\documentclass[useAMS,usenatbib]{mn2e}

\usepackage{graphicx}
\usepackage{scalefnt}
\usepackage{amssymb}

\title[Simulating disc galaxy evolution in a group environment. I. The influence 
of the global tidal field]{Simulating the evolution of disc galaxies in a group 
environment. I. The influence of the global tidal field}
\author[\'A. Villalobos, G. De Lucia, S. Borgani, and G. Murante]{\'A. 
Villalobos$^{1}$\thanks{villalobos@oats.inaf.it}, G. De Lucia$^{1}$, 
S. Borgani$^{1,2,3}$, and G. Murante$^{4}$
\\$^{1}$INAF - Astronomical Observatory of Trieste, via G.B. Tiepolo 11, I-34143 
Trieste, Italy\\$^{2}$Astronomy Unit, Department of Physics, University of Trieste, 
via G.B. Tiepolo 11, I-34131 Trieste, Italy\\$^{3}$INFN - Istituto Nazionale di 
Fisica Nucleare, Trieste, Italy\\$^{4}$INAF - Istituto Nazionale di Astrofisica 
- Osservatorio Astronomico di Torino, Str. Osservatorio 25, I-10025, Pino Torinese, 
Torino, Italy}

\begin{document}

\pagerange{\pageref{firstpage}--\pageref{lastpage}} \pubyear{---}

\maketitle

\label{firstpage}

\begin{abstract}
We present the results of a series of numerical simulations aimed to study 
the evolution of a disc galaxy within the global tidal field of a group 
environment. Both the disc galaxy and the group are modelled as 
multi-component, collision-less, $N$-body systems, composed by both dark 
matter and stars. In our simulations, the evolution of disc galaxies is 
followed after they are released from the group virial radius, and as 
their orbits sink towards the group centre, under the effect of dynamical 
friction. We explore a broad parameter space, covering several aspects of 
the galaxy-group interaction that are potentially relevant to galaxy 
evolution. Namely, prograde and retrograde orbits, orbital eccentricities, 
disc inclination, role of a central bulge in discs, internal disc kinematics, 
and galaxy-to-group mass ratios. We find that significant disc transformations 
occur only after the mean density of the group, measured within the orbit 
of the galaxy, exceeds $\sim$0.3--1 times the central mean density of 
the galaxy. The morphological evolution of discs is found to be strongly 
dependent on the initial inclination of the disc with respect to its orbital 
plane. That is, discs on face-on and retrograde orbits are shown to retain 
longer their disc structures and kinematics, in comparison to prograde discs. 
This suggests that after interacting with the global tidal field alone, a 
significant fraction of disc galaxies should be found in the central regions 
of groups. Prominent central bulges are not produced, and pre-existing bulges 
are not enhanced in discs after the interaction with the group. Assuming that 
most S0 are formed in group environments, this implies that prominent 
bulges should be formed mostly by young stars, created only after a galaxy 
has been accreted by a group. Finally, contrary to some current 
implementations of tidal stripping in semi-analytical models of galaxy 
evolution, we find that more massive galaxies suffer more tidal stripping. 
This is because dynamical friction brings them faster to the group centre, in 
comparison to their lower mass counterparts.

\end{abstract}


\section{Introduction}

Galaxies in high density environments, such as clusters or groups, have 
properties that are consistently different from those of galaxies in 
regions of the Universe with ``average'' density. For instance, in 
comparison to field galaxies, the population of galaxies within groups and 
clusters is characterised by a higher fraction of red galaxies with lower 
average star formation rate, as well as by a lower fraction of disc galaxies
\citep[e.g.,][]{dressler1997,lewis2002,gomez2003,balogh2004,weinmann2006,mcgee2008}. 
Although the precise mechanisms responsible for such differences remain 
unknown, these observational findings do highlight the key role played by 
environment in the evolution of galaxies.

In the framework of the currently favoured cosmological model (the 
$\Lambda$CDM model), galaxy clusters are formed at relatively late times, 
through a series of mergers and accretion events of both dark matter and 
baryonic material, ranging from small halos, isolated galaxies to large 
galaxy groups. In such a context, a galaxy that resides in a cluster at 
present-time has experienced several different environments during its 
lifetime \citep[e.g.,][]{lacey1994,zhao2003,berrier2009,mcgee2009}. In each 
of these environments, a galaxy might have been affected by different 
physical processes, such as, strangulation, ram-pressure stripping, as well 
as tidal forces against the background potential, and close encounters with 
other galaxies and substructures orbiting within the same dark matter halo. 
The combined effect of these processes likely played a significant role in 
shaping the observed physical properties of galaxies residing in groups and 
clusters at present time \citep[see the review by][]{weinmann2011-2}. Thus, 
in order to better understand the origin of these properties, it is crucial 
to characterise and quantify the relative effect that those physical 
processes have on galaxies, in different environments.

From a theoretical point of view, a number of studies have used numerical 
$N$-body simulations to investigate the influence of environments on galaxy 
evolution. Most of them, however, have been focused either on cluster mass 
scales \citep[e.g.,][]{moore1999,gnedin2003,mastropietro2005}, or on Milky 
Way-like mass scales 
\citep[e.g.,][]{mayer2001,mayer2001-2,klimentowski2009,kazantzidis2011}. 
For example, \citeauthor{mastropietro2005} studied the influence of tidal 
stripping and galaxy harassment on the evolution of disc galaxies within 
a cluster environment. In their study, the authors extracted a cluster 
environment from a cosmological simulation, and replaced 20 of its particles 
(randomly chosen at redshift $z$=0.5) with high-resolution multi-component 
galaxy models. The authors found that the cluster environment induces a 
series of strong bar formation episodes that transform the galaxy 
morphologies from late-type discs to dwarf spheroidals. The galaxies 
retain, however, a significant fraction of their rotational motion. Other 
authors have taken advantage of hydrodynamic simulations, using either 
smoothed-particle hydrodynamics (SPH) or grid codes, to study the removal 
of a galaxy's gas disc during its motion through the intracluster medium 
\citep[][]{quilis2000,roediger2006}. At the lower end of the mass spectrum, 
in a recent study, \citet{kazantzidis2011} examined the transformation of 
dwarf disc galaxies into dwarf spheroidals, driven by tidal interactions 
with a Milky Way-size host galaxy, using $N$-body simulations of 
self-consistent stellar discs placed within cosmologically motivated 
dark matter (DM) halos.

On the group mass scale, some studies have explored the role played 
by ram pressure stripping in the evolution of galaxies that move through 
an intragroup medium \citep[e.g.,][]{marcolini2003,roediger2005,tecce2011}, 
while recent work has focused on the evolution of dwarf galaxies in the 
Local Group \citep{sawala2011}, and on the transformation of spiral 
galaxies into S0 galaxies via repeated close encounters with other group 
galaxies \citep{bekki2011}. Overall, surprisingly little work has been 
dedicated so far to systems with velocity dispersions typical of galaxy 
groups, even though they arguably represent the most common 
environment that galaxies experience. Indeed, it has long been claimed 
that a significant fraction of the cluster galaxy population might have 
been ``pre-processed'' in groups, that later merged onto clusters 
\citep{zabludoff1998}. This appears to be confirmed by
\citet[][but see also \citeauthor{berrier2009}~\citeyear{berrier2009}]{mcgee2009}, 
who used merger tree information from semi-analytic catalogues to 
explore the accretion of galaxies onto clusters and groups. The authors 
found that clusters at redshift epochs ranging from $z$=0 to $z$=1.5 have 
accreted a significant fraction of galaxies through groups. Specifically, 
they estimated that $\sim$40 per cent of galaxies with 
$M_{\rm stellar}>10^9 M_{\sun}$ were accreted onto clusters with 
$M_{\rm cluster}$=$10^{14.5} M_{\sun}$ through groups with 
$M_{\rm group}>10^{13} M_{\sun}$. 
In a recent paper, \citet{delucia2011} confirm that a significant
fraction of the galaxies that currently reside in a cluster have spent
time as satellites of a lower mass system before being accreted onto the
final parent halo. This fraction is largest for lowest mass galaxies. In
addition, by comparing their theoretical predictions with observational
estimates of the passive fraction as a function of the parent halo mass 
and cluster-centric radius, they argue that satellite galaxies become
passive after they have spent a relatively long time ($\sim 5-7$~Gyr) in
halos more massive than $\sim 10^{13} M_{\sun}$. It is therefore crucial 
to understand the influence of different physical processes on galaxy 
evolution for scales typical of galaxy groups.

The aim of this paper is to help alleviate the aforementioned lack 
of numerical studies by exploring the evolution of disc galaxies within 
a group environment, covering a broad parameter space. We use 
high-resolution $N$-body simulations and focus specifically on the effect 
of the global tidal forces on galaxy evolution. This approach allows us 
to isolate this effect, distinguishing it from those of other relevant 
processes, such as, low-velocity close encounters with other group members. 
Ultimately, we plan a better quantification of the relative importance that 
different physical process have on galaxy evolution, as they are gradually 
introduced in the next papers of this series. Because of this approach, 
our group environments are built from idealised initial conditions, do not 
account for hierarchical growth, and do not contain substructures. Thus, 
by design, the simulations presented in this study do not include the effect 
of close encounters on galaxy evolution.

The layout of this paper is as follows: \S \ref{sec-setup} describes the 
set-up of our experiments; \S \ref{sec-descrip} describes the results of 
our simulations; in \S \ref{sec-discuss} we discuss our results and in 
\S \ref{sec-conclusion} we summarise our conclusions. At the end of each 
subsection in \S \ref{sec-descrip} we include short summaries for readers 
who prefer to skip the detailed description of the simulations.

\section{Set-up of numerical experiments}
\label{sec-setup}

\begin{table*}
\centering
\begin{minipage}{65mm}
 \caption{Properties of group environments.}
 \label{group-param}
 \begin{tabular}{@{}lllr@{}}
  \hline
                                    & ``$z$=0''            & ``$z$=1''            &               \\
  \hline
  \hline
  DM Halo                           &                      &                      &               \\
  \hline
  Virial mass                       & $9.9 \times 10^{12}$ & $9.9 \times 10^{12}$ & ($M_{\sun}$)  \\
  Virial radius                     & 555.94               & 329.19               & (kpc)         \\
  Concentration                     & 9.74                 & 4.87                 &               \\
  Circular velocity                 & 276.97               & 360.07               & (km s$^{-1}$) \\
  Number particles                  & $1.1 \times 10^{6}$  & $1.1 \times 10^{6}$  &               \\
  Softening                         & 0.55                 & 0.32                 & (kpc)         \\
  \hline
  Stellar spheroid                  &                      &                      &               \\
  \hline
  Mass                              & $10^{11}$            &  $10^{11}$           & ($M_{\sun}$)  \\
  Scale radius                      & 3.24                 &  1.91                & (kpc)         \\
  Number particles                  & $5 \times 10^{5}$    &  $5 \times 10^{5}$   &               \\
  Softening                         & 0.1                  &  0.06                & (kpc)         \\
  \hline
  \hline
 \end{tabular}
\end{minipage}
\end{table*}

\begin{table*}
\centering
\begin{minipage}{170mm}
 \caption{Properties of disc galaxies.}
 \label{disk-galaxies-param}
 \begin{tabular}{@{}lllllllr@{}}
  \hline
                                    & \multicolumn{3}{c}{``$z$=0''}                                            & \multicolumn{3}{c}{``$z$=1''}                                       &               \\
  \hline
  Label                             & \tiny{REF,BUL$^{(1)}$,CIR,}    & \tiny{MR1:5} & \tiny{MR1:20}        & \tiny{REF,BUL$^{(1)}$,CIR,}    & \tiny{MR1:10} & \tiny{MR1:40}          &               \\
  experiment                        & \tiny{RAD,RET,COL$^{(2)}$,FON} &              &                      & \tiny{RAD,RET,COL$^{(2)}$,FON} &               &                        &               \\
  \hline
  DM Halo                           &                       &                       &                      &                       &                        &                        &               \\
  \hline
  Virial mass                       & $10^{12}$             & $2 \times 10^{12}$    &  $5 \times 10^{11}$  & $5.07 \times 10^{11}$ & $10^{12}$              & $2.53 \times 10^{11}$  & ($M_{\sun}$)  \\
  Virial radius                     & 258.91                & 326.2                 &  205.5               & 122.22                & 153.98                 & 97                     & (kpc)         \\
  Concentration                     & 13.12                 & 11.99                 &  14.36               & 6.56                  & 6                      & 7.18                   &               \\
  Circular velocity                 & 129.17                & 162.76                &  102.51              & 133.93                & 168.77                 & 106.29                 & (km s$^{-1}$) \\
  Number particles                  & $5 \times 10^{5}$     & $5 \times 10^{5}$     &  $5 \times 10^{5}$   & $10^{6}$$^{(3)}$      & $10^{6}$               & $10^{6}$               &               \\
  Softening                         & 0.35                  & 0.44                  &  0.28                & 0.41                  & 0.6                    & 0.6                    & (kpc)         \\
  \hline
  Stellar disc                      &                       &                       &                      &                       &                        &                        &               \\
  \hline
  Disc mass                         & $2.8 \times 10^{10}$  &  $5.6 \times 10^{10}$ & $1.4 \times 10^{10}$ & $1.42 \times 10^{10}$ & $2.84 \times 10^{10}$  & $7.09 \times 10^{9}$   & ($M_{\sun}$)  \\
  Scale-length                      & 3.5                   &  4.4                  & 2.7                  & 1.6                    & 1.6                   & 1.6                    & (kpc)         \\
  Scale-height                      & 0.35                  &  0.44                 & 0.27                 & 0.16                   & 0.16                  & 0.16                   & (kpc)         \\
  $Q$                               & 2                     &  2                    & 2                    & 2.5$^{(3)}$            & 2                     & 3                      &               \\
  Number particles                  & $10^{5}$              &  $10^{5}$             & $10^{5}$             & $10^{5}$               & $10^{5}$              & $10^{5}$               &               \\
  Softening                         & 0.05                  &  0.06                 & 0.04                 & 0.012                  & 0.012                 & 0.012                  & (kpc)         \\
  \hline
  \hline
 \end{tabular}
\\(1): A stellar bulge is added at the centre of the disc, with $M_{\rm bulge}$=$0.3 M_{\rm disc}$ and $a_{\rm bulge}$=$0.2 R_{\rm D}$.
(2): Disc kinematically cold, with $Q$=$1.25$.
(3): Values are larger, with respect to ``$z$=0'', to enhance the stability of the disc against the formation of a bar.
\end{minipage}
\end{table*}
        
\begin{table*}
\centering
\begin{minipage}{125mm}
 \caption{List of experiments at redshift ``$z$=0'' [and ``$z$=1''].}
 \label{list-exper}
 \begin{tabular}{@{}lcccccl@{}}
  \hline
  Label               & Orbit      & $\theta$$^a$ & $(V_r,V_{\phi})$$^a$ & $e$$^a$  & $M_{\rm galaxy}$:$M_{\rm group}$ & Notes                      \\
                      & $^{(1)}$   & $^{(2)}$     & $^{(3)}$             & $^{(4)}$ &  $^{(5)}$                        & $^{(6)}$                   \\
  \hline
  \hline REF          & Prograde   & 0$\degr$     & (0.9,0.6)            & 0.86     & 1:10 [1:20]                      & Reference experiment       \\
  \hline BUL          & Prograde   & 0$\degr$     & (0.9,0.6)            & 0.86     & 1:10 [1:20]                      & Bulge added to disc        \\
  \hline CIR          & Prograde   & 0$\degr$     & (0.6,1.1)            & 0.6      & 1:10 [1:20]                      & More ``circular'' infall   \\
  \hline RAD          & Prograde   & 0$\degr$     & (1.2,0.3)            & 0.97     & 1:10 [1:20]                      & More ``radial'' infall     \\
  \hline RET          & Retrograde & 180$\degr$   & (0.9,0.6)            & 0.86     & 1:10 [1:20]                      & Retrograde orbital infall  \\
  \hline FON          & -          & 90$\degr$    & (0.9,0.6)            & 0.86     & 1:10 [1:20]                      & ``Face-on'' infall         \\
  \hline COL          & Prograde   & 0$\degr$     & (0.9,0.6)            & 0.86     & 1:10                             & ``Colder'' disc kinematics \\
  \hline MR1:5        & Prograde   & 0$\degr$     & (0.9,0.6)            & 0.86     & 1:5                              & More massive galaxy        \\
         $[$MR1:10$]$ &            &              &                      &          & [1:10]                           &                            \\
  \hline MR1:20       & Prograde   & 0$\degr$     & (0.9,0.6)            & 0.86     & 1:20                             & Less massive galaxy        \\
         $[$MR1:40$]$ &            &              &                      &          & [1:40]                           &                            \\
  \hline \hline
 \end{tabular}
\\(1) Sense of the orbital infall with respect to the disc rotation. (2) Initial angle between the orbital and intrinsic angular momentum of the disc.
(3): Radial and tangential components of the initial velocity of the disc, in units of the virial circular velocity of the group. (4):
Initial orbital eccentricity. (5): Initial galaxy-to-group mass ratio. (6): Differences with respect to the reference experiment.
($a$): Same values used at both redshift epochs.
\end{minipage}
\end{table*}

We have carried out a total of 17 experiments in order to study the general
evolution of disc galaxies within a group environment. Our basic strategy is 
to release a single disc galaxy at a time, from the virial radius of the 
group, studying the evolution of its morphology and kinematics as it orbits 
within the group. Each experiment explores a single aspect of the 
galaxy-group interaction (see below), in order to estimate its role in the 
general evolution of galaxies. We run our simulations for 12 Gyr, 
independently of whether the galaxy reaches the centre of the group, it is 
destroyed or it survives.

The group environment is modelled as a $N$-body (i.e. ``live'') DM halo, 
following a NFW density profile \citep{navarro1997}:
\begin{equation} \label{nfw}
\rho_{\rm{halo}}(r) = \frac{\rho_{\rm s}}{(r/r_{\rm s})(1+r/r_{\rm s})^2} ,
\end{equation}
where $\rho_{\rm s}$ is a characteristic scale density and $r_{\rm s}$ a 
scale radius. The simulated group also includes a spherically symmetric 
stellar component at its centre to account for a central galaxy, that is 
assumed to follow a Hernquist density profile \citep{hernquist1990}:
\begin{equation}
\rho_{\rm *}(r) = \frac{M_{\rm *}}{2\pi} \frac{a_{\rm *}}{r(r+a_{\rm *})^3} ,
\label{bulgeprof}
\end{equation}
where $M_{\rm *}$ is the stellar mass and $a_{\rm *}$ is the scale radius.
Galaxies are modelled as multi-component systems composed by a stellar disc 
embedded in a DM halo. Stellar discs are built according to the following 
density profile:
\begin{equation} \label{diskprof}
\rho_{\rm disc}(R,z) = \frac{M_{\rm disc}}{8\pi R_{\rm D}^2 z_{\rm D}}
\exp\left(-\frac{R}{R_{\rm D}}\right)\ \textrm{sech}^2\left(\frac{z}{2z_{\rm D}}\right) ,
\end{equation}
where  $M_{\rm disc}$ is the disc mass, $R_{\rm D}$ is the exponential 
scale-length, and $z_{\rm D}$ is the exponential scale-height. Depending on 
the experiment, the stellar discs can also contain a central stellar bulge, 
following a Hernquist profile. Galaxy DM halos follow a NFW density profile
and are initially spherical, do not rotate, and the structure of their inner 
region has been adiabatically contracted to account for the growth of the 
baryonic component(s). We refer the interested reader to 
\citet{villalobos-helmi2008} for a complete description of the procedure 
followed to generate the self-consistent initial conditions used for this 
study.

Our experiments probe an ample range of the parameter space, exploring 
different aspects of the galaxy-group interaction that are potentially 
relevant to the evolution of galaxies. In particular, we study the effect 
of: prograde and retrograde infalls (with respect to the sense of rotation 
of the disc), different orbital eccentricities, the presence of a central 
stellar bulge in the disc, different internal kinematics of the disc, and 
different galaxy-to-group mass ratios. The initial orbital parameters of 
discs are chosen to be consistent with distributions of orbital parameters 
of infalling substructures, obtained from cosmological simulations \citep{benson2005}.
In order to cover those distributions, we choose for our experiments the most 
likely eccentricity of infall, and both the least and the most eccentric 
orbits.

We place our simulations at two different redshift epochs, ``$z$=0'' and 
``$z$=1''\footnote{The quotation marks denote that our group environments 
are not evolved with cosmic time throughout the simulations.}. Thus, we are 
able to also explore possible dependencies of disc evolution on the group 
structure. We choose to use a fixed value for the mass of groups at both 
``$z$=0'' and ``$z$=1'' (10$^{13} M_{\sun}$, that is within the mass range 
reported by \citeauthor{mcgee2009} \citeyear{mcgee2009} where significant 
environmental effects must take place), only evolving with redshift both 
the group concentration and virial radius.

At each redshift, we define our ``reference'' galaxy models as those with 
bulge-less discs, on prograde infalling orbits with the most likely orbital 
eccentricity. At ``$z$=0'' the reference galaxy resembles a bulge-less, 
slightly less massive Milky Way (MW), while at ``$z$=1'' it is a smaller 
version of the MW, obtained by scaling down its properties as in \citet*{mo1998}.
This scaling involves defining both the disc mass and the disc scale-length 
as constant fractions of the virial mass and virial radius of the galaxy halo, 
respectively, at each redshift. Therefore, our reference galaxy at 
``$z$=1'' is half as massive as that at ``$z$=0''. Tables~\ref{group-param} 
and \ref{disk-galaxies-param} list the structural parameters of our groups 
and disc galaxies, respectively, and Table~\ref{list-exper} provides the 
complete list of experiments and their characteristics. Throughout the paper 
we refer to each experiment by its label or its main characteristic 
interchangeably.

We have also performed simulations of all disc galaxies in isolation, evolving 
them as long as the full simulations, 12~Gyr. Our galaxies are found to be 
stable, in terms of their structure and kinematics, and show a negligible 
evolution in the conservation of both their total energy and total angular 
momentum.

The typical computational cost required to complete a single simulation is 
between 432 and 768 core processor hours, using 32 tasks (i.e., 16 cores in 
simultaneous multi-threading mode) of the IBM P575 Power 6 (``SP6'') 
supercomputer hosted by CINECA\footnote{http://hpc.cineca.it/.}. The 
simulations were run using GADGET-3, a non-public evolution of the GADGET-2 
code. We refer the interested reader to \citet{springel2005}, for a complete 
description of the latter. The main improvement of GADGET-3 over GADGET-2 
resides in its multiple domain decomposition. Domain decomposition in the 
GADGET code is based on the computation of the Peano-Hilbert space-filling 
curve connecting all the particles. This curve is then split into $M$ segments, 
to be assigned to different processors. These segments are then ranked 
according to their estimated computational cost. In the GADGET-3 code, each 
of the $N$ ($<$$M$) tasks is allowed to receive more than one disjoined 
segment of the Peano-Hilbert curve, with the first task receiving the most 
and the least expensive, and so on, thus providing a more efficient 
work-load balance among the tasks. This kind of decomposition allows a 
substantial increase in speed, especially when a few massive objects dominate 
the processing cost, as in the simulations presented here.

\section{Results}
\label{sec-descrip}

\subsection{Disc orbital evolution}
\label{subsec-orbital}

\begin{figure*}
\begin{center}
\includegraphics[width=71mm]{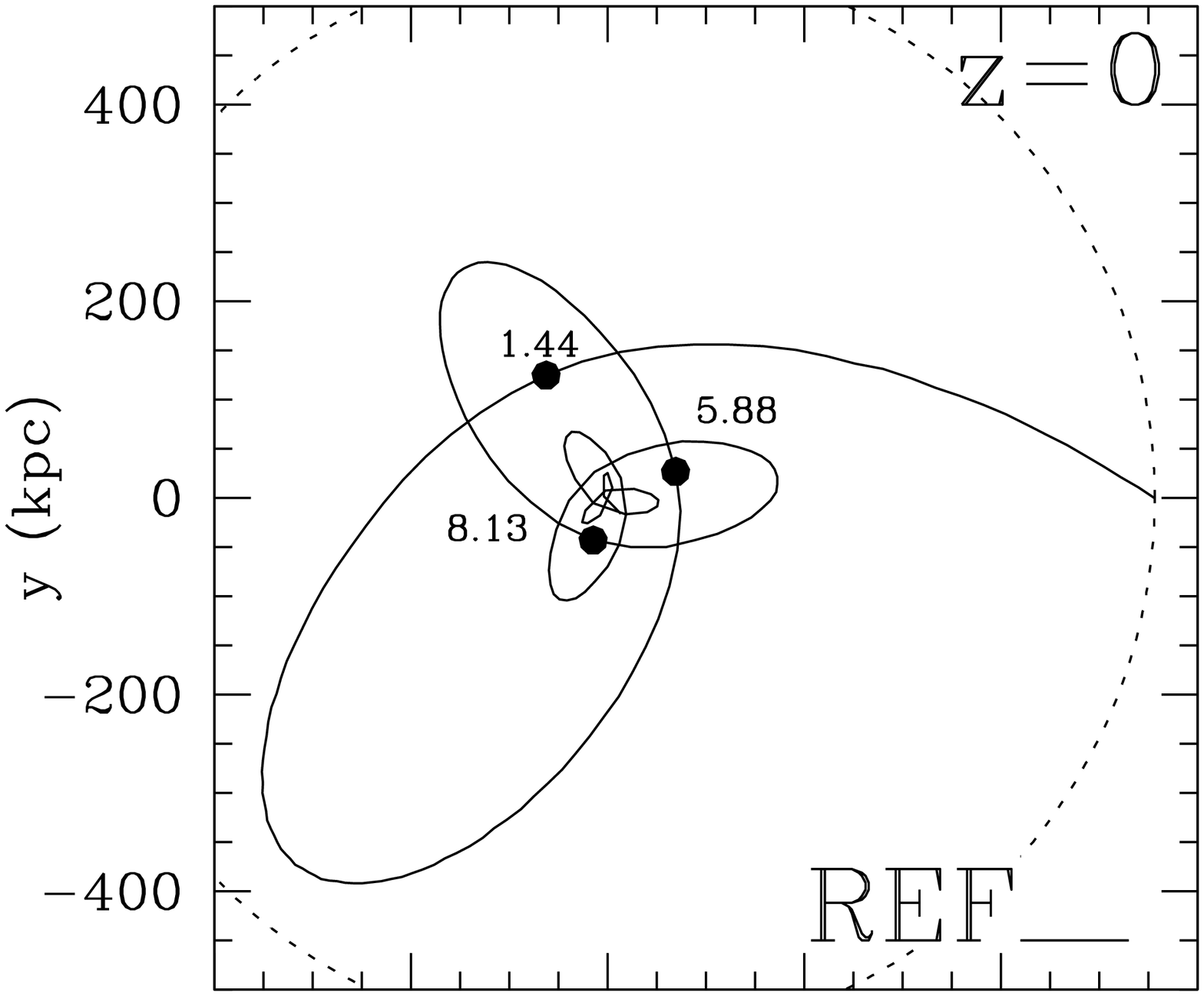}\hspace*{-1.68cm}
\includegraphics[width=71mm]{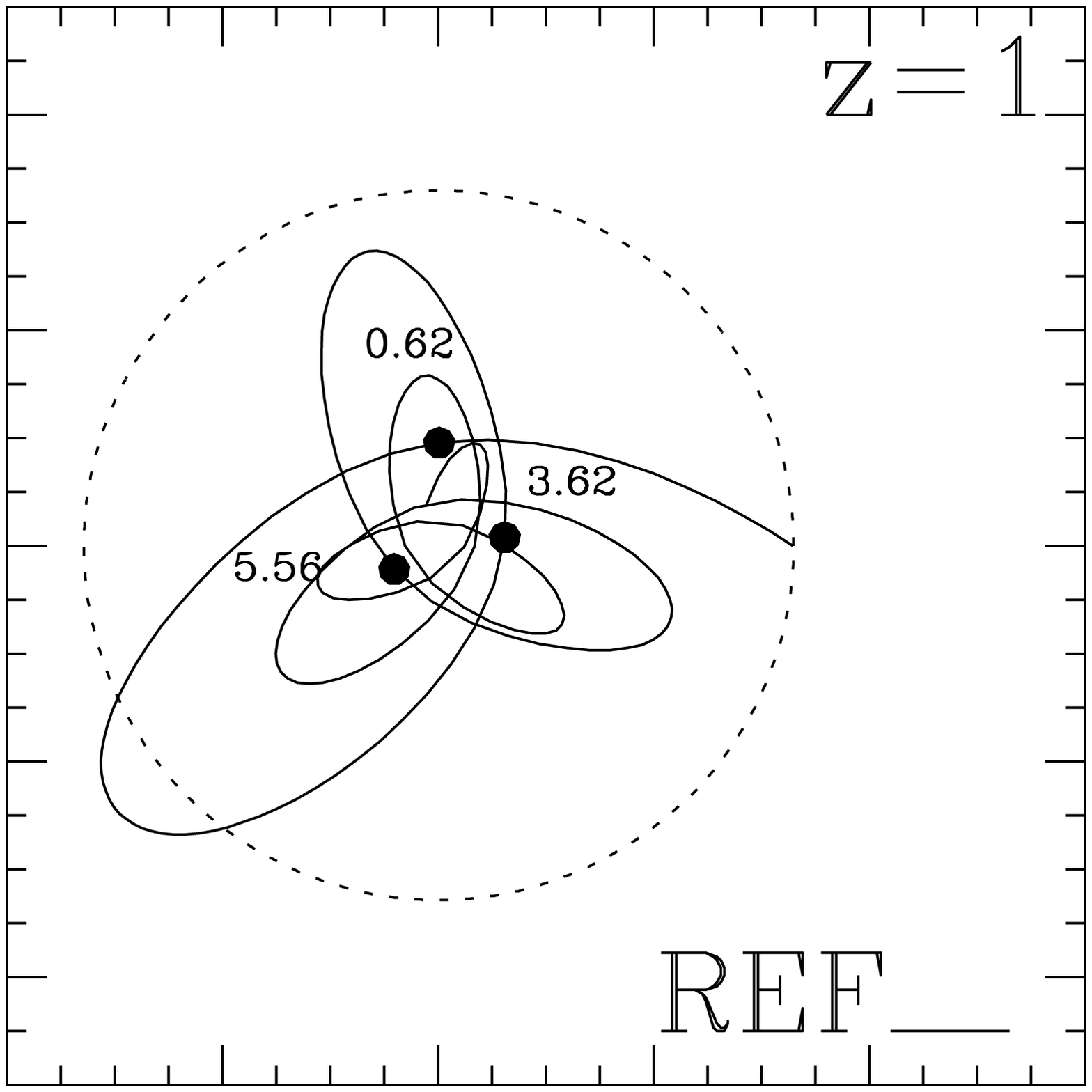}\vspace*{-1.61cm}\\
\includegraphics[width=71mm]{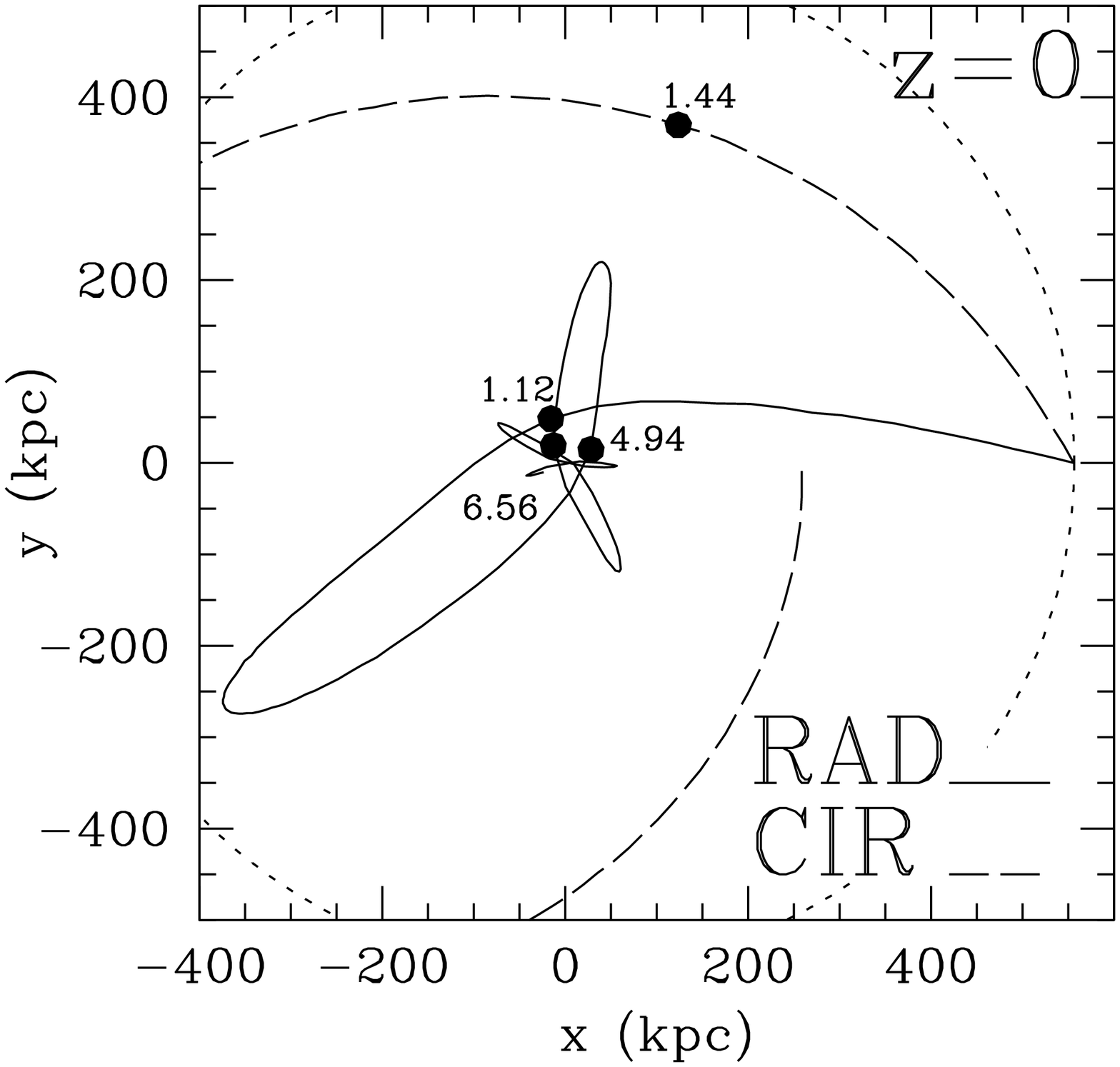}\hspace*{-1.68cm}
\includegraphics[width=71mm]{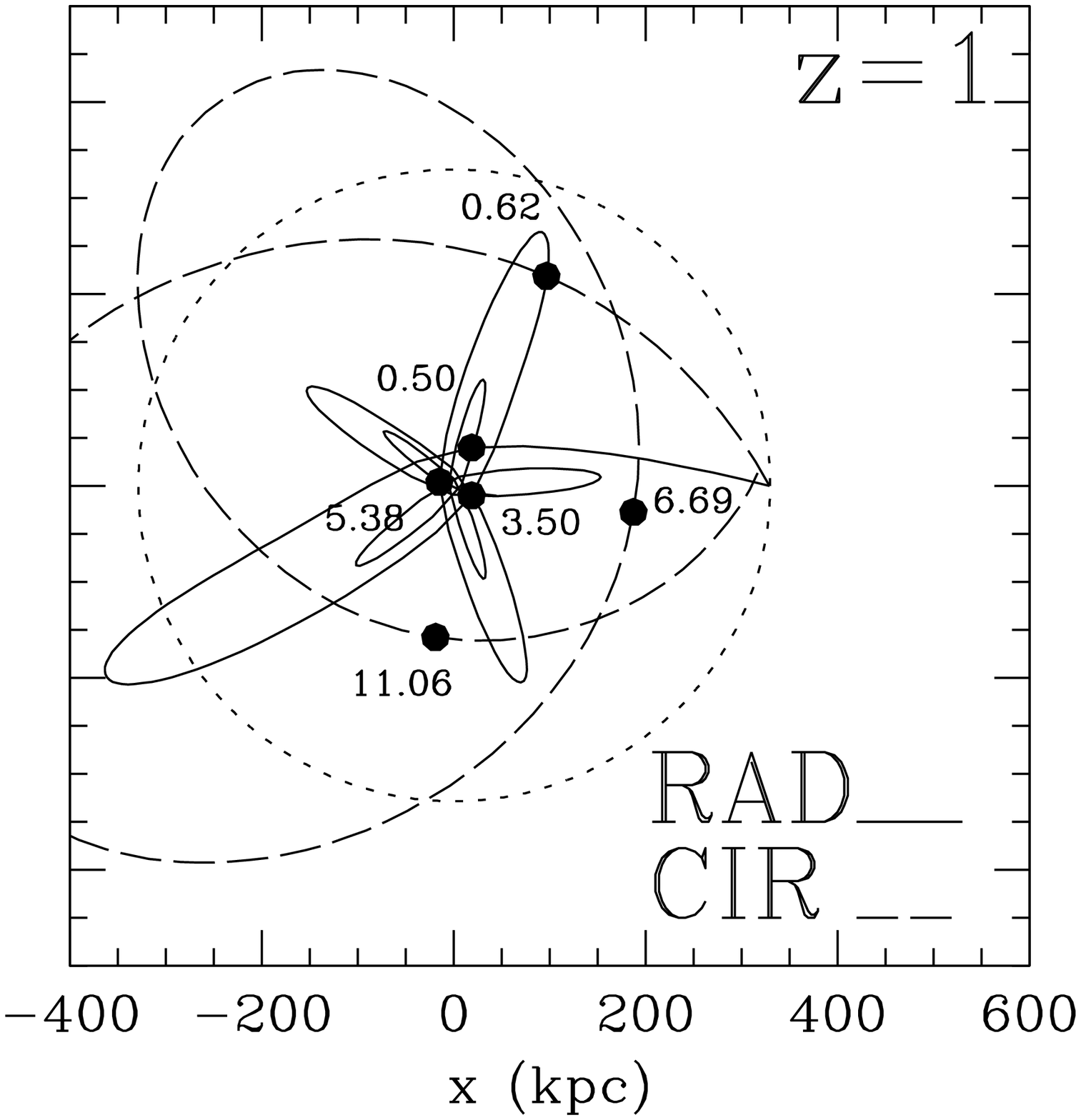}
\end{center}
\caption{Trajectories of the centres of mass of disc galaxies with initial 
orbital eccentricities $e$=0.86, 0.6 and 0.97 (REF, CIR and RAD experiments, 
respectively), as they spiral in towards the centre of the group at redshift 
``$z$=0'' and ``$z$=1''. The coordinate system is centred on the centre of 
mass of the group and the XY projections are on the orbital plane of the disc. 
The first pericentric passages (solid dots), and their respective times (in 
units of Gyr), are included as reference, as well as the initial virial 
radius of the group (dotted lines).}
\label{trayectories}
\end{figure*}

\begin{figure*}
\begin{center}
\includegraphics[width=141mm]{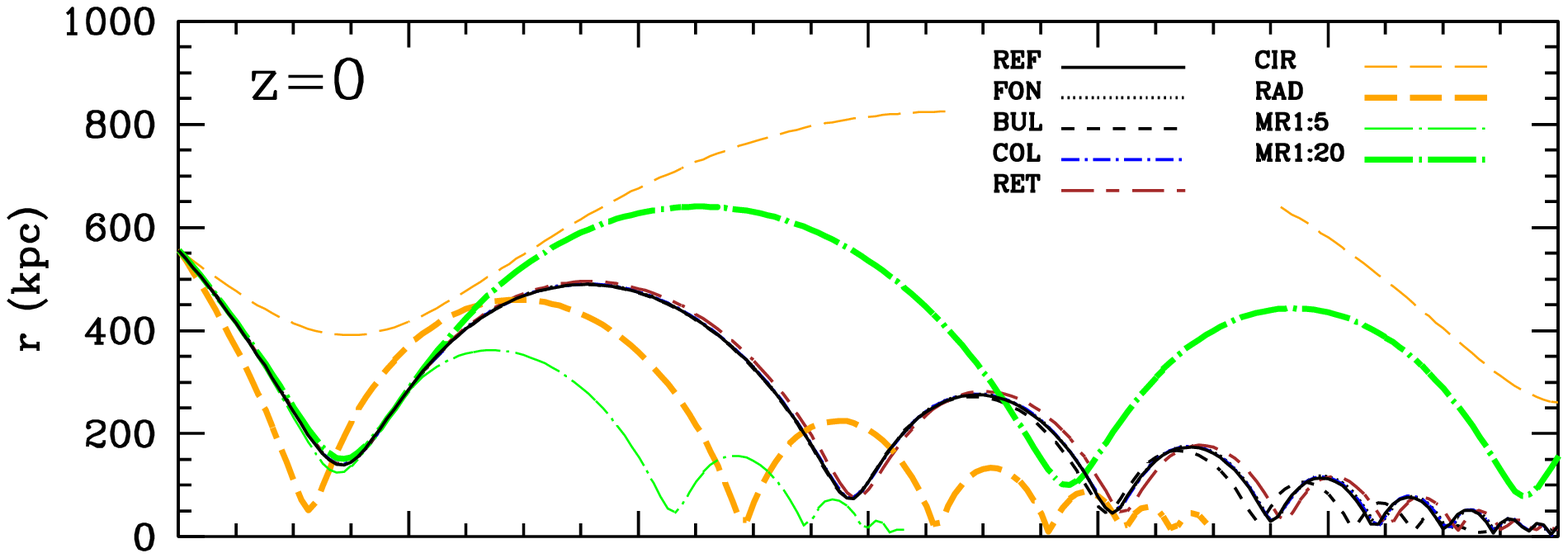}\vspace*{-9.7cm}\\
\includegraphics[width=141mm]{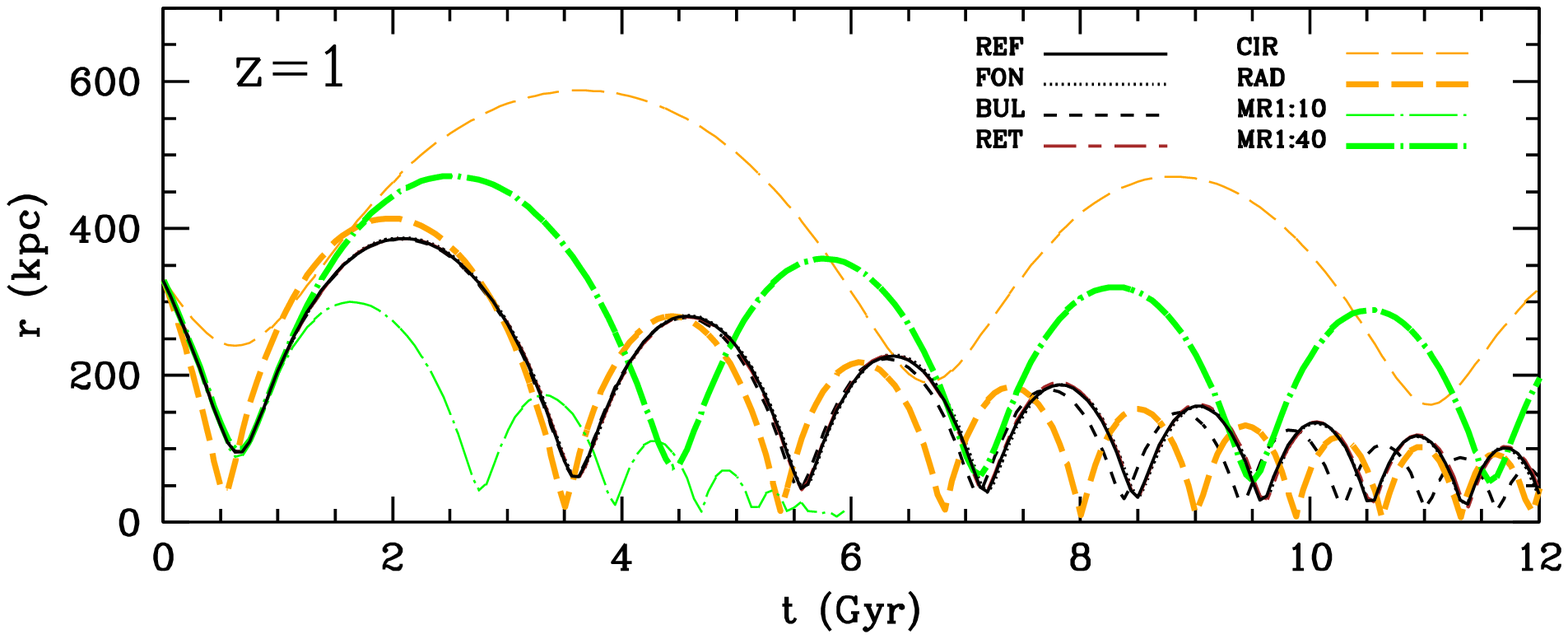}\vspace*{-8cm}
\end{center}
\caption{Evolution of the distance between the centres of mass of the disc 
and that of the group for all our experiments, at redshift epochs ``$z$=0'' 
and ``$z$=1''.}
\label{separation-evol}
\end{figure*}

\begin{figure*}
\begin{center}
\includegraphics[width=85mm]{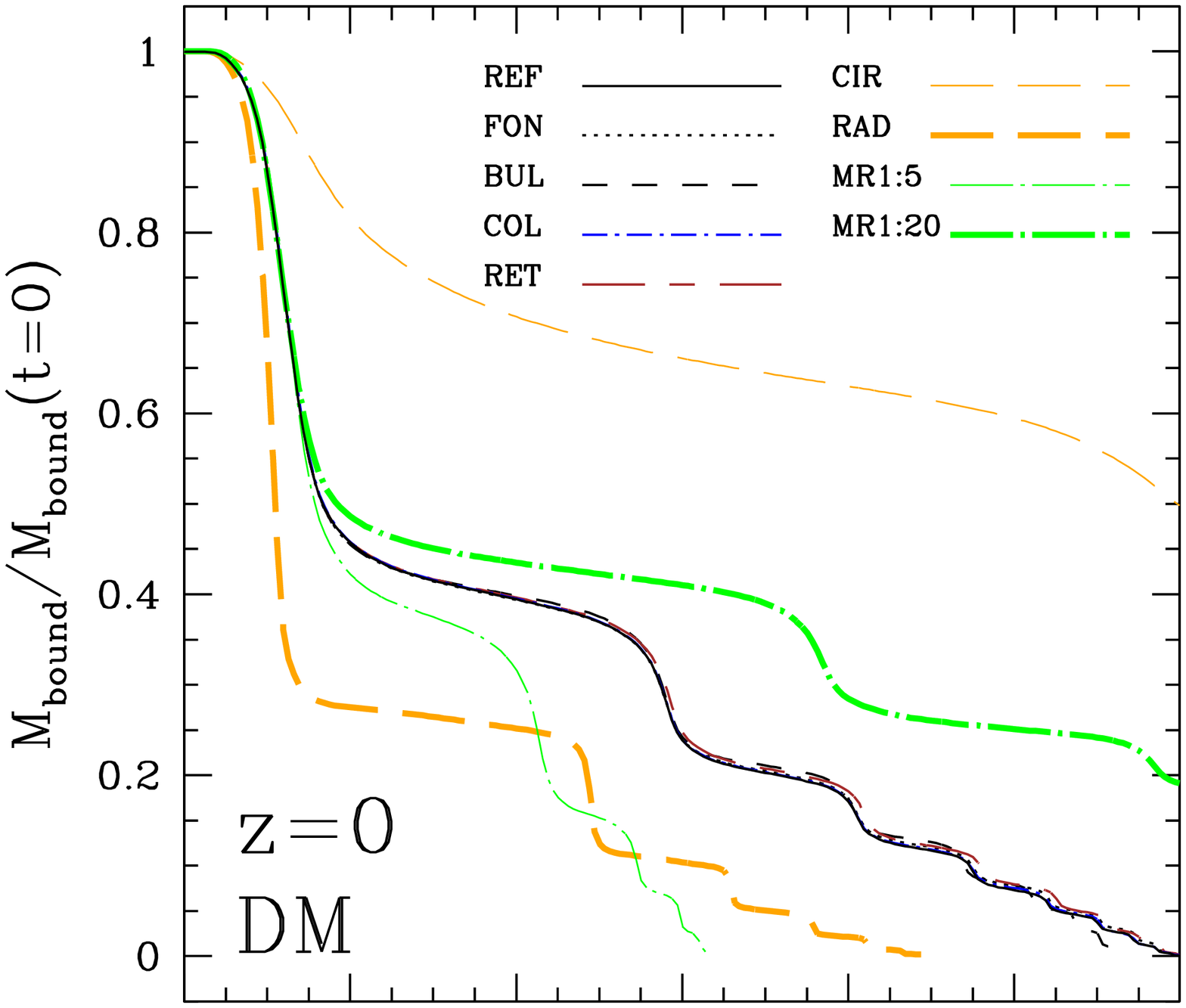}\hspace*{-2cm}
\includegraphics[width=85mm]{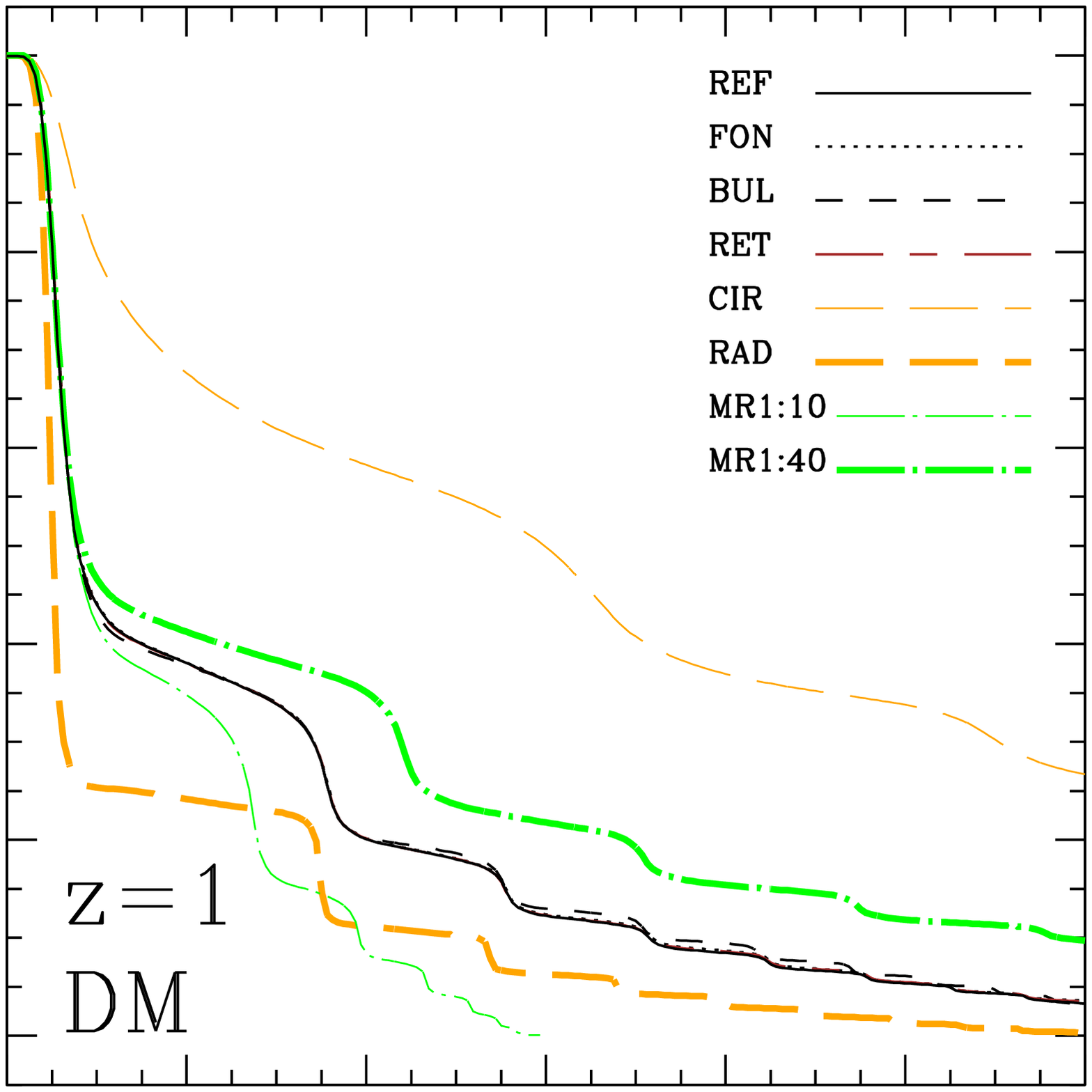}\vspace*{-1.92cm}\\
\includegraphics[width=85mm]{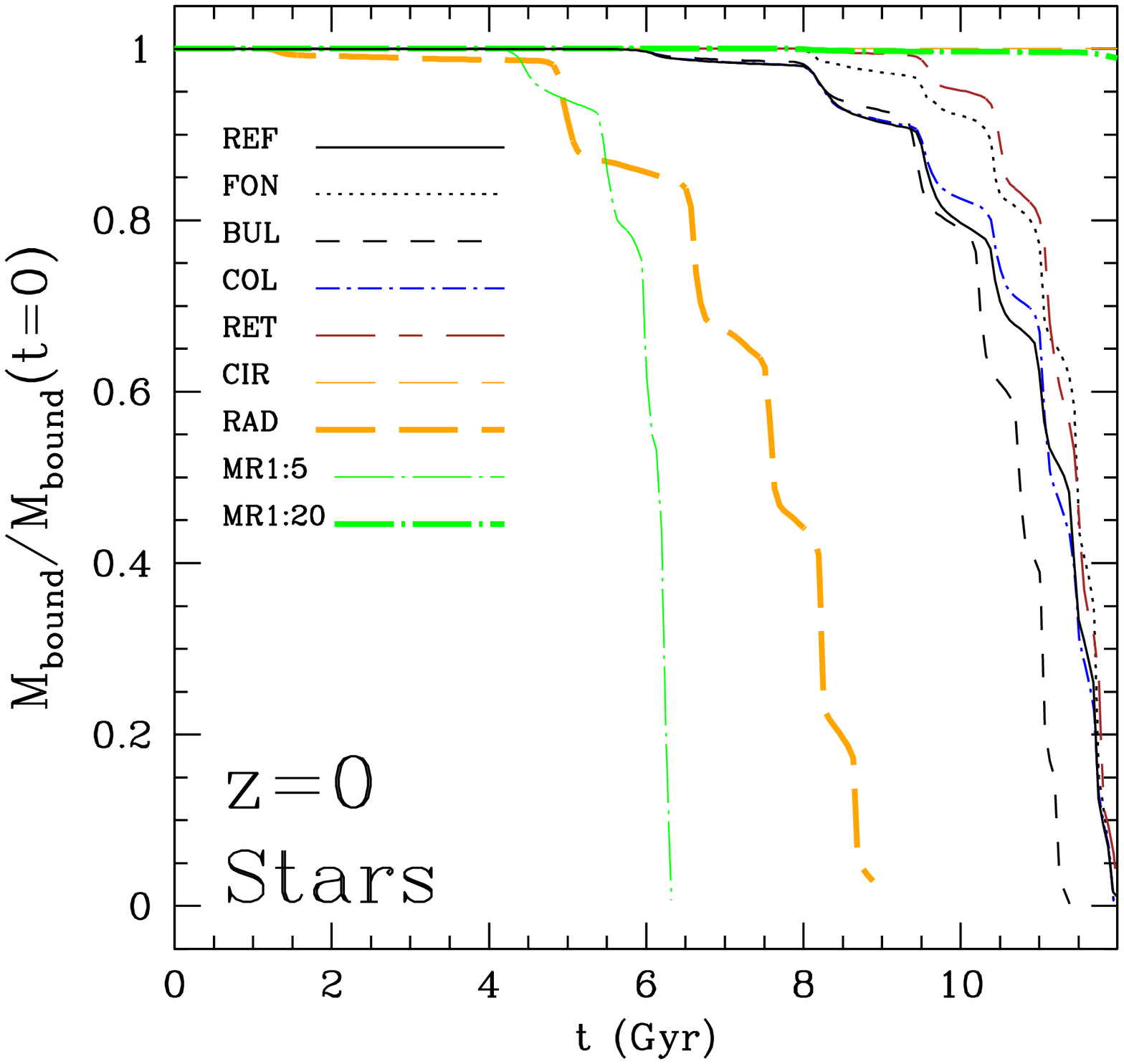}\hspace*{-2cm}
\includegraphics[width=85mm]{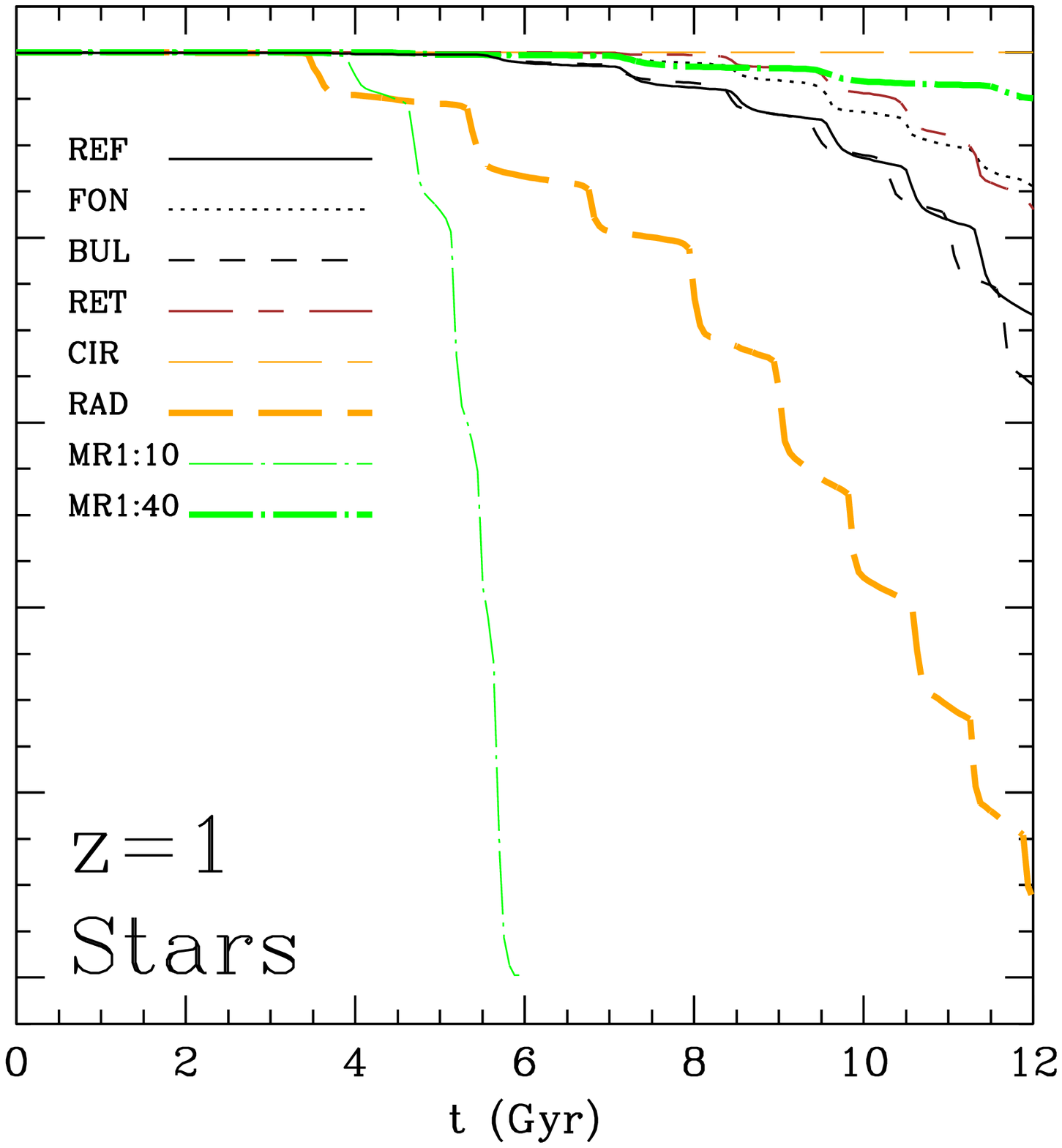}
\end{center}
\caption{Evolution of the mass fraction (both DM and stars) that remains 
bound to the galaxy during the galaxy-group interaction, for all our 
experiments at redshift epochs ``$z$=0'' and ``$z$=1''.}
\label{massbound-evol}
\end{figure*}

In our simulations, the evolution of the disc galaxies takes place from the 
moment they are released on an infalling orbit at the virial radius of the 
group environment. Figure~\ref{trayectories} shows the trajectories, over 
12 Gyr of infall, of the centre of mass of the disc galaxy for our REF 
(reference), RAD (most radial infall) and CIR (most circular infall) 
experiments for the redshift epochs ``$z$=0'' and ``$z$=1''. This figure 
offers a direct comparison between the most likely orbit of infall, according 
to velocity distributions from cosmological simulations, and the extremes of 
those distributions. At a given redshift epoch, the most likely orbit seems 
more similar to the most radial infall than to the most circular one. This 
can also be noted in terms of the initial eccentricities of the orbits (see 
Table~\ref{list-exper}). However, in the most radial orbit, the disc galaxy 
penetrates deeper in the central region of the group and on shorter 
timescales than in the case of the most likely orbit. Thus, the disc galaxy 
on a more radial orbit is affected faster by stronger tidal shocks against 
the central potential of the group. This leads to a more dramatic evolution 
in the disc, especially with respect to a more circular orbit in which the 
disc galaxy experiences fewer pericentric passages, generally closer to the 
virial radius of the group than to its centre.

The disc galaxy on the most likely infalling orbit at ``$z$=1'' experiences 
more pericentric passages in comparison to the same case at ``$z$=0''. This 
is expected as a consequence of the smaller galaxy-to-group mass ratio at 
``$z$=1'' (1:20 vs. 1:10), which is inversely proportional to the merging 
timescale driven by dynamical friction. In general, in our experiments at 
``$z$=1'', the disc galaxy spends more time away from the denser central 
region of the group, as opposed to ``$z$=0'', meaning that the disc is 
exposed to a broader range of mostly lower densities within the group. In 
fact, for the most circular orbit at ``$z$=1'', the galaxy spends a 
significant amount of time beyond the (initial) virial radius of the group.

The strong influence of dynamical friction on the infalling disc galaxy, 
that progressively shrinks its orbit over time, is evident in 
Figure~\ref{trayectories}. Dynamical friction is usually understood as the 
dragging force exerted on an object that moves through a homogeneous medium, 
which in $N$-body simulations is usually modelled by numerous less massive 
particles. Such dragging force comes from the gravitational pull produced by 
medium particles, that cluster behind the moving object. As it can be seen 
in our experiments, such dragging force continuously reduces the galaxy 
tangential velocity (and angular momentum), causing it to spiral-in towards 
the centre of the group. This can also be seen in Figure~\ref{separation-evol} 
as a monotonic decrease of both the apocentric and pericentric distances, over 
subsequent passages of the disc galaxy around the centre of the group. 
Figure~\ref{separation-evol} also shows the dependence of dynamical friction 
on the initial orbital parameters of the disc galaxy and on the 
galaxy-to-group mass ratio. Dynamical friction is less efficient for orbits 
that are initially more circular (CIR vs. REF vs. RAD), and also for lower 
galaxy-to-group mass ratio (MR1:20 vs. REF vs. MR1:5, for ``$z$=0''; MR1:40 
vs. REF vs. MR1:10, for ``$z$=1''). The first trend can be understood in 
terms of the radial mass distribution of the group. A disc galaxy initially 
on a more circular infalling orbit spends more time at larger distances 
from the centre of the group, where the medium surrounding the galaxy is 
less dense. Therefore, it is less efficient at dragging the disc galaxy and 
reducing its angular momentum. The second trend, as mentioned above, accounts 
for the inverse proportionality between the galaxy-to-group mass ratio and 
the timescale of dynamical friction. Interestingly, at ``$z$=1'' there is 
little difference in the evolution of the orbit between the most likely 
infall (REF) and the most radial one (RAD), compared to the corresponding 
experiments at ``$z$=0''. This indicates that the effect of the dynamical 
friction is more similar in these experiments at ``$z$=1''. This can be 
understood as a consequence of the different structure of the group DM halo 
at different redshifts. In general, DM halos are found to be less concentrated 
at higher redshift, for a fixed virial mass. That is, its mass content is 
more evenly distributed with radii in halos at higher redshift. Then, in 
those halos, the efficiency of dynamical friction is expected to show less 
variation with radius, in comparison to more concentrated halos. Thus, one 
can also expect less differences in the orbital evolution of galaxies having 
different initial orbital eccentricities.

Summarising, galaxies in more eccentric orbits and/or with higher 
galaxy-to-group mass ratios experience stronger dynamical friction as they 
orbit in a group, which makes them decay faster towards the group centre. At 
higher redshift, galaxies with different initial eccentricities show less 
differences in the evolution of their orbits, at fixed halo mass. This can 
be linked to the less centrally concentrated mass distribution of groups at 
higher redshift, which causes weaker radial variations in the dynamical 
friction acting upon galaxies. 

\subsection{Mass loss}
\label{sec-massloss}

In order to quantify how much bound mass (both DM and stars) galaxies retain 
as they orbit within a group environment, the following algorithm was used:
(i) Consider all DM and stellar particles of the galaxy that were bound at 
the previous snapshot as bound at the current snapshot. If the current 
snapshot is the initial one, then all galaxy particles are considered bound 
by construction;
(ii) Compute the total bound mass of the galaxy and the velocity of its 
centre of mass;
(iii) Compute the binding energy of the particles that are considered bound 
using the updated velocity of the galaxy's centre of mass;
(iv) Retain only those particles that are still bound (i.e., with negative 
binding energy), and recompute the total mass of the galaxy;
(v) If the total bound mass from (ii) and (iv) has converged, then record it 
and go back to (i) for the next snapshot. If the total bound mass has not 
converged, then go back to (ii) using the bound particles found at step (iv).

Figure~\ref{massbound-evol} shows the evolution of the number of particles 
that remain bound in both the DM halo and the stellar disc of the galaxy 
during its infall within the group environment. In general, the evolution is 
characterised by alternating phases of fast and slow mass loss rates, which 
are closely related to the orbital evolution of the galaxy. Phases of fast 
mass loss take place when the galaxy is near one of its pericentric passages,
where DM particles and disc stars in the galaxy suffer significant impulsive 
accelerations by the group potential. On the other hand, slow phases of mass 
loss occur when the galaxy approaches the apocentres.
In general, galaxies lose DM mass much earlier than stellar mass. In 
most cases, more than half of the initial DM mass is lost after the first 
pericentric passage of the galaxy, while stellar discs are able to retain 
half of their initial mass after several pericentric passages. This can be 
explained by the fact that DM halos are less concentrated and (initially) 
more radially extended than stellar discs. As galaxies orbit within the group, 
the barely-bound external particles of halos are quickly removed as they 
encounter the high density central region of the group, where they experience 
strong impulsive accelerations. On the other hand, disc stars sit at the 
bottom of the potential well of the galaxy, where they require a significant 
increase in their kinetic energies to be removed from the discs, therefore 
remaining bound to the galaxy for longer times.

In our simulations, the extreme cases in the evolution of mass loss are found 
for the experiments that explore different orbital eccentricities and 
galaxy-to-group mass ratios. Namely, at ``$z$=0'', the CIR; RAD; MR1:5 and 
MR1:20 experiments, and at ``$z$=1'', the CIR; RAD; MR1:10 and MR1:40 experiments.
With respect to the REF experiment, more DM mass remains bound to the galaxy 
halo in the CIR; MR:1:20 and MR1:40 experiments, that have in common the fact 
of being affected by less efficient dynamical friction 
(see \S\ref{subsec-orbital}), which keeps the galaxies away from the central 
region of the group where strong accelerations at pericentric passages are more 
intense. In these experiments, discs are capable to retain most of their mass 
($>$95 per cent) over 12 Gyr of evolution within the group environment. On the 
other hand, in the RAD; MR1:5; and MR1:10 experiments, the dynamical friction
is more efficient and galaxies are dragged towards the central region of the 
group early on during the infall. In these cases, both DM and stellar mass are 
stripped faster than in the REF experiment. Typically, DM halos only retain 
$\sim$30-40 per cent of their mass after the first pericentric passage, while
discs can delay the start of their mass loss until the third pericentric passage.

The rest of the experiments (i.e., FON, BUL\footnote{Throughout the paper 
the outcome of our BUL experiments is analysed exclusively from the point of 
view of the evolution of the disc. Only disc stars are included in all 
measurements, excluding stars belonging to the bulge by construction.}, COL, 
RET) show an evolution of DM mass loss that is basically identical to that in 
the REF experiment. This is to be expected since those experiments explore 
different configurations related only to the galaxies' discs. On the other 
hand, the stellar mass loss in those experiments show very interesting variations.
Discs infalling with different inclinations with respect to their orbital 
planes (REF vs. FON vs. RET) lose stellar mass in distinct ways during the 
infall within the group. RET experiments (where discs have their intrinsic and 
orbital angular momentum vectors antiparallel) are shown to retain up to 
$\sim$15 per cent more stellar mass along their orbits, with respect to the REF 
experiments (which have the aforementioned angular momentum vectors 
parallel)\footnote{Note that our definition of retrograde infall differs from 
that used in \citet{murante2010}, where the sense of the orbit was defined 
with respect to the primary halo/disc spin.}. Interestingly, the FON experiments, 
in which discs infall face-on (i.e., their intrinsic and orbital angular 
momentum vectors are orthogonal), show an intermediate behaviour with respect 
to the REF and RET experiments, where discs infall edge-on. This suggests that 
the inclination of the stellar discs with respect to their orbits is an 
important factor for galaxy evolution during the infall within groups (see 
Section~\ref{sec-morph}). Note also that, at ``$z$=0'', the discs in the REF, 
FON and RET experiments have all lost almost completely their stellar content 
by the end of the simulations ($\sim$12 Gyr), independently of differences in 
their past evolution. This shows that the longer the discs retain their mass 
during intermediate stages of the infall, the faster they will lose it at later 
times. This can be understood in terms of the role played by dynamical friction 
during the infall. The more mass a disc retains, the more efficient the influence 
of dynamical friction is, which brings the disc close to the centre of the group
on shorter timescales. Here, it can be affected by stronger impulsive 
accelerations, leading to a faster mass loss rate. In a similar vein, the 
presence of a central stellar bulge in the disc in the BUL experiment (with 
1/3 of the disc's mass), deepens the potential well enough to allow the disc 
to retain slightly more bound mass during the first part of the infall, in 
comparison to the REF experiment. Similarly to the FON and RET experiments, 
this retention of extra bound mass makes the disc in the BUL experiment to 
lose mass faster at later stages of the infall. This type of behaviour is seen 
yet again for the COL experiment, where the colder kinematics of the disc 
(i.e., with a lower $Q$ parameter) keeps the stars closer to the midplane, which 
makes them less likely to get stripped during the early stages of the evolution. 
Again, the disc is able to temporarily retain more bound mass, only to lose it 
faster by the end of the infall, compared to the REF experiment.

Note that, in general, the fraction of bound stellar mass decreases more 
slowly in the experiments at ``$z$=1'', compared to those  at ``$z$=0''. This can 
be understood as a consequence of the more compact discs at ``$z$=1'', and of the 
shallower radial variation of the global potential well, due to the less 
concentrated mass distribution of the group at higher redshift. This leads to 
weaker and less destructive impulsive accelerations affecting the disc at each 
pericentric passage, which helps preserving for a longer time the mass originally 
bound to the discs. This highlights the key role played by the structure of the 
environment on the evolution of disc galaxies.

Summarising, we find that galaxies with more eccentric initial orbits and 
those with higher galaxy-to-group mass ratios experience a faster loss of 
bound mass, in comparison to the reference experiment. This occurs as a 
consequence of the more efficient dynamical friction affecting the galaxies 
in the first cases, which brings them faster within the dense central region 
of the group, exposing them to stronger impulsive accelerations. Additionally, 
the initial inclination of disc galaxies, with respect to their orbital 
planes, is found to be an important factor in the evolution of their mass loss. 
In particular, galaxies with infalling inclinations of 90\degr and 180\degr 
are found to be able retain increasingly more mass during intermediate stages 
of the orbit, compared to galaxies with 0\degr inclination. We also find that 
when galaxies retain more mass during intermediate stages of the infall, they 
end up losing it faster at later times. This is also related to the more 
efficient dynamical friction acting on more massive galaxies. Finally, less 
concentrated groups, in experiments at higher redshift, are found to be linked 
to significantly less violent decreases in the fraction of bound stellar mass 
of galaxies, given that they provide a relatively weaker global force field.

\subsection{Disc morphology}
\label{sec-morph}

\begin{figure*}
\begin{center}
\includegraphics[width=34.9mm]{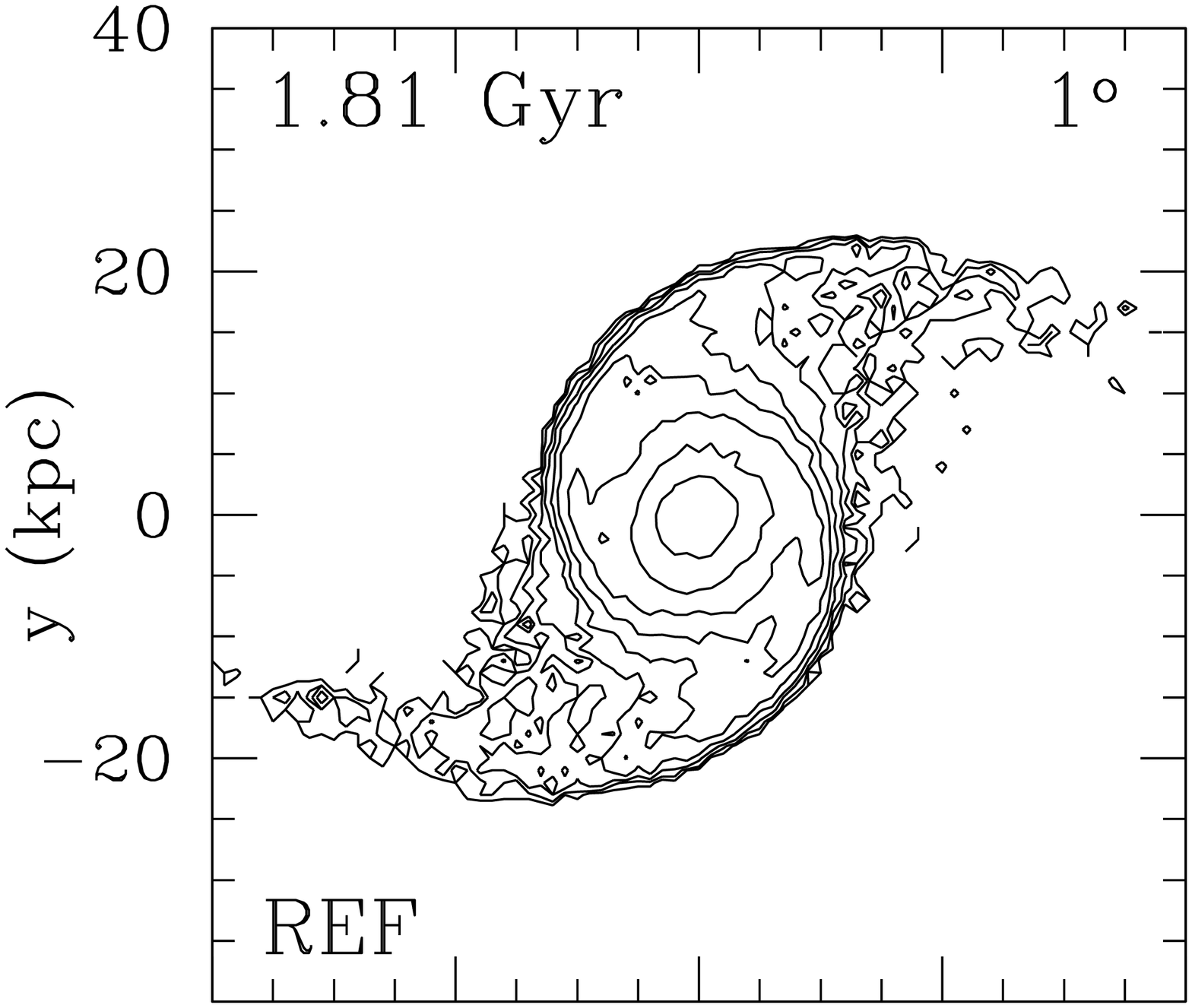}\hspace*{-0.87cm}
\includegraphics[width=34.9mm]{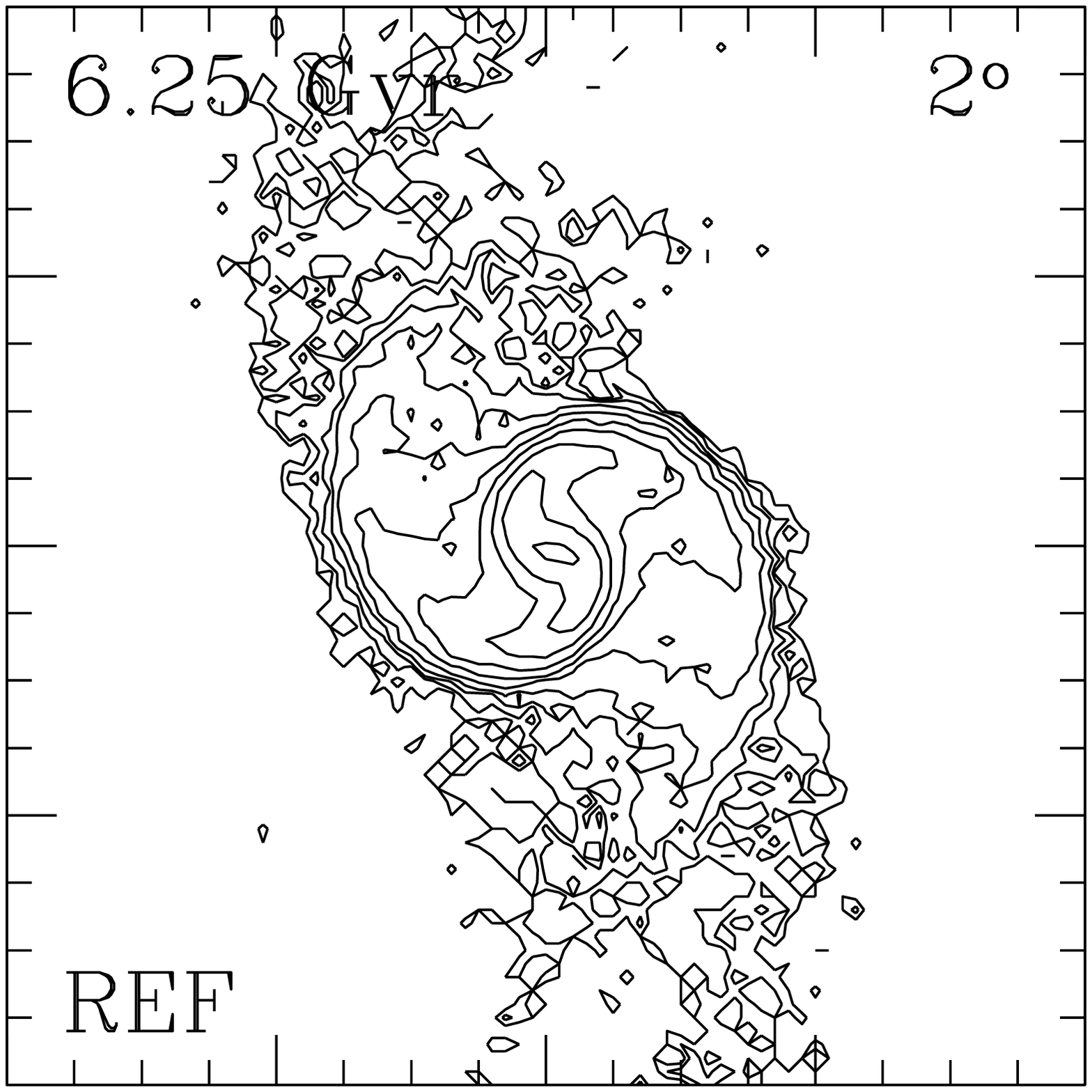}\hspace*{-0.87cm}
\includegraphics[width=34.9mm]{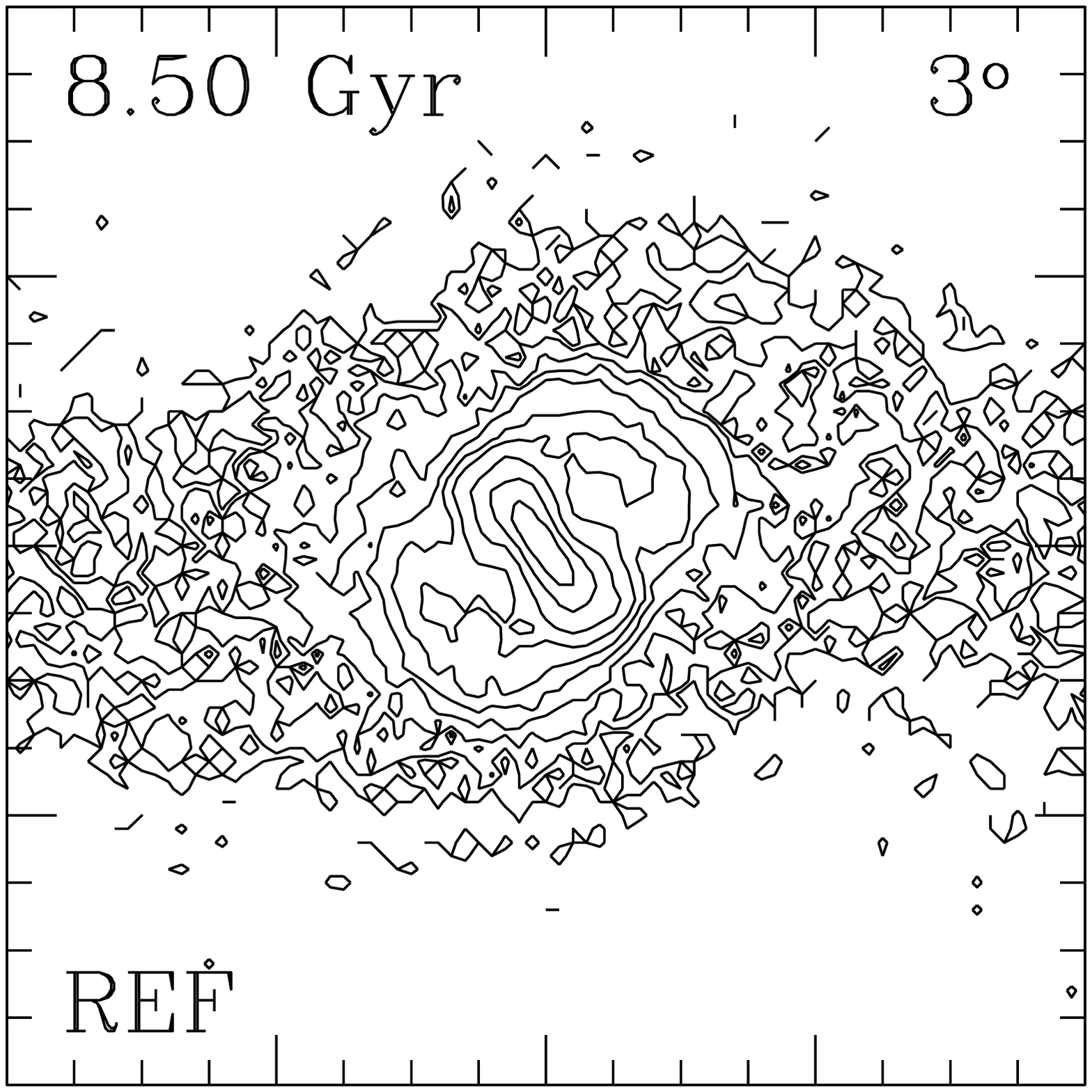}\hspace*{-0.87cm}
\includegraphics[width=34.9mm]{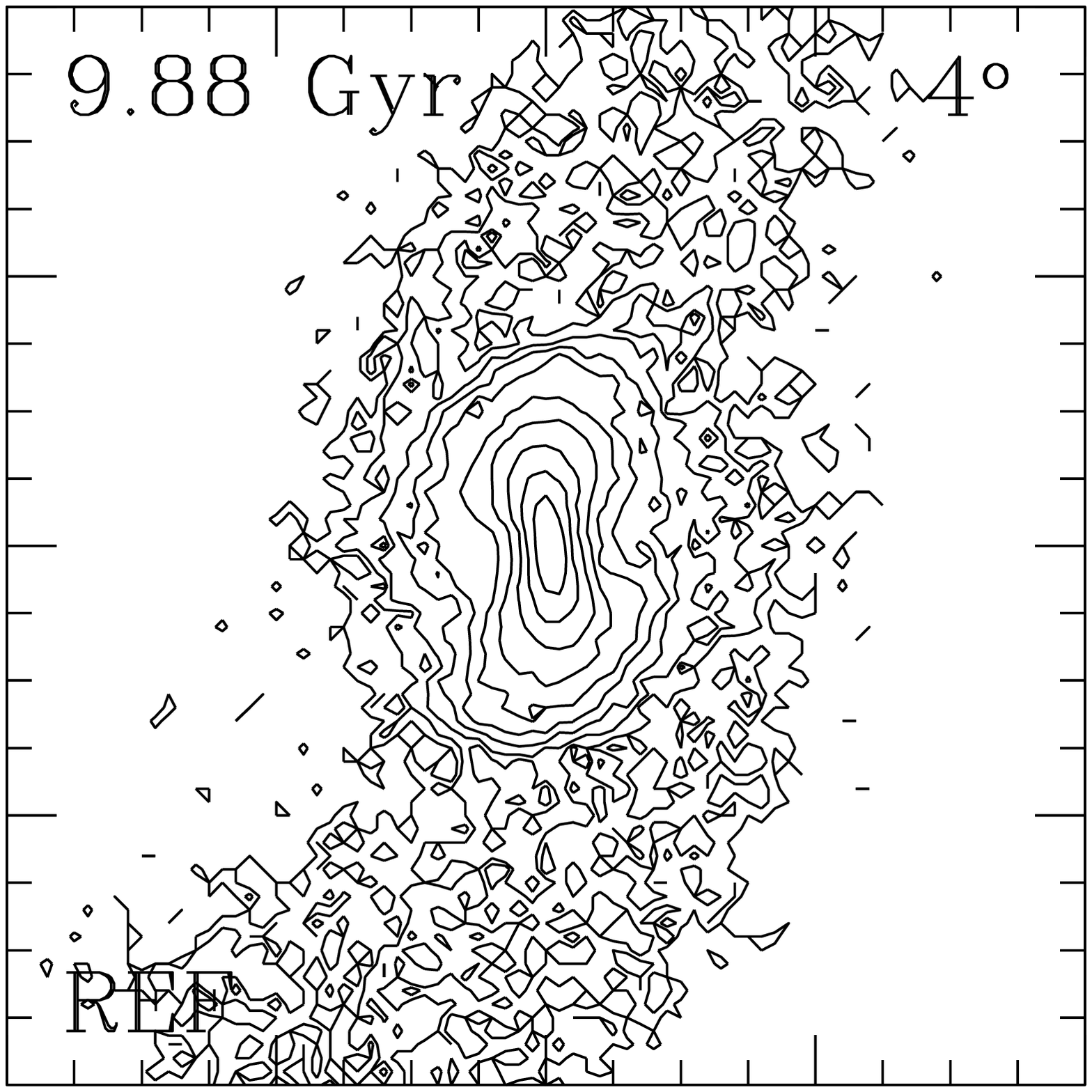}\hspace*{-0.87cm}
\includegraphics[width=34.9mm]{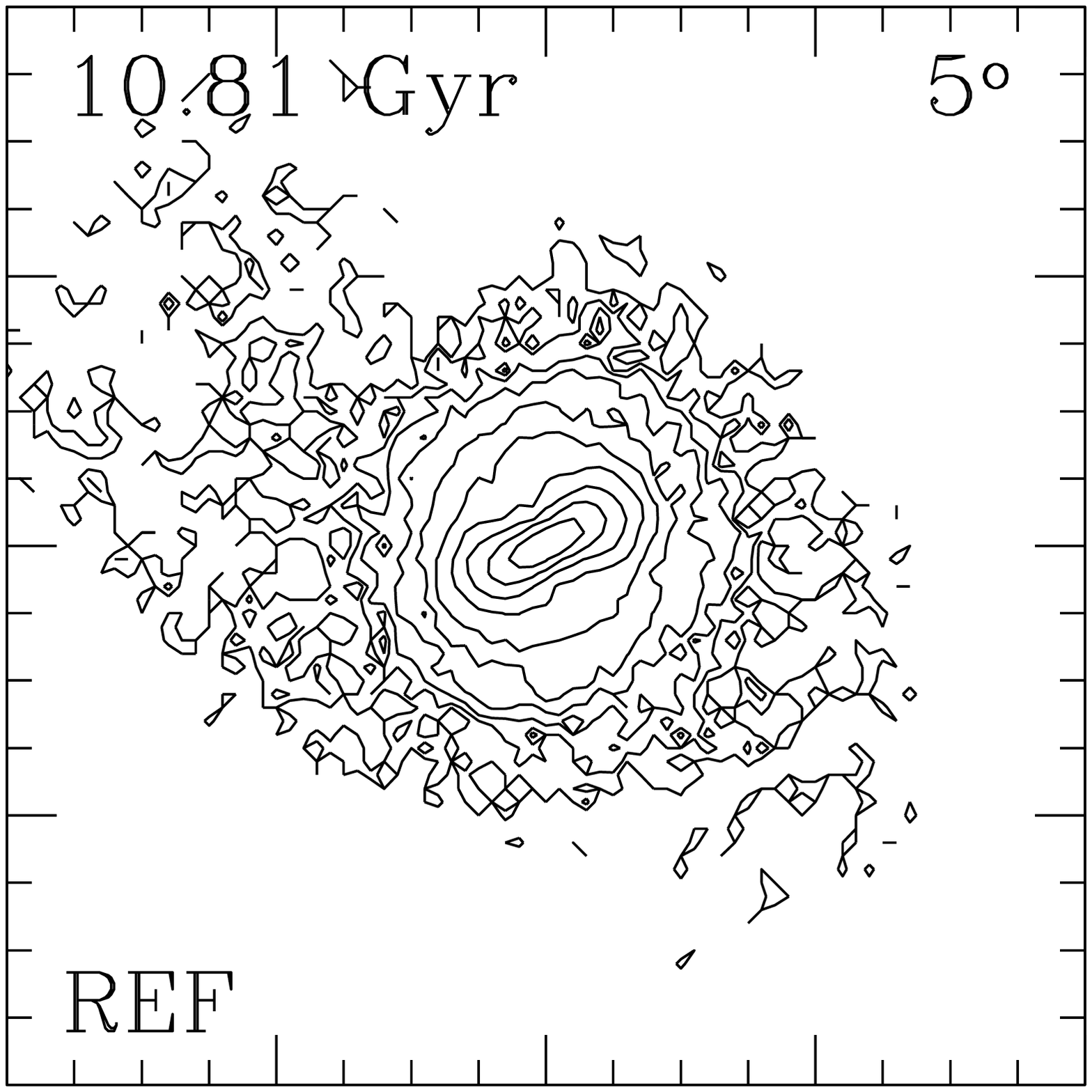}\hspace*{-0.87cm}
\includegraphics[width=34.9mm]{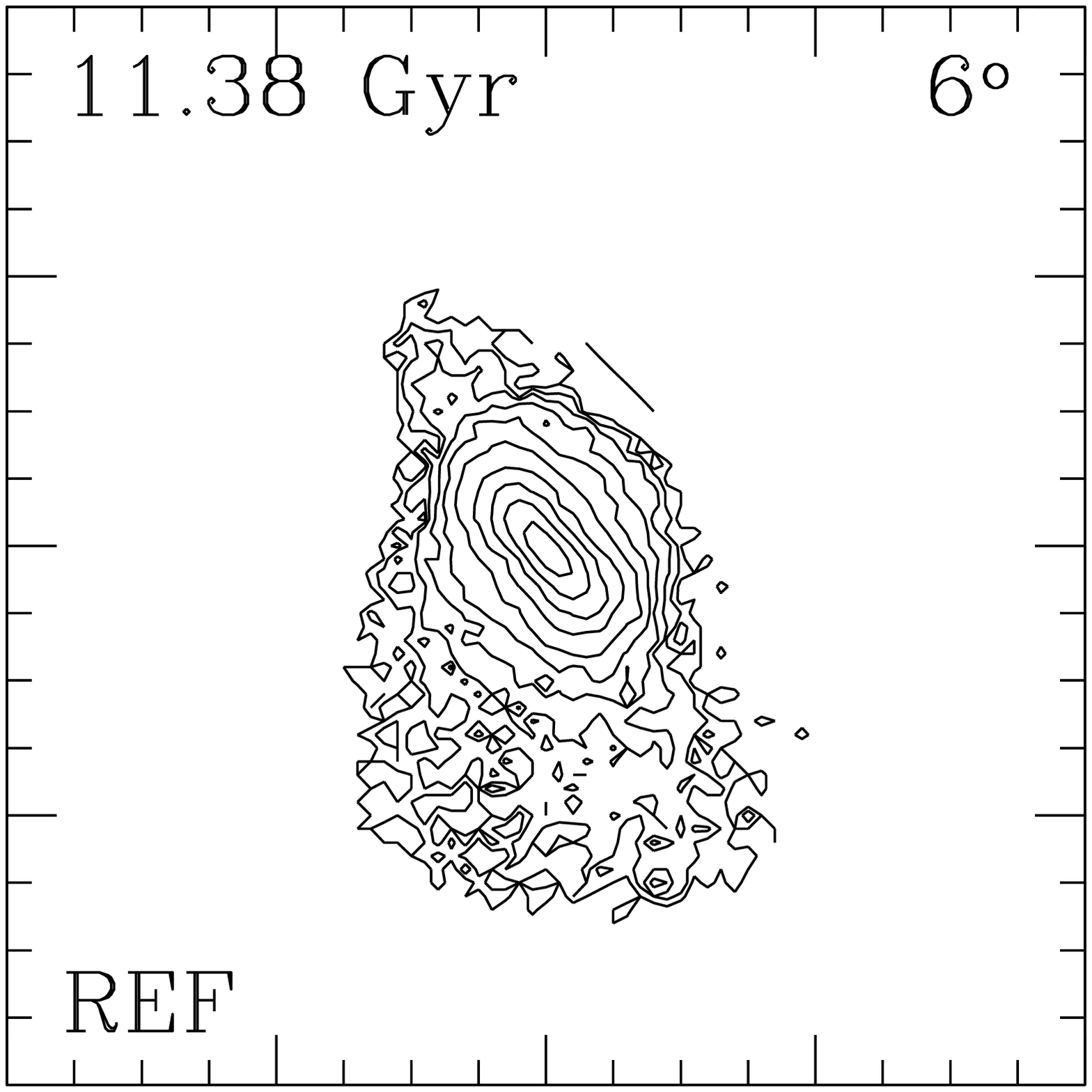}\vspace*{-0.81cm}\\
\includegraphics[width=34.9mm]{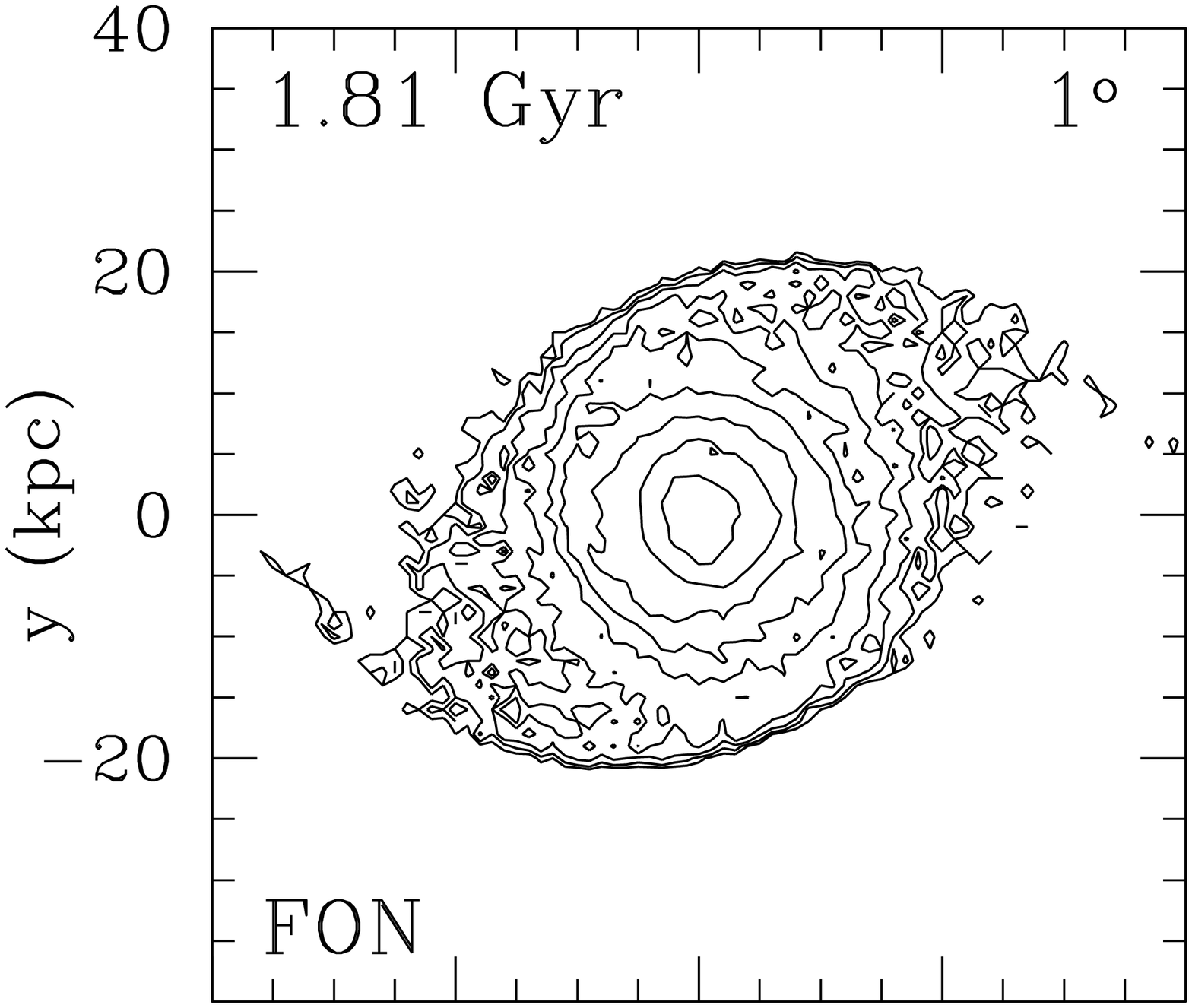}\hspace*{-0.87cm}
\includegraphics[width=34.9mm]{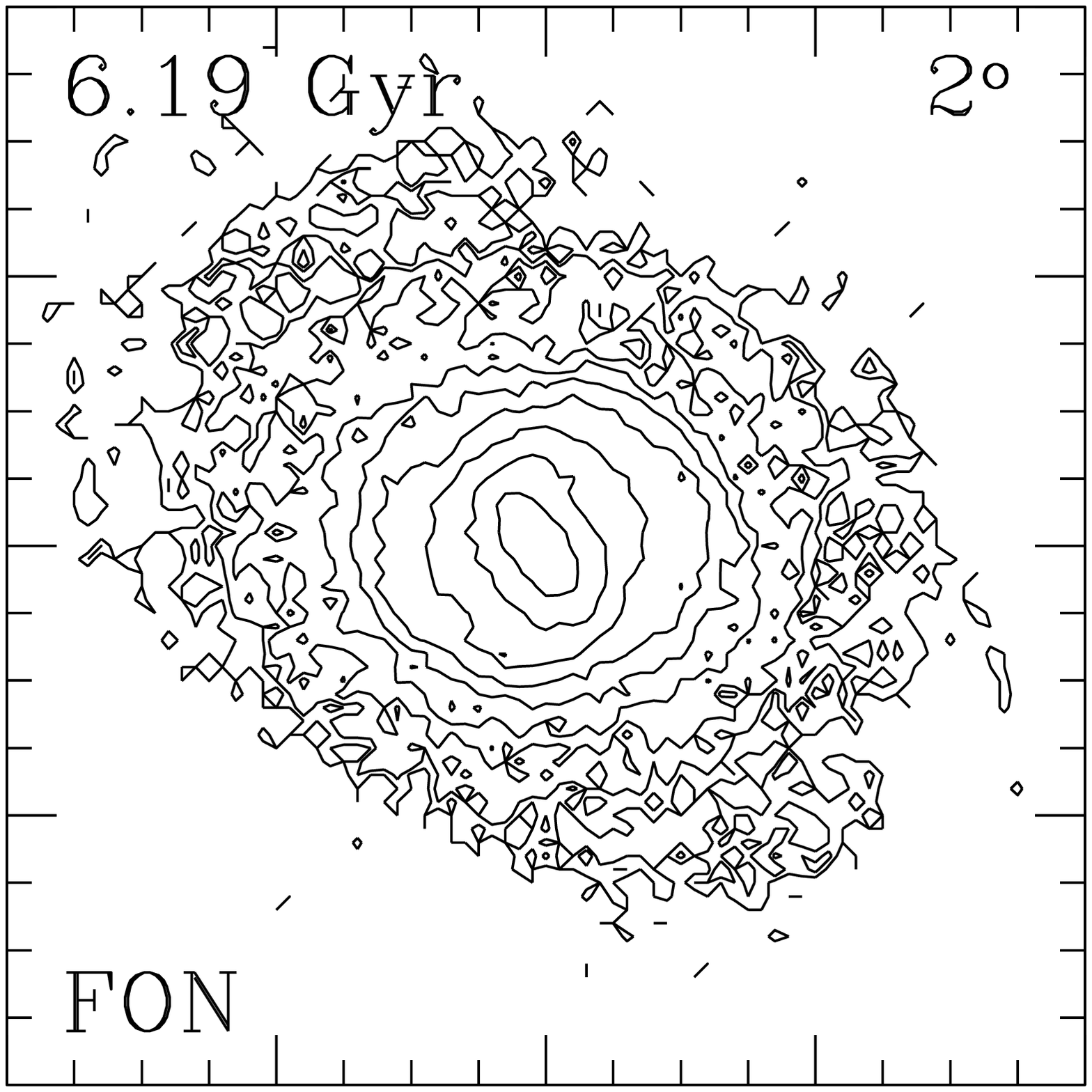}\hspace*{-0.87cm}
\includegraphics[width=34.9mm]{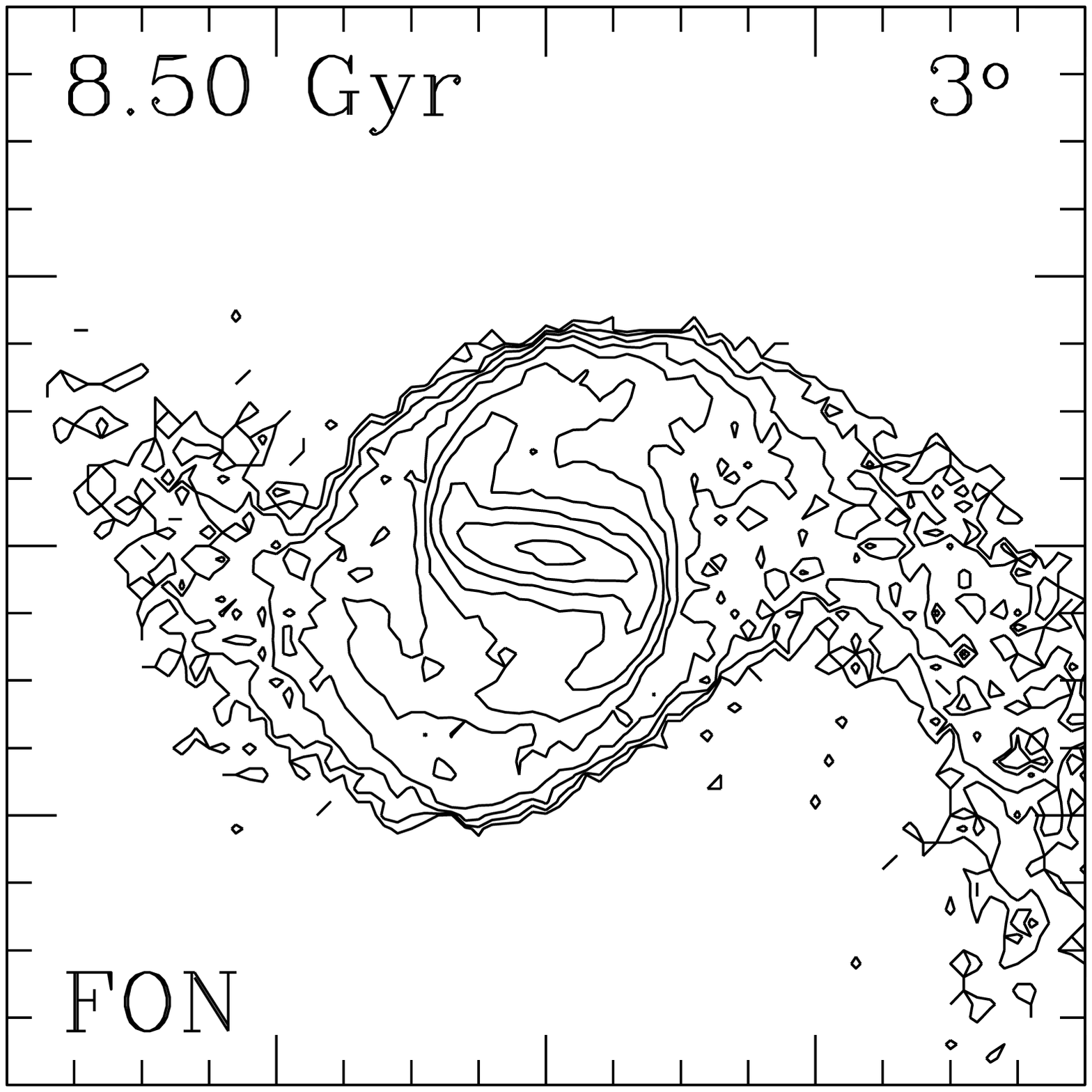}\hspace*{-0.87cm}
\includegraphics[width=34.9mm]{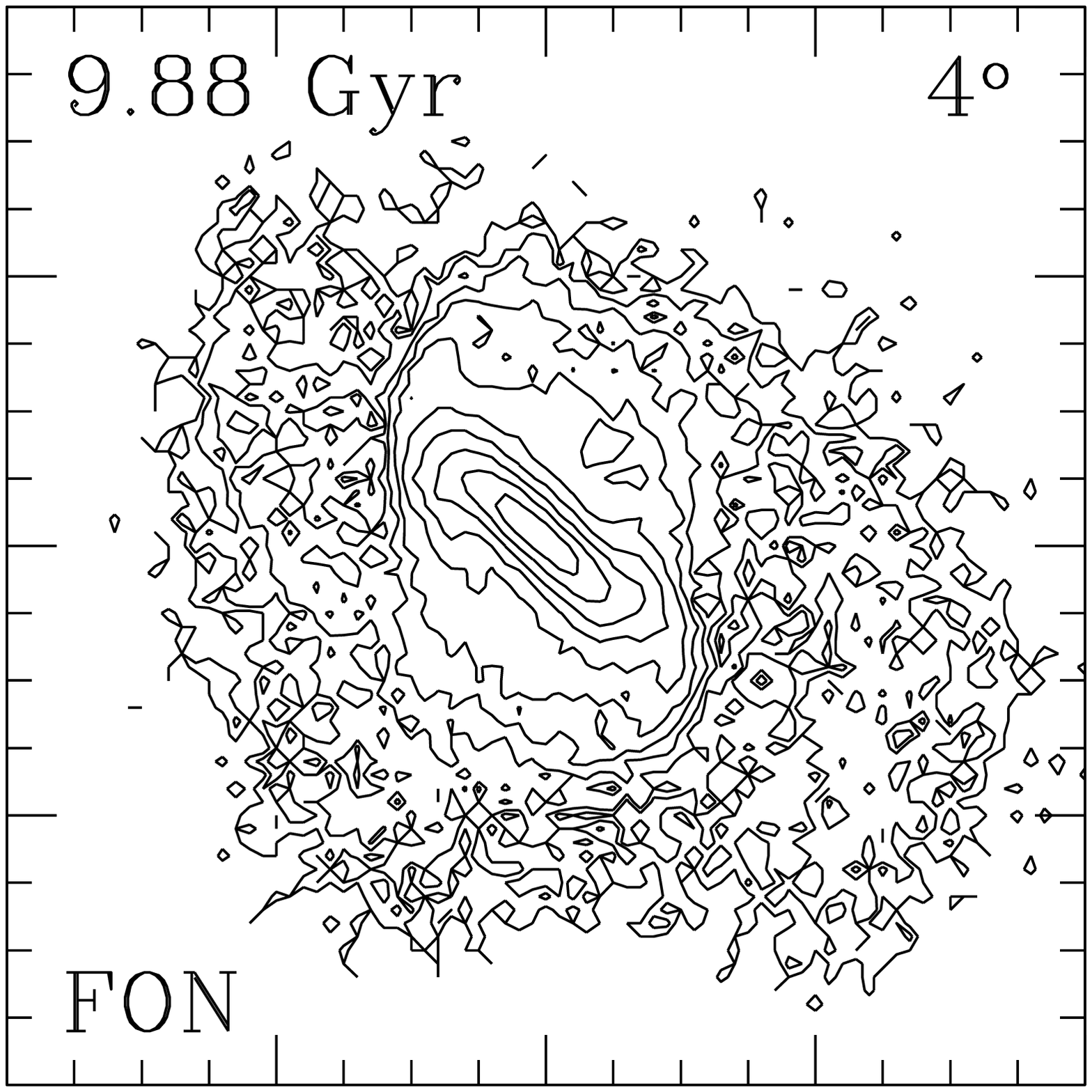}\hspace*{-0.87cm}
\includegraphics[width=34.9mm]{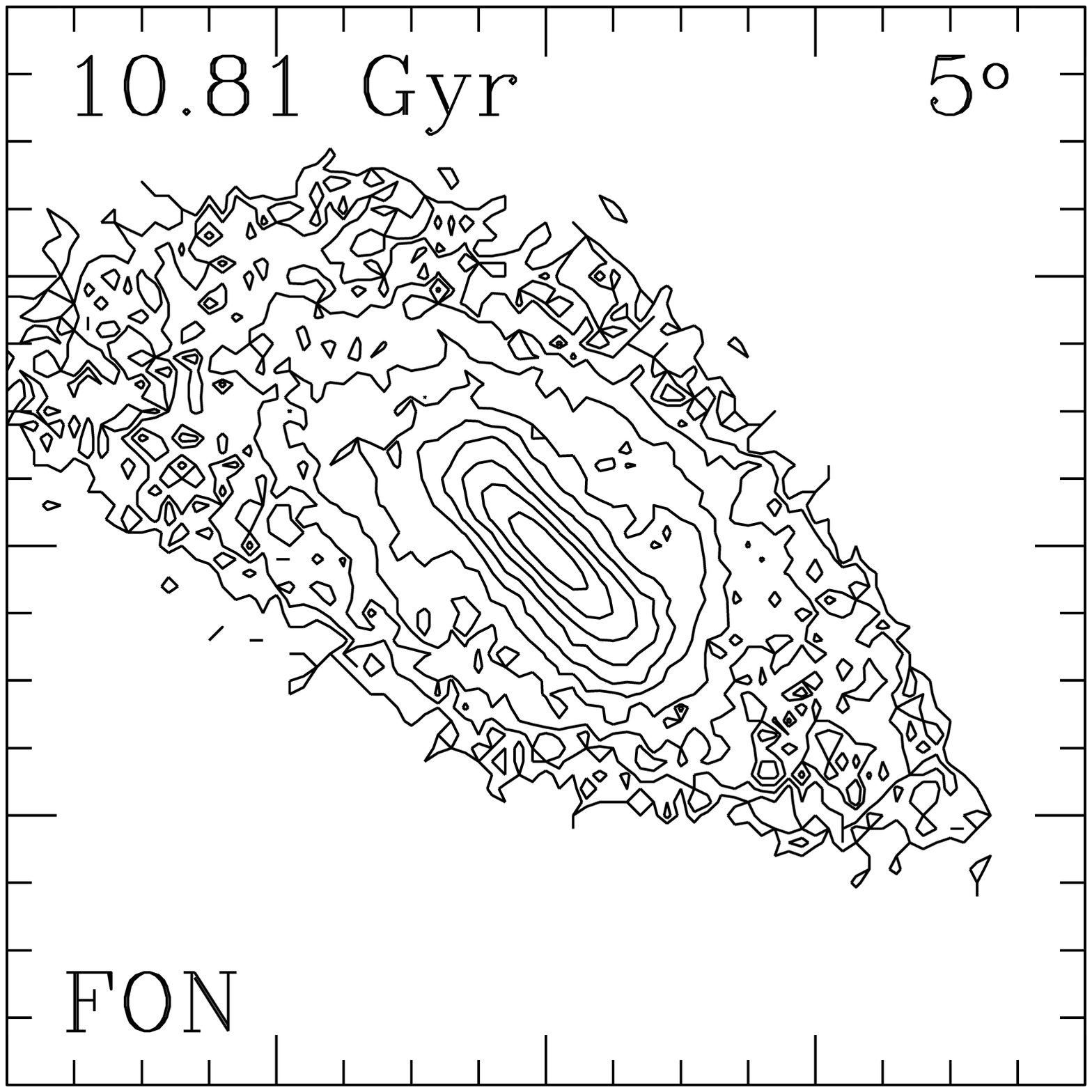}\hspace*{-0.87cm}
\includegraphics[width=34.9mm]{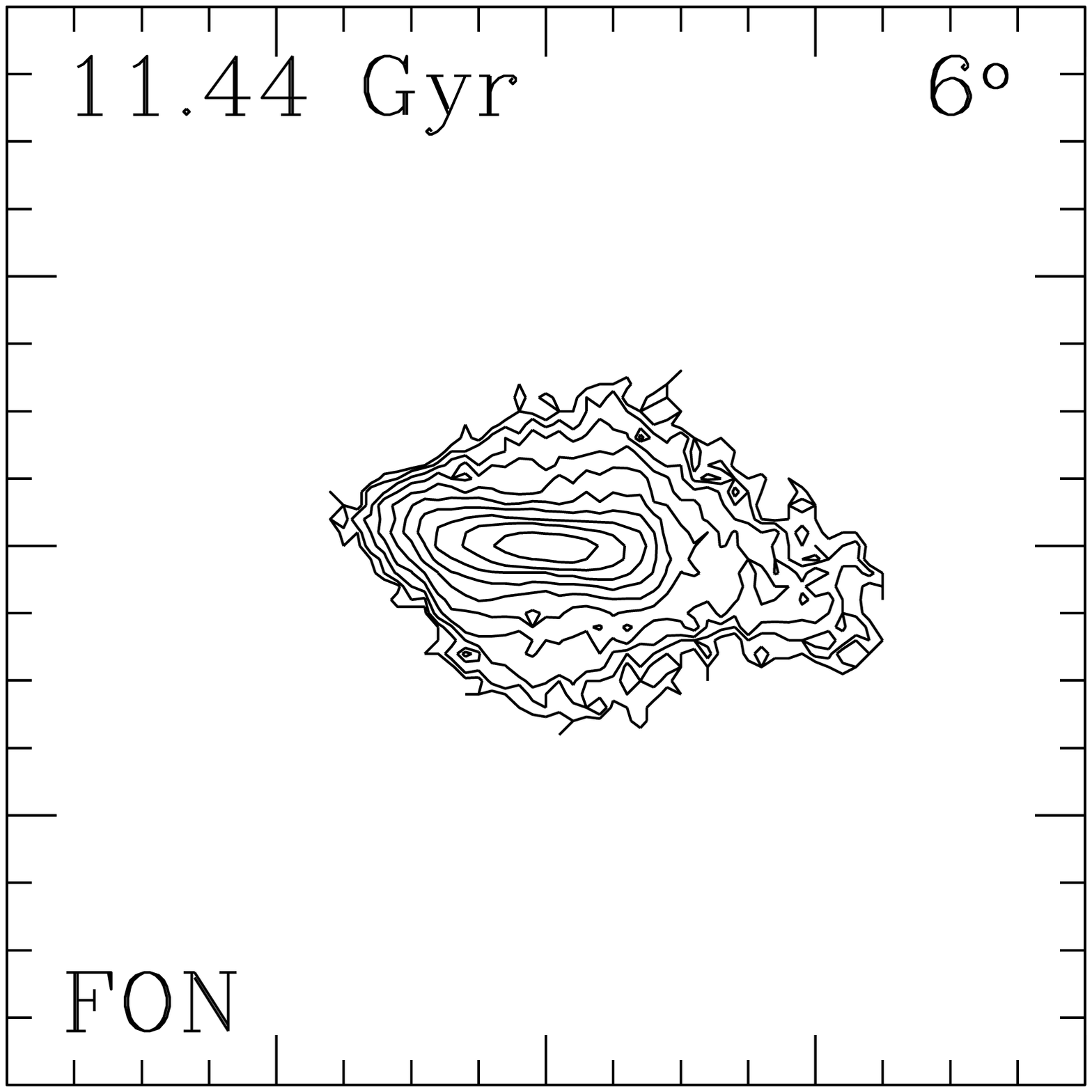}\vspace*{-0.81cm}\\
\includegraphics[width=34.9mm]{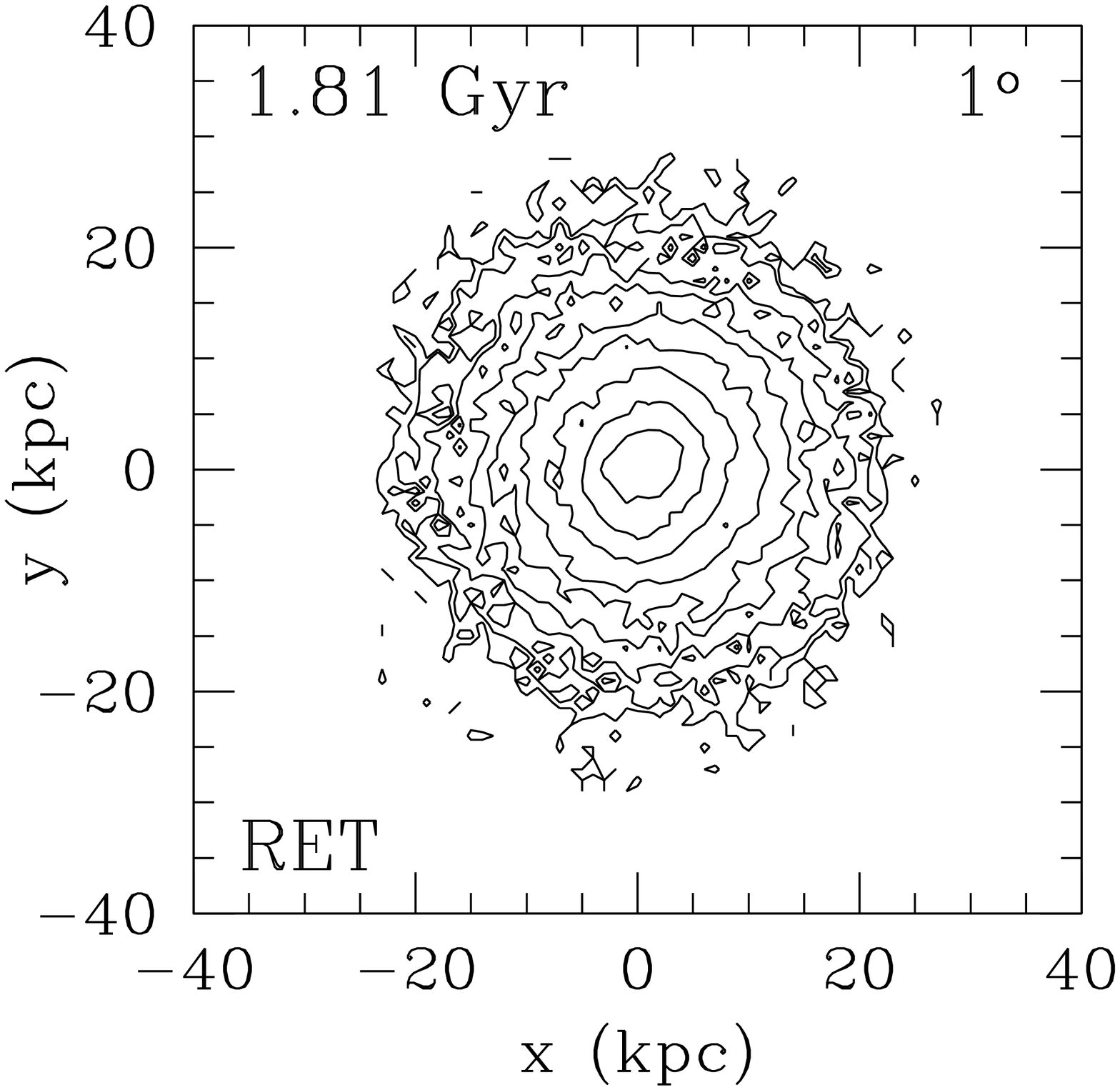}\hspace*{-0.87cm}
\includegraphics[width=34.9mm]{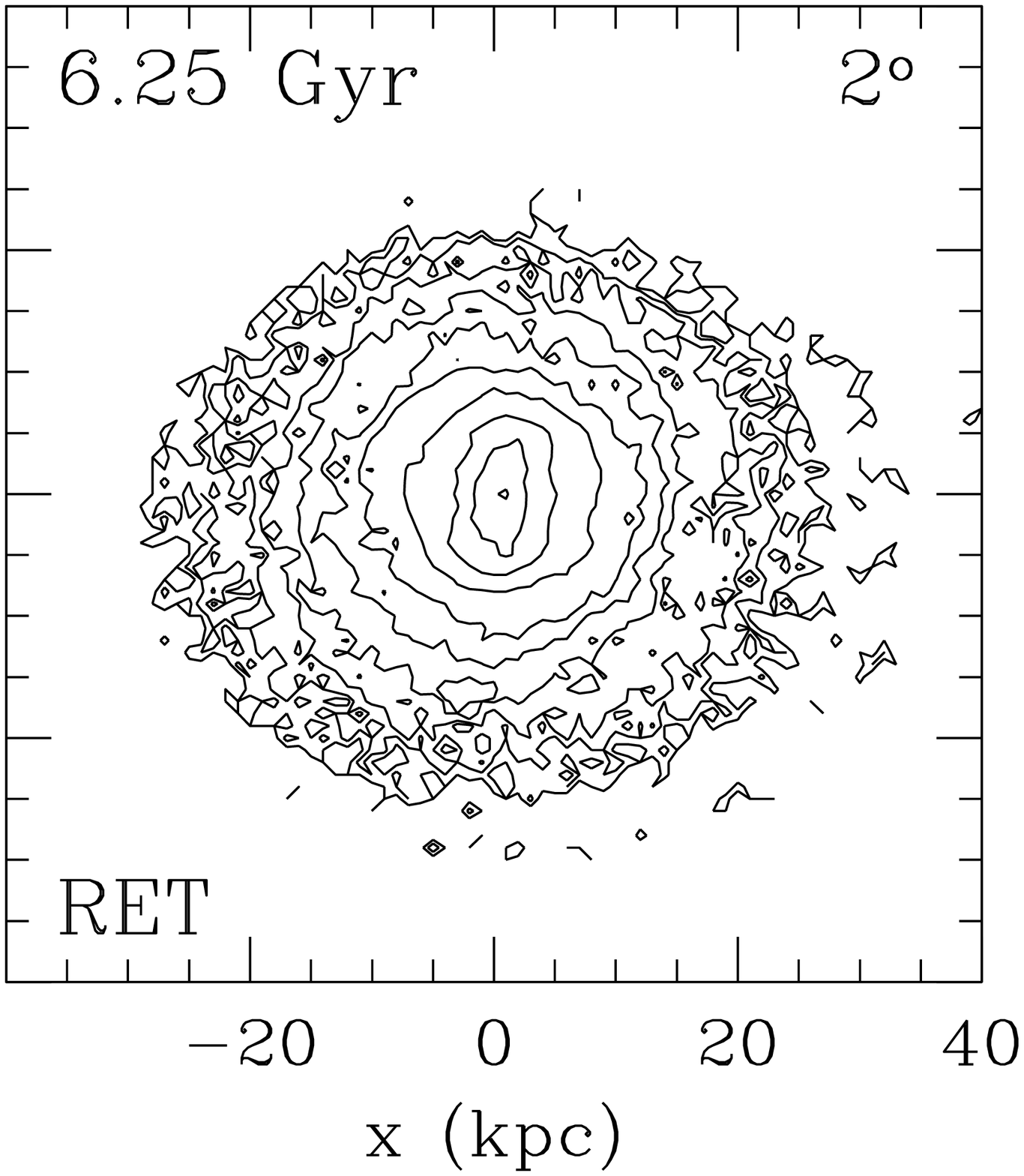}\hspace*{-0.87cm}
\includegraphics[width=34.9mm]{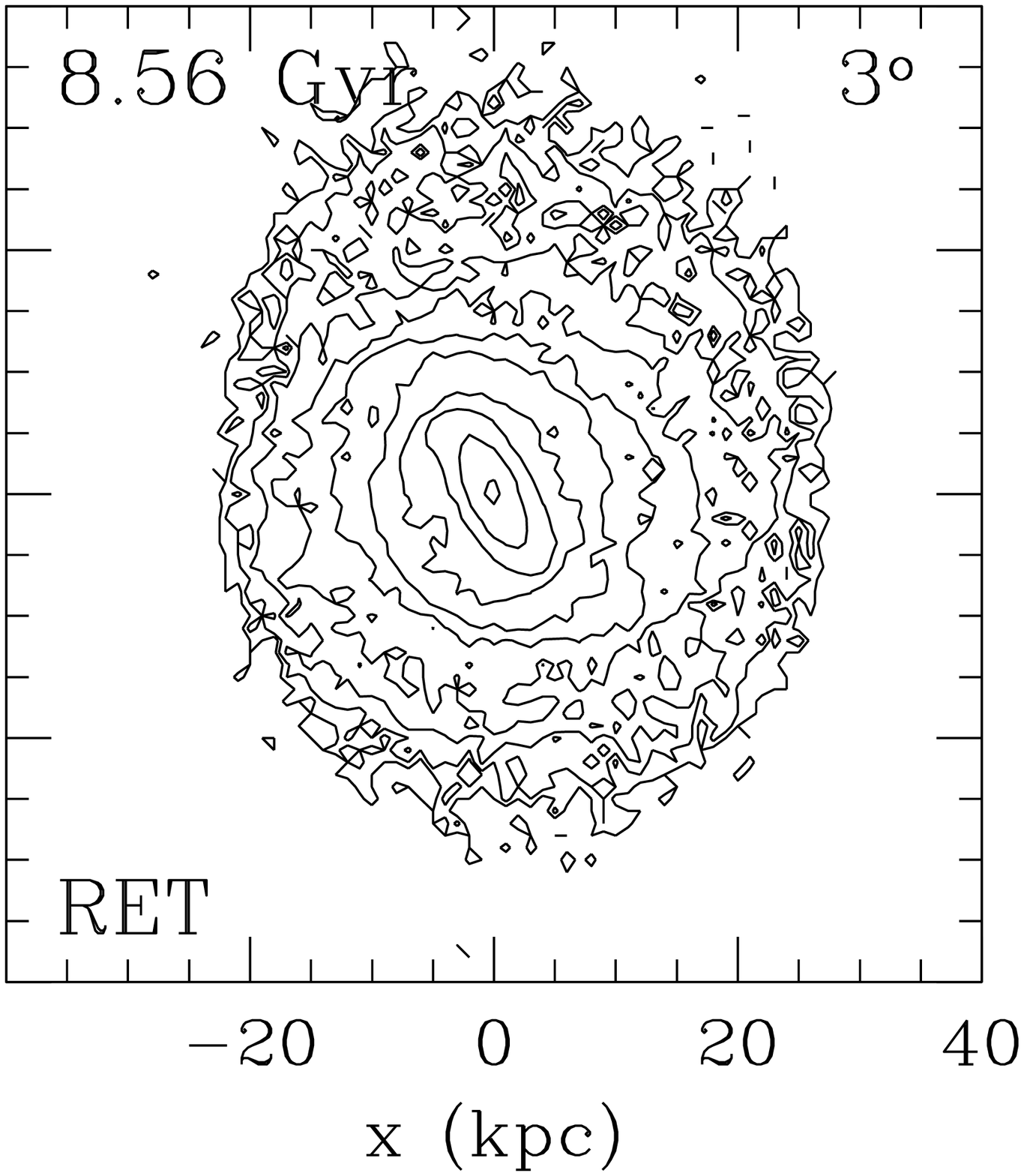}\hspace*{-0.87cm}
\includegraphics[width=34.9mm]{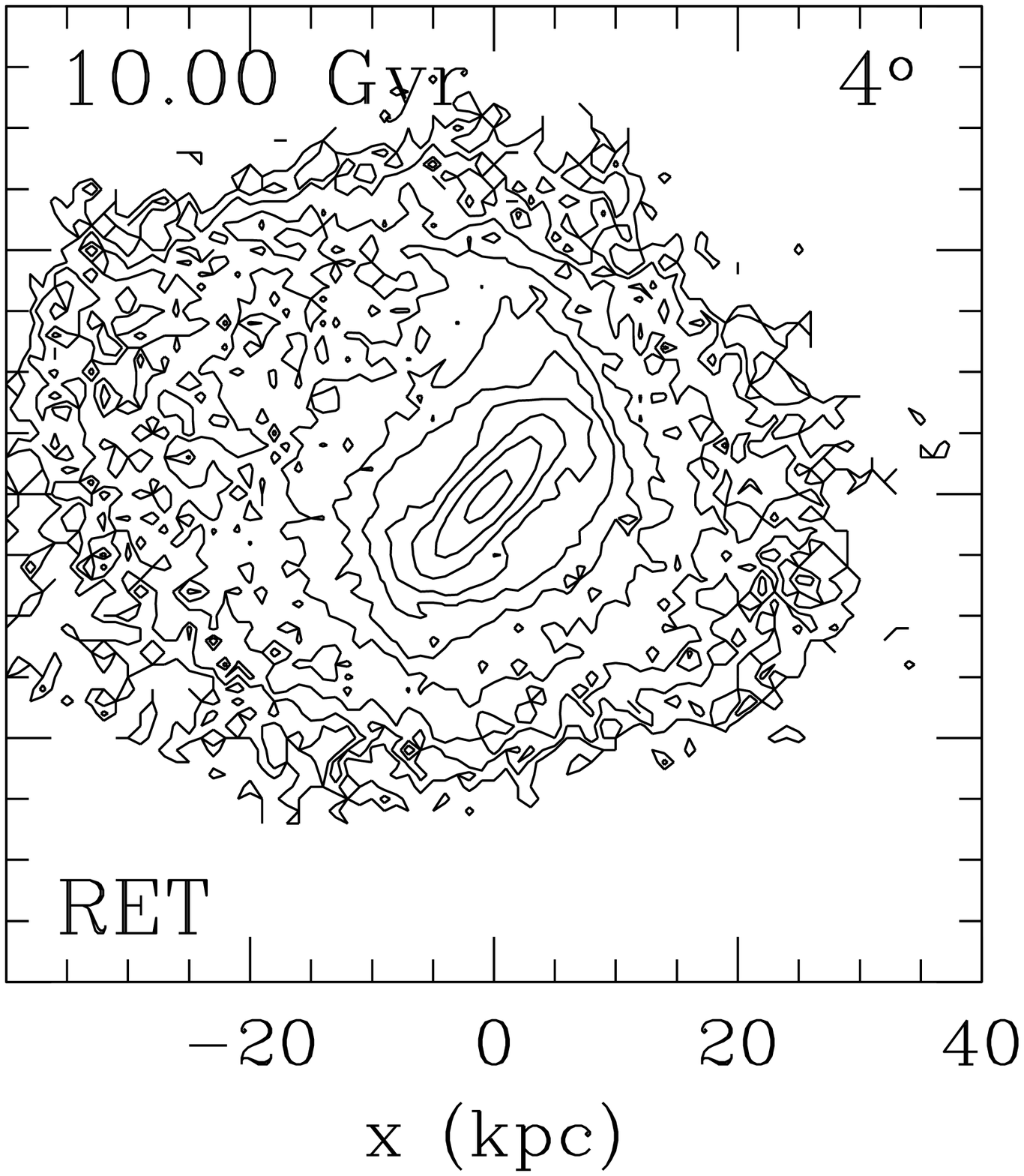}\hspace*{-0.87cm}
\includegraphics[width=34.9mm]{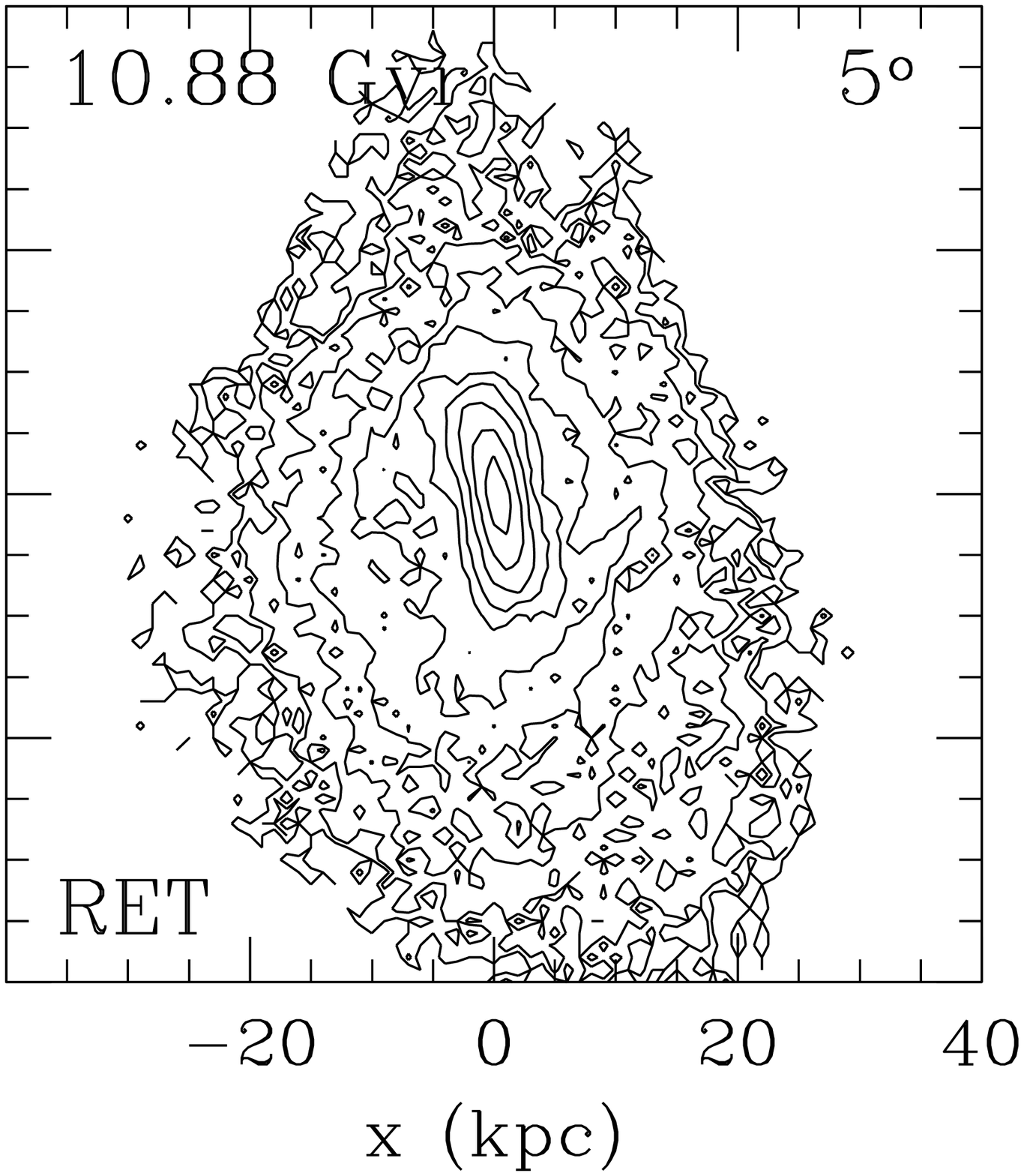}\hspace*{-0.87cm}
\includegraphics[width=34.9mm]{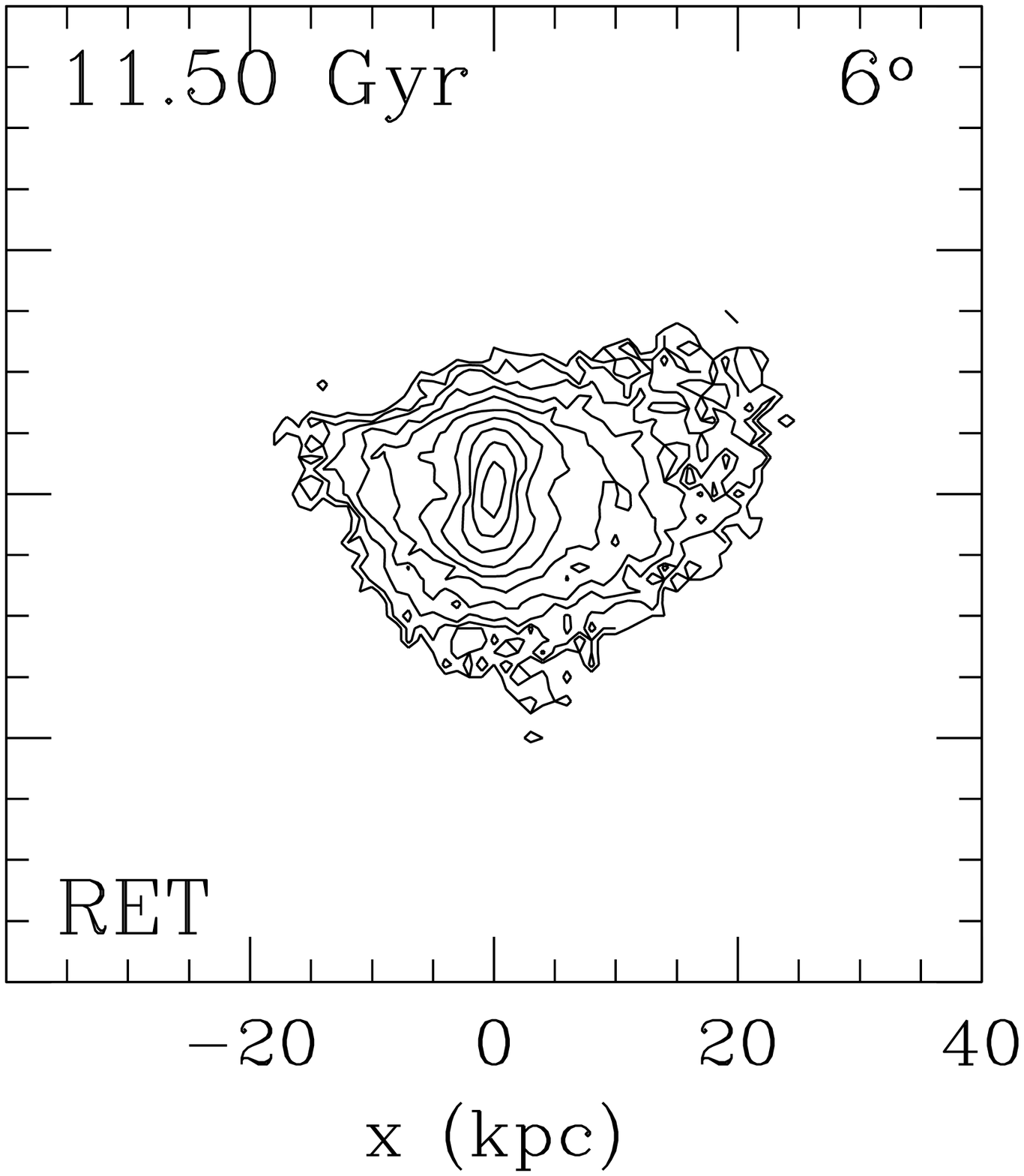}
\end{center}
\caption{Face-on views of the morphological evolution of disc galaxies for our 
REF (reference), FON (face-on infall) and RET (retrograde infall) experiments 
at ``$z$=0'', after each of their first six pericentric passages. Number density
contours are drawn at the same levels to ease the comparison, and only stars 
that remain bound to the disc, at a given time, are considered.}
\label{contours-xy-z0}
\end{figure*}

\begin{figure*}
\begin{center}
\includegraphics[width=34.9mm]{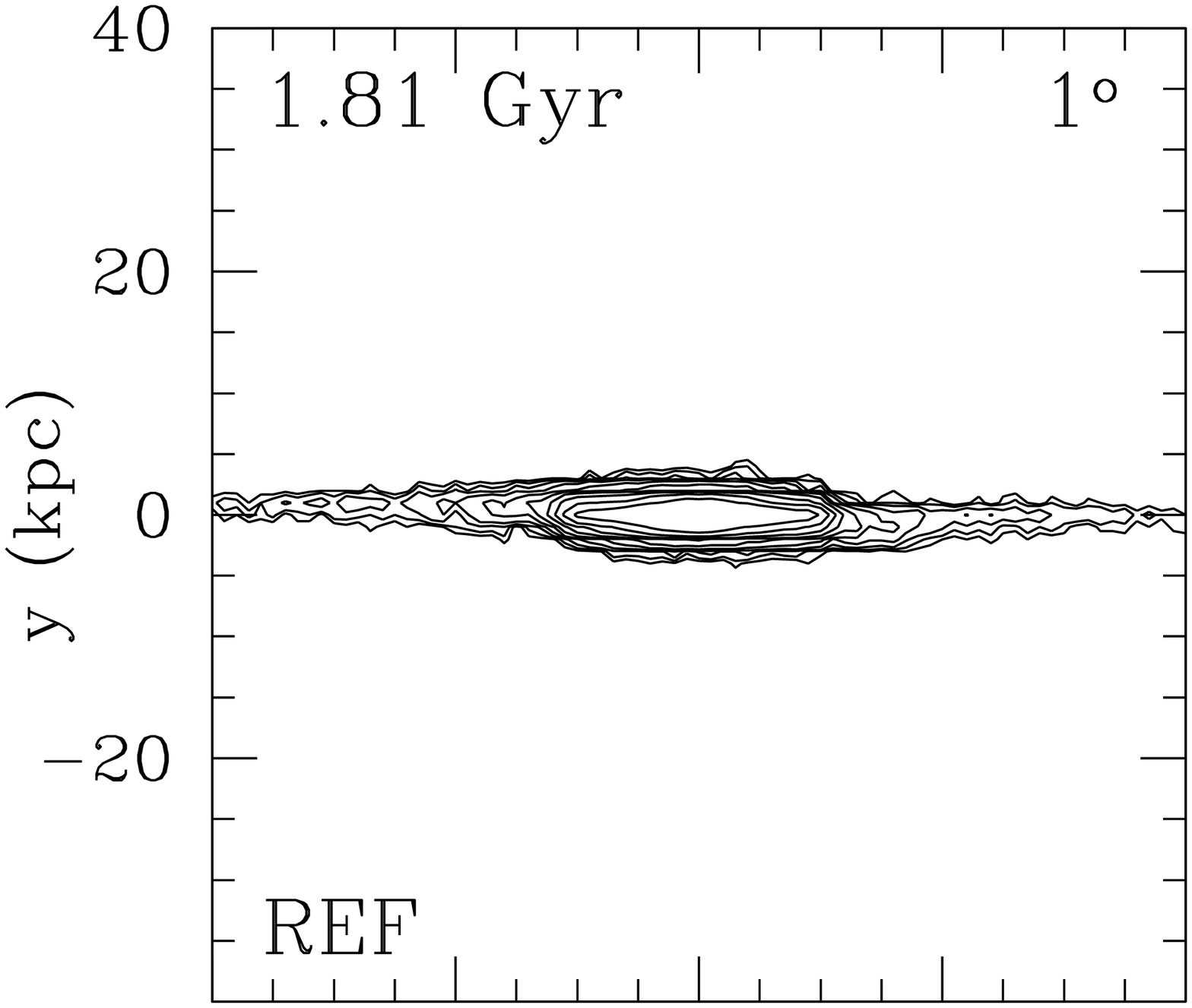}\hspace*{-0.87cm}
\includegraphics[width=34.9mm]{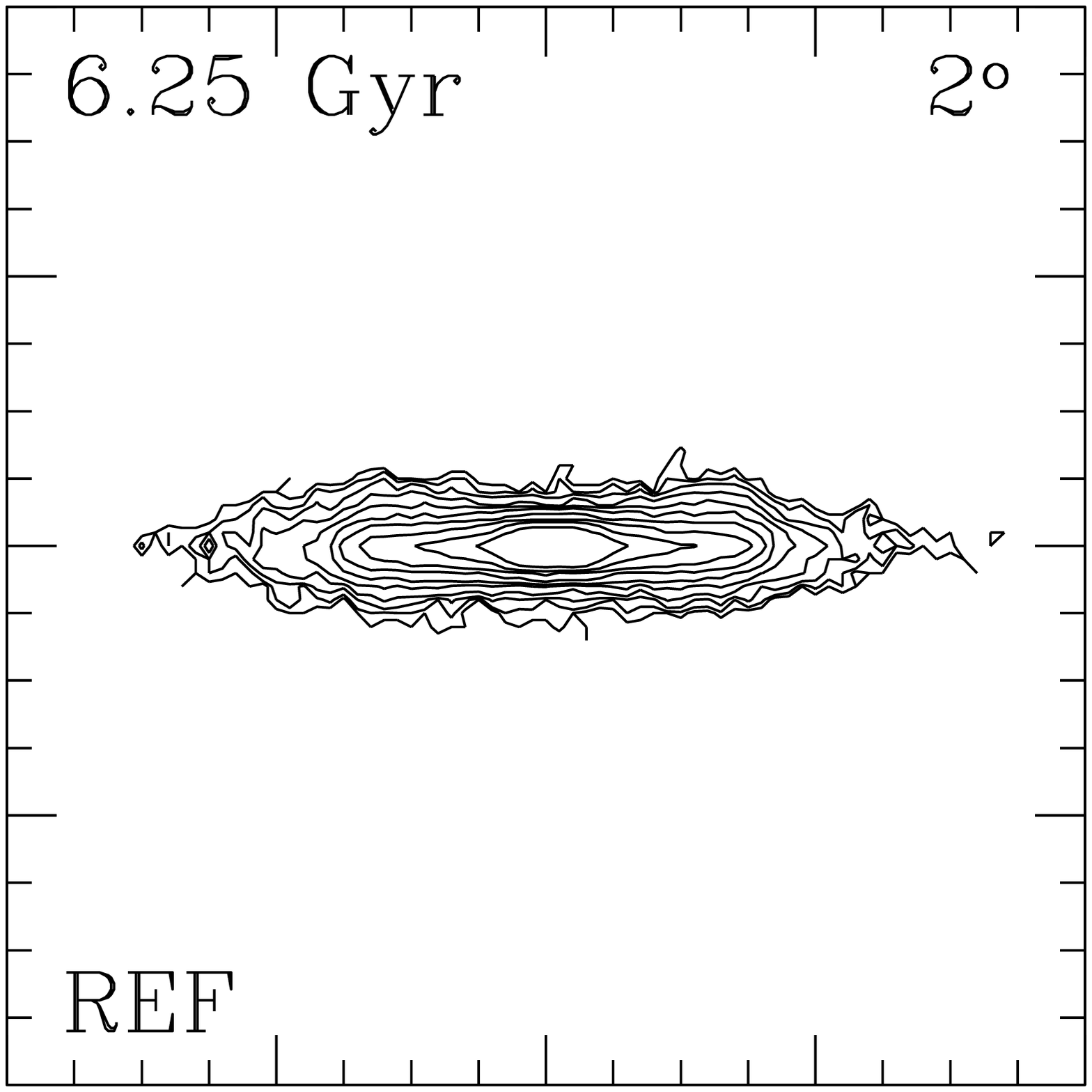}\hspace*{-0.87cm}
\includegraphics[width=34.9mm]{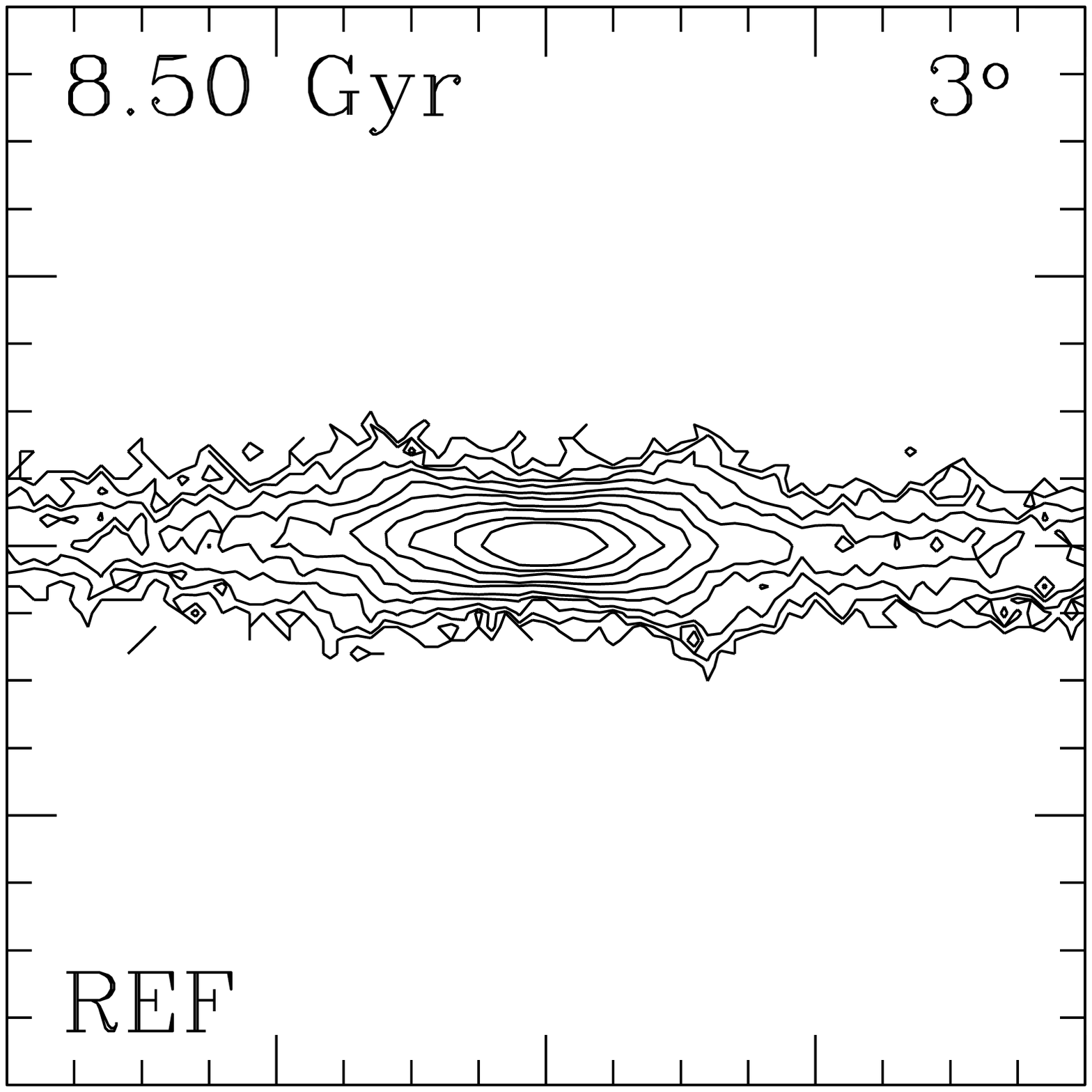}\hspace*{-0.87cm}
\includegraphics[width=34.9mm]{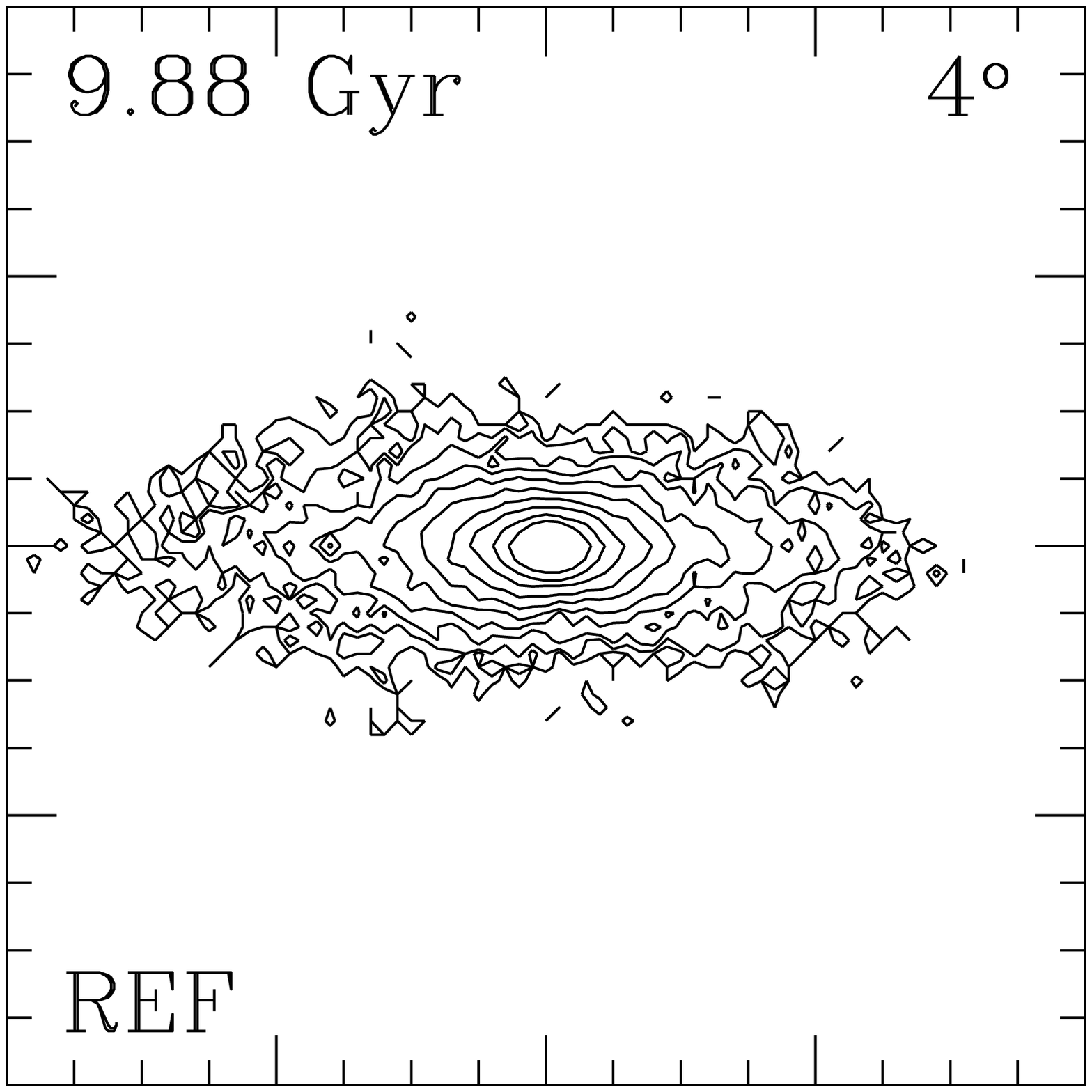}\hspace*{-0.87cm}
\includegraphics[width=34.9mm]{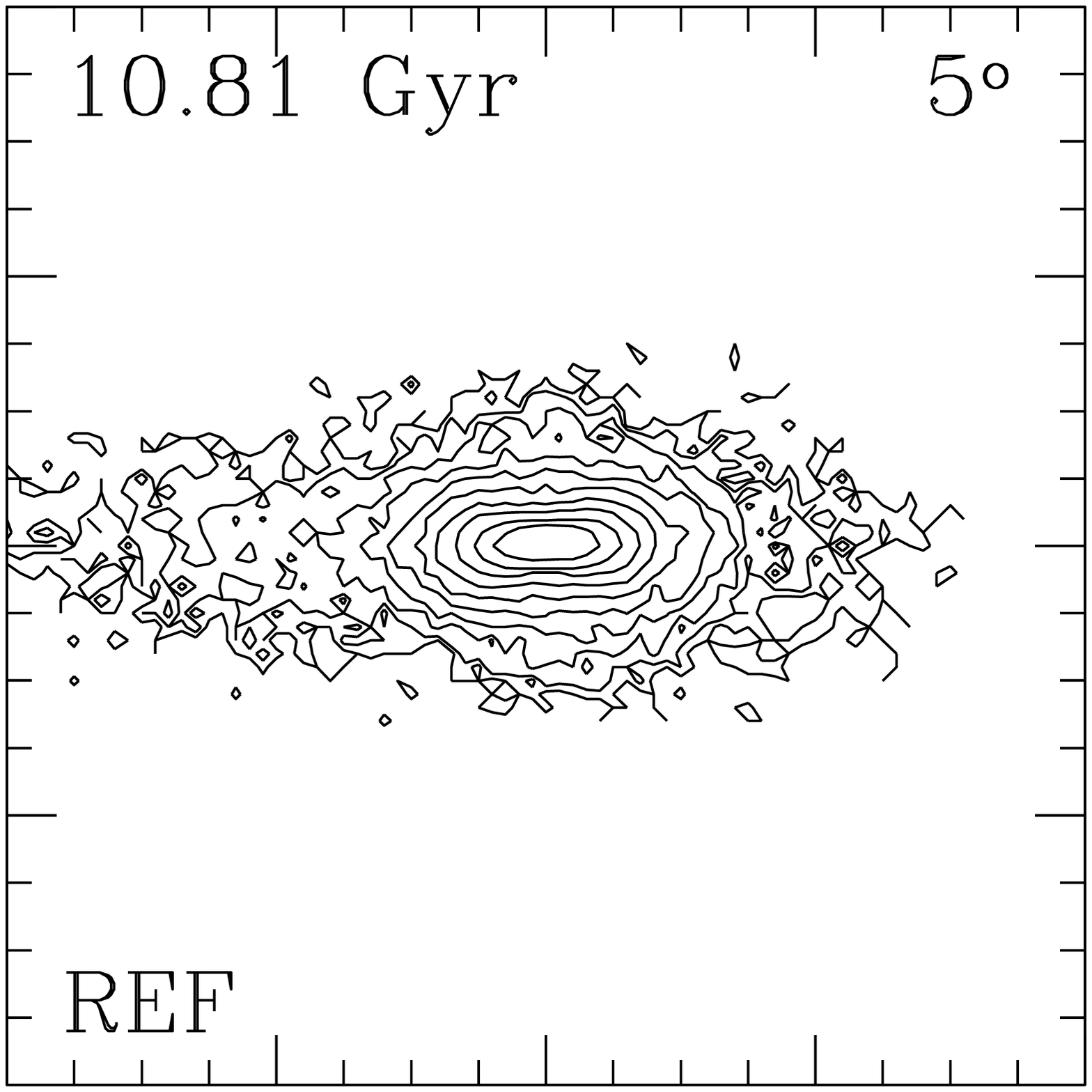}\hspace*{-0.87cm}
\includegraphics[width=34.9mm]{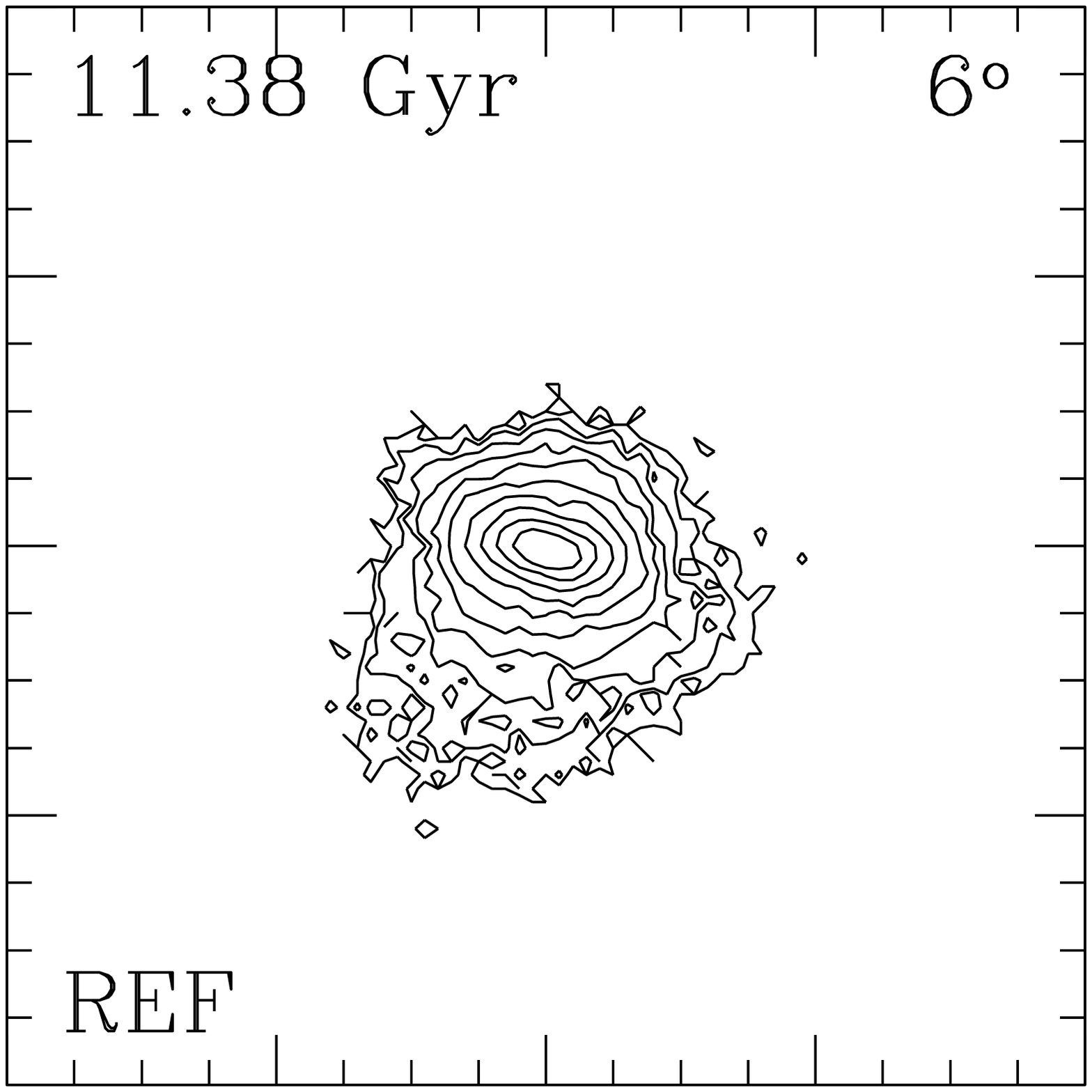}\vspace*{-0.81cm}\\
\includegraphics[width=34.9mm]{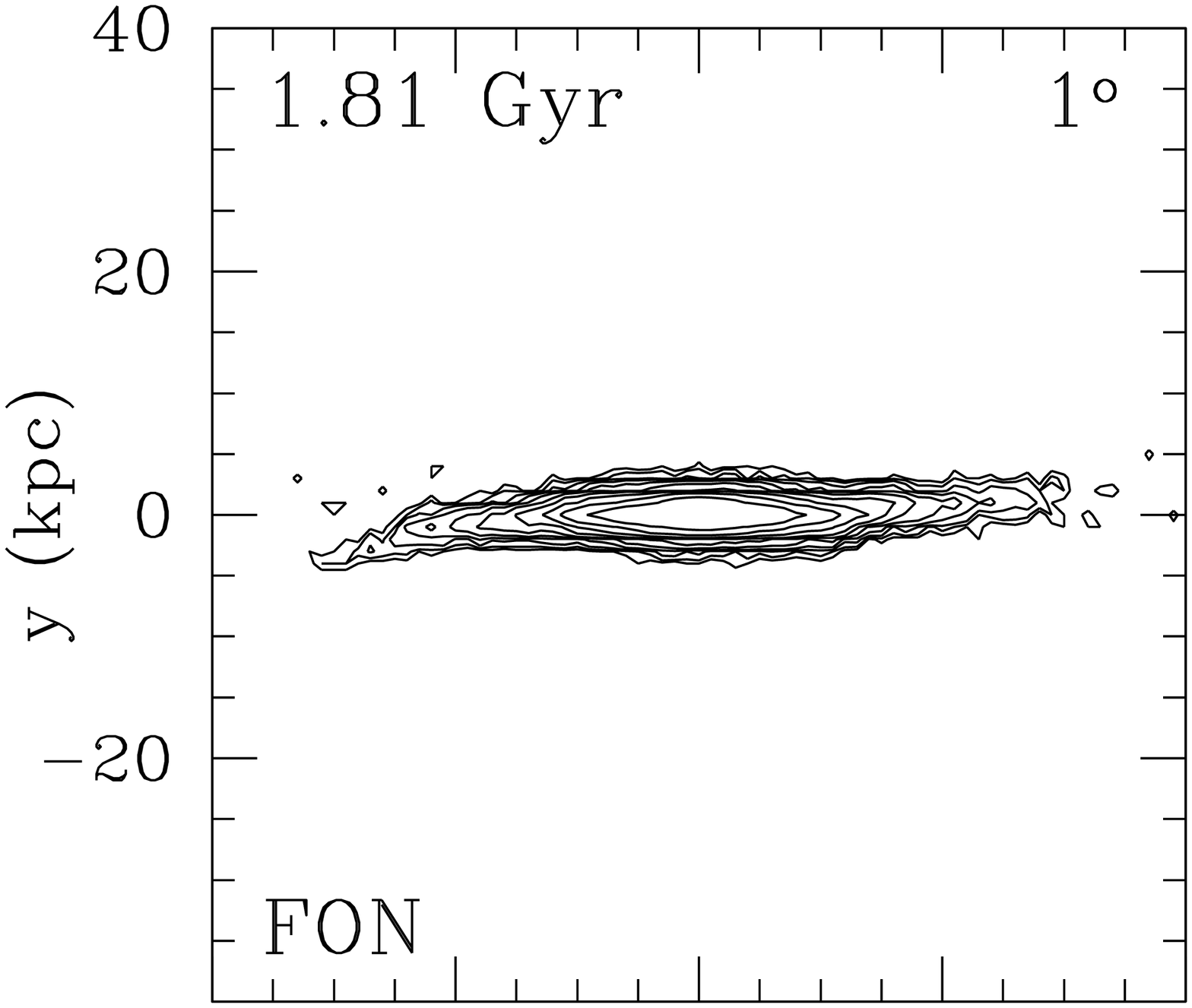}\hspace*{-0.87cm}
\includegraphics[width=34.9mm]{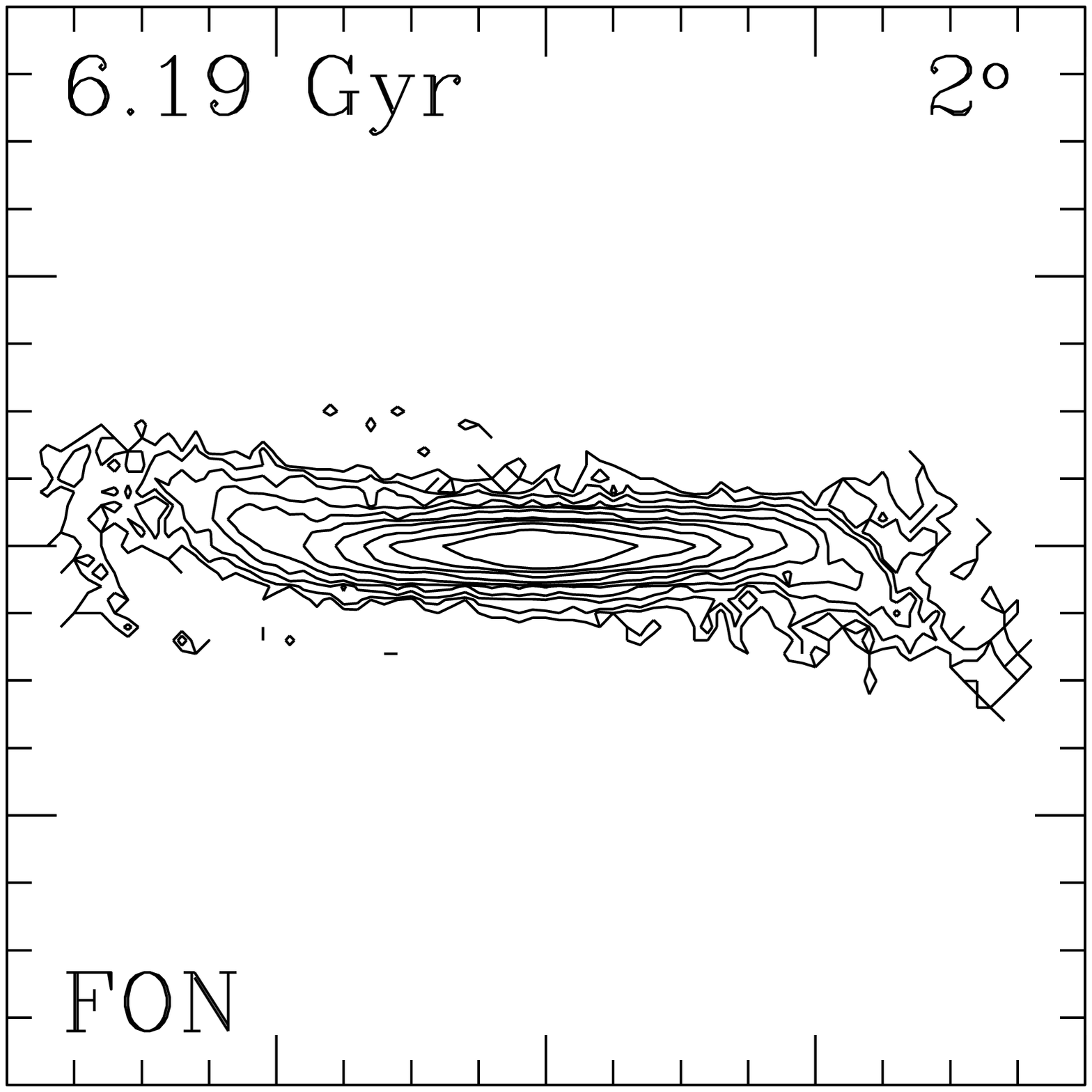}\hspace*{-0.87cm}
\includegraphics[width=34.9mm]{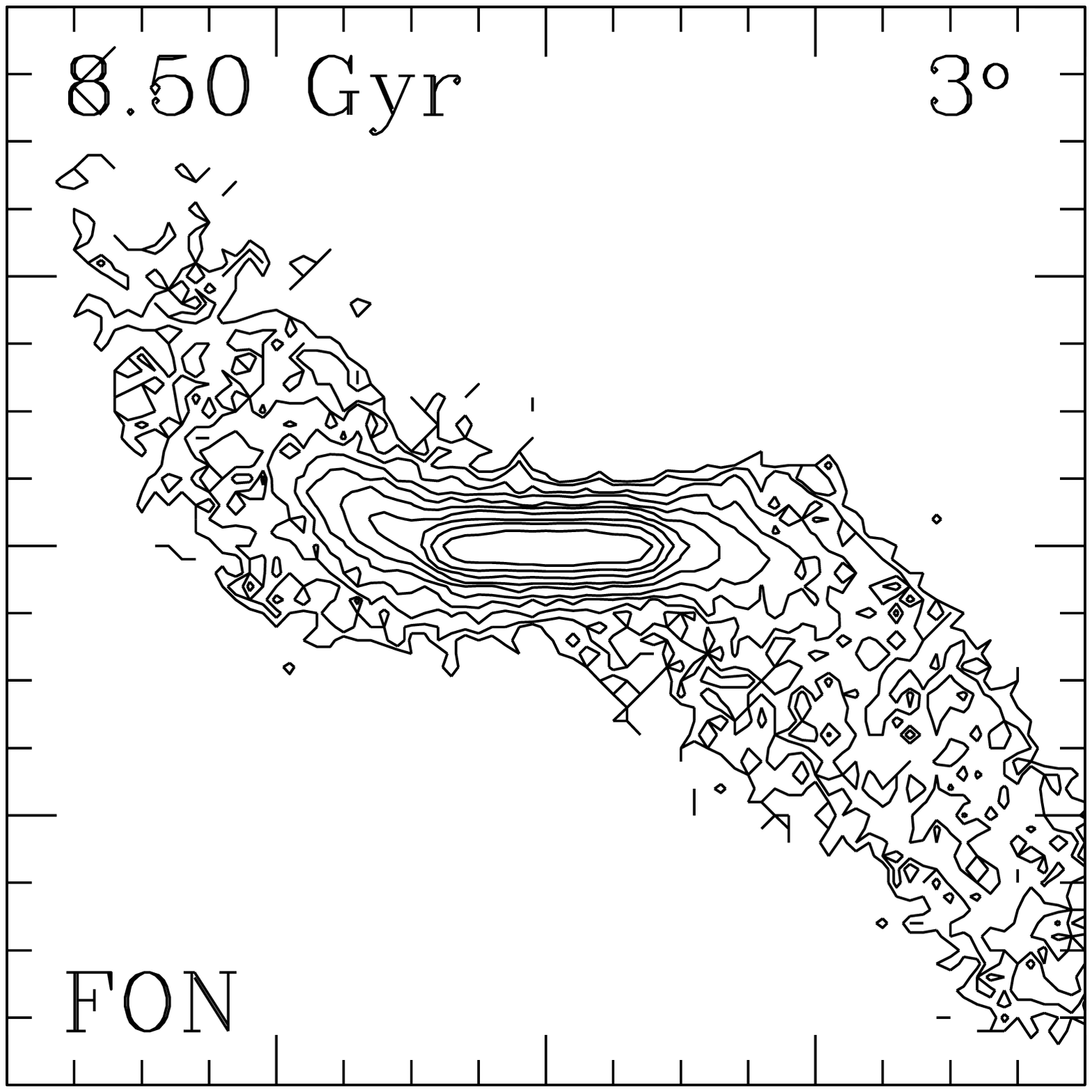}\hspace*{-0.87cm}
\includegraphics[width=34.9mm]{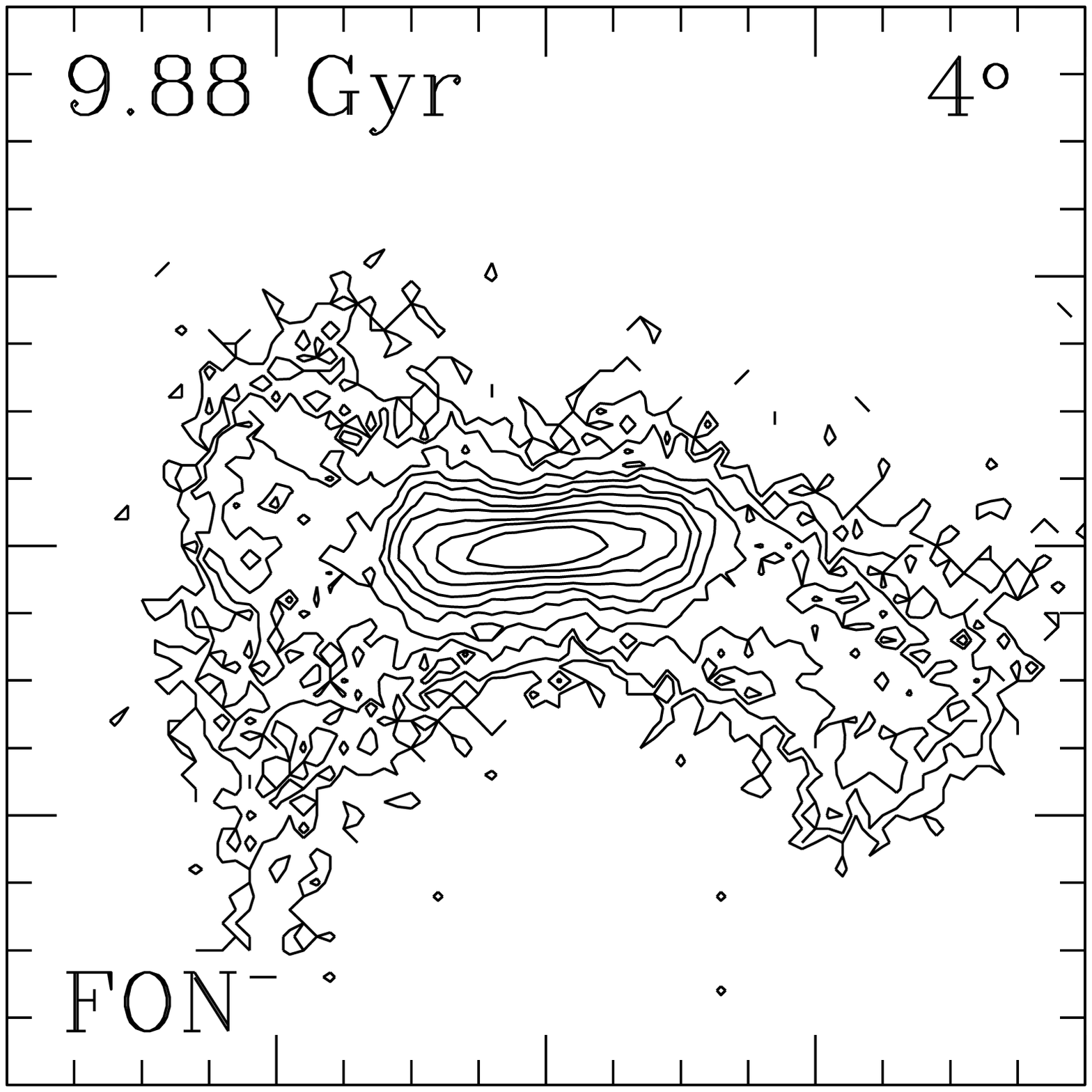}\hspace*{-0.87cm}
\includegraphics[width=34.9mm]{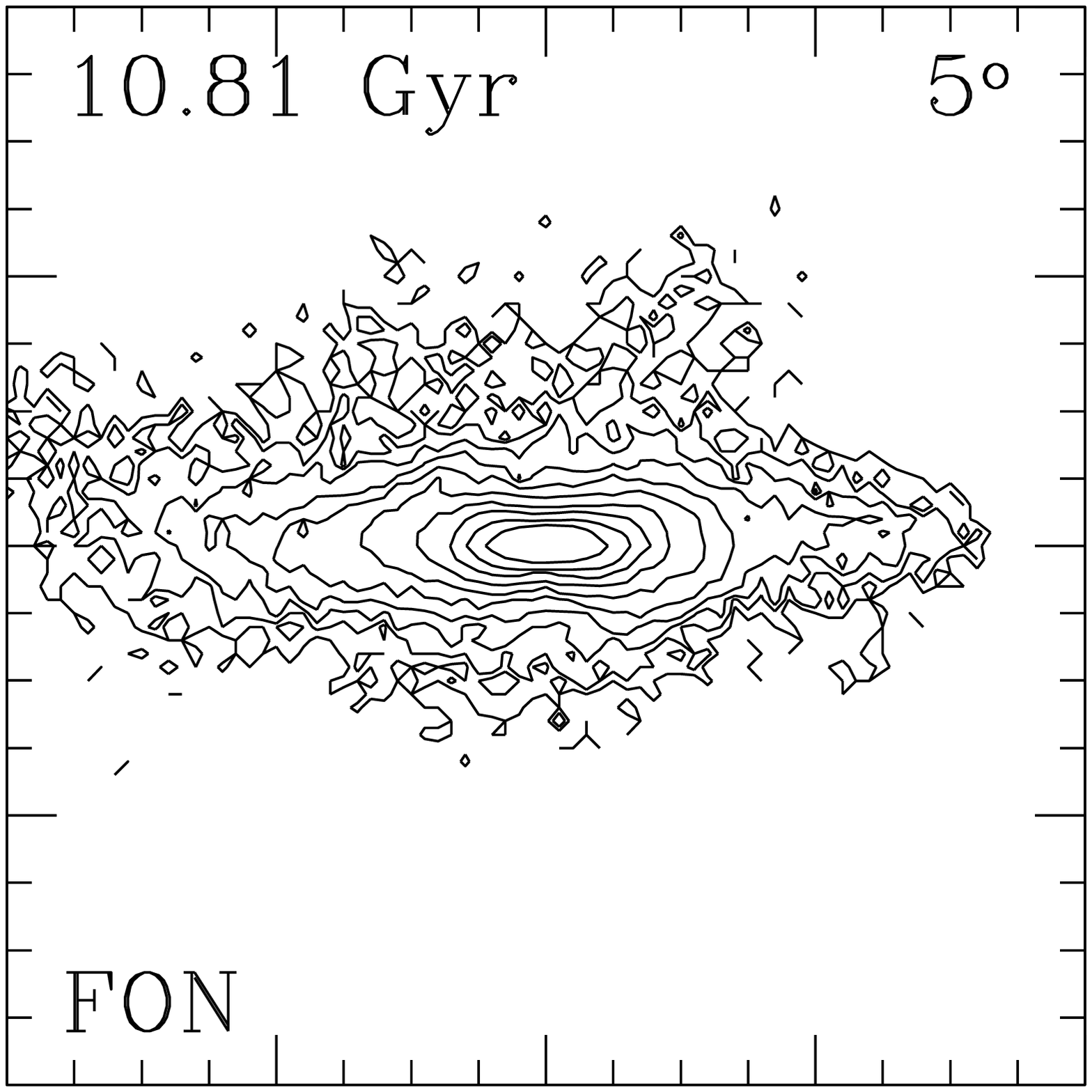}\hspace*{-0.87cm}
\includegraphics[width=34.9mm]{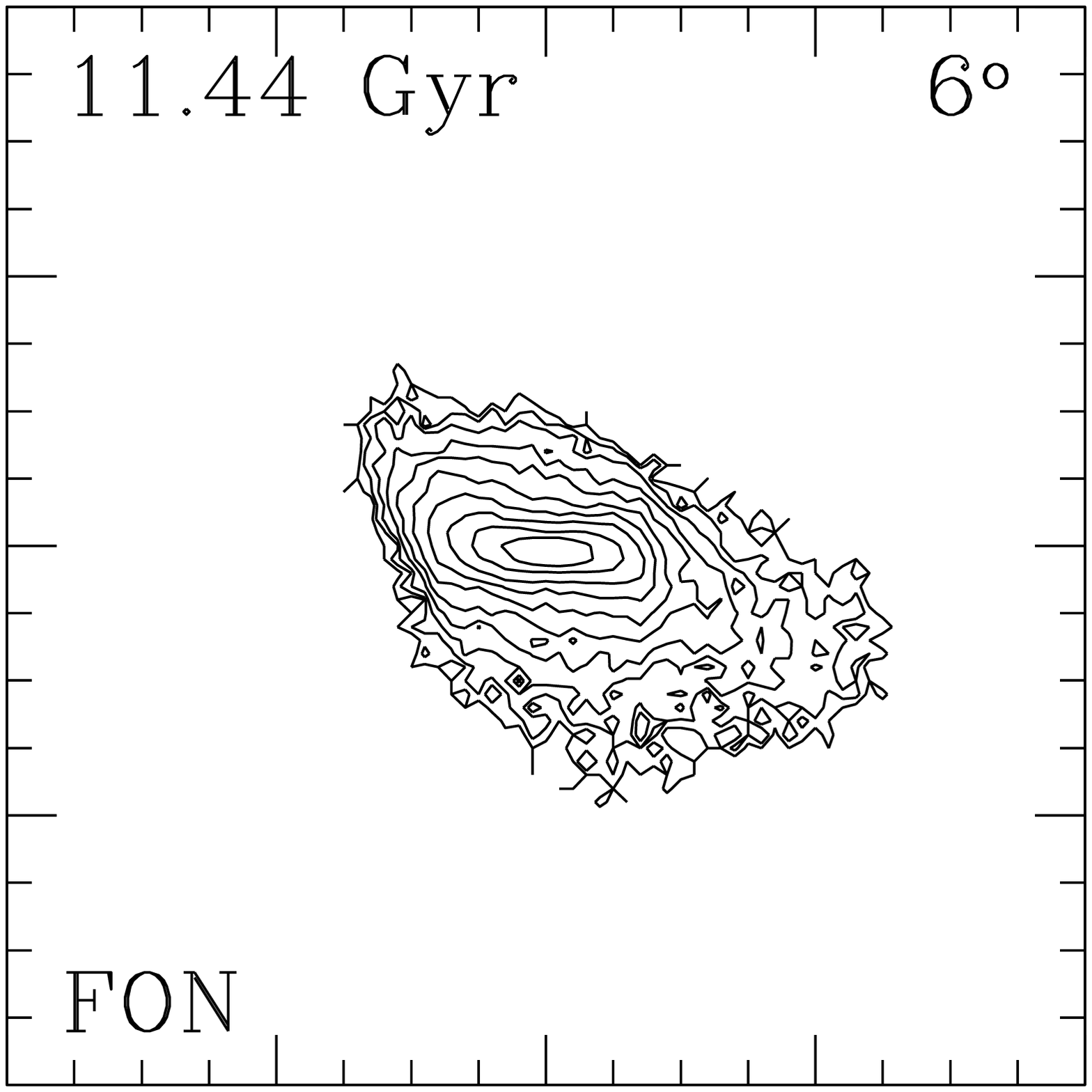}\vspace*{-0.81cm}\\
\includegraphics[width=34.9mm]{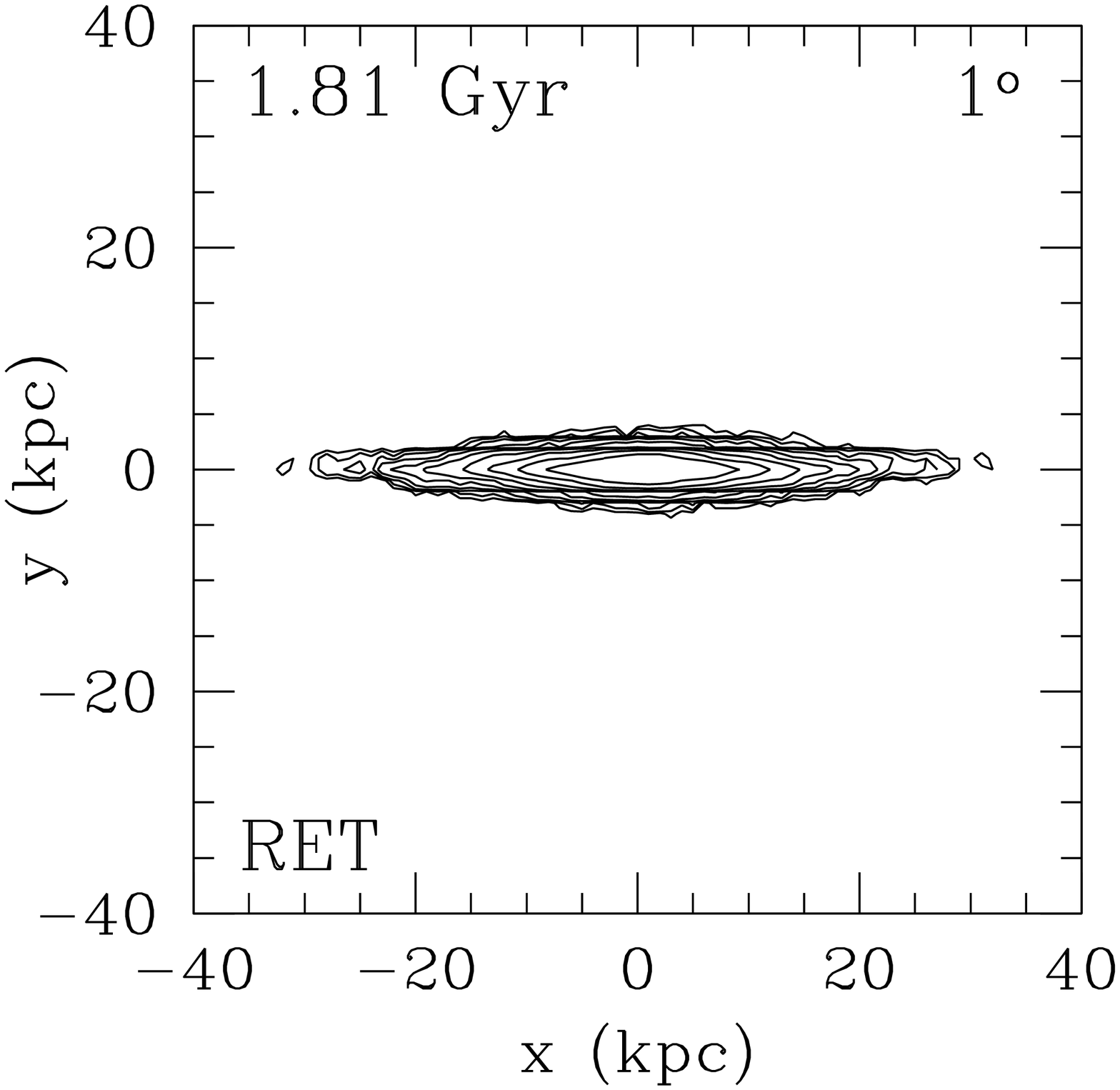}\hspace*{-0.87cm}
\includegraphics[width=34.9mm]{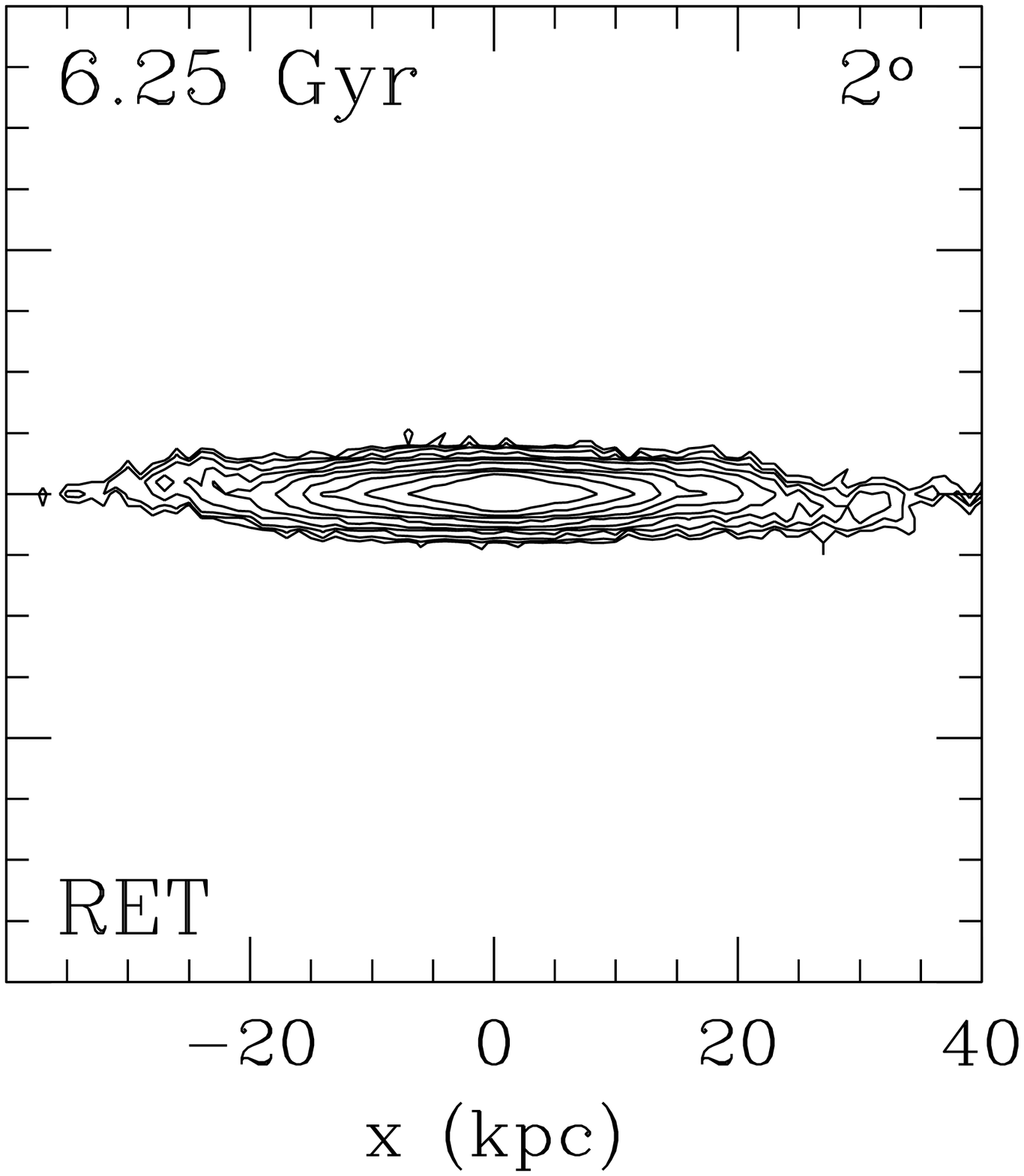}\hspace*{-0.87cm}
\includegraphics[width=34.9mm]{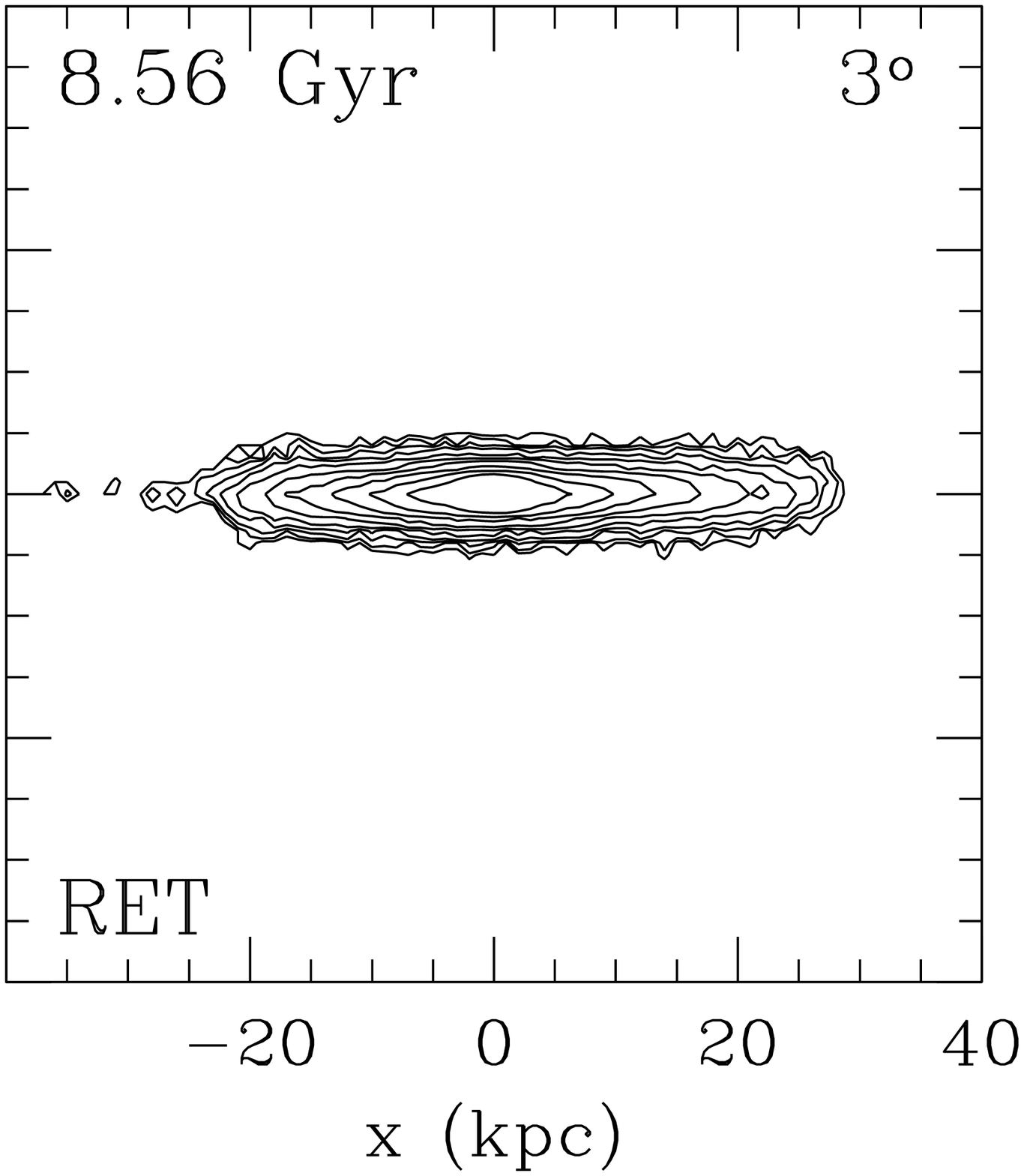}\hspace*{-0.87cm}
\includegraphics[width=34.9mm]{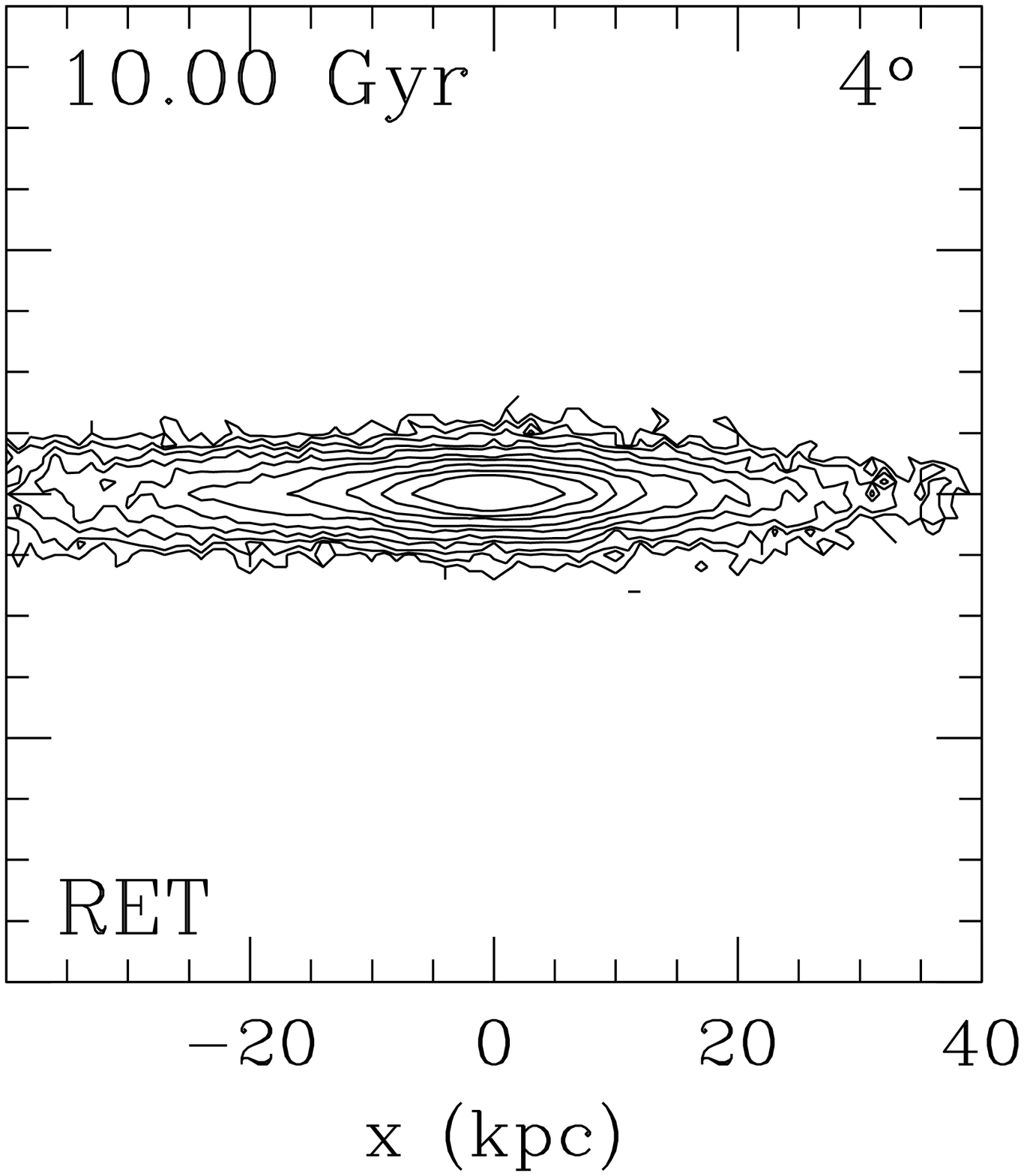}\hspace*{-0.87cm}
\includegraphics[width=34.9mm]{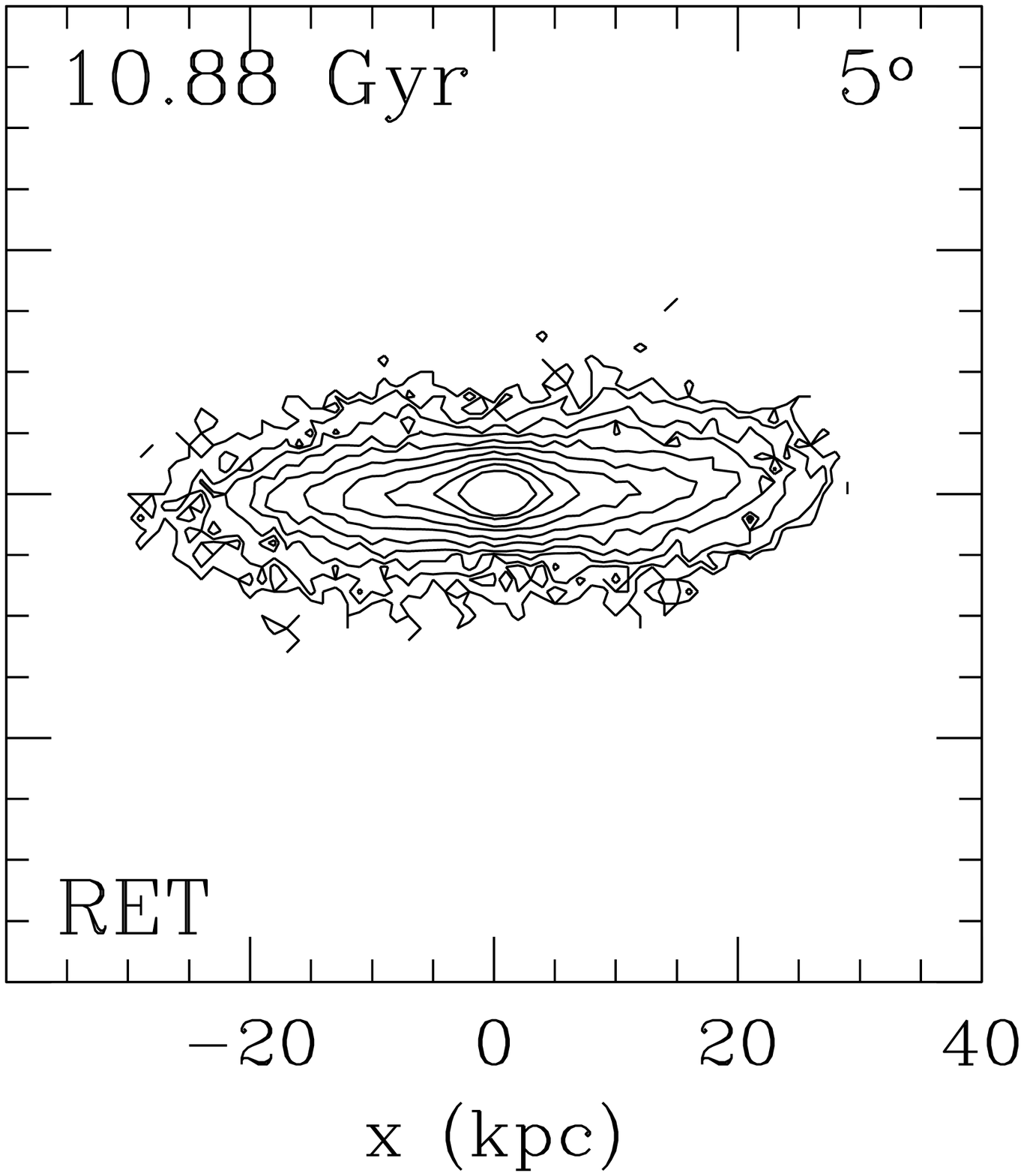}\hspace*{-0.87cm}
\includegraphics[width=34.9mm]{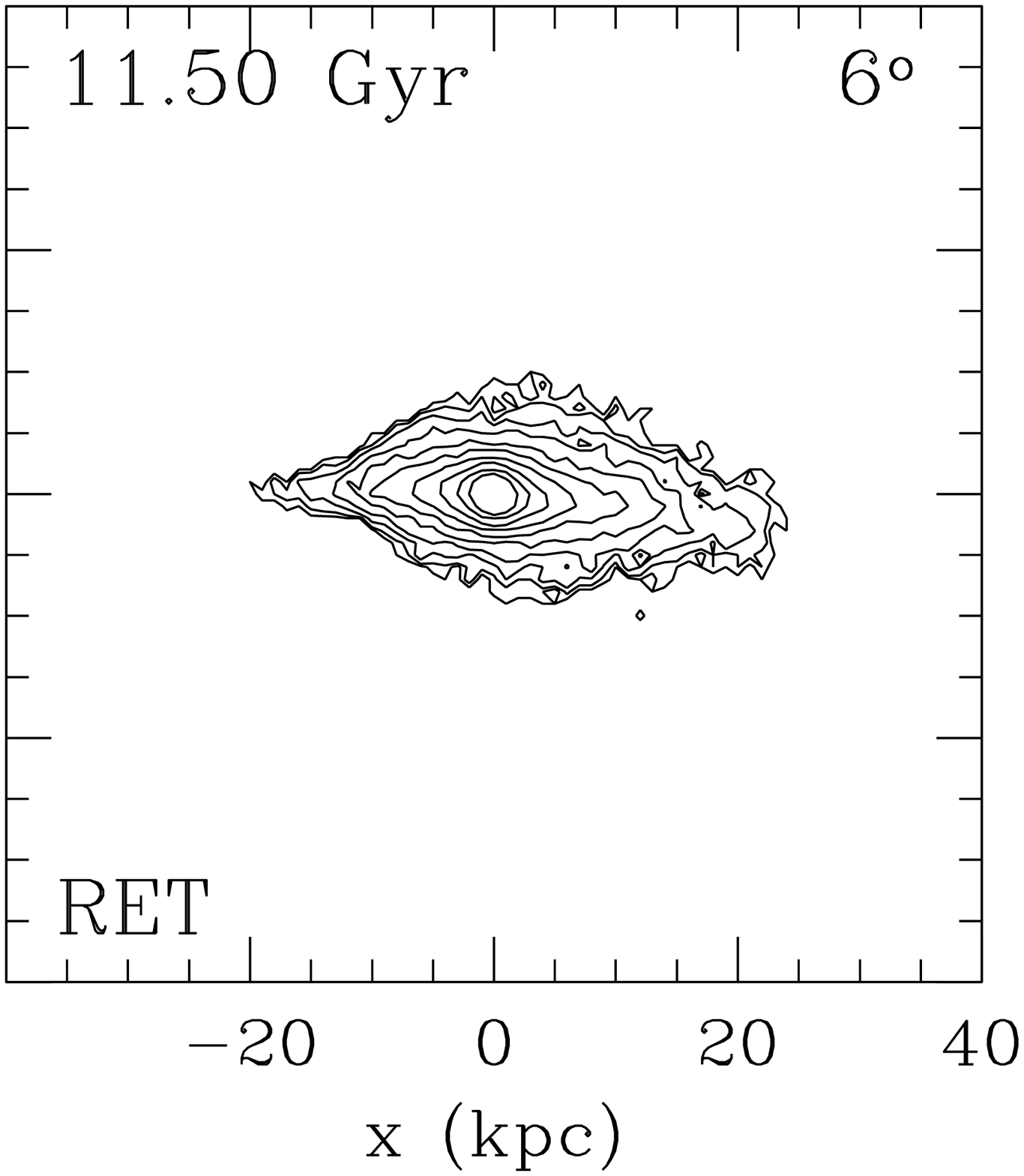}
\end{center}
\caption{Same as Fig.~\ref{contours-xy-z0} but in edge-on view.}
\label{contours-xz-z0}
\end{figure*}

\begin{figure*}
\begin{center}
\includegraphics[width=34.9mm]{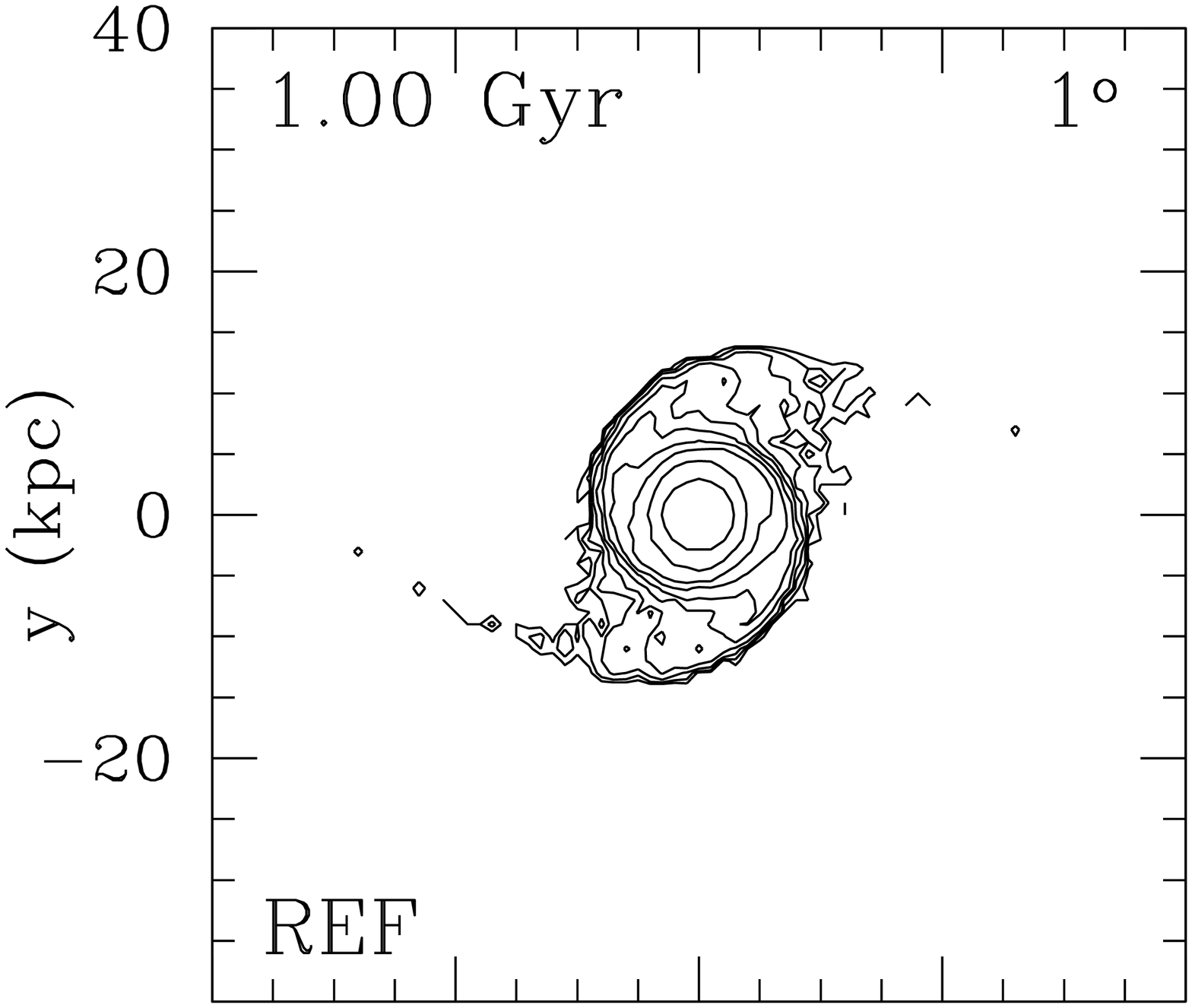}\hspace*{-0.87cm}
\includegraphics[width=34.9mm]{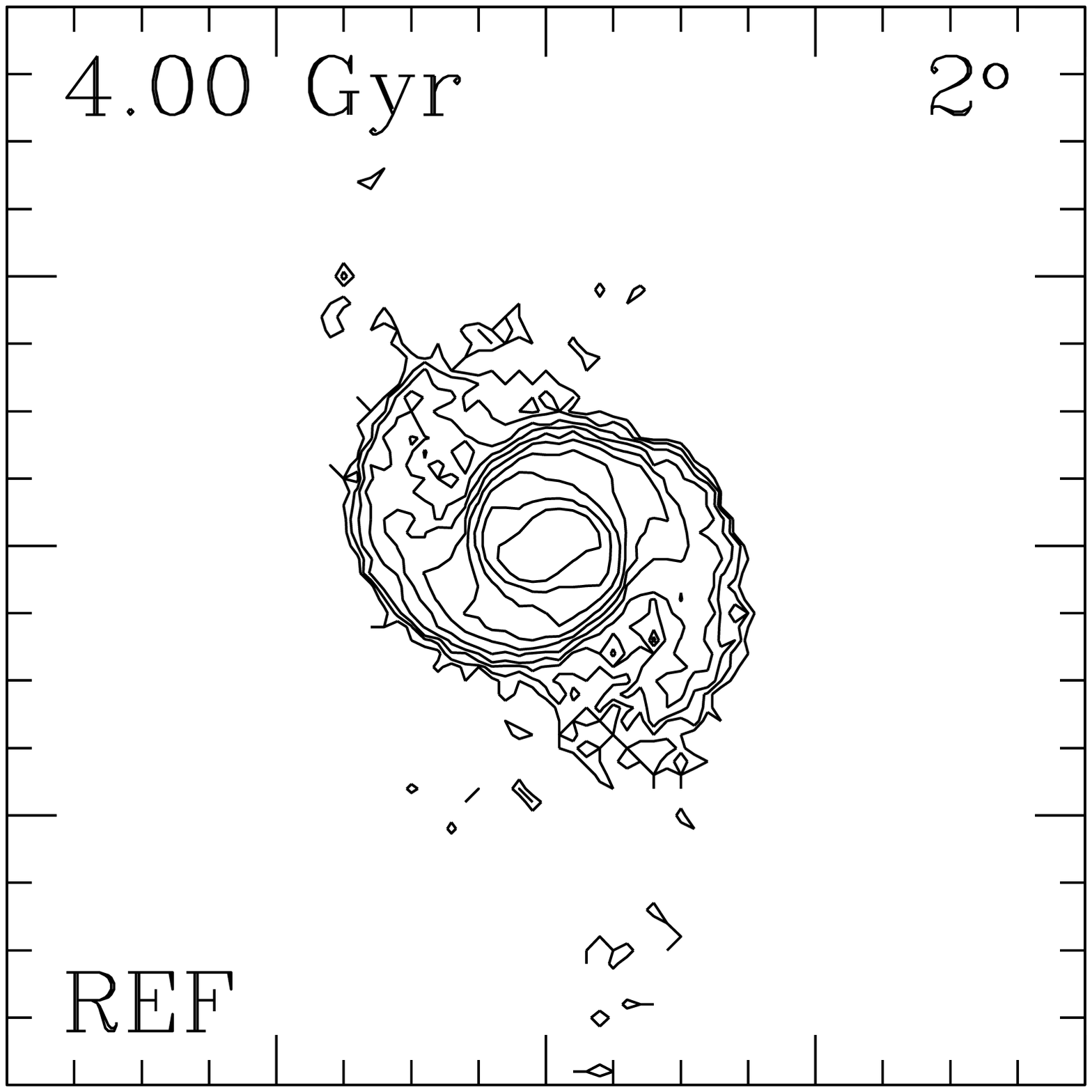}\hspace*{-0.87cm}
\includegraphics[width=34.9mm]{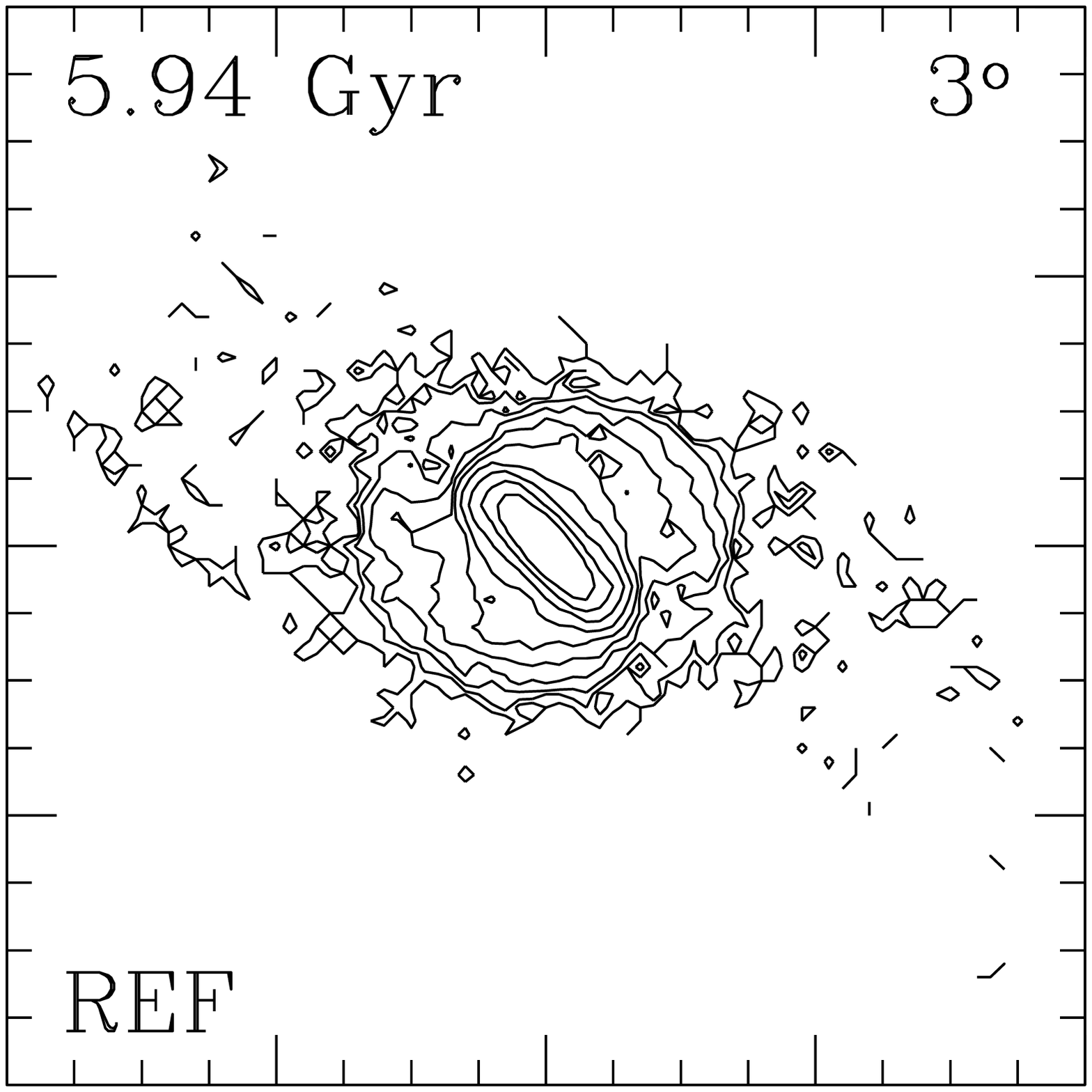}\hspace*{-0.87cm}
\includegraphics[width=34.9mm]{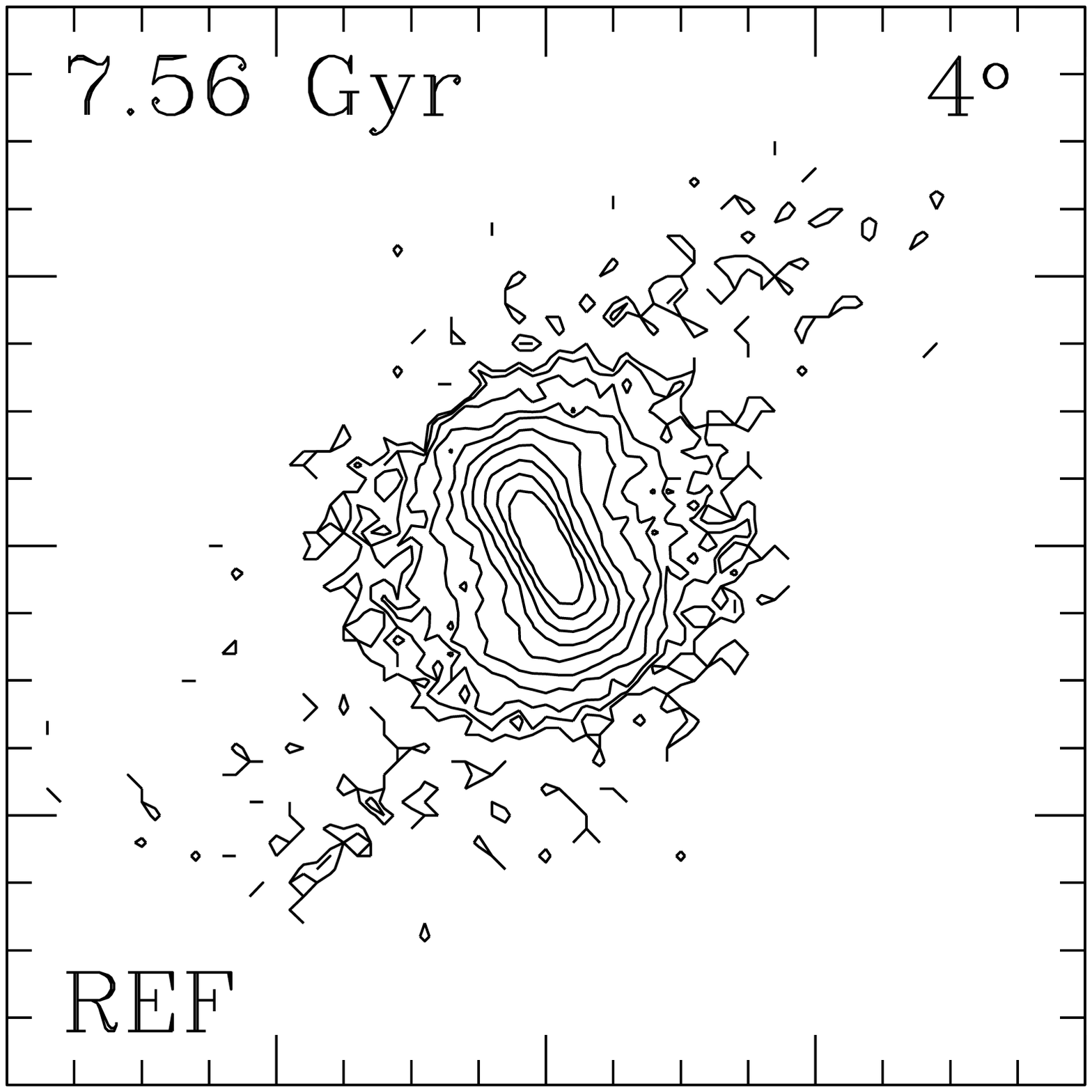}\hspace*{-0.87cm}
\includegraphics[width=34.9mm]{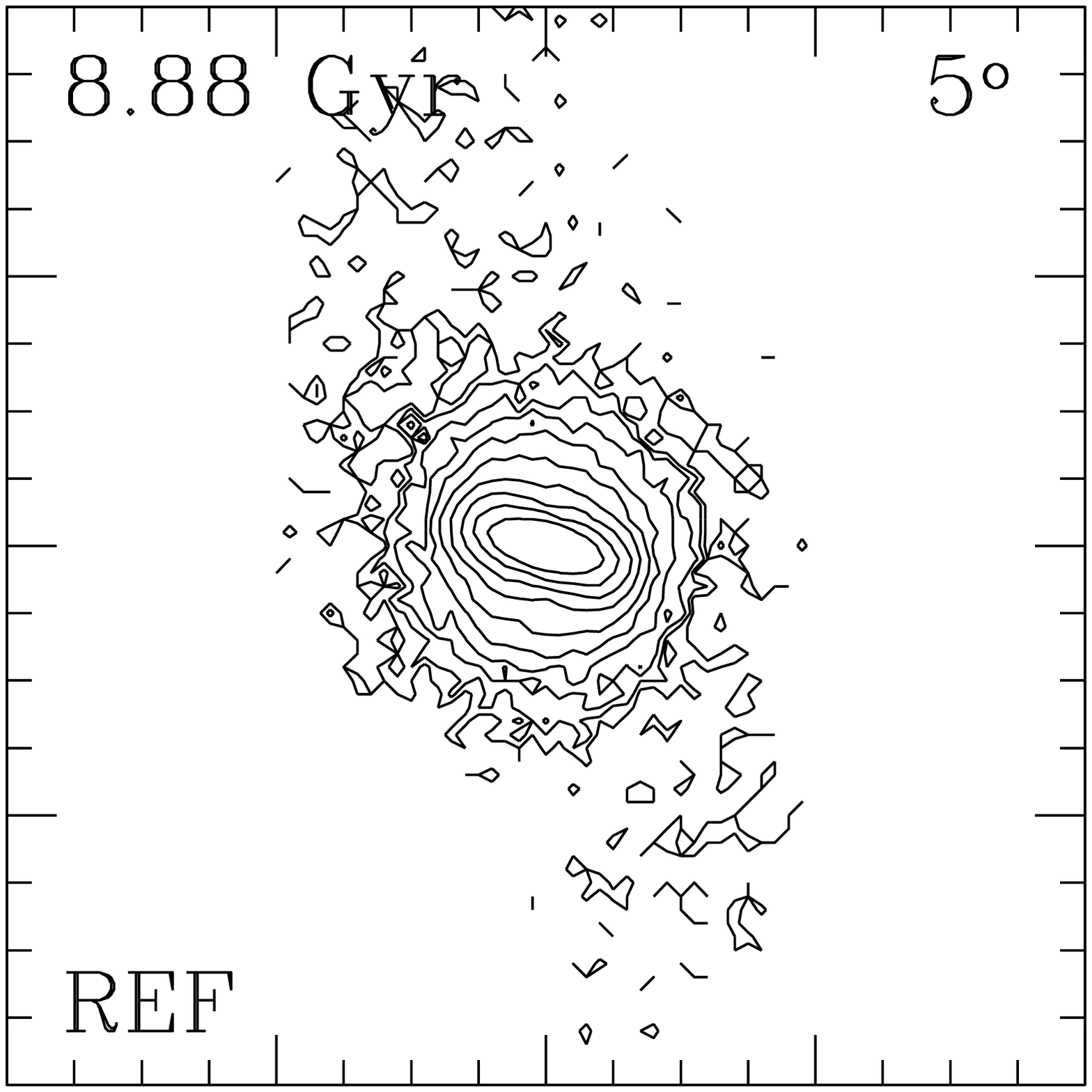}\hspace*{-0.87cm}
\includegraphics[width=34.9mm]{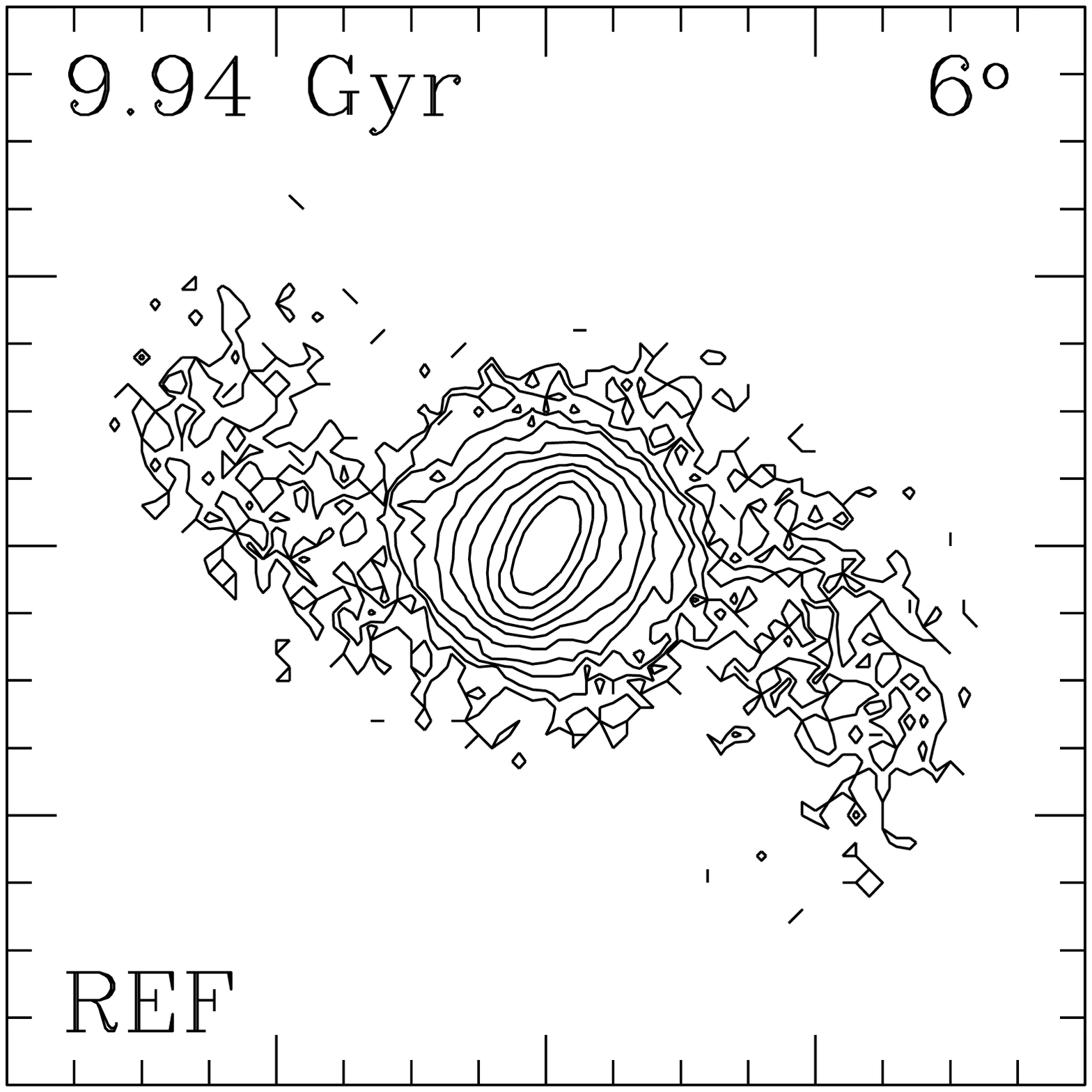}\vspace*{-0.81cm}\\
\includegraphics[width=34.9mm]{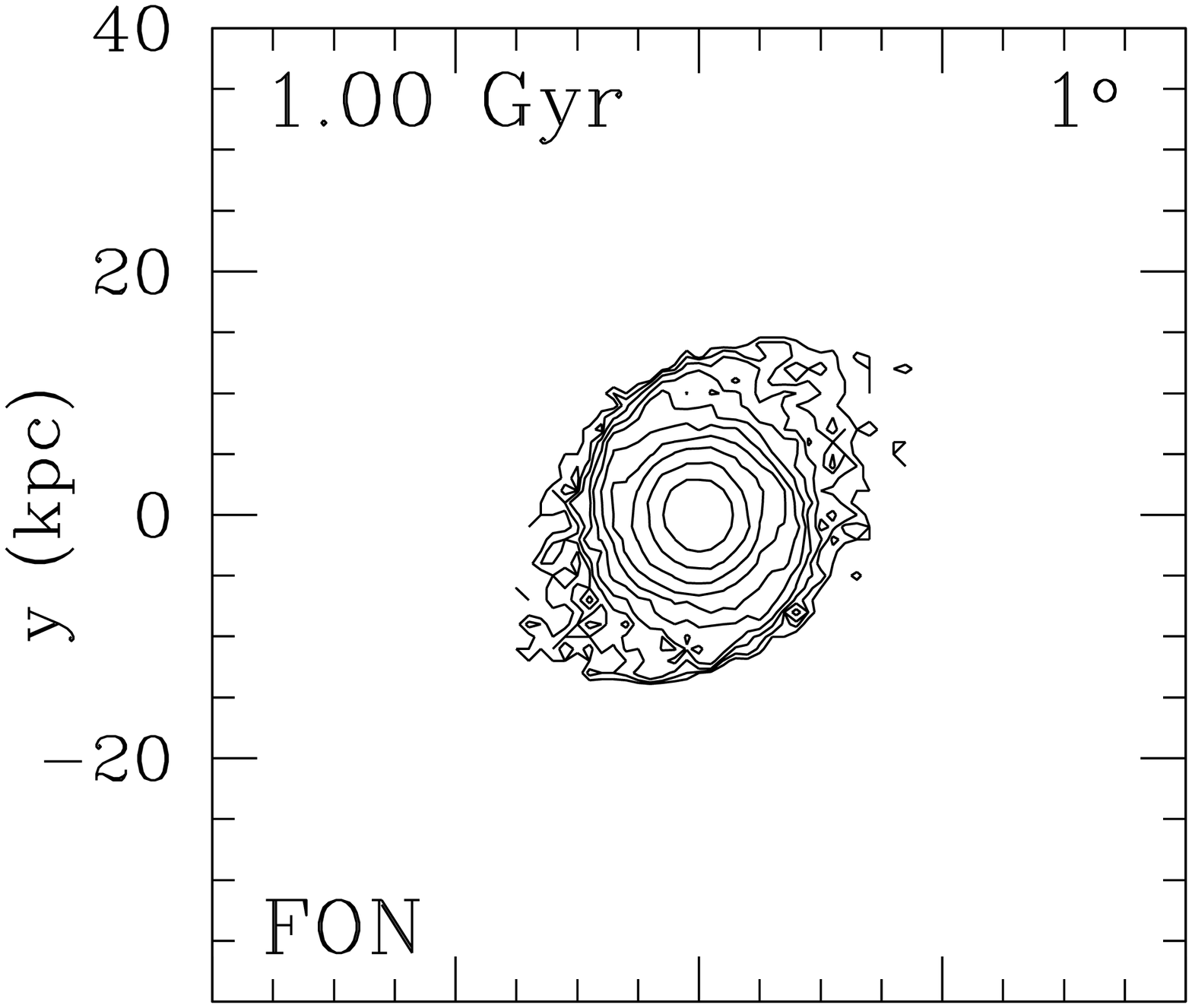}\hspace*{-0.87cm}
\includegraphics[width=34.9mm]{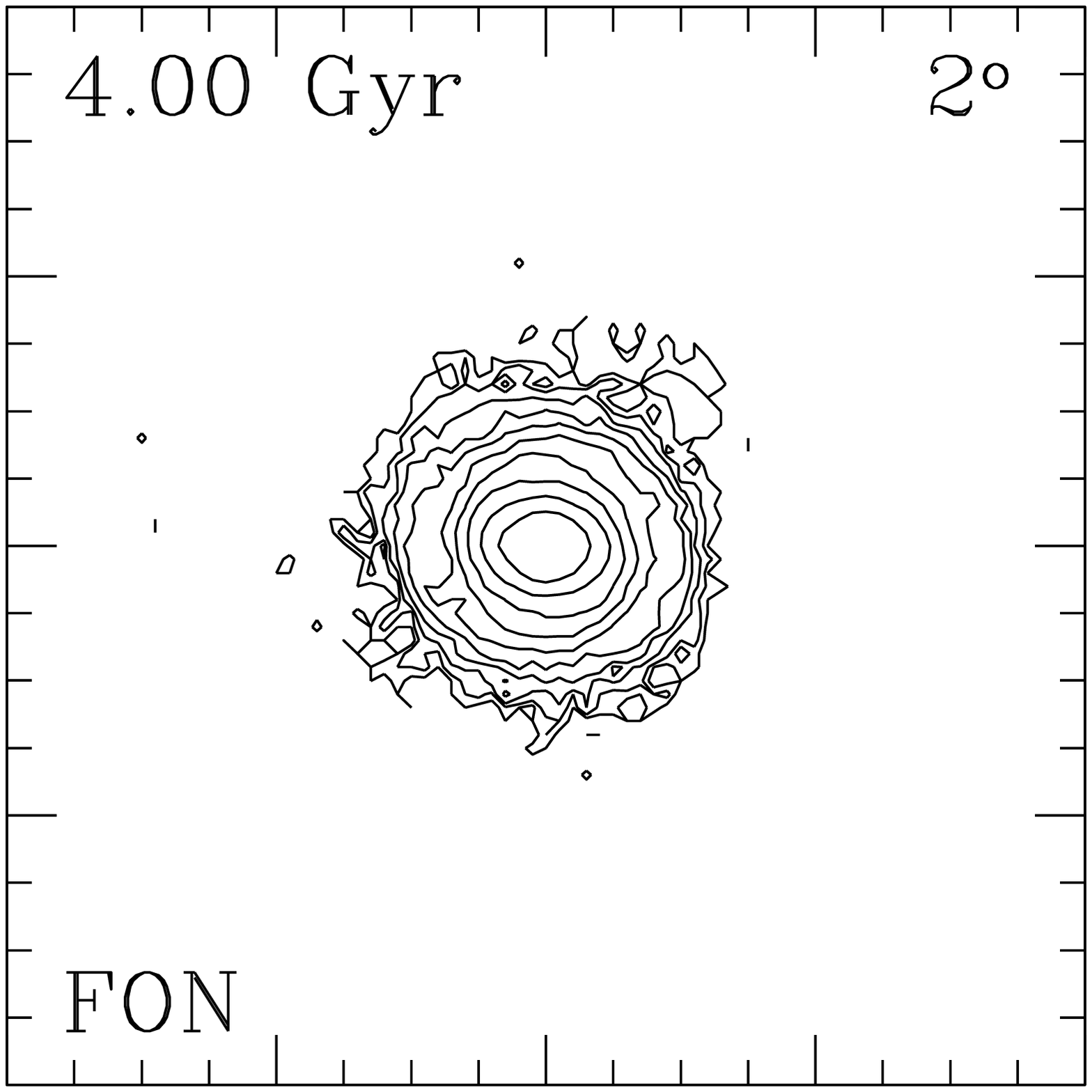}\hspace*{-0.87cm}
\includegraphics[width=34.9mm]{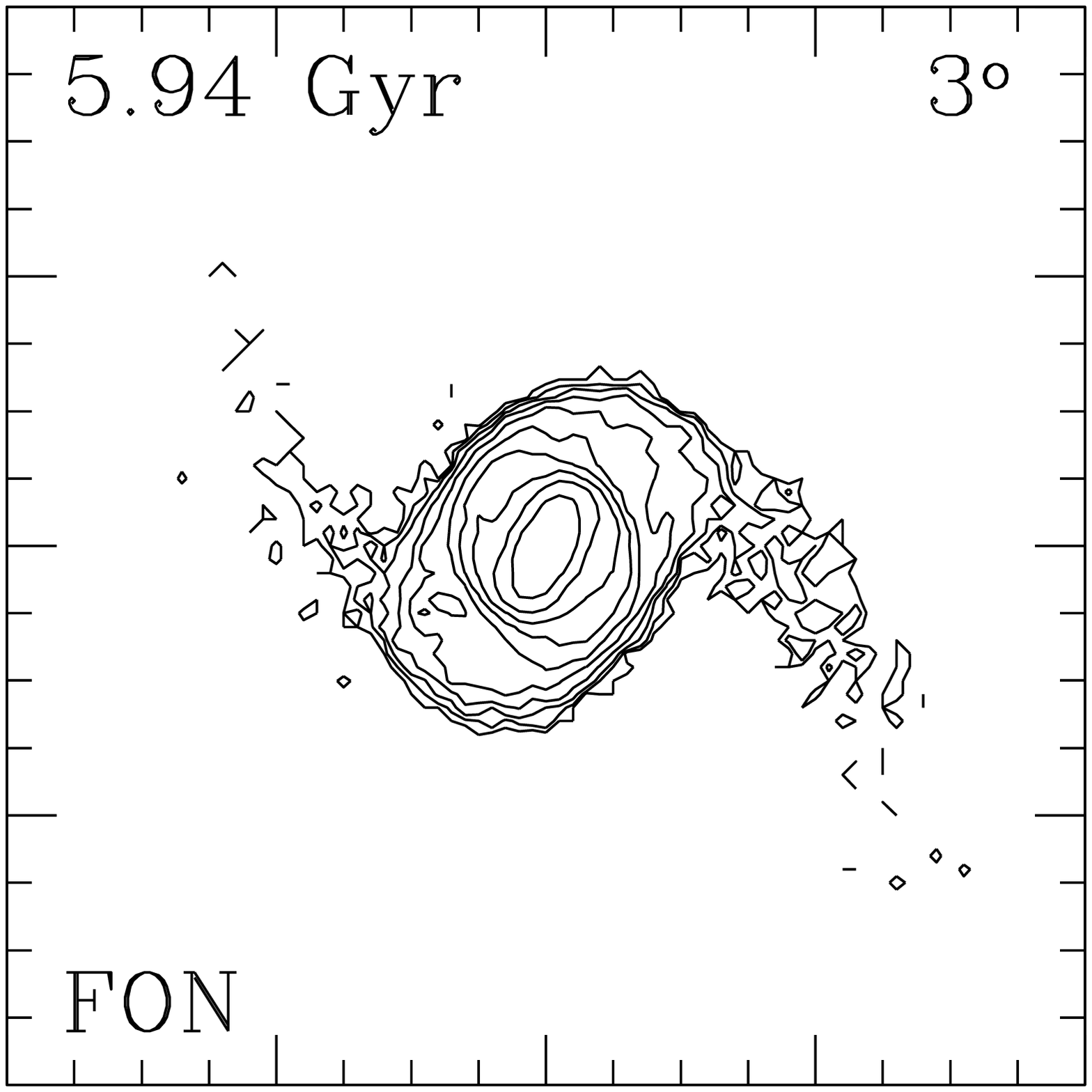}\hspace*{-0.87cm}
\includegraphics[width=34.9mm]{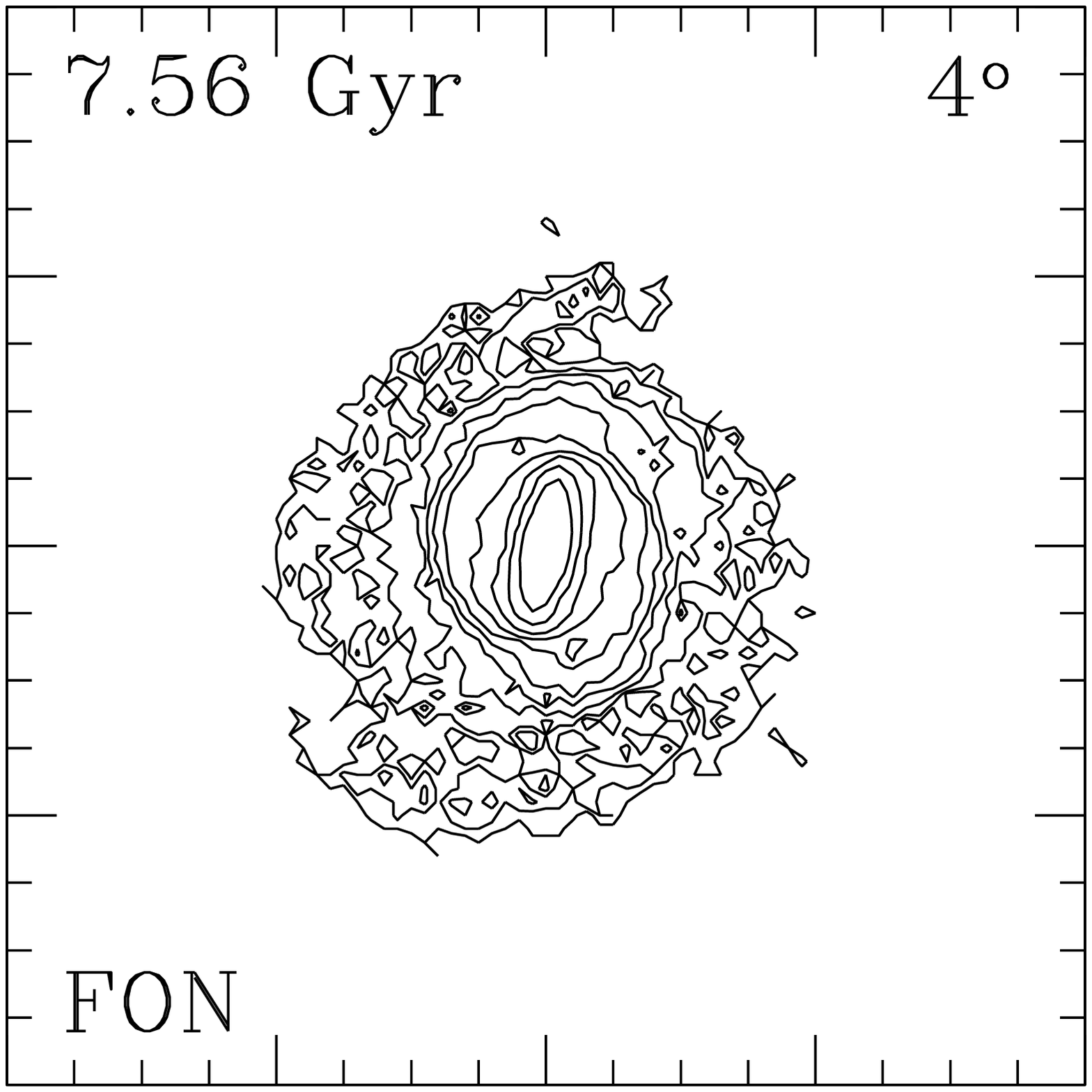}\hspace*{-0.87cm}
\includegraphics[width=34.9mm]{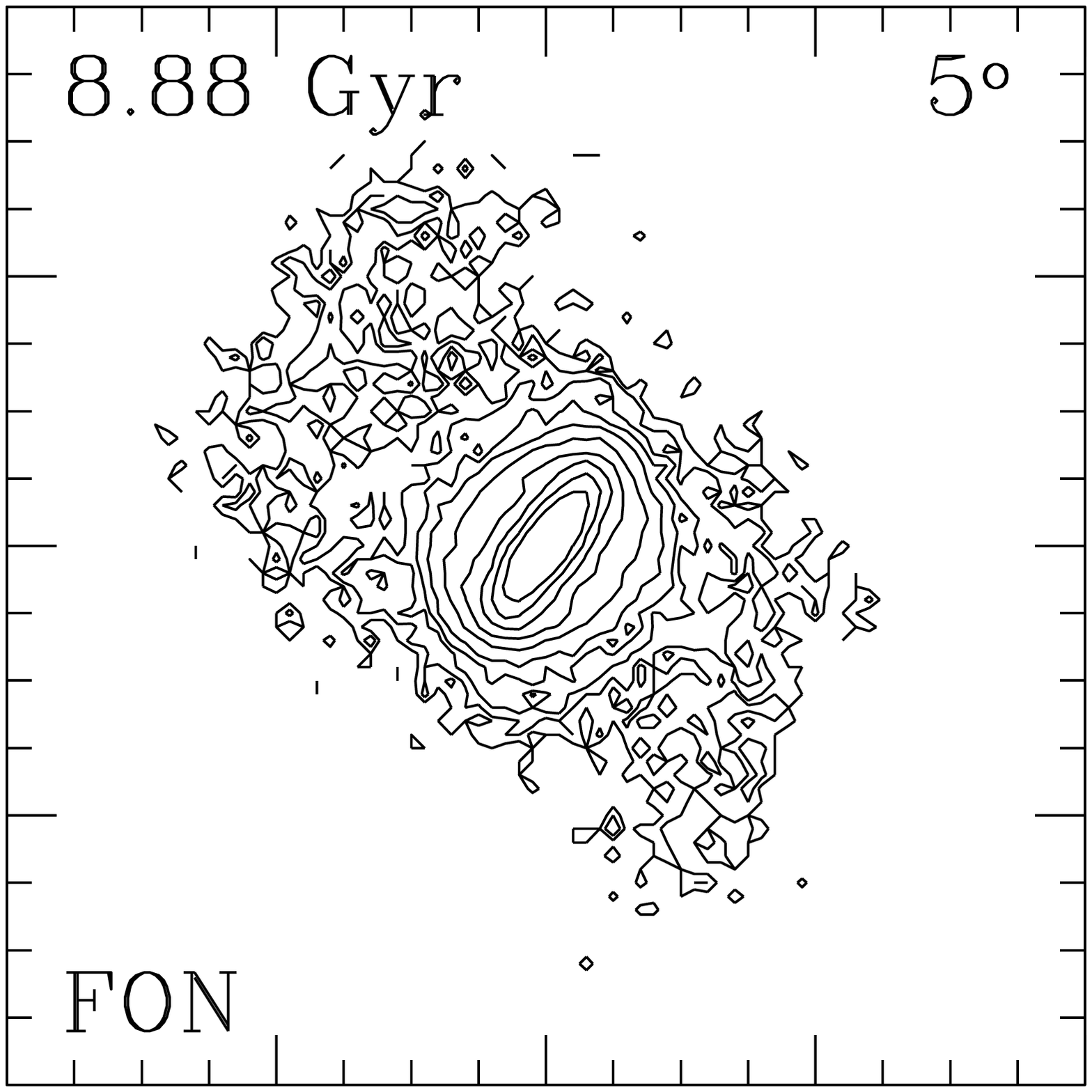}\hspace*{-0.87cm}
\includegraphics[width=34.9mm]{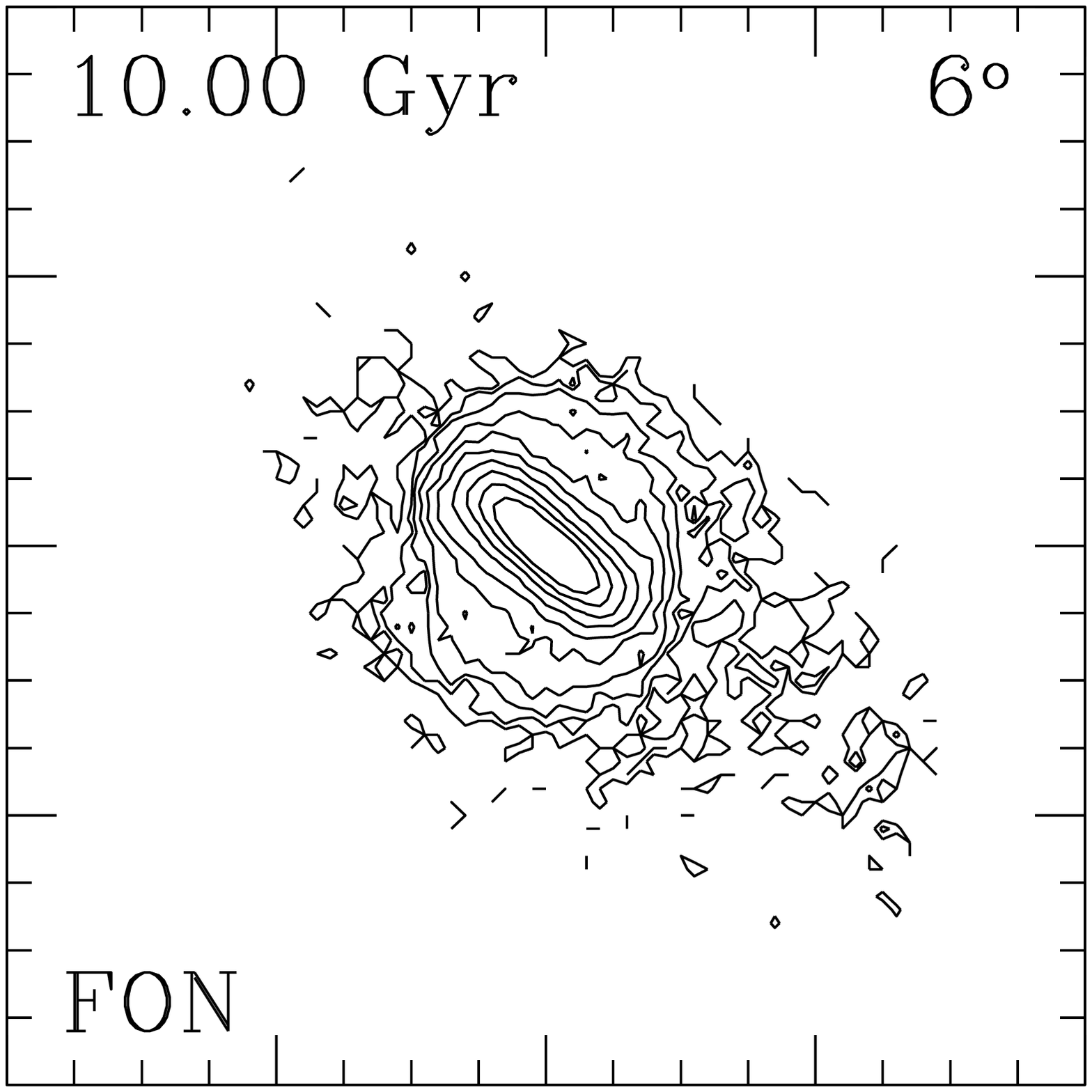}\vspace*{-0.81cm}\\
\includegraphics[width=34.9mm]{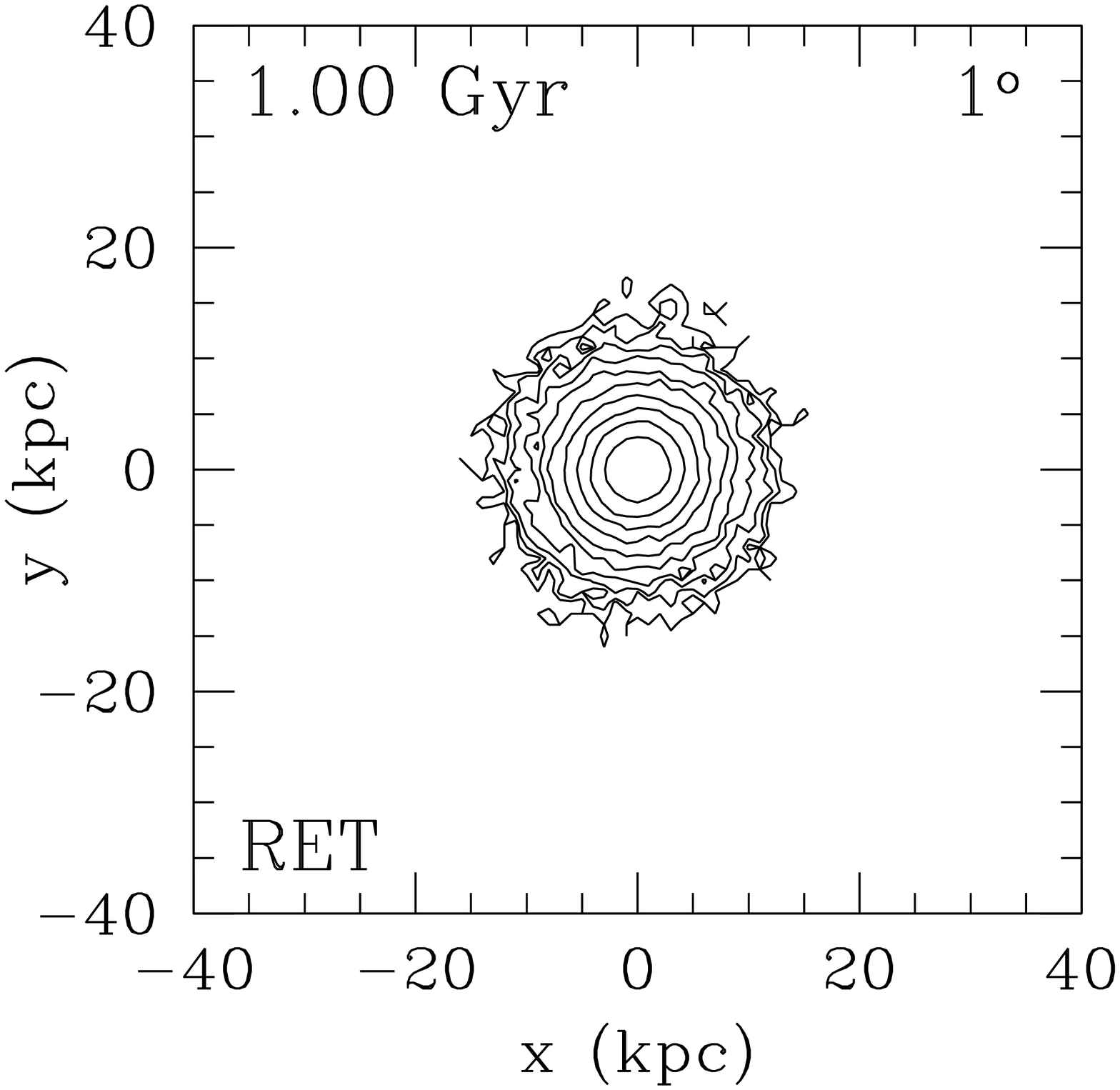}\hspace*{-0.87cm}
\includegraphics[width=34.9mm]{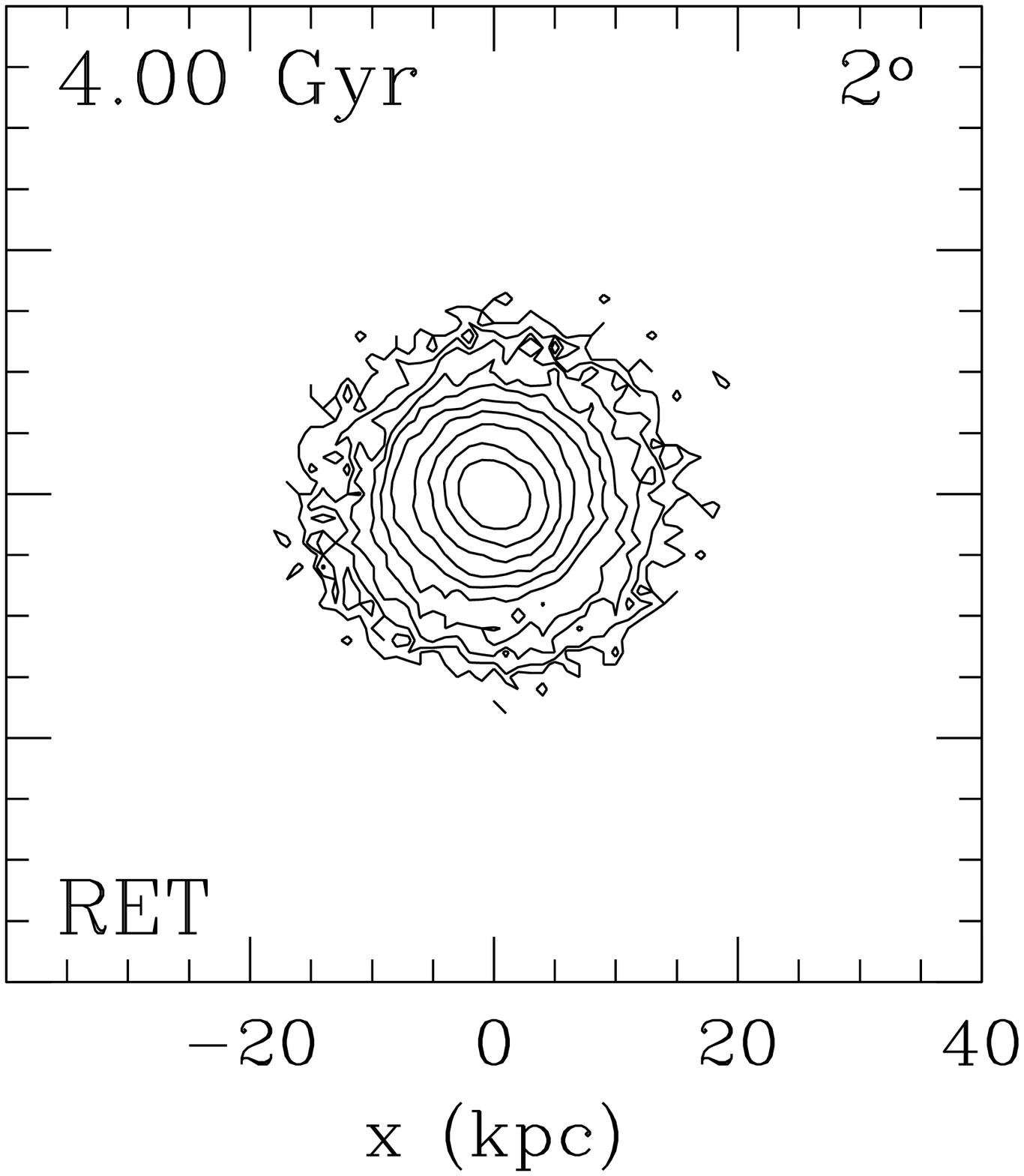}\hspace*{-0.87cm}
\includegraphics[width=34.9mm]{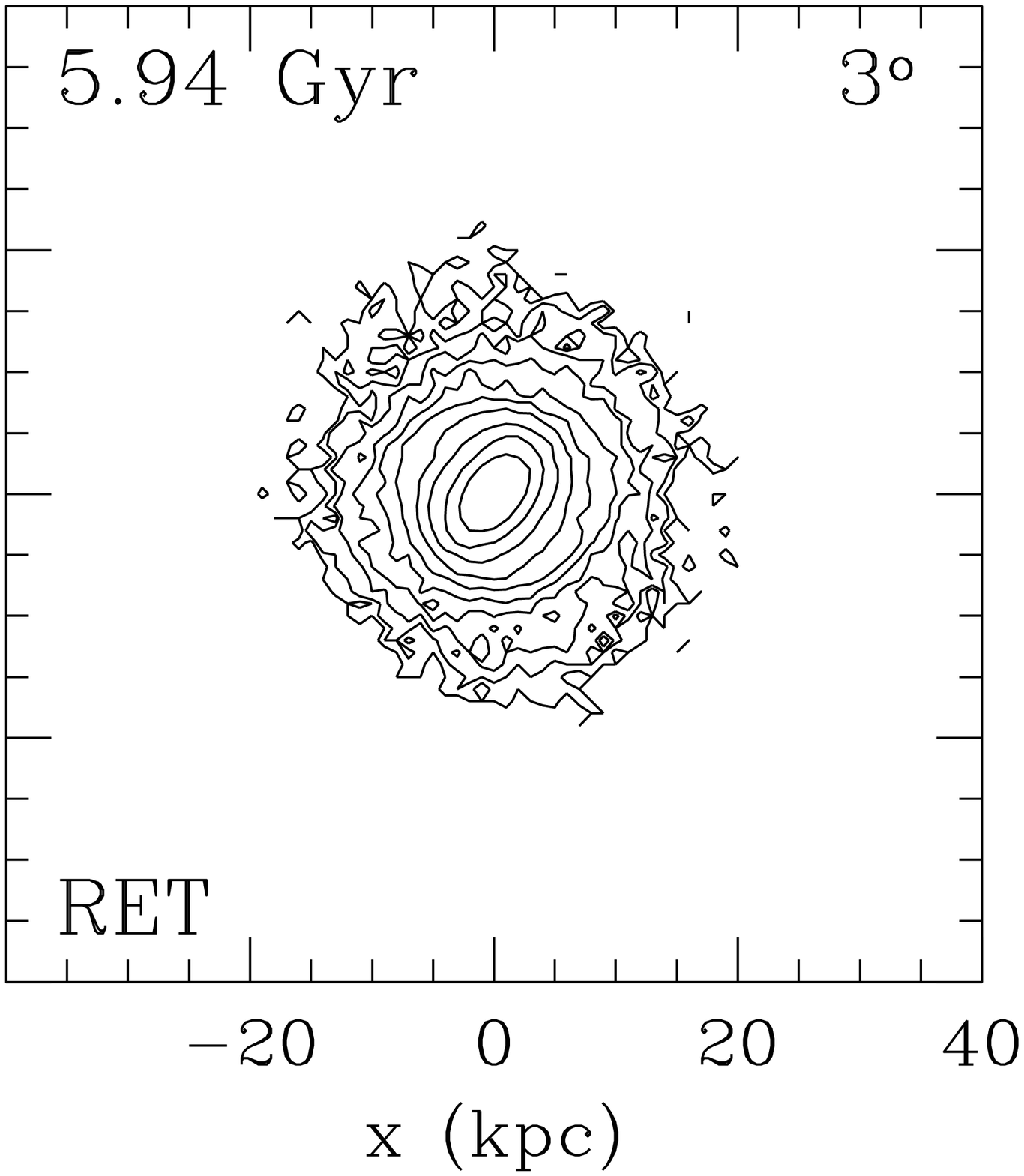}\hspace*{-0.87cm}
\includegraphics[width=34.9mm]{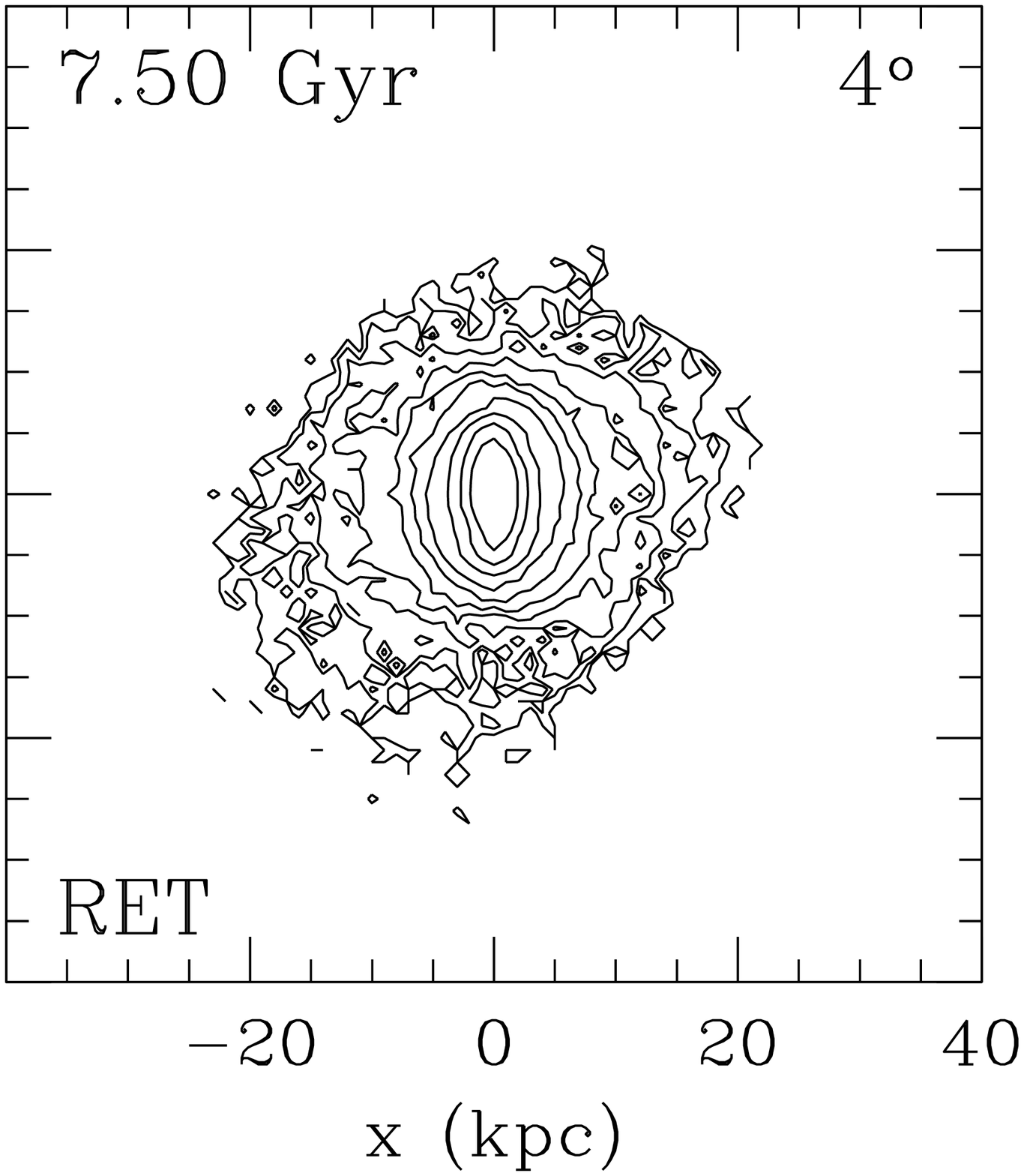}\hspace*{-0.87cm}
\includegraphics[width=34.9mm]{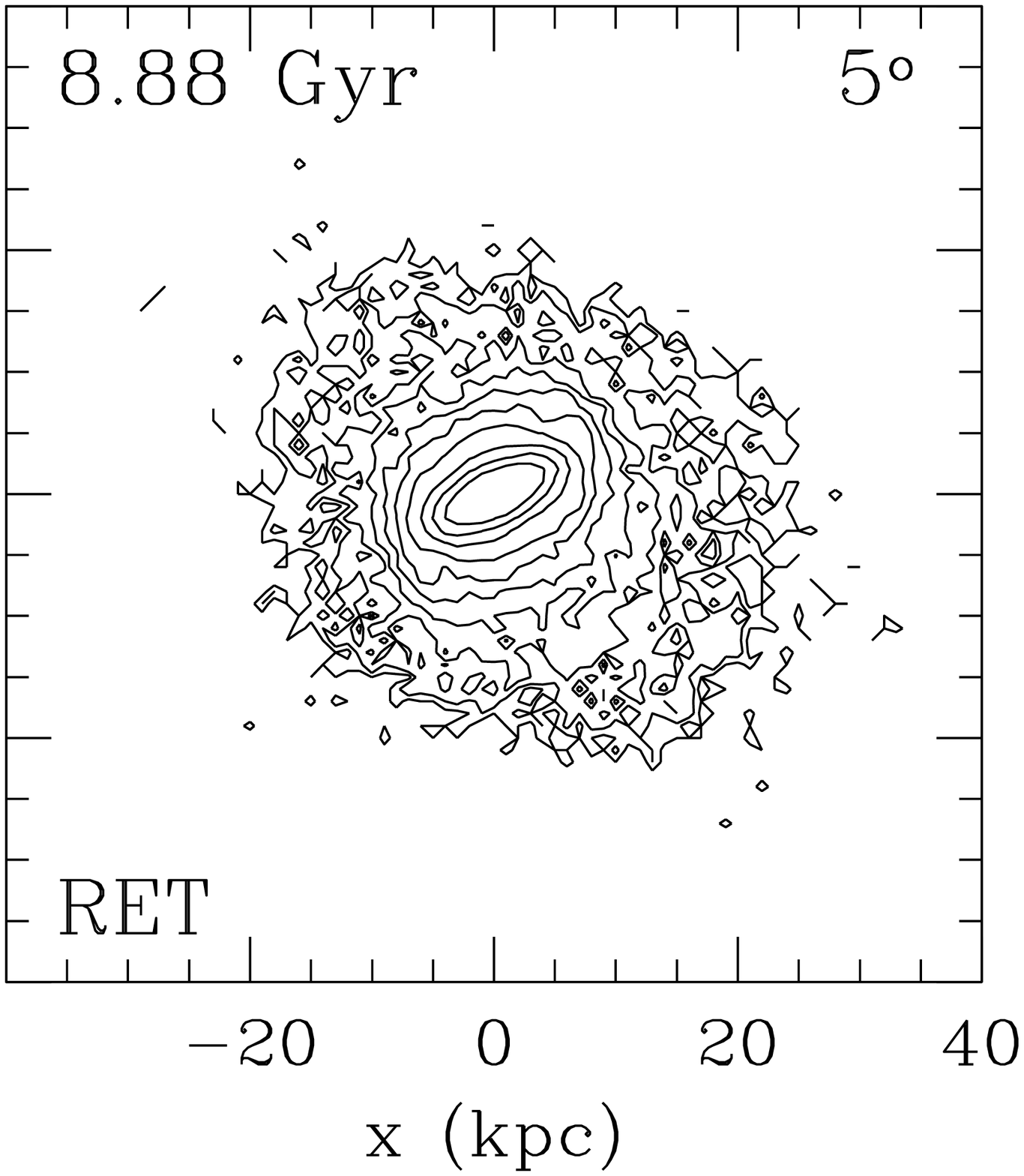}\hspace*{-0.87cm}
\includegraphics[width=34.9mm]{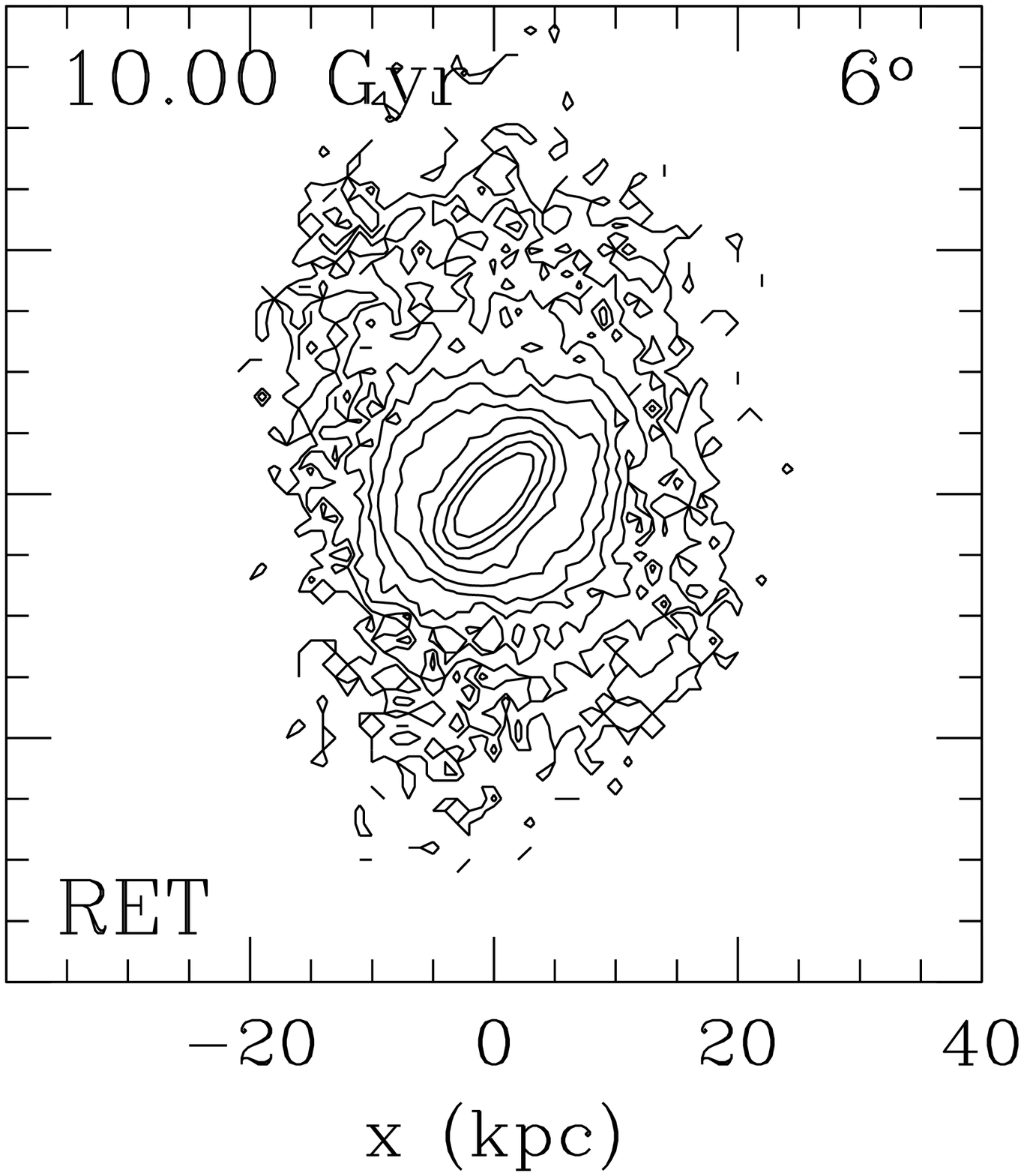}
\end{center}
\caption{Face-on views of the morphological evolution of disc galaxies for 
our REF (reference), FON (face-on infall) and RET (retrograde infall) 
experiments at ``$z$=1'', after each of their first six pericentric passages. 
Number density contours are drawn at the same levels to ease the comparison, 
and only stars that remain bound to the disc, at a given time, are considered.}
\label{contours-xy-z1}
\end{figure*}

\begin{figure*}
\begin{center}
\includegraphics[width=34.9mm]{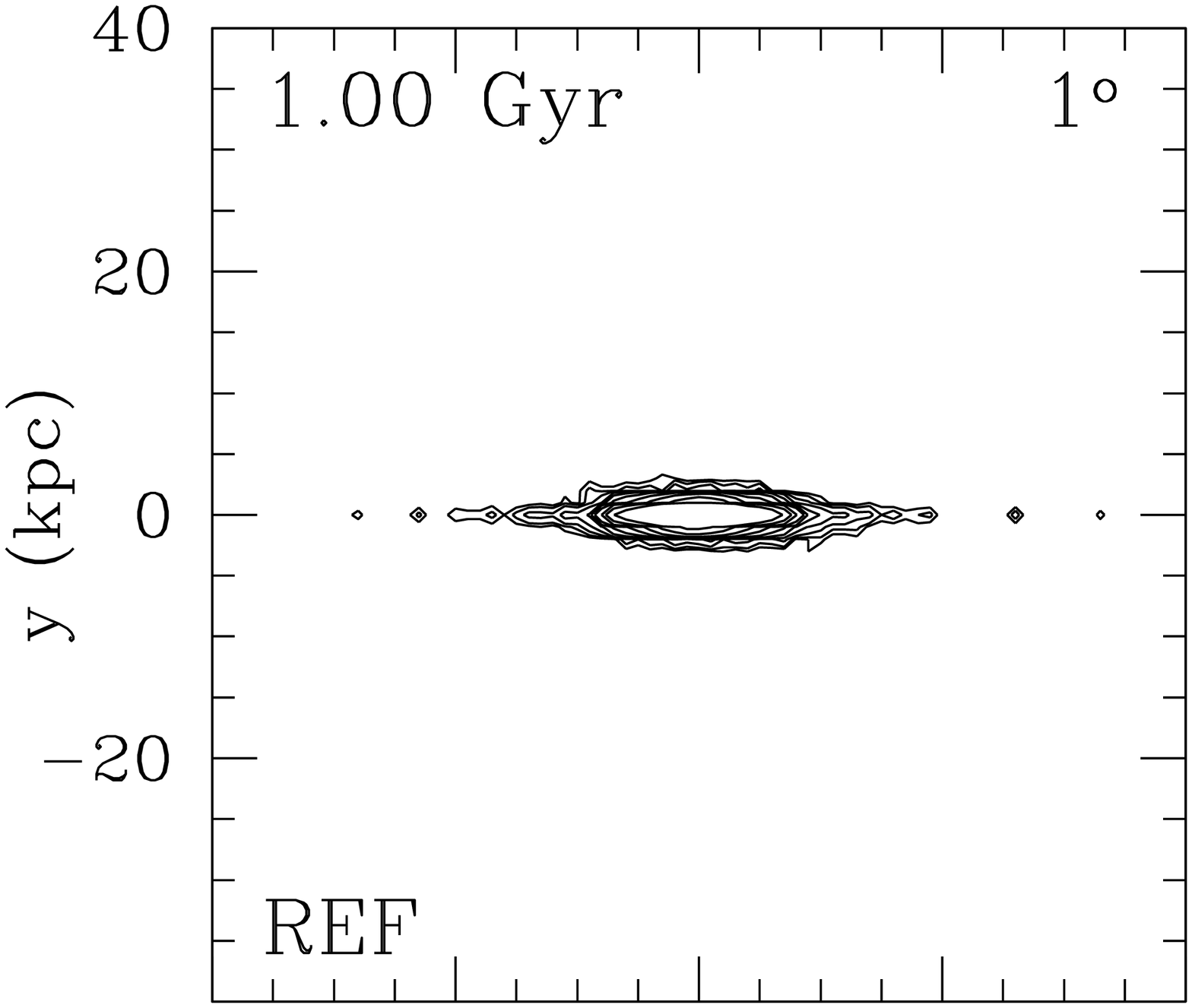}\hspace*{-0.87cm}
\includegraphics[width=34.9mm]{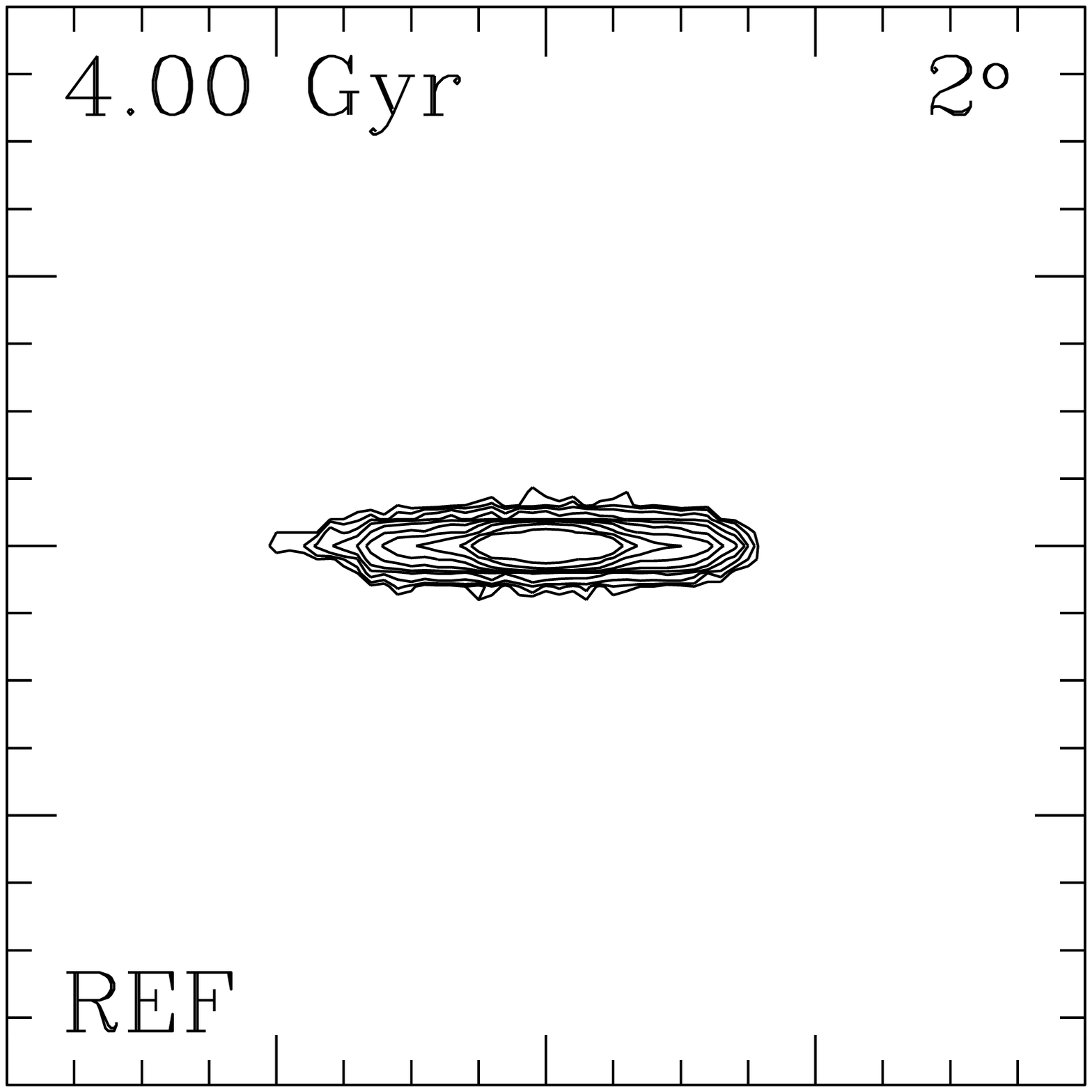}\hspace*{-0.87cm}
\includegraphics[width=34.9mm]{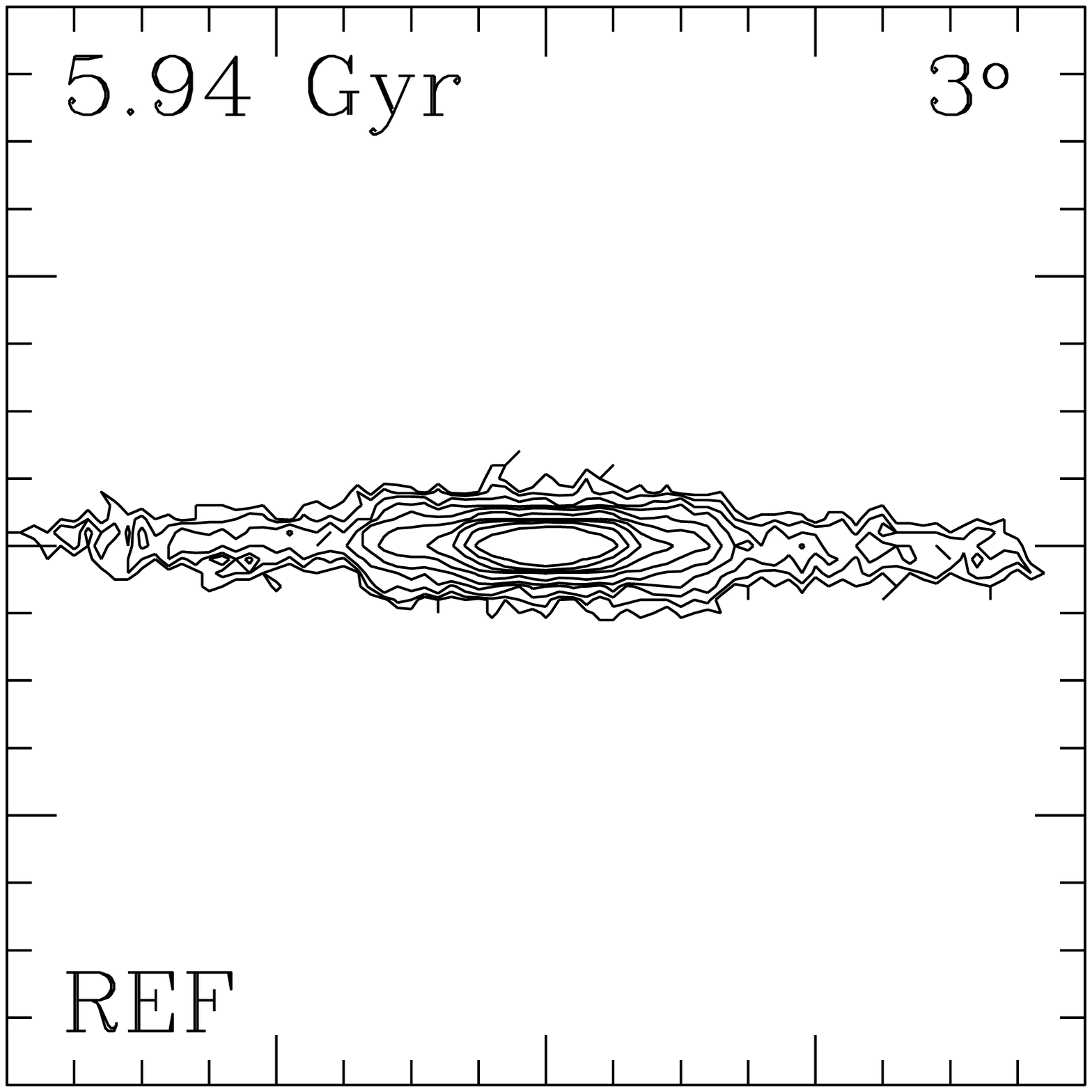}\hspace*{-0.87cm}
\includegraphics[width=34.9mm]{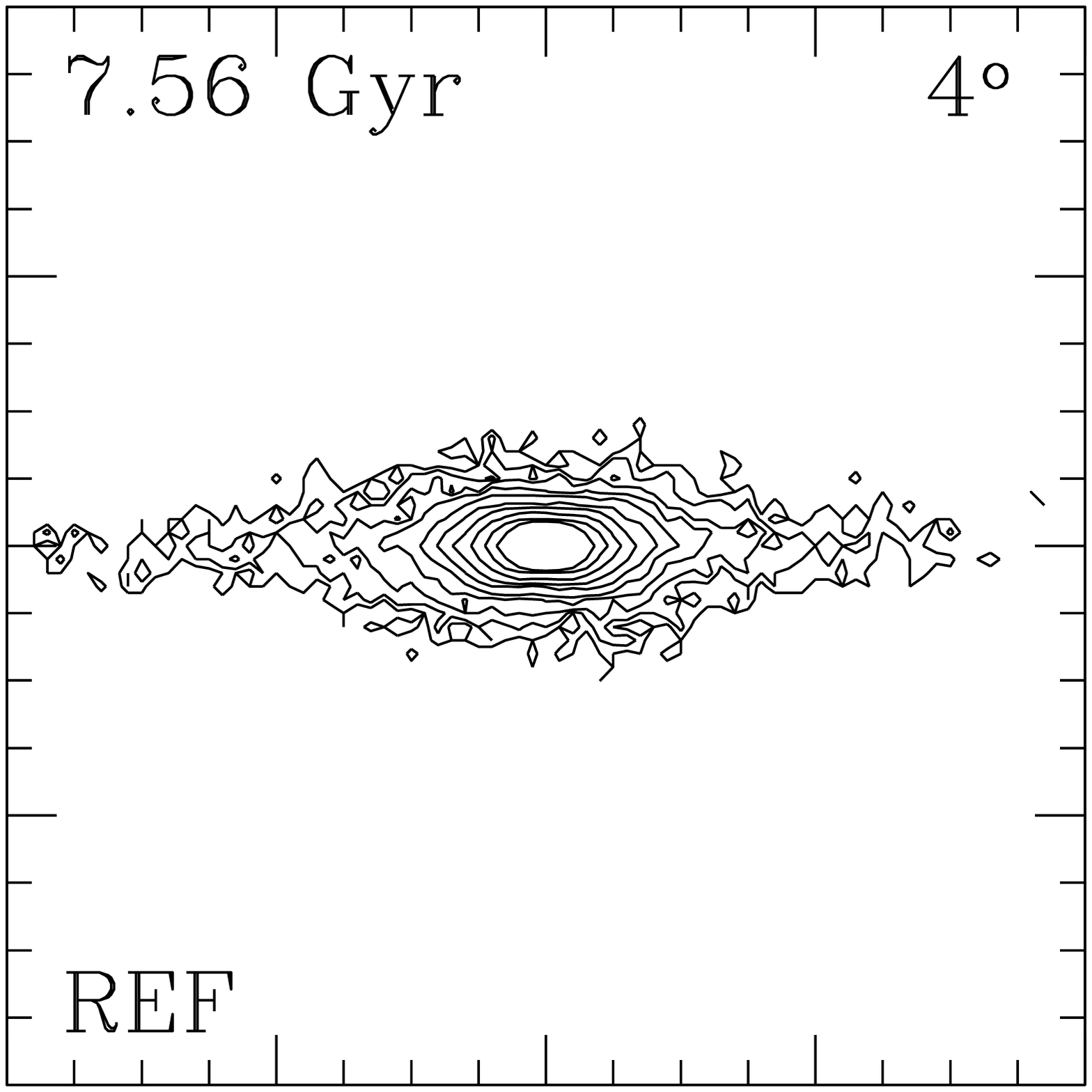}\hspace*{-0.87cm}
\includegraphics[width=34.9mm]{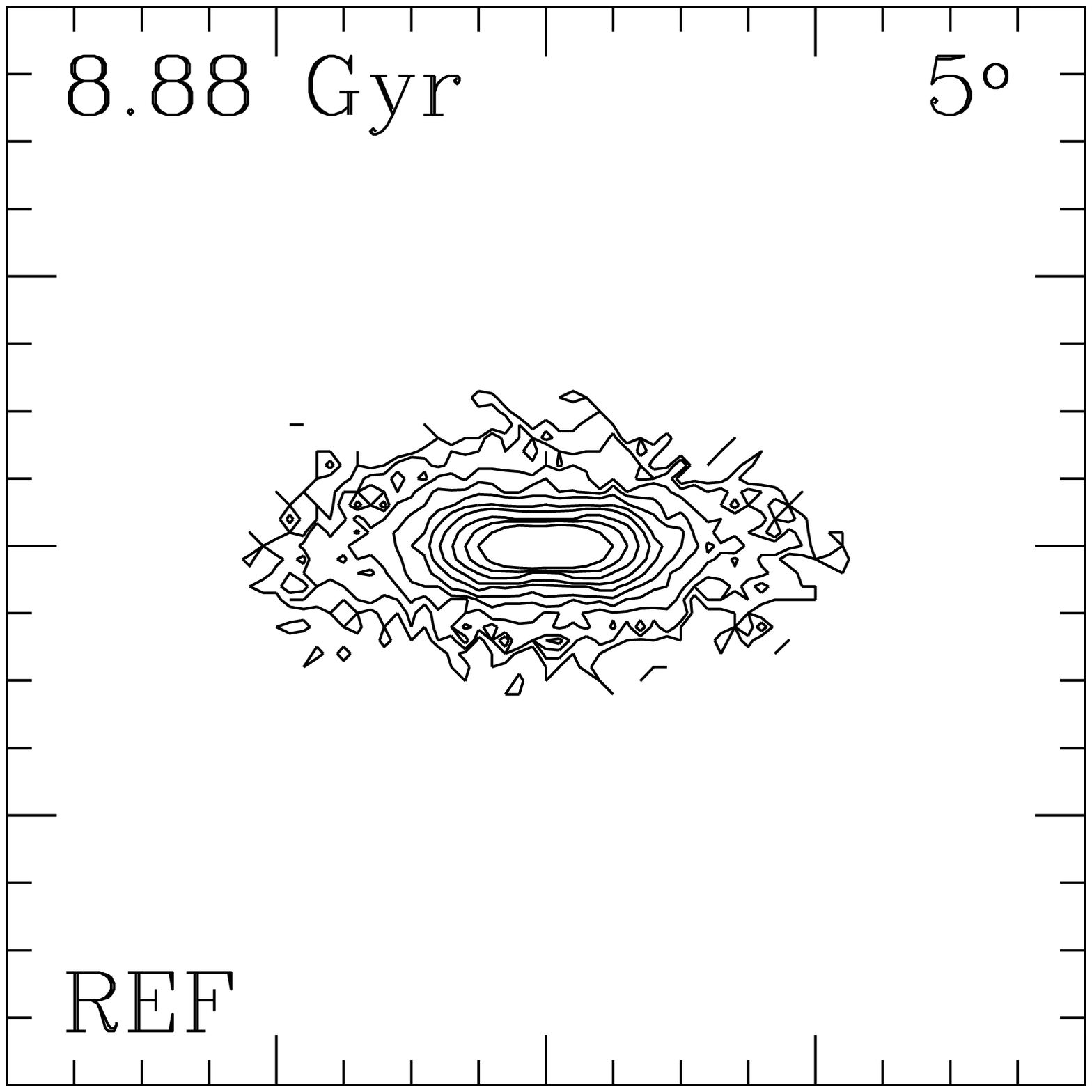}\hspace*{-0.87cm}
\includegraphics[width=34.9mm]{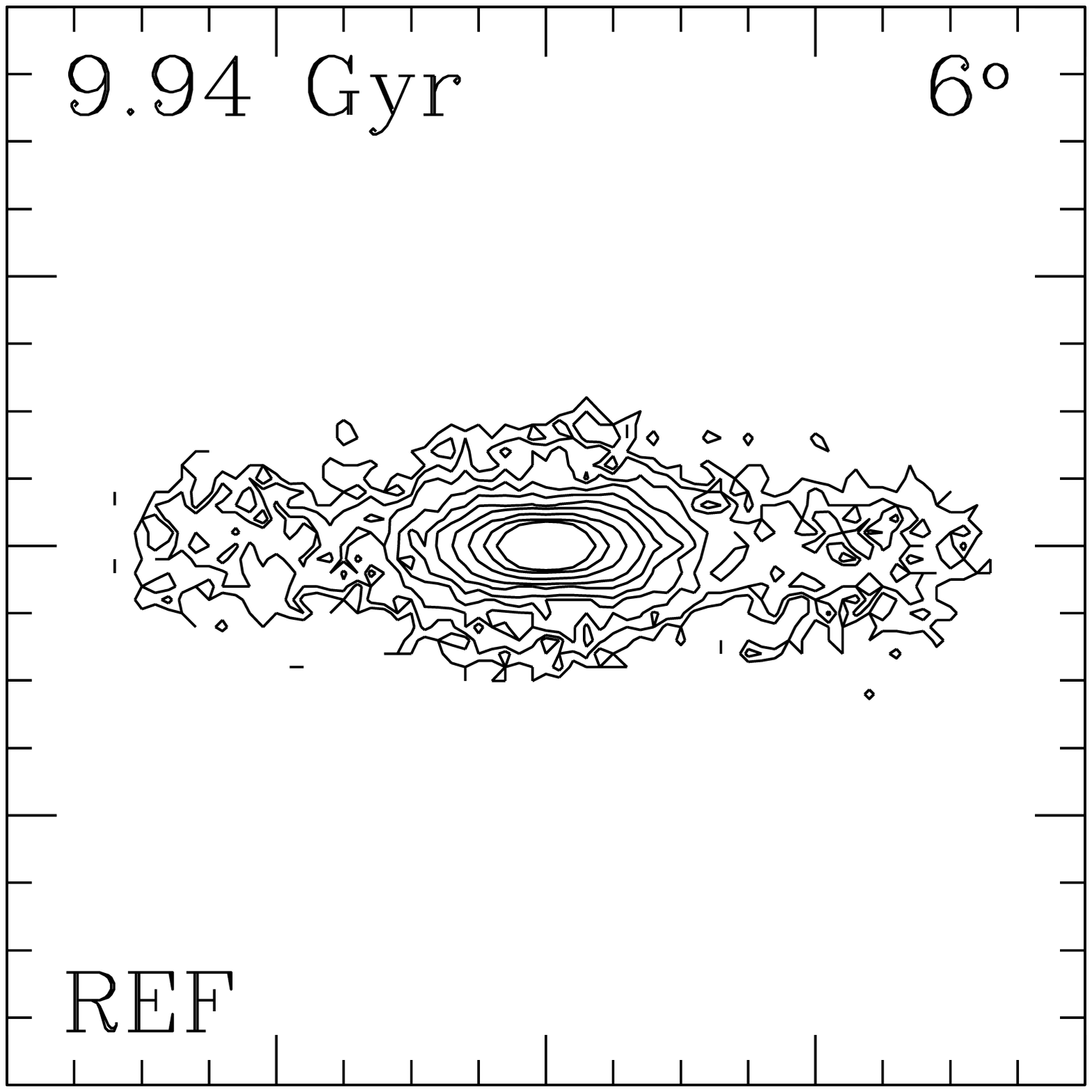}\vspace*{-0.81cm}\\
\includegraphics[width=34.9mm]{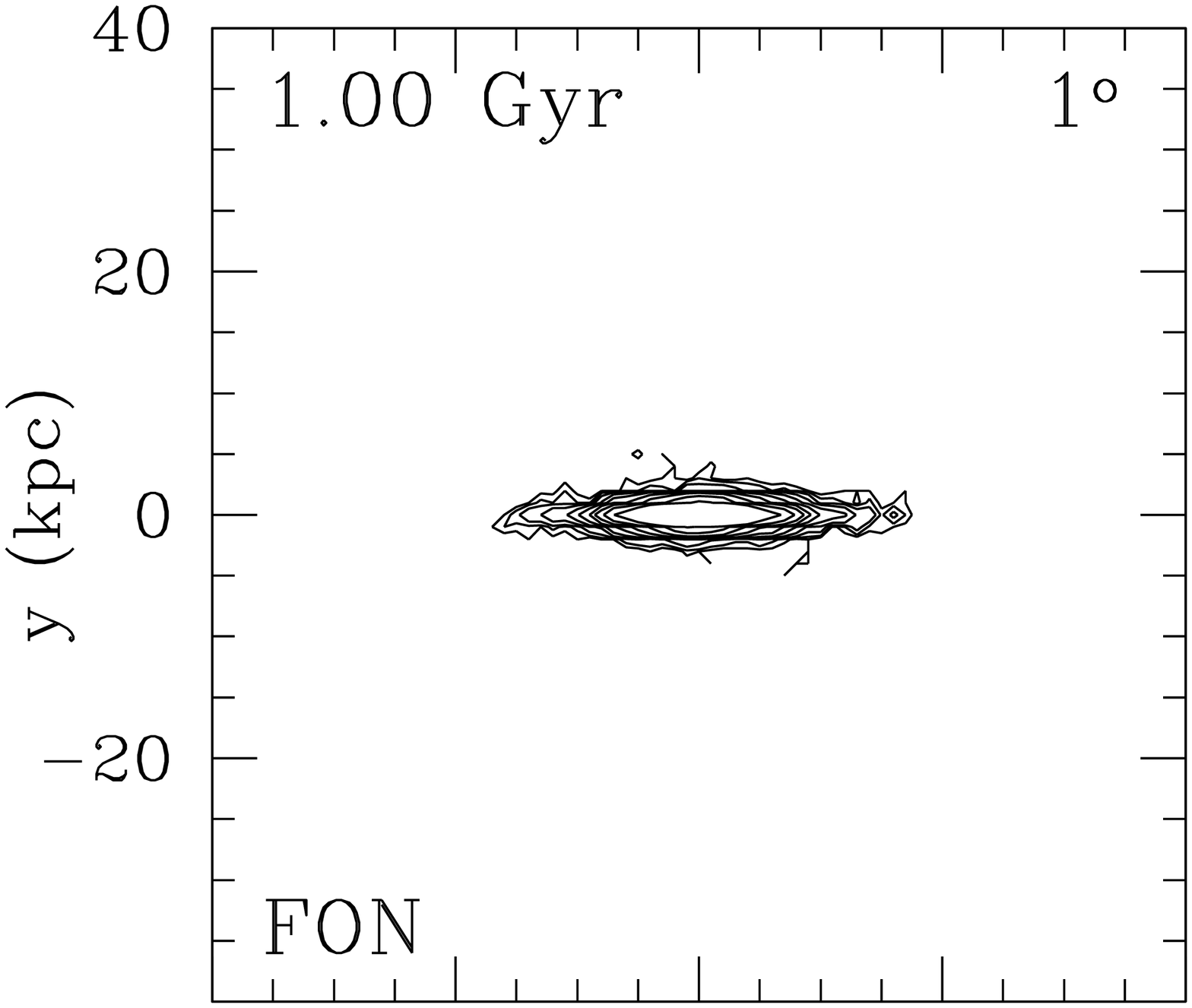}\hspace*{-0.87cm}
\includegraphics[width=34.9mm]{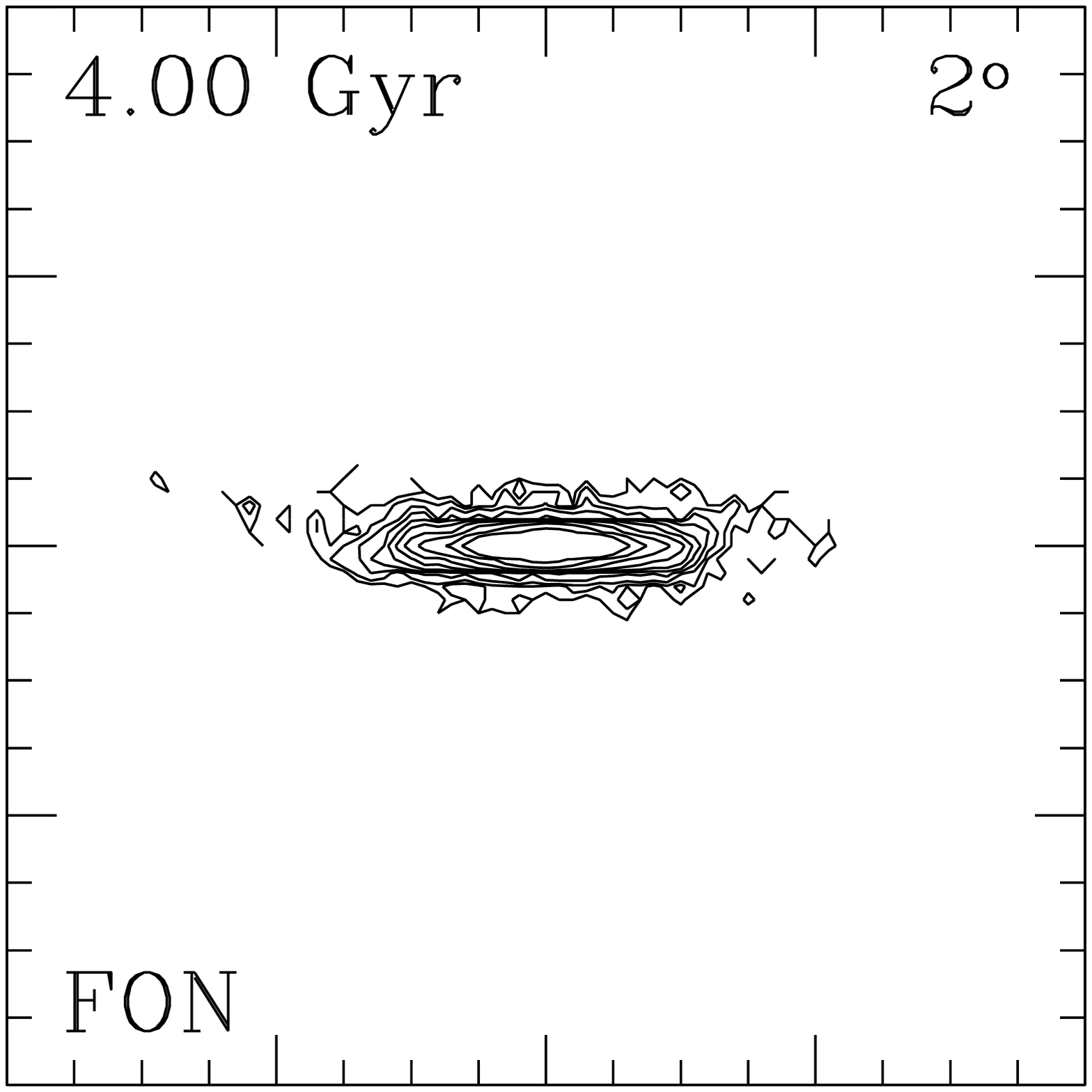}\hspace*{-0.87cm}
\includegraphics[width=34.9mm]{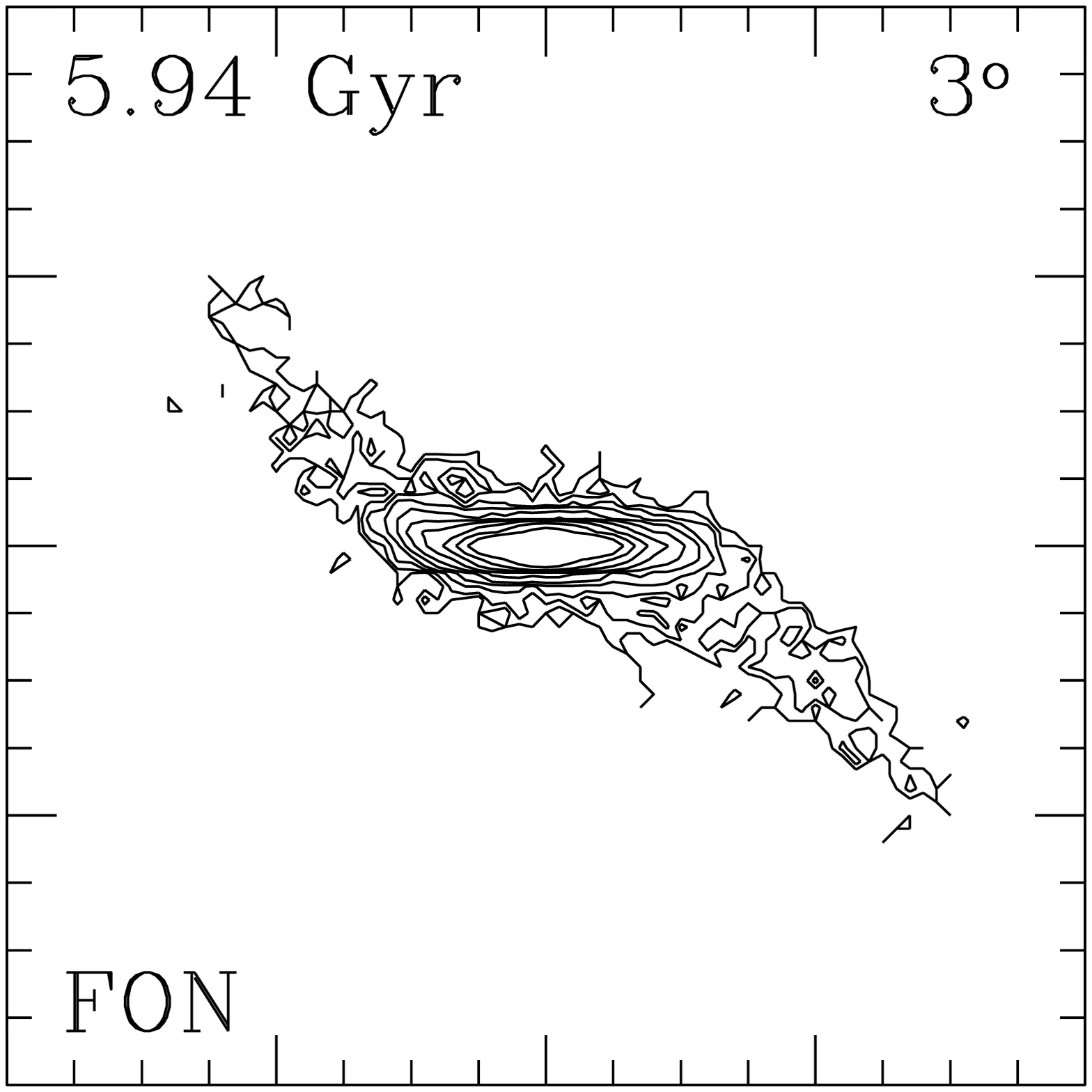}\hspace*{-0.87cm}
\includegraphics[width=34.9mm]{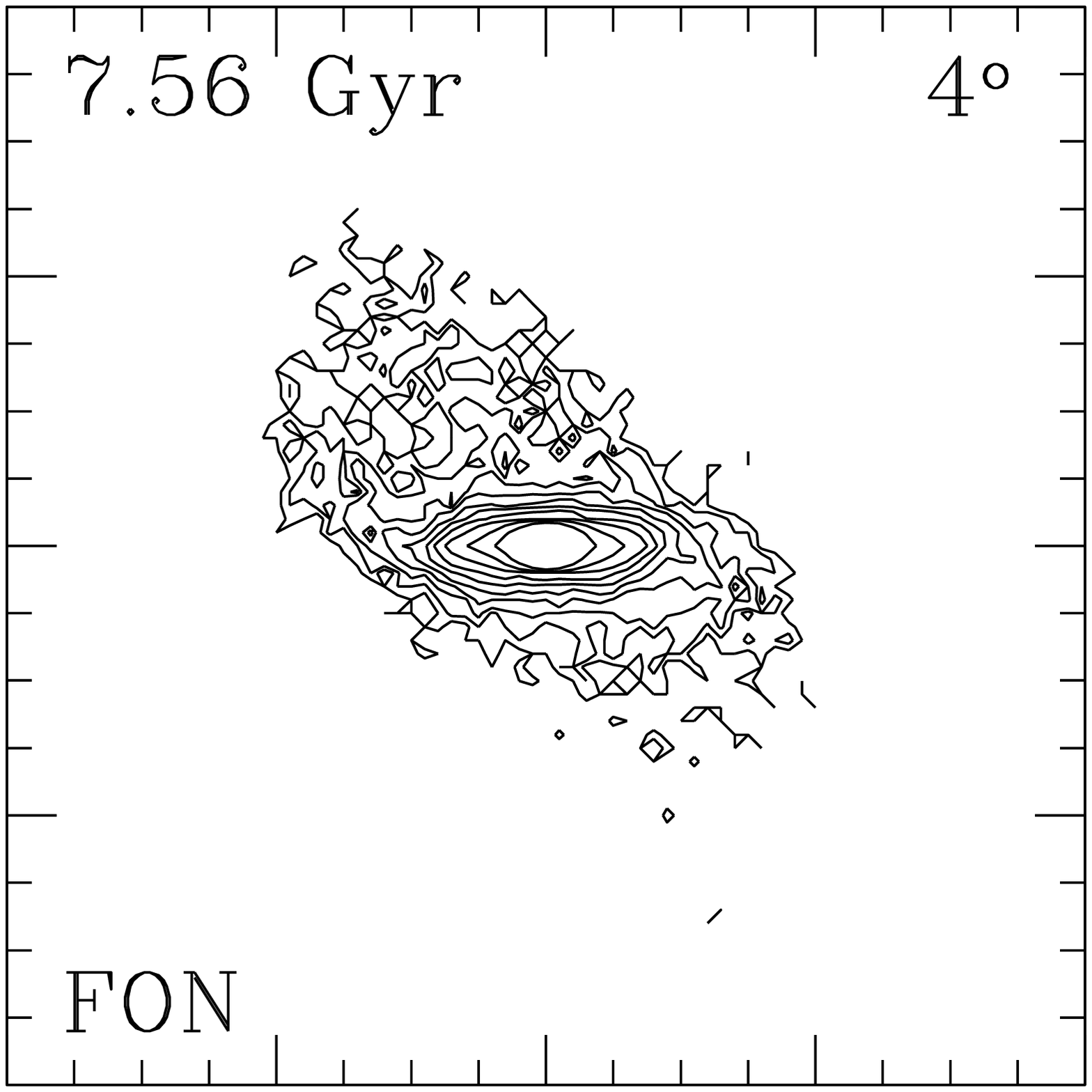}\hspace*{-0.87cm}
\includegraphics[width=34.9mm]{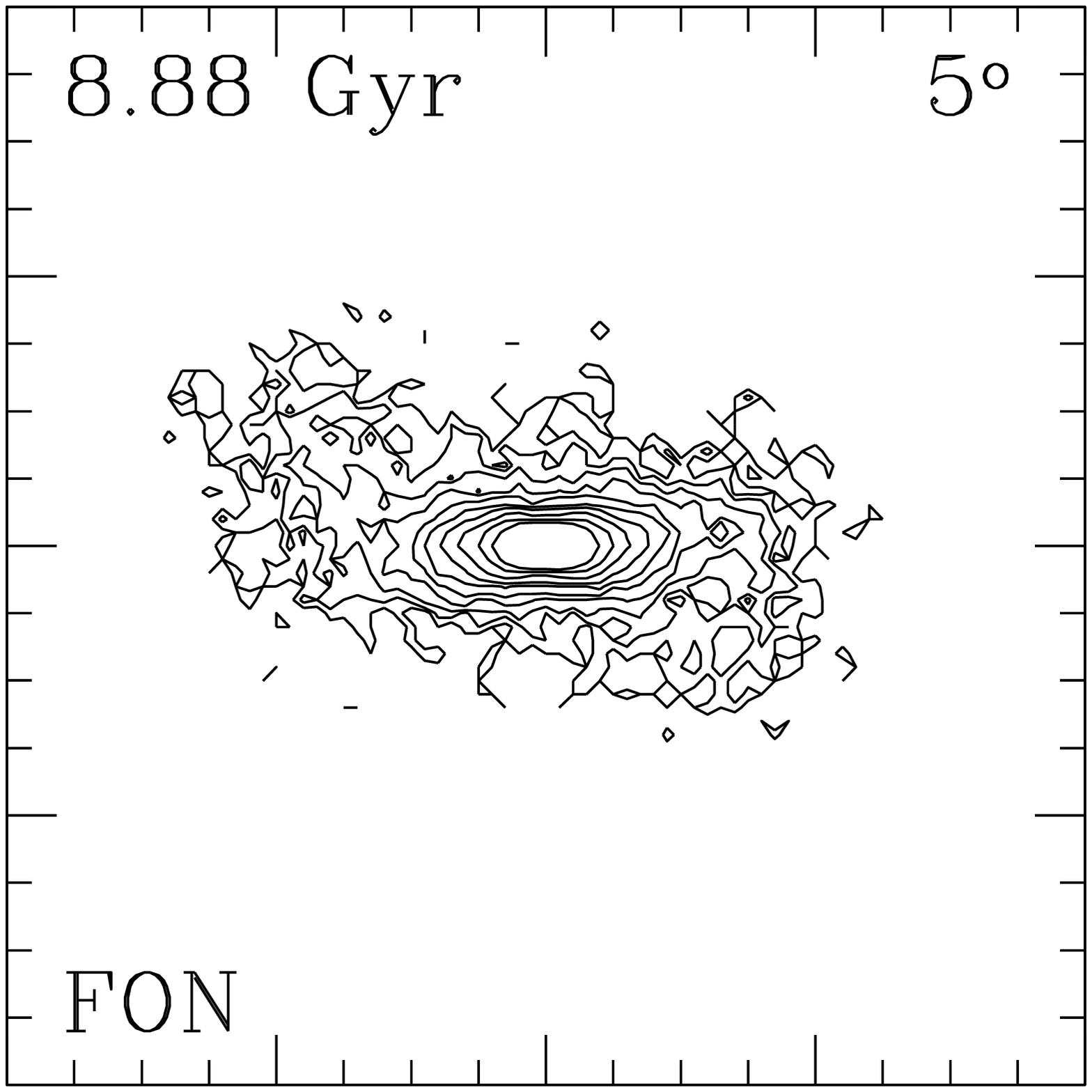}\hspace*{-0.87cm}
\includegraphics[width=34.9mm]{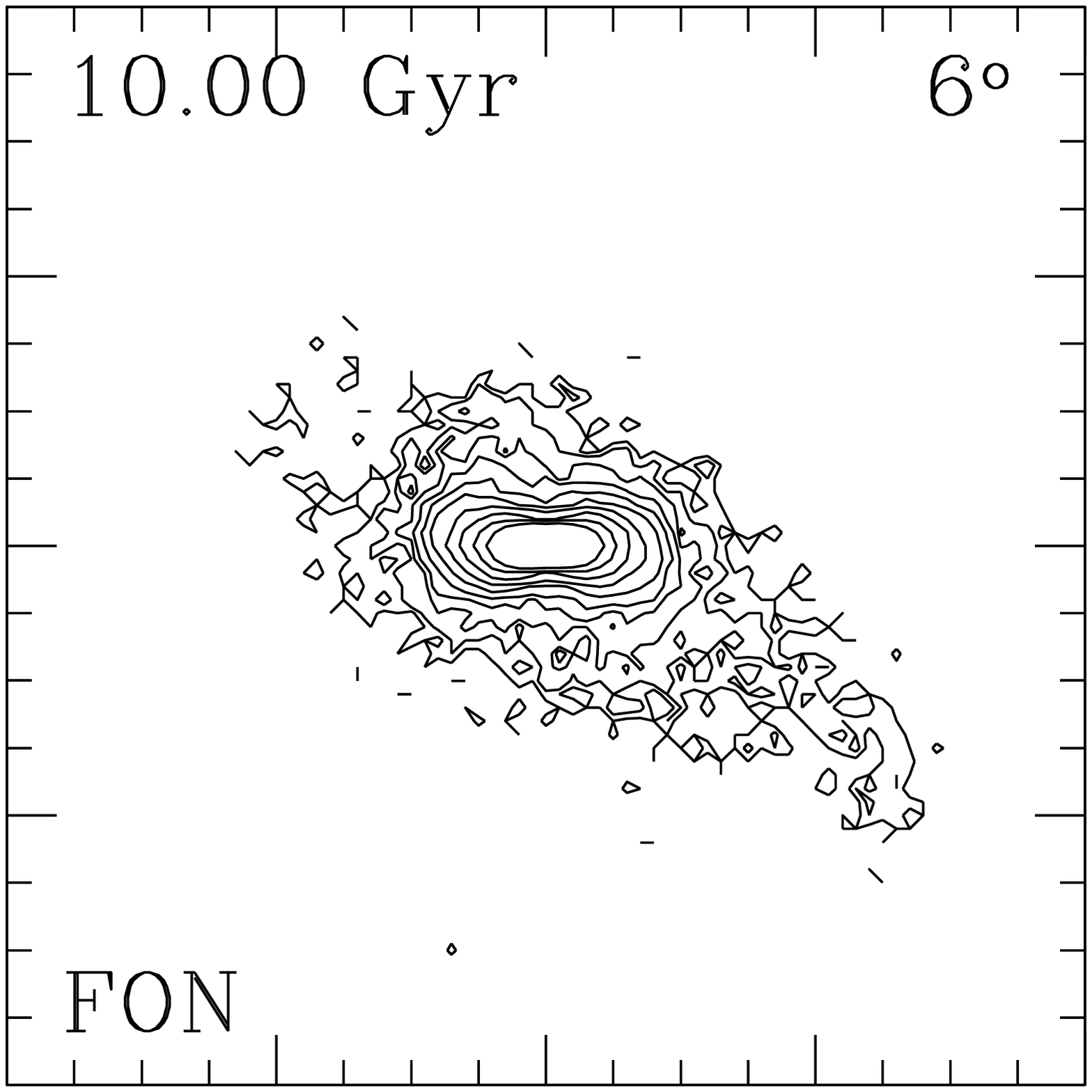}\vspace*{-0.81cm}\\
\includegraphics[width=34.9mm]{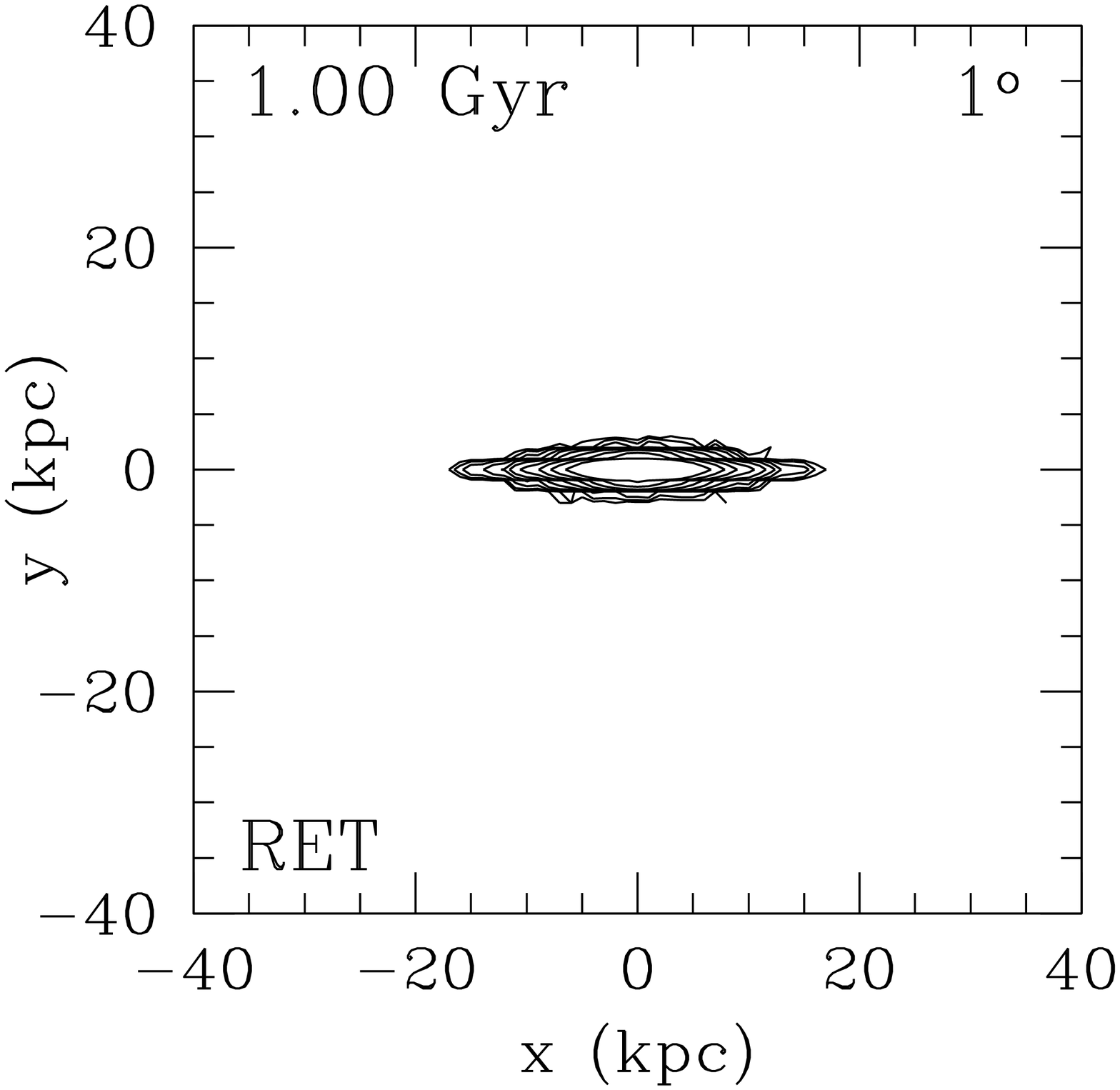}\hspace*{-0.87cm}
\includegraphics[width=34.9mm]{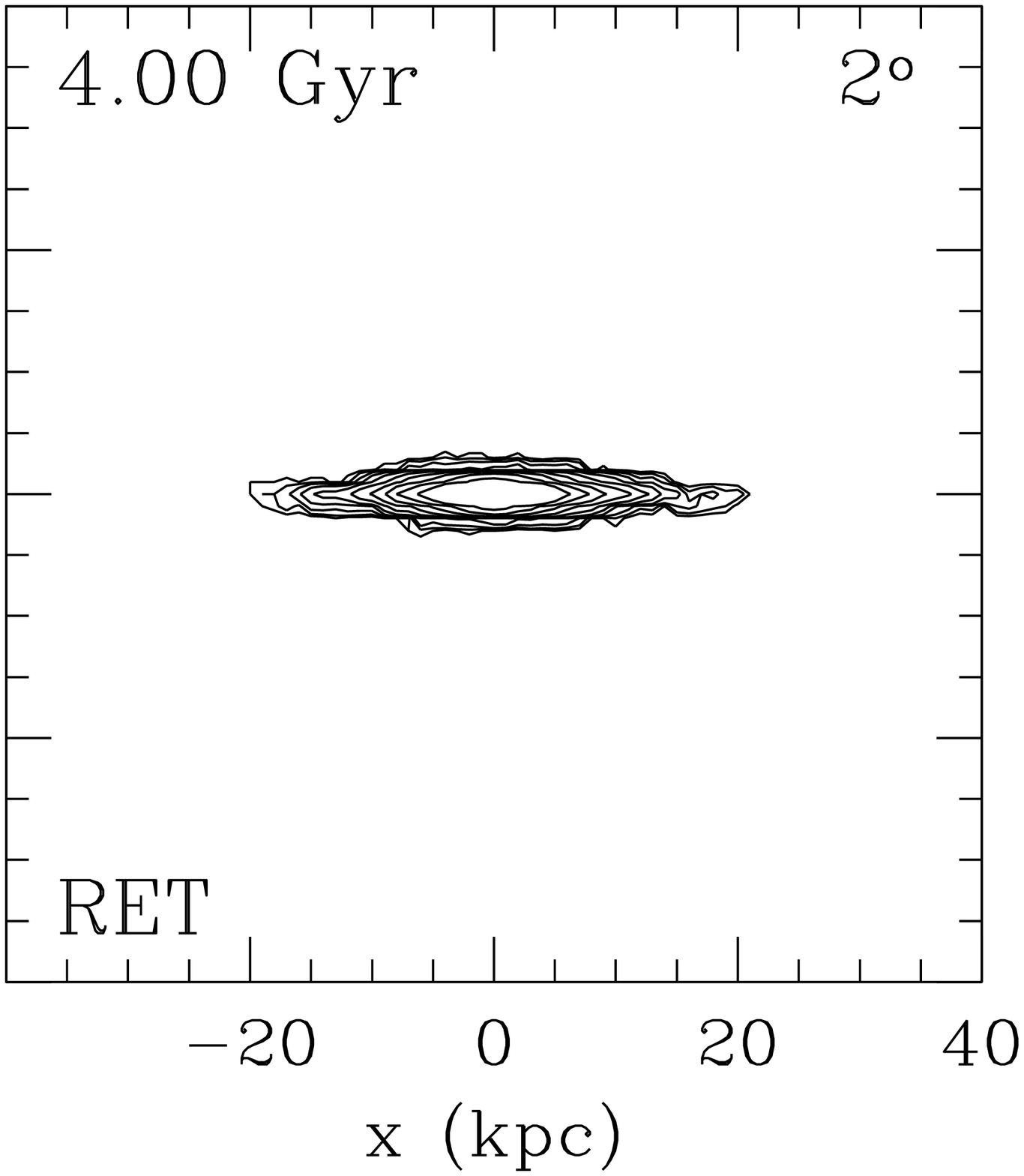}\hspace*{-0.87cm}
\includegraphics[width=34.9mm]{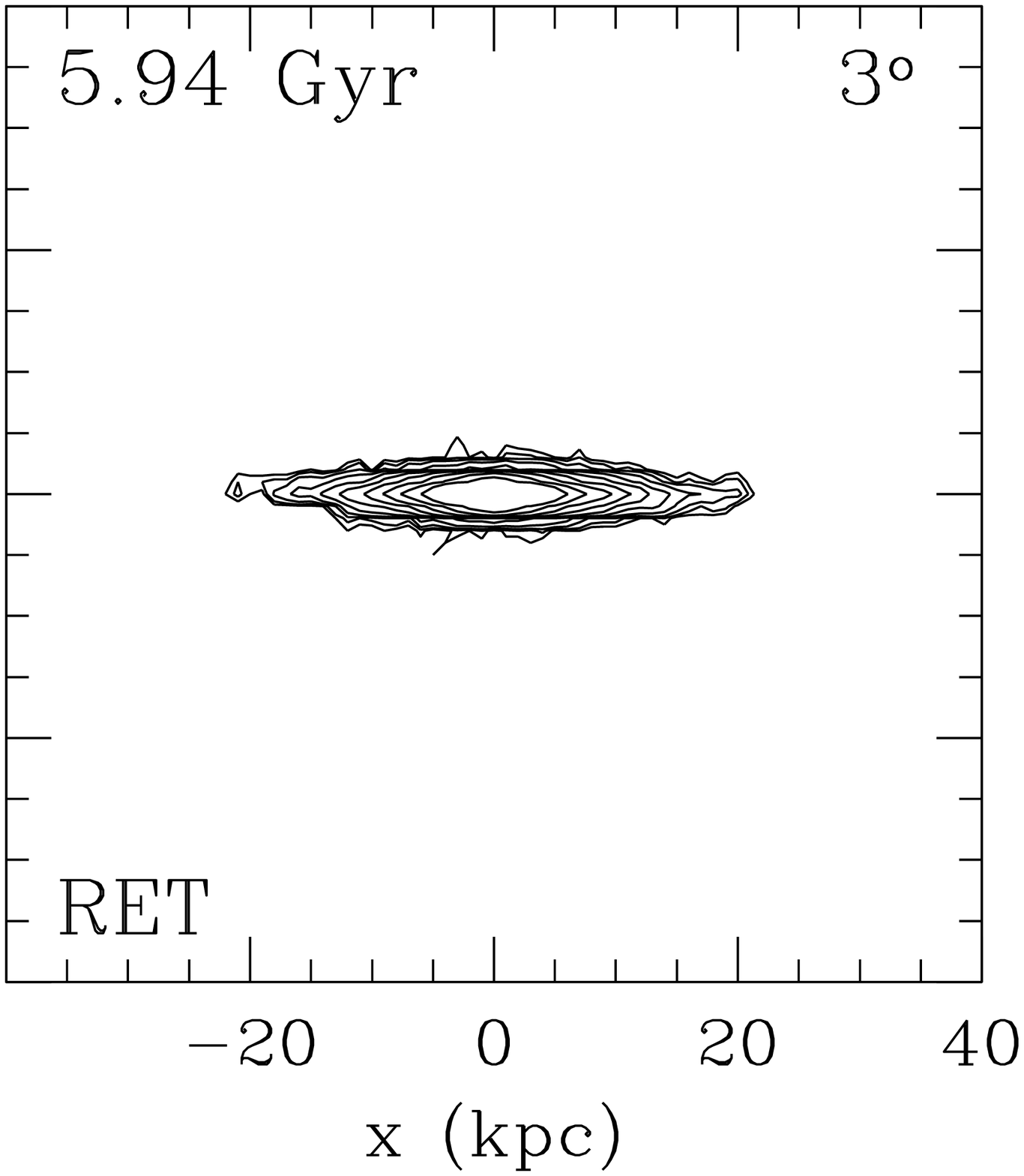}\hspace*{-0.87cm}
\includegraphics[width=34.9mm]{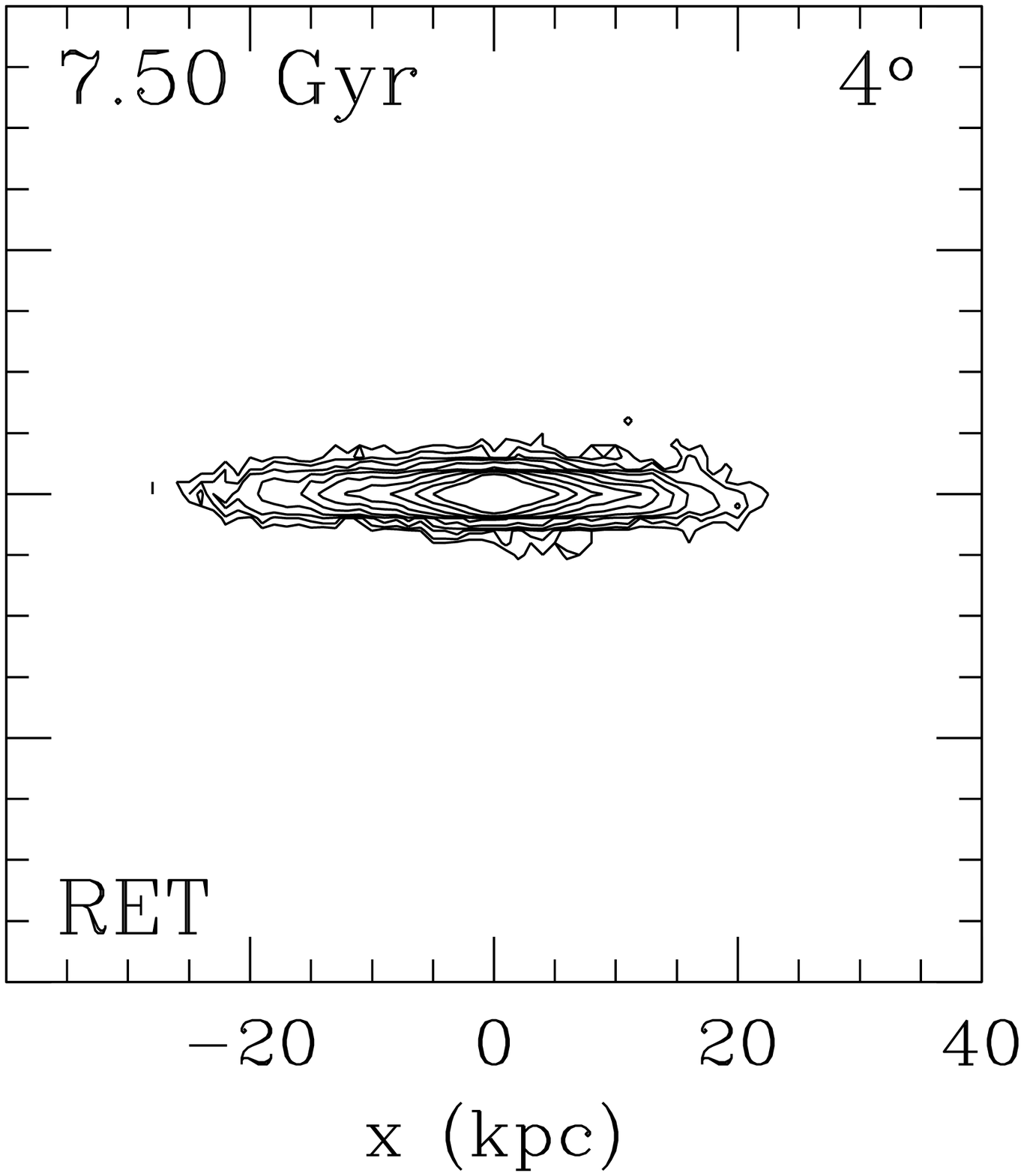}\hspace*{-0.87cm}
\includegraphics[width=34.9mm]{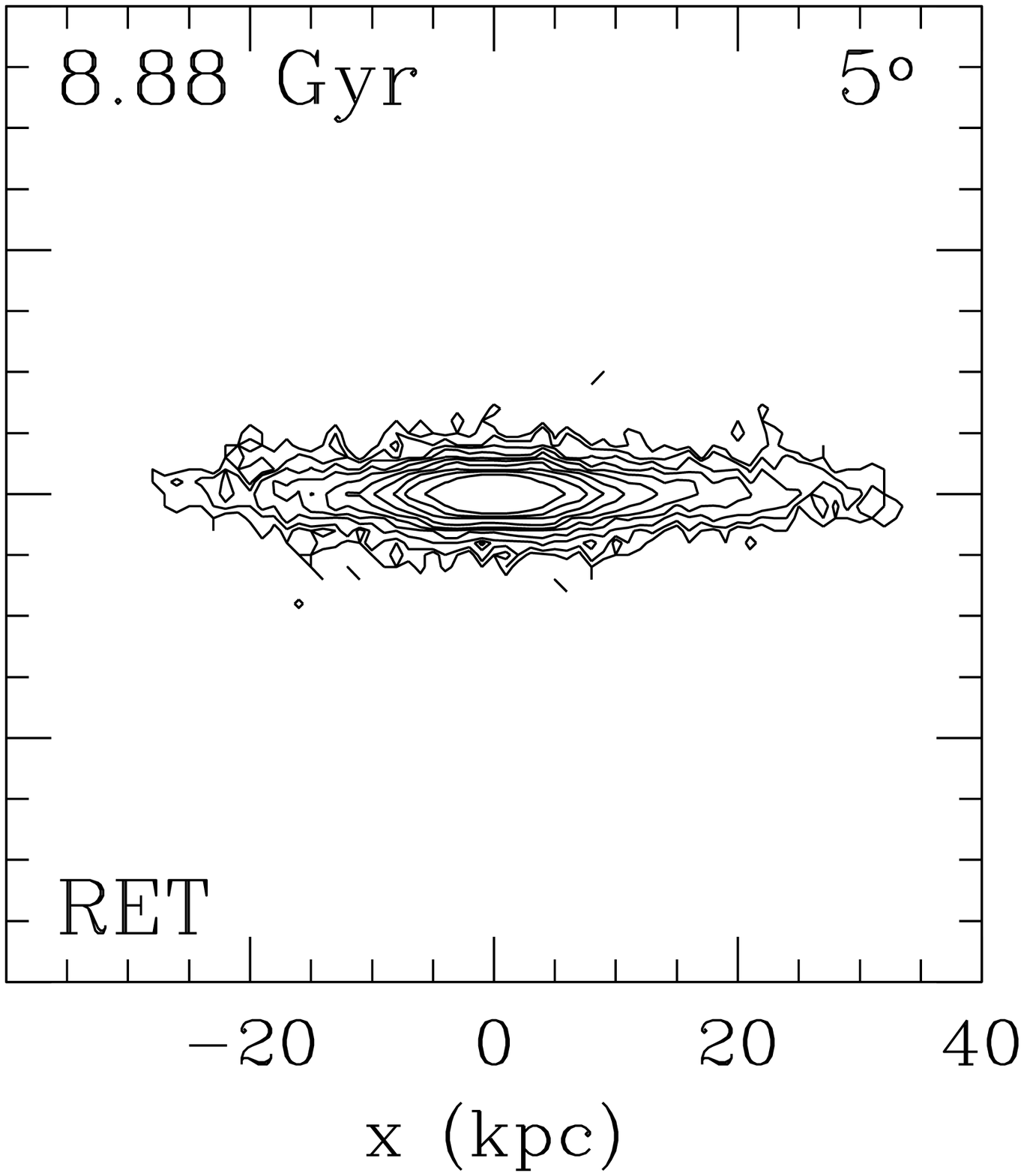}\hspace*{-0.87cm}
\includegraphics[width=34.9mm]{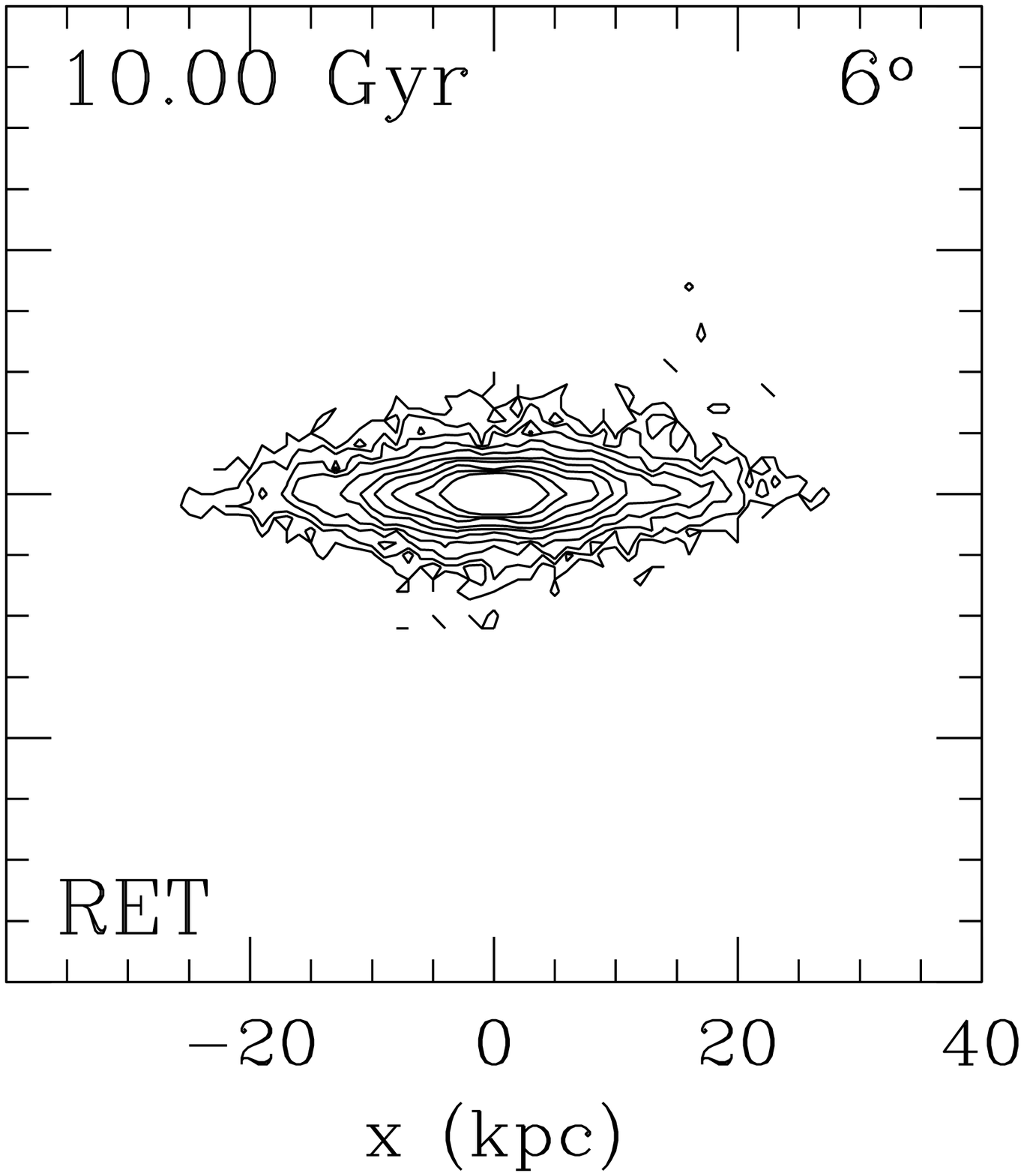}
\end{center}
\caption{Same as Fig.~\ref{contours-xy-z1} but in edge-on view.}
\label{contours-xz-z1}
\end{figure*}

\begin{figure*}
\begin{center}
\includegraphics[width=75mm]{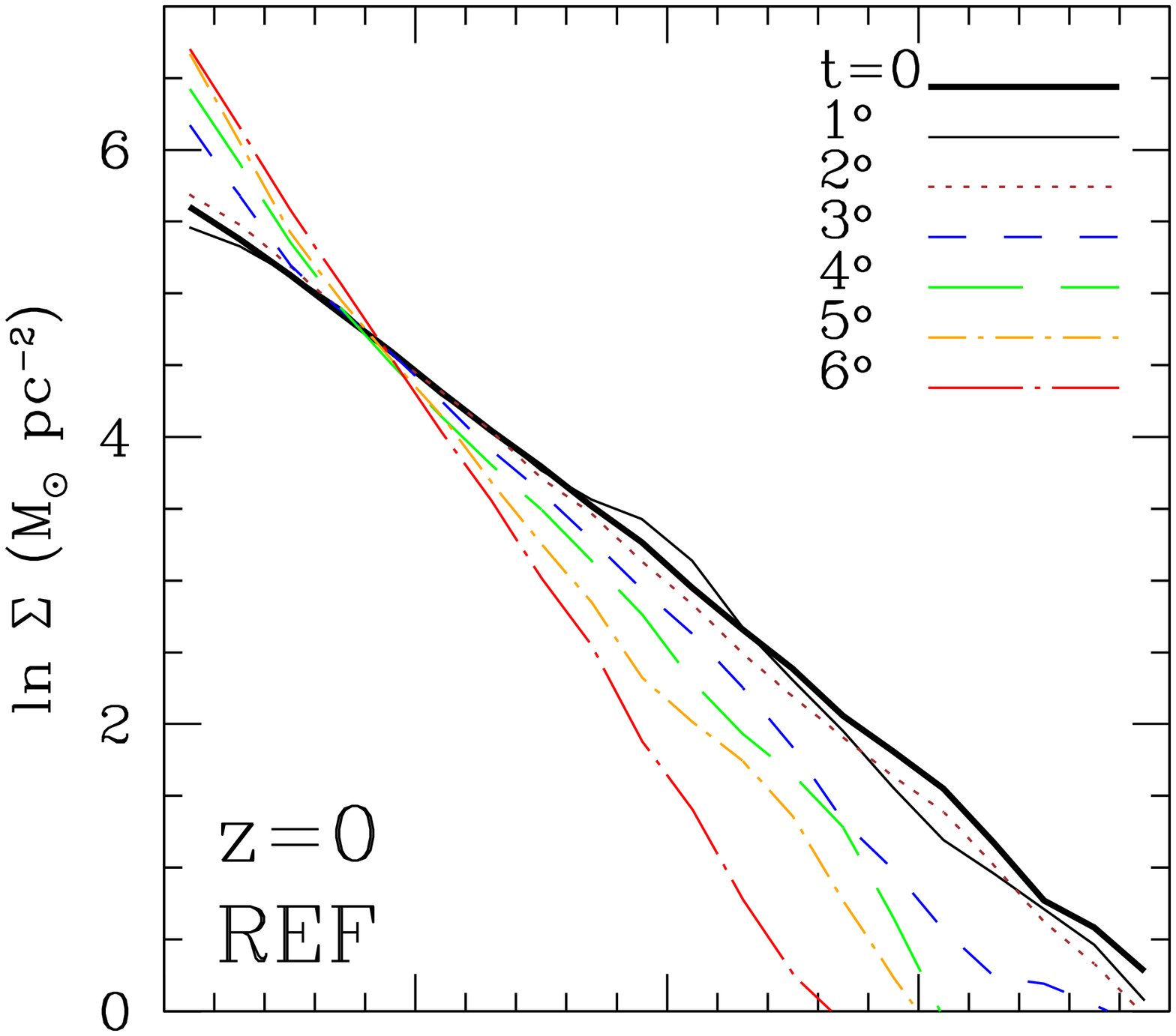}\hspace*{-1.76cm}
\includegraphics[width=75mm]{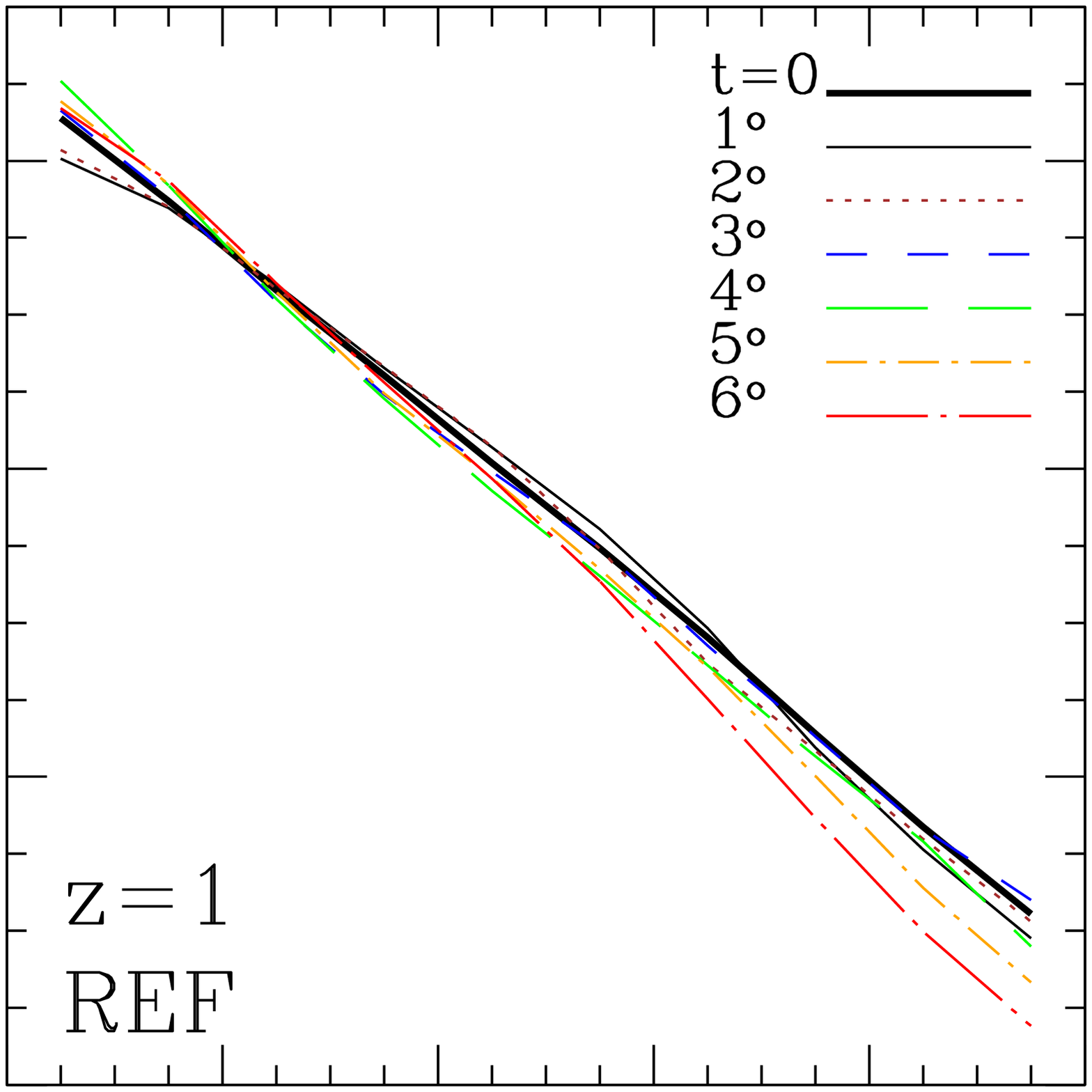}\vspace*{-1.7cm}\\
\includegraphics[width=75mm]{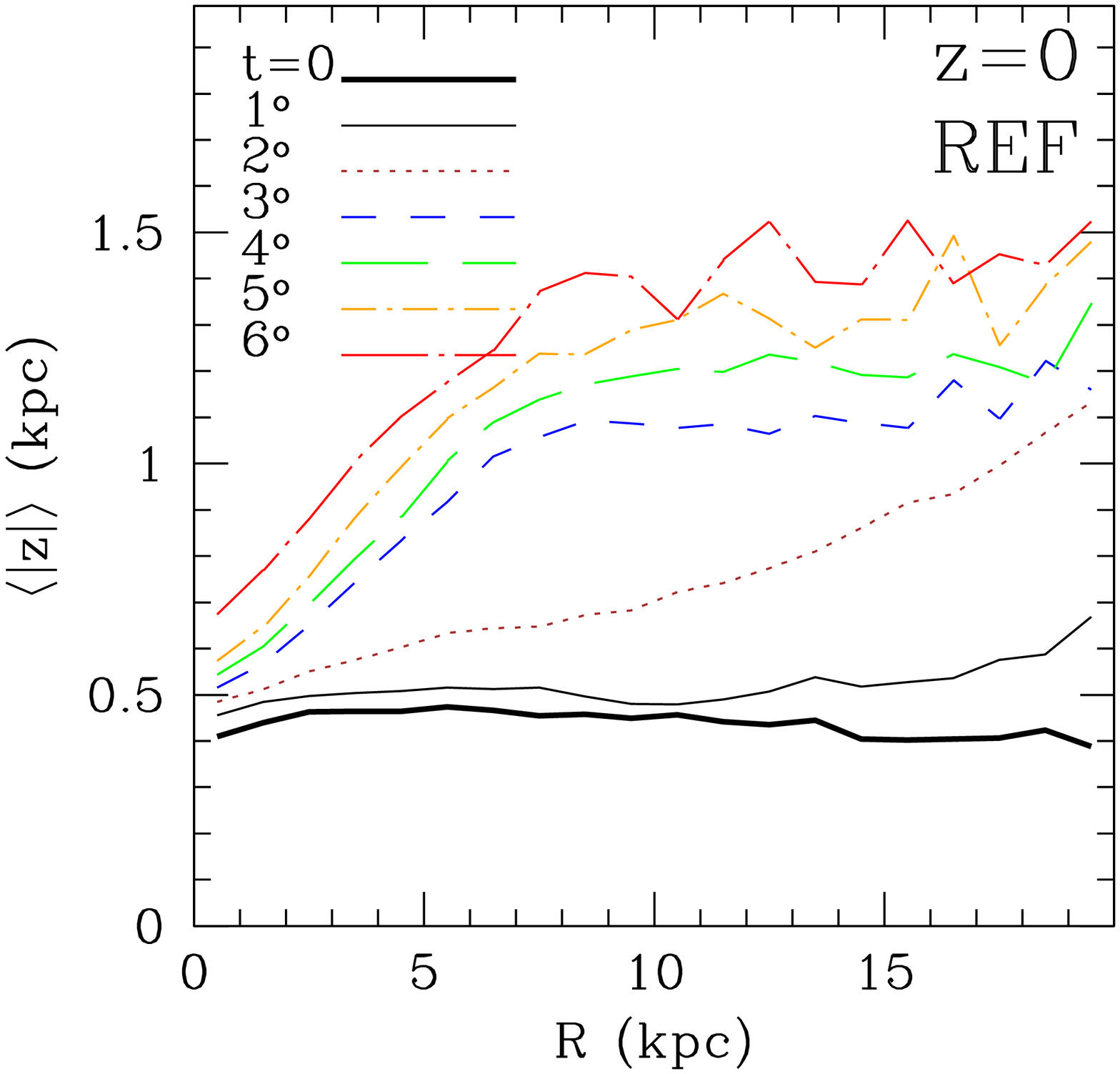}\hspace*{-1.76cm}
\includegraphics[width=75mm]{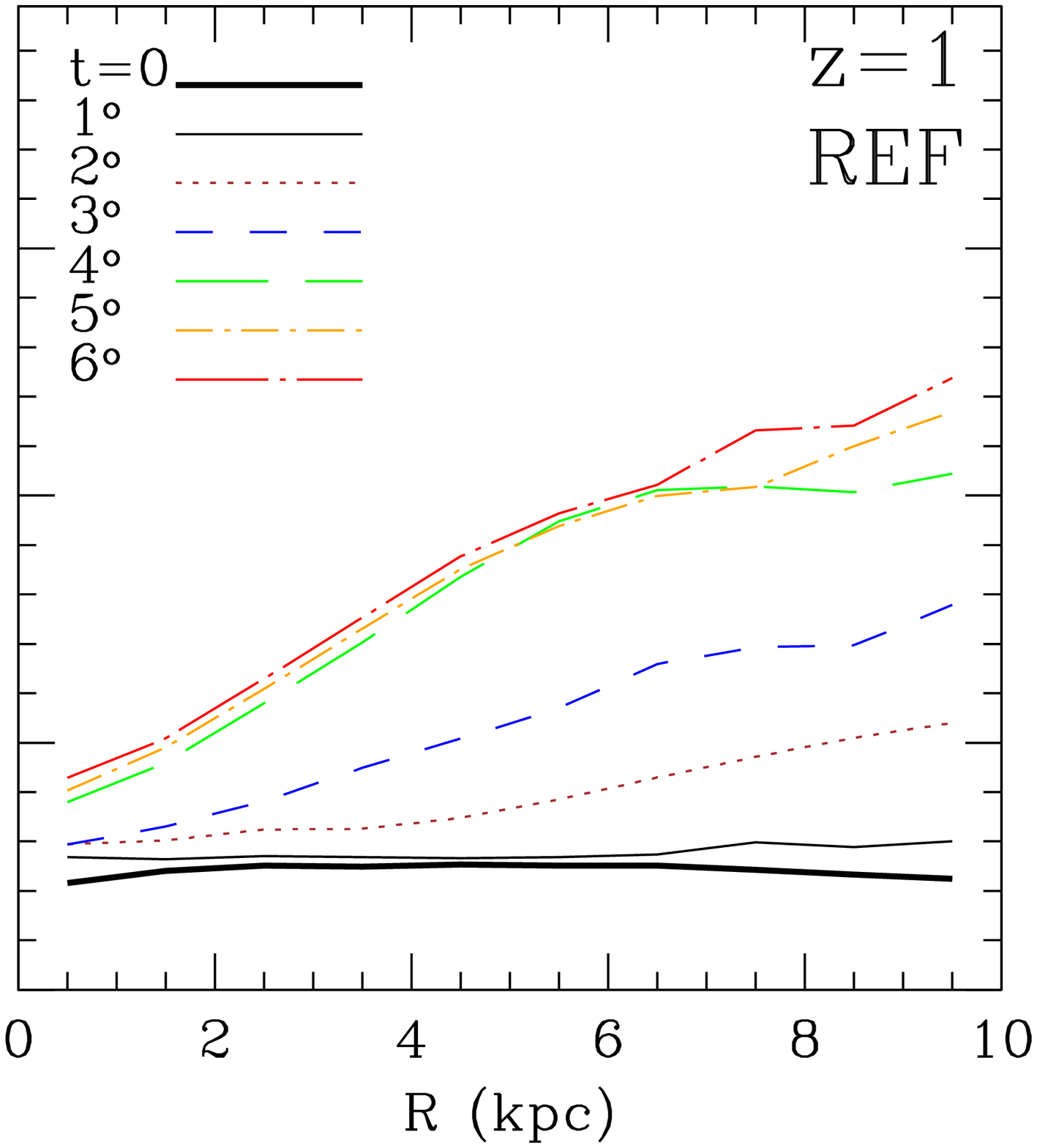}
\end{center}
\caption{Evolution of the mass surface density and thickness profiles of 
disc galaxies, for our REF (reference) experiments at ``$z$=0'' and ``$z$=1'', 
after each of their first six pericentric passages. The profiles have been 
computed in concentric rings, 1~kpc wide, considering only stars that remain 
bound and are located within 3~kpc from the midplane. At each radius the 
thickness of discs is computed as $\langle |z| \rangle$.
}
\label{lnsigma-thickness-evol-ref-z0-z1}
\end{figure*}

\begin{figure*}
\begin{center}
\includegraphics[width=88mm]{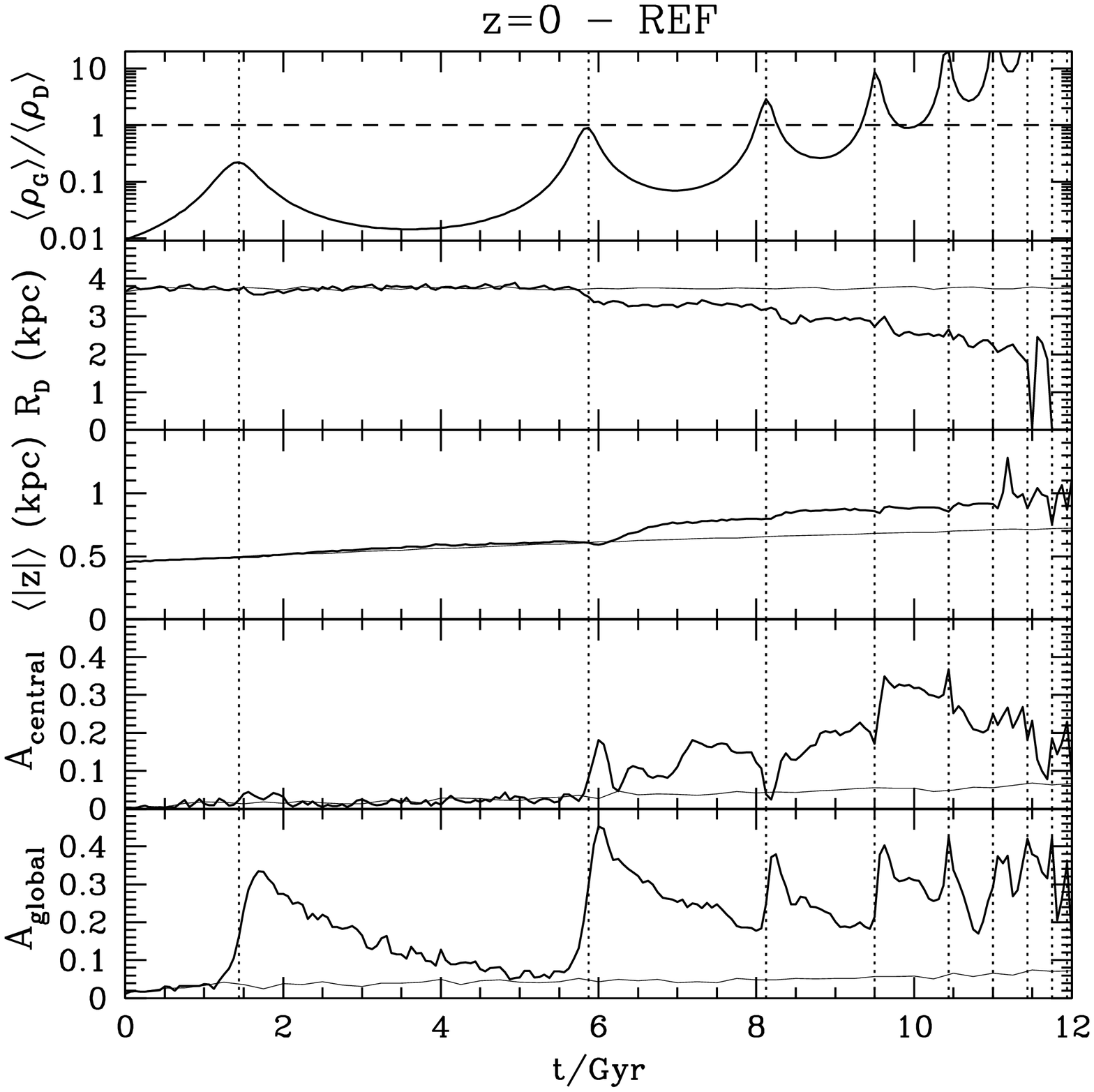}
\includegraphics[width=88mm]{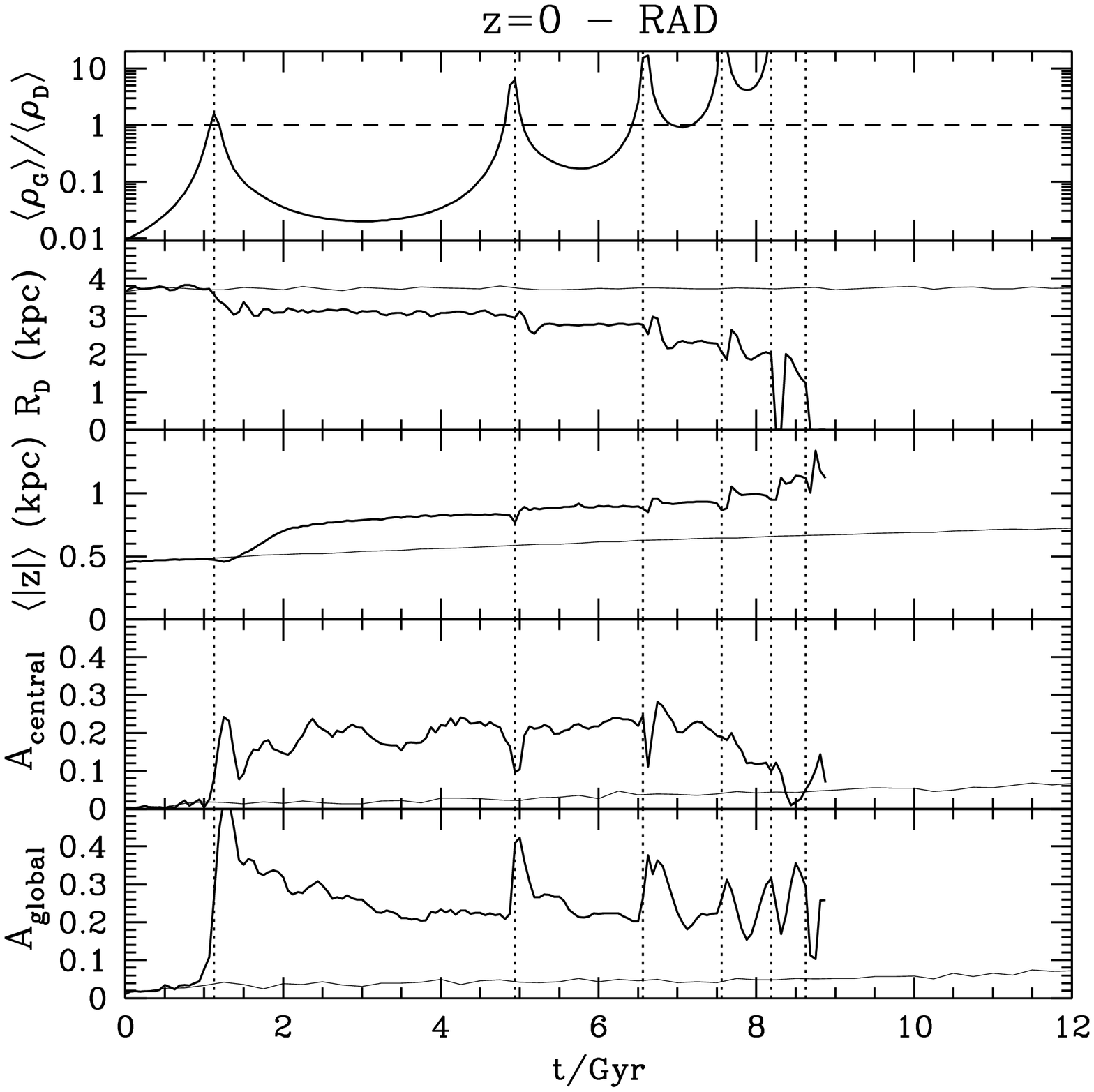}
\includegraphics[width=88mm]{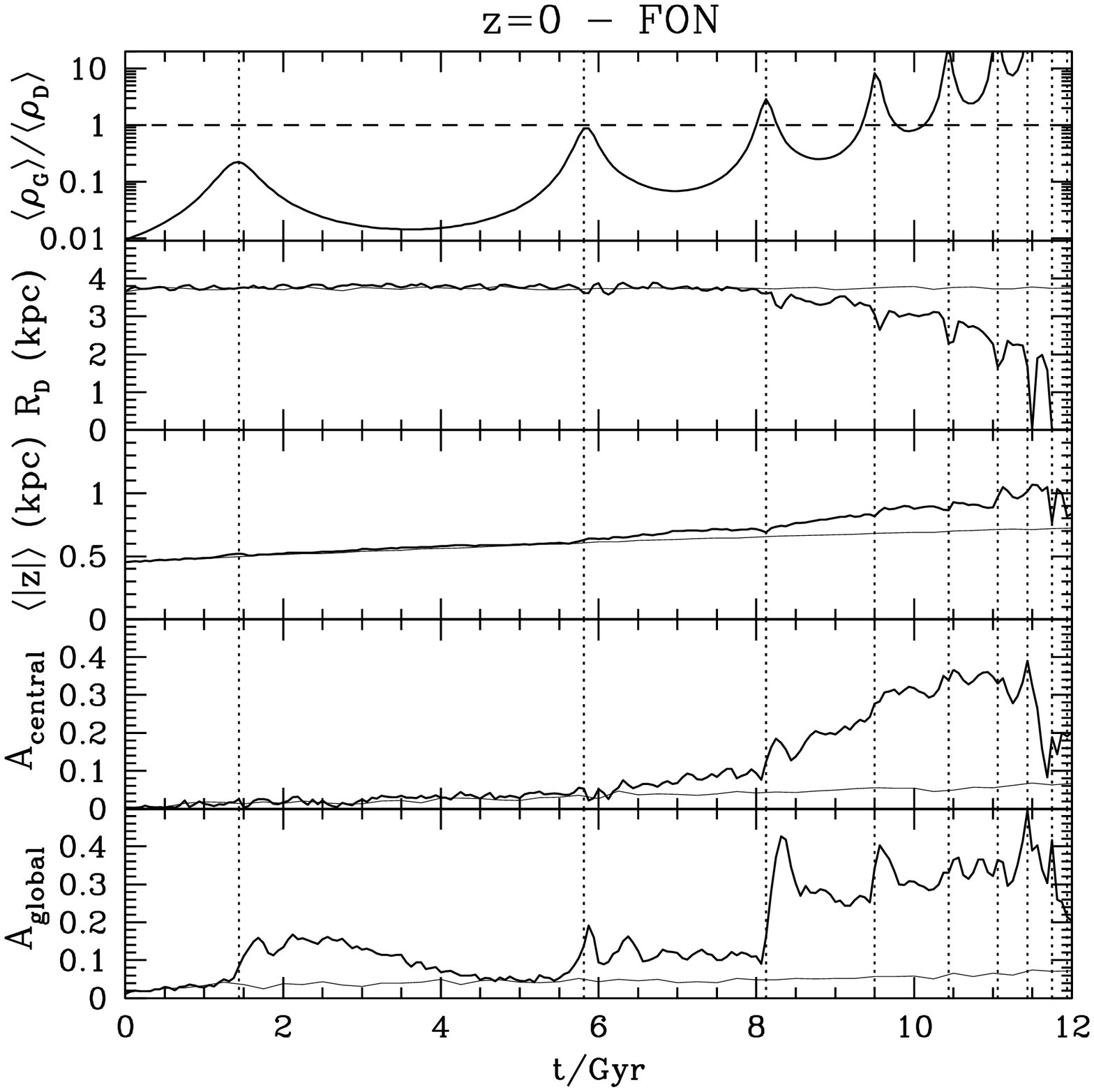}
\includegraphics[width=88mm]{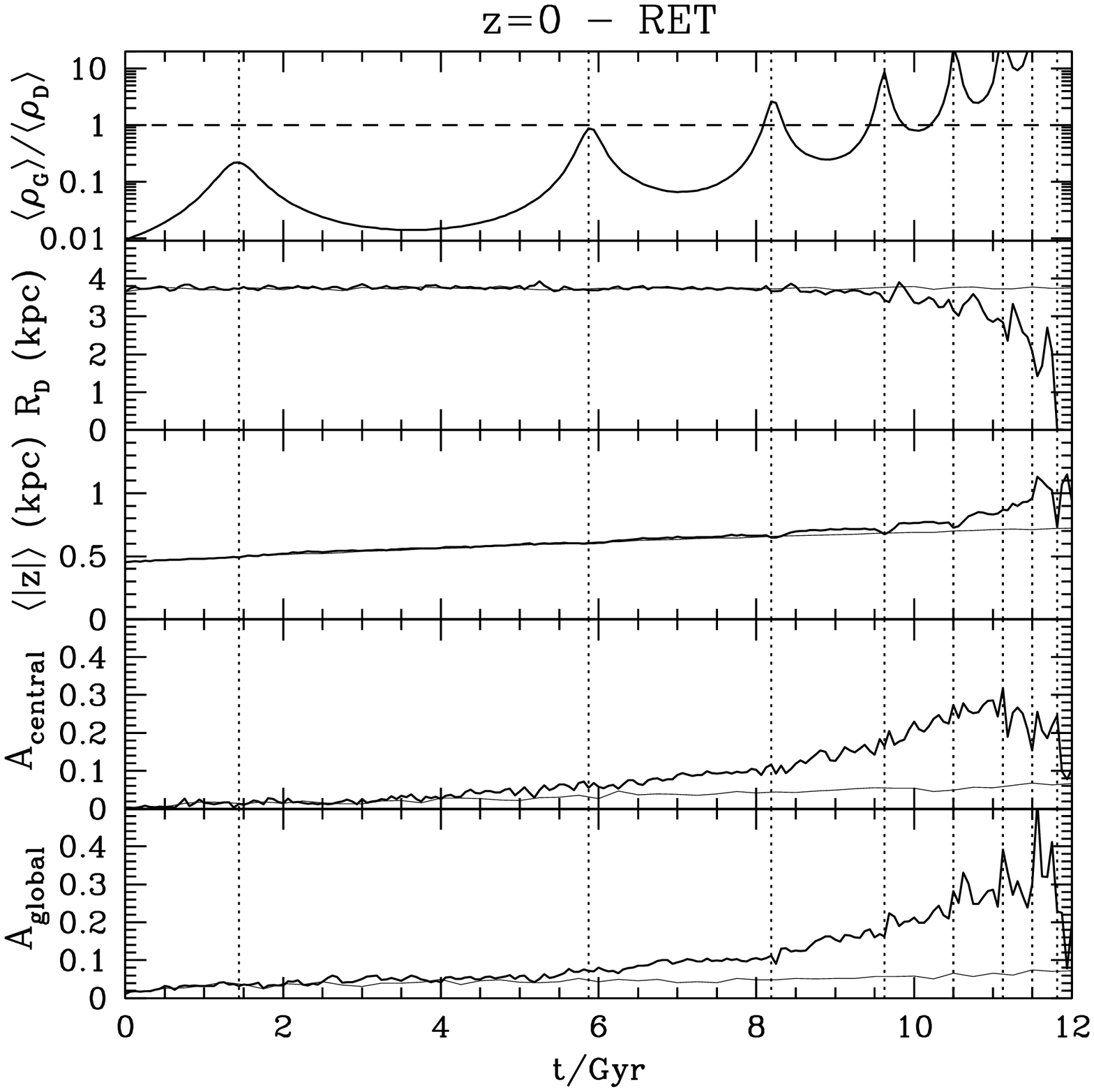}
\end{center}
\caption{
Evolution of the structure of disc galaxies in terms of their scale-lengths, 
$R_{\rm D}$, mean thickness, $\langle|z|\rangle$, and both central and global 
$m$=2 Fourier amplitudes, $A_{\rm central}$ and $A_{\rm global}$. The evolution 
of discs is shown along with the strength of the global tidal force acting upon 
them along their orbits, estimated as the group-to-galaxy mean density ratio, 
$\langle \rho_{\rm group} \rangle / \langle \rho_{\rm galaxy} \rangle$ (see 
details in the text). For comparison, the evolution of the structure of isolated 
disc galaxies (thin lines), and the times of pericentric passages (vertical 
dotted lines), are also included. For clarity, only the REF (reference), RAD 
(more eccentric infall), FON (face-on infall) and RET (retrograde infall) 
experiments at ``$z$=0'' are shown, while the rest of the experiments can be 
found in the Appendix.
}
\label{struct-correlations-z0}
\end{figure*}

\begin{figure*}
\begin{center}
\includegraphics[width=88mm]{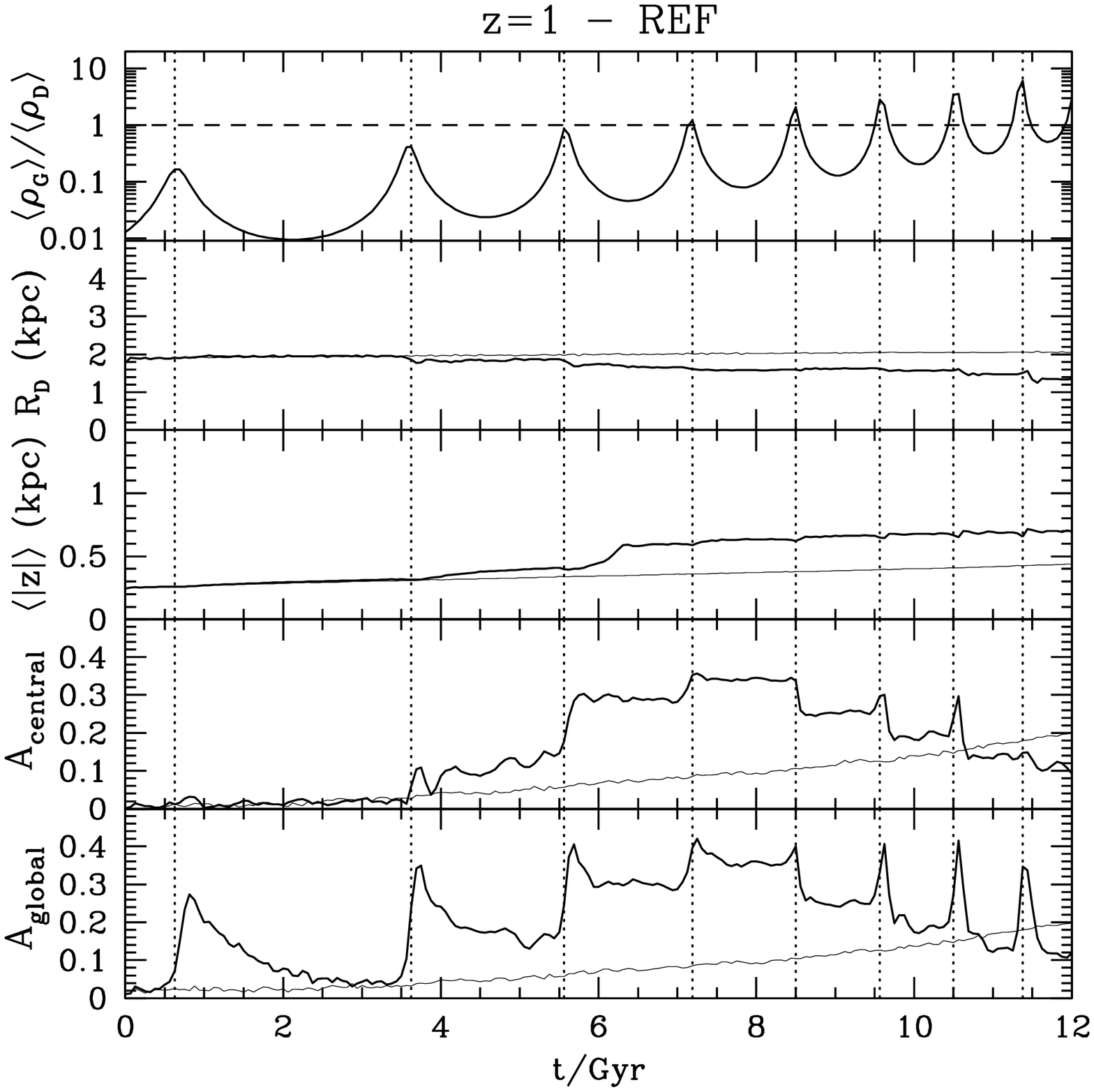}
\includegraphics[width=88mm]{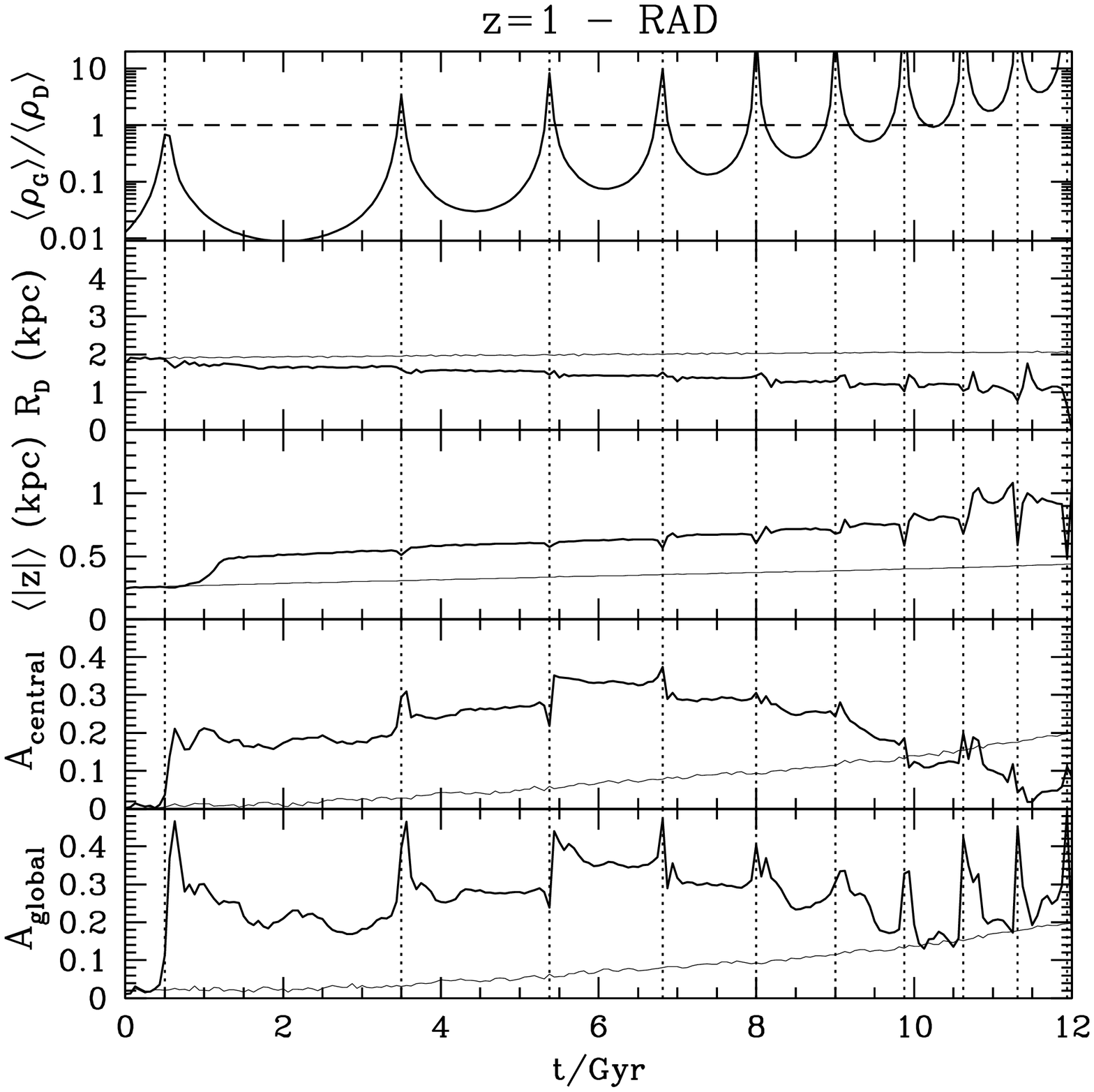}
\includegraphics[width=88mm]{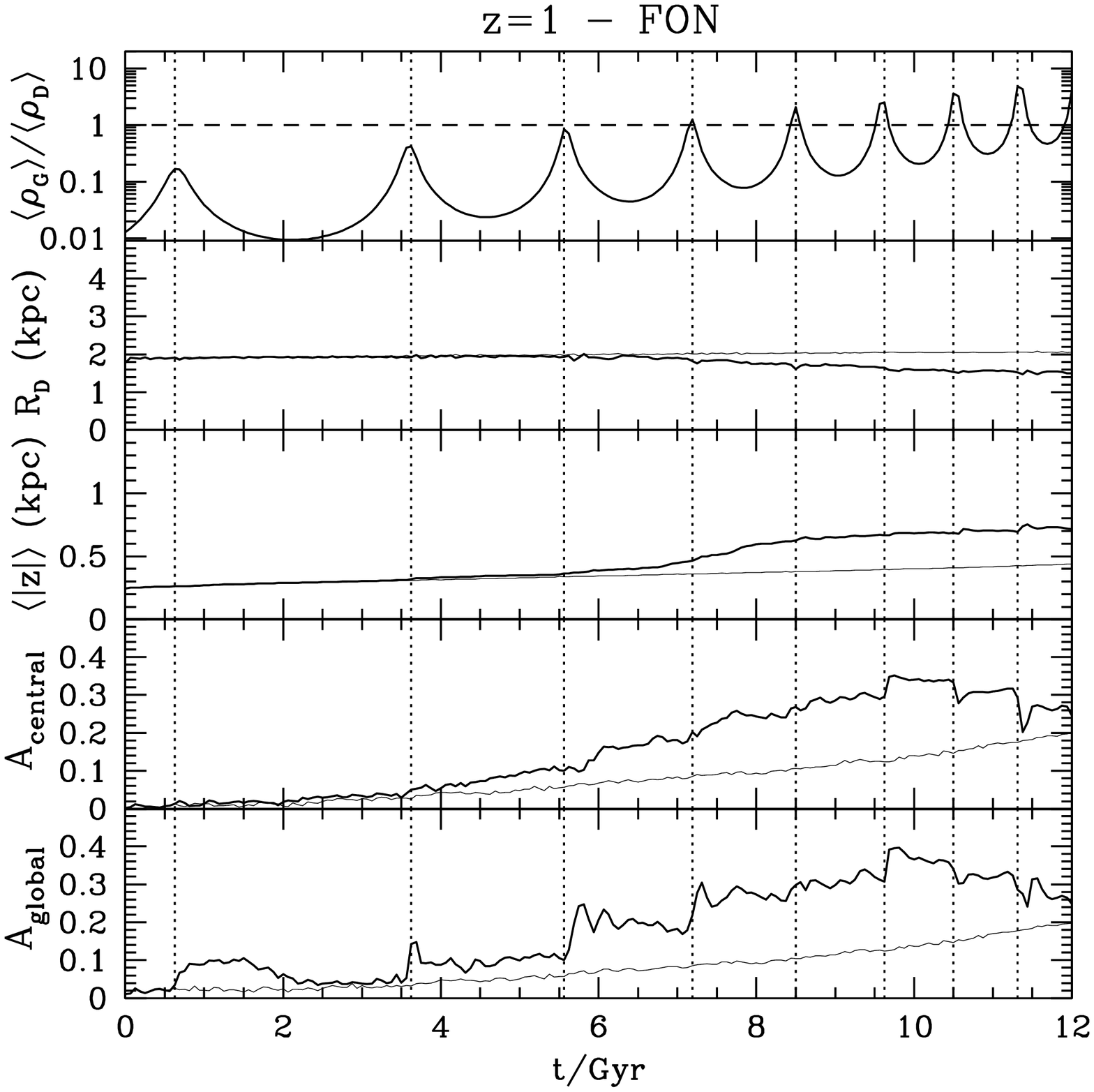}
\includegraphics[width=88mm]{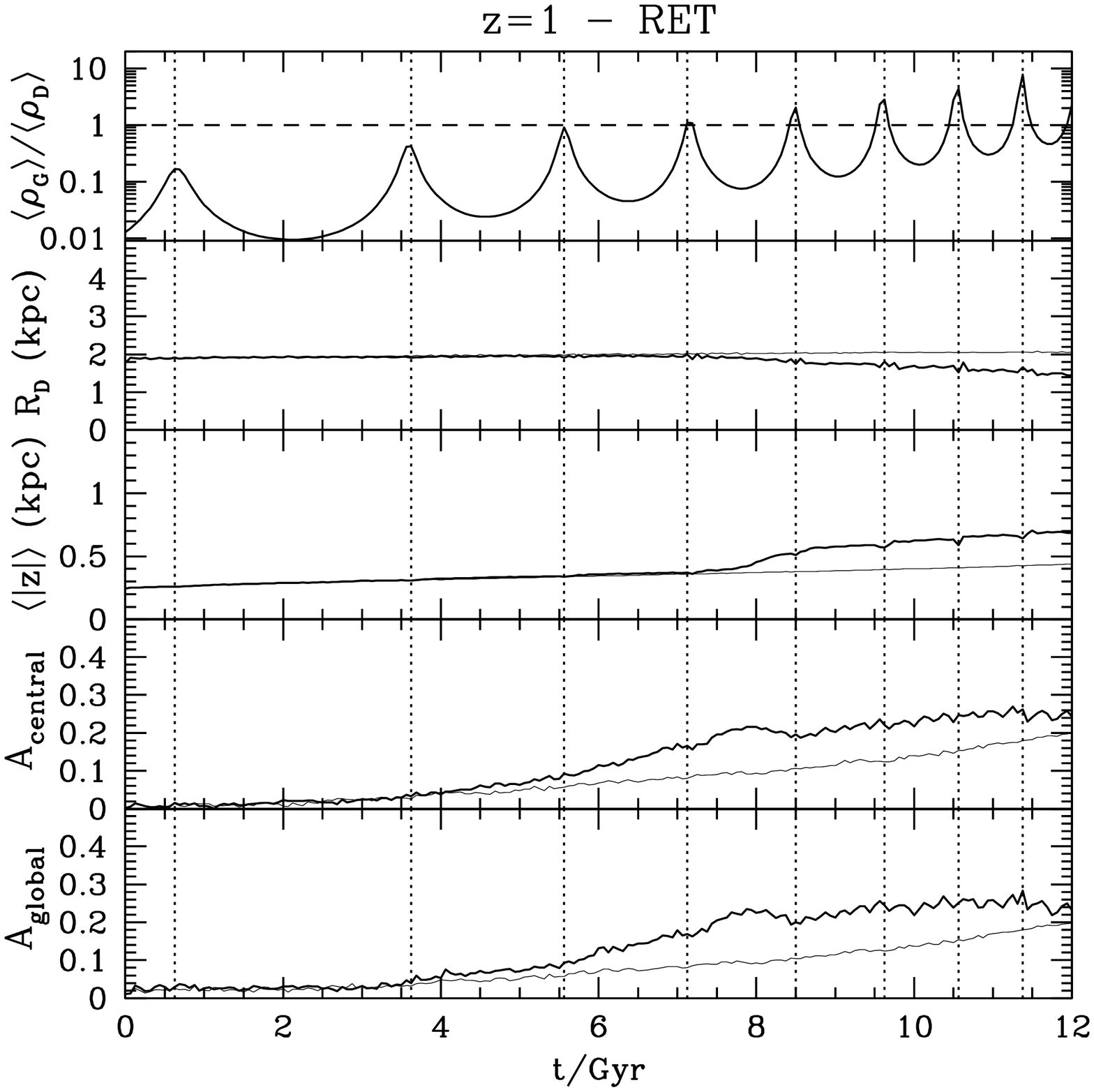}
\end{center}
\caption{Same as Fig.~\ref{struct-correlations-z0}, but for experiments at 
``$z$=1''.}
\label{struct-correlations-z1}
\end{figure*}

\begin{figure*}
\begin{center}
\includegraphics[width=88mm]{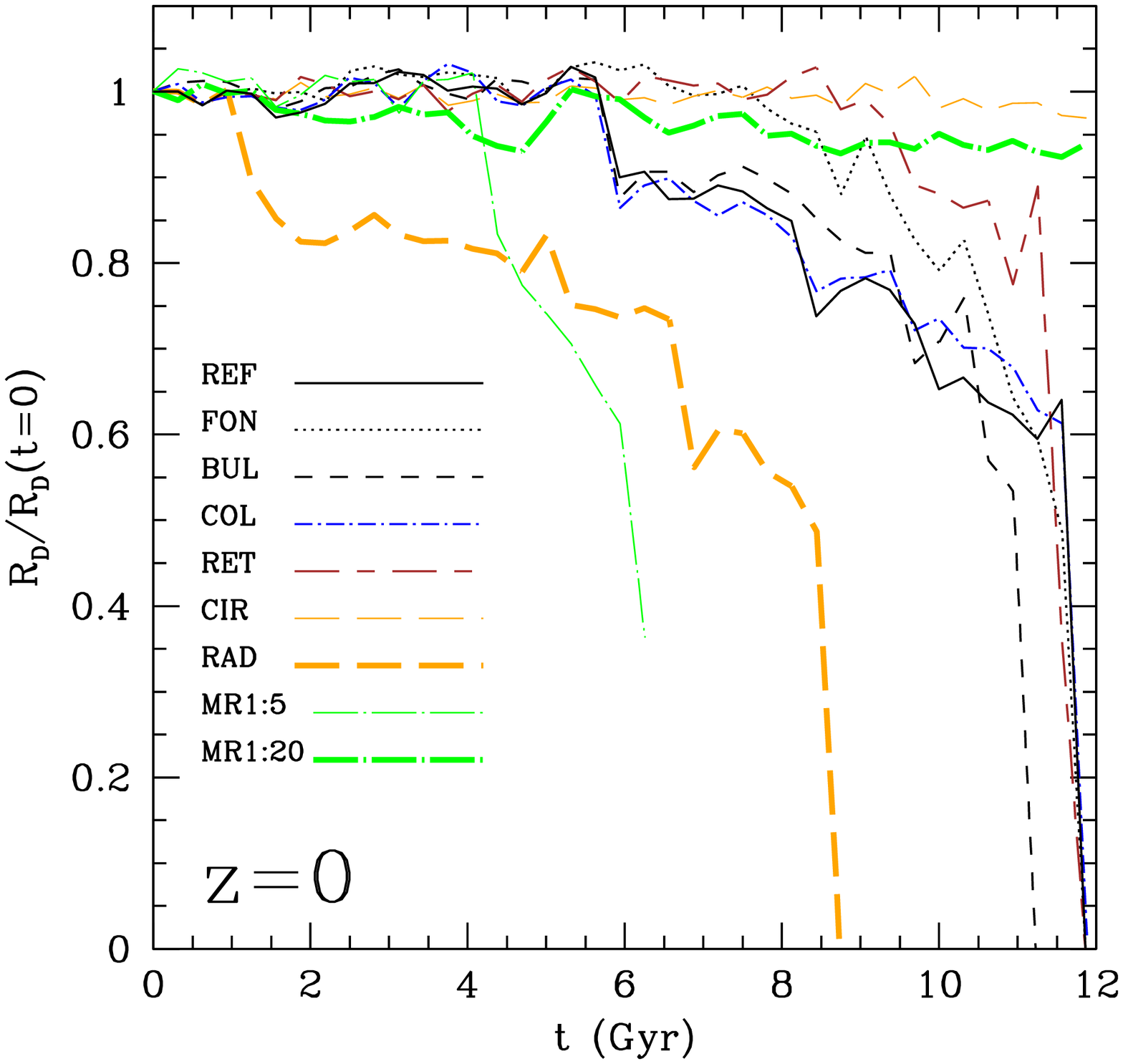}
\includegraphics[width=88mm]{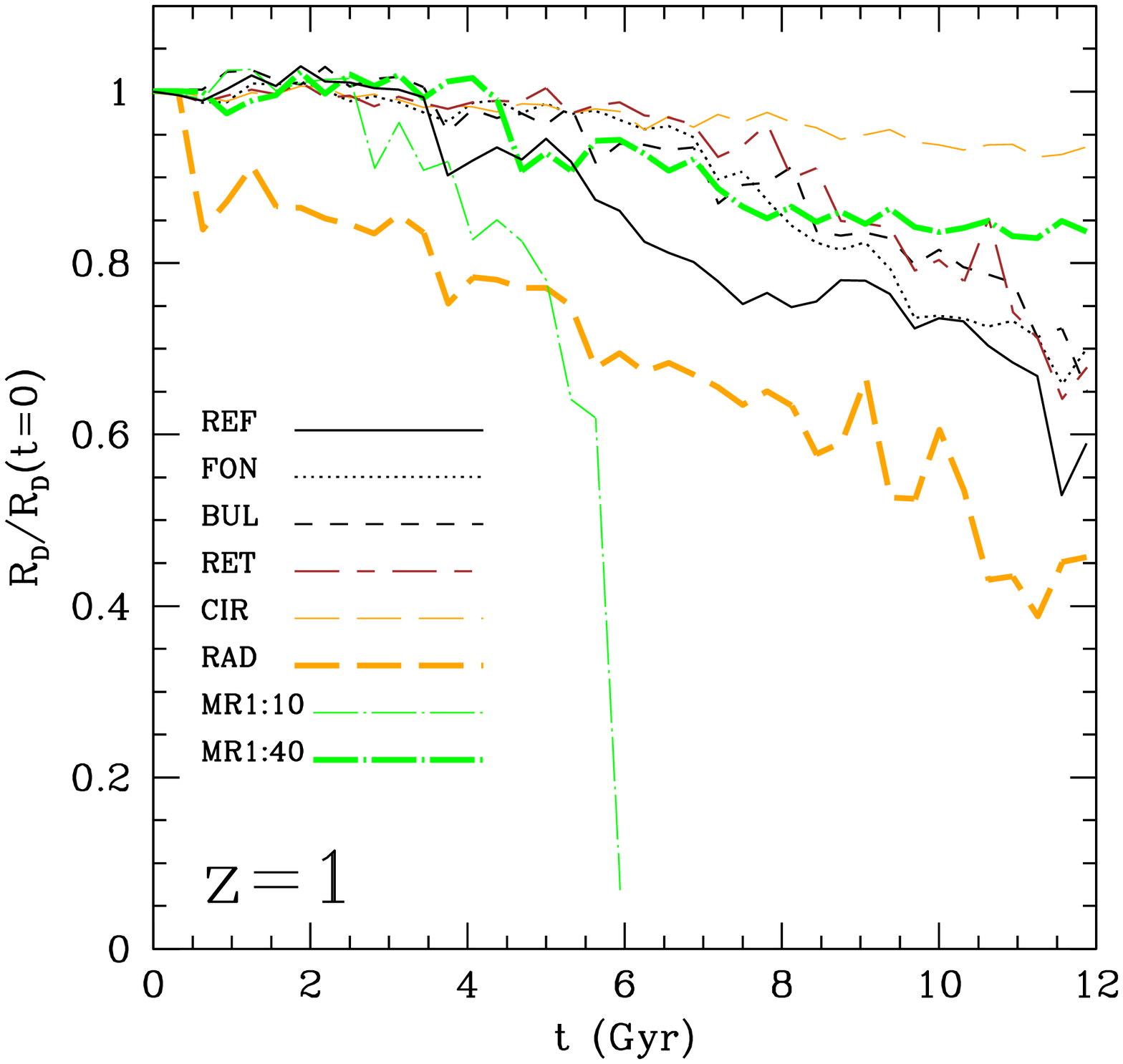}
\includegraphics[width=88mm]{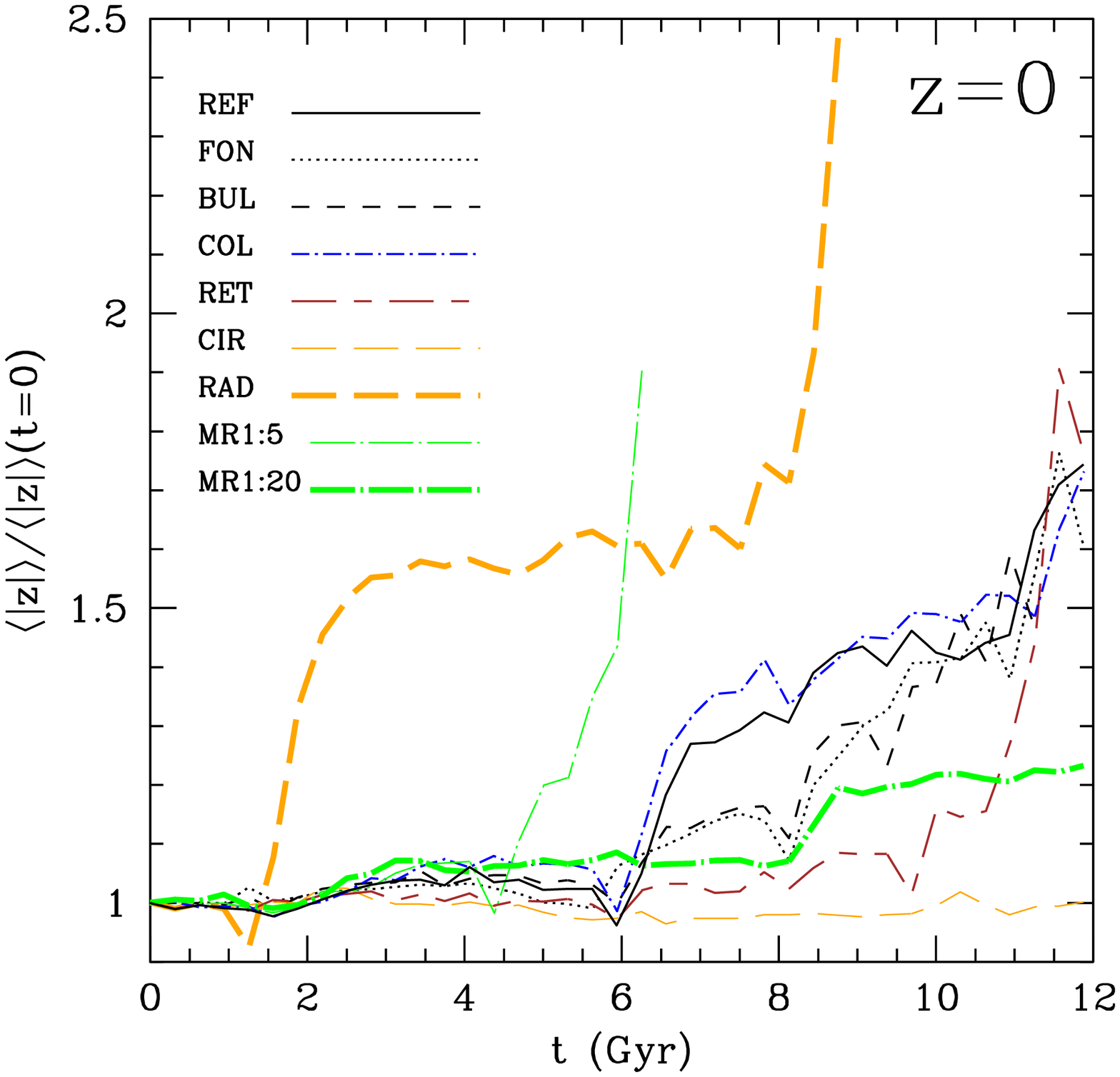}
\includegraphics[width=88mm]{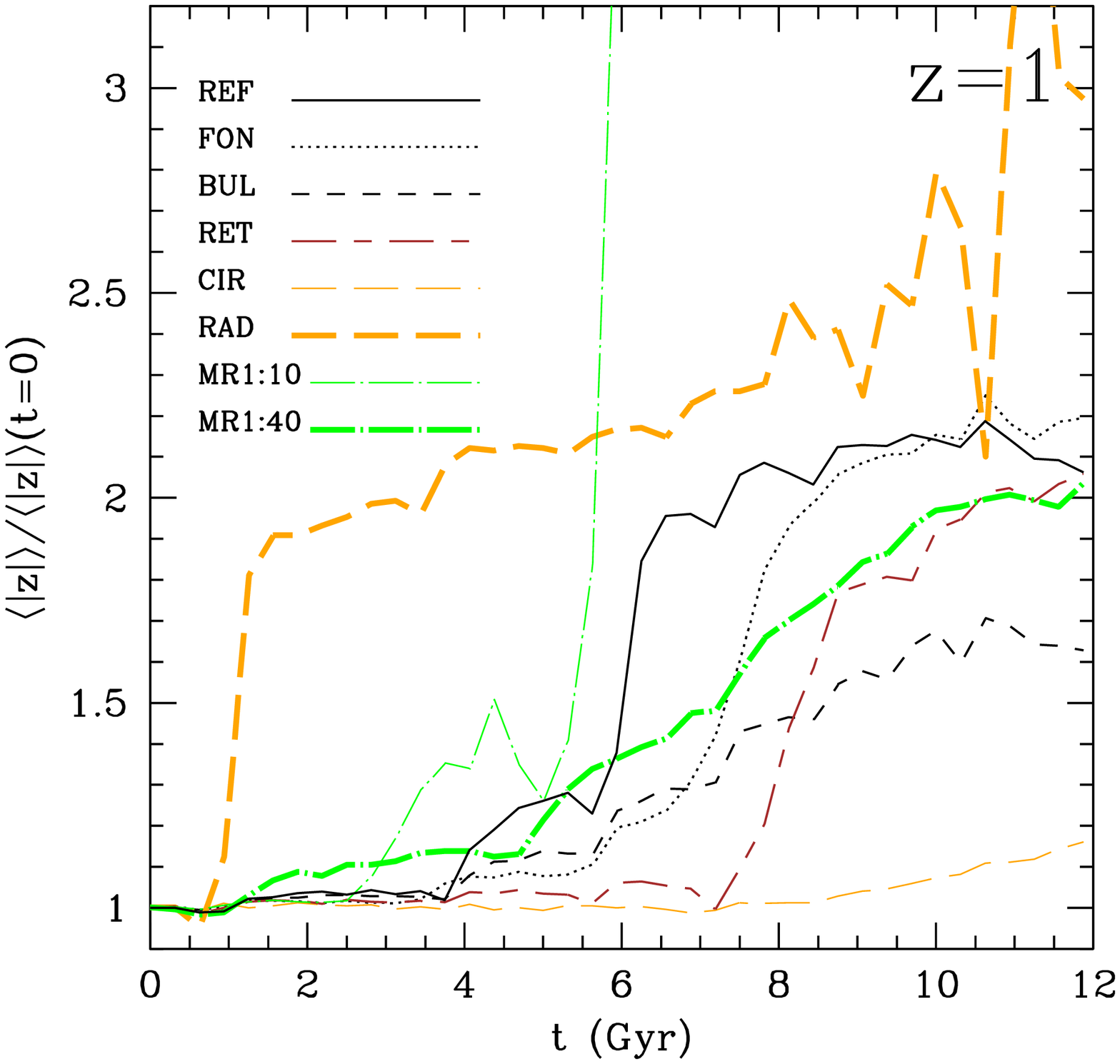}
\end{center}
\caption{Evolution of the scale-lengths and the mean thickness of discs 
galaxies for all our experiments at redshift epochs ``$z$=0'' and ``$z$=1''. 
Both the scale-lengths and mean thickness have been normalised by their 
corresponding initial values, and have been corrected by the evolution
of the respective disc in isolation.}
\label{structure-evol-z0-z1}
\end{figure*}

As shown in the previous section, disc galaxies go through significant 
evolution in their stellar content as they orbit within a group environment.

Figures~\ref{contours-xy-z0}--\ref{contours-xz-z1} show the evolution of the 
morphology of disc galaxies in the REF, FON and RET experiments at ``$z$=0'' 
and ``$z$=1''. These experiments explore different initial inclinations of 
the discs with respect to their orbital planes. Changes in the morphology 
are illustrated in terms of (equally-spaced) number density contours, 
considering only disc stars that remain bound at the respective pericentric 
passage. 

The Figures show striking differences in the morphology of discs after the 
first pericentric passage. Disc galaxies in the REF experiments are characterised 
by the formation of strong tidal arms, while in the RET experiments discs 
maintain their initial circular shape. Interestingly, discs in the FON 
experiments are an intermediate case, showing milder tidal arms than in REF, 
but conserving most of their initial shape, as in RET. Note that the first 
pericentric passage occurs after the same infall time and at the same 
distance from the group centre, for all three experiments at a given redshift. 
Therefore, all three discs are affected by the same global force. A very 
similar evolution is observed up to the fourth (sixth) pericentric passages 
at ``$z$=0'' (``$z$=1''). In general, discs in the FON and RET experiments
show a remarkable increasing resilience against the formation of tidal arms, 
with respect to discs in the REF experiments.

The different evolution of discs in prograde and retrograde orbits, regarding 
the formation of tidal arms, can be explained by the different net tangential 
velocities of stars on opposite sides of the disc during a pericentric passage.
In a prograde orbit, disc stars that are closer to the group centre have lower 
angular momenta than the disc centre of mass, with respect to the centre of 
the group, while stars on the opposite side of the disc have higher angular 
momenta. The combined effect of this different distribution of angular momenta 
and the global tidal field stretches the disc in opposite directions, inducing 
the formation of tidal arms. On the other hand, in a retrograde orbit, disc 
stars closer to the group centre have higher angular momenta than the disc 
centre of mass, and those on the opposite side of the disc have lower angular 
momenta, with respect to the group centre. In this case, the global tidal 
field is less efficient at stretching the disc, and the formation of tidal 
arms is inhibited.

Although the results discussed above are valid for disc galaxies orbiting 
in a group-size halo, it is interesting to note that the presence of spiral 
arms induced by the tidal field seems to be consistent with observations of 
spiral arms in unrelaxed populations of early-type dwarf galaxies in the 
Virgo cluster \citep[e.g., see][]{lisker2007}.

Figures~\ref{contours-xy-z0}--\ref{contours-xz-z1} also show a quantitative 
difference in the central bars formed in each case. They appear stronger for 
discs in the REF experiments, and progressively milder in the FON and RET 
experiments. It is interesting to notice that the relative strength of the 
central bars seems to correlate with the thickening generated in the discs 
after subsequent pericentric passages. In general, discs in the RET 
experiments remain significantly thinner than those in the REF experiments, 
while in the FON experiments the initial vertical structure of discs is 
considerably altered by a noticeable warping during the first pericentric 
passages, especially after the third one.

Finally, regarding the rest of the experiments, our simulations show that all 
disc galaxies on prograde orbits suffer morphological changes that are rather 
similar to those seen in the REF experiments. This evolution is found to be 
independent of the galaxy-to-group mass ratio, initial orbital parameters, 
presence of bulge or disc kinematics. The timescales at which these morphological 
changes occur, however, are directly related to the efficiency of the dynamical 
friction affecting the disc in each case. That is, the morphological evolution 
is found to be faster for galaxies with more eccentric initial orbits and/or 
with larger galaxy-to-group mass ratios.

Summarising, the morphological changes experienced by disc galaxies orbiting 
within a group are characterised by the formation of tidal arms and a central 
bar, induced by the global tidal field during pericentric passages. Our 
simulations show that the significance of these morphological features is 
closely related to the initial inclination of the discs with respect to their 
orbital planes. The morphological evolution is found to be stronger for discs 
with inclination 0\degr, and weaker for discs with inclination 180\degr. In 
the latter case, discs are able to retain their initial morphology (especially 
their thickness) after several pericentric passages. The morphology of disc 
galaxies with initial inclination 90\degr are typically affected by a strong 
warping during the first pericentric passages. All disc galaxies in prograde 
orbits are found to suffer similar morphological changes, and the timescales 
of these changes are directly linked to the efficiency of the dynamical 
friction acting on the discs.

\subsection{Disc structure}
\label{sec-struc}

In order to study the structural changes of disc galaxies, as they orbit 
within a group environment, we have first centred the discs and aligned them 
in such a way that the Z-axis is defined by their rotation axes. Then, their 
structural properties have been computed in concentric rings, 1~kpc wide (out 
to 20~kpc from the disc centre for the ``$z$=0'' experiments, and out to 10~kpc
for the ``$z$=1'' experiments), considering only stars that remain bound to the 
disc galaxy at a given time, and that are located within 3~kpc from the 
midplane\footnote{No significant variations in either the structural or 
the kinematical properties of discs are found for different (reasonable) 
vertical limits.}.

Figure~\ref{lnsigma-thickness-evol-ref-z0-z1} shows the evolution of the 
surface mass density profiles $\Sigma(R)$ of discs in the REF experiments at 
redshift ``$z$=0'' and ``$z$=1'', after each of their first six pericentric 
passages. The profiles show that discs start losing stars from the outskirts 
first. This is expected since those stars have lower binding energies and can 
be more easily stripped off, especially after strong tidal interactions 
between the disc and the central region of the group during pericentric passages.
On the other hand, there is an increase of the surface mass density in the 
central region of the discs (most notably at ``$z$=0''), due to the accumulation 
of mass that forms a central non-axisymmetric feature. Interestingly, the 
surface mass density profiles stay fairly exponential (seen as a straight 
line in log-linear scale) during the evolution of the discs, in spite 
of the central feature formed at later times. In our experiments, we have 
computed the scale-length of discs by applying a linear fit to $\ln \Sigma(R)$. 
We have considered only bound stars out to 0.75$R_{\rm D}$ (where $R_{\rm D}$ 
is the initial disc scale-length), in order to avoid spurious measurements 
due to tidal arms. The evolution in the radial structure of the stellar discs, 
during their infall within a group environment, is characterised by a steady 
decrease in the scale-length of their mass distributions. Note that the 
decrease in the surface mass density at larger radii appears more significant 
at ``$z$=0'' than at ``$z$=1''. This is consistent with the faster mass loss 
rate observed during pericentric passages at ``$z$=0'', as described in
Section~\ref{sec-massloss}.

Figure~\ref{lnsigma-thickness-evol-ref-z0-z1} also shows the evolution in the 
thickness of the stellar discs as a function of radius, after each of their 
first six pericentric passages, for the REF experiments at redshift ``$z$=0'' 
and ``$z$=1''. At each radius, the thickness is expressed as $\langle |z| \rangle$.
As discs orbit within the group environment, they are subject to significant 
thickening that is not uniform with radius, but more pronounced towards the 
outskirts. In the REF experiments, most of the evolution in the thickening
of the discs is seen before the third (fourth) pericentric passage at ``$z$=0'' 
(``$z$=1''). In fact, we shall see that the pericentric passage at which the 
first significant change in the thickening takes place is clearly correlated 
to the appearance of a central non-axisymmetric feature in the disc. Since the 
thickening increases with radius, we have adopted a single \emph{global} 
thickening in order to compare discs of different experiments. This global 
thickening has been defined as a ``mean'' thickening averaged over radial bins, 
and weighted by the number of stars in each bin.

Interestingly, our simulated discs resemble the unrelaxed populations of 
early-type dwarf galaxies observed in the Virgo cluster, which have exponential 
brightness profiles and oblate intrinsic shapes \citep{lisker2007}. However, as 
mentioned above, our simulations might not be directly comparable to those 
observations.

In the context of our simulations, the evolution of disc galaxies is exclusively 
driven by tidal interactions with the global potential of the group environment. 
This motivates us to examine the structural changes in the discs with respect to 
the strength of the global tidal forces acting upon them. 
Figures~\ref{struct-correlations-z0} and \ref{struct-correlations-z1} show the 
evolution of the scale-length and scale-height of discs, and of the amplitudes 
of their central and global non-axisymmetries, comparing them to the evolution 
of the global tidal force on the disc galaxies along their orbits. This is shown 
for the following representative experiments: REF, RAD, FON and RET, at 
redshift epochs ``$z$=0'' and ``$z$=1''. Also included are the times at which 
the pericentric passages take place, as well as the structural evolution of 
the discs in isolation (scale-length, scale-height and amplitudes of 
non-axisymmetries), that is, without the interaction with the group environment.
The strength of the global tidal force acting on discs along their orbits is 
estimated as the group-to-galaxy mean density ratio, 
$\langle \rho_{\rm group} \rangle / \langle \rho_{\rm galaxy} \rangle$.
Both $\langle \rho_{\rm group} \rangle$ and  $\langle \rho_{\rm galaxy} \rangle$ 
are computed considering both DM and stellar mass, the former within the 
disc's orbit from the centre of the group, and the latter within 10$R_{\rm D}$ 
from the disc centre. The amplitudes of the $m$=2 non-axisymmetries are 
computed by binning the spatial distribution of bound disc stars in cylindrical 
shells of 1~kpc width, out to $\sim$6$R_{\rm D}$. In each bin, the second 
harmonics of angular distribution are computed as:
\begin{equation} \label{fourier-comp}
a_2 = \frac{1}{N} \sum^{N}_{i=1} \sin (2\phi_i), \qquad b_2 = \frac{1}{N} \sum^{N}_{i=1} \cos (2\phi_i),
\end{equation}
where in each bin, $N$ and $\phi_i$ are the number of stars and their angular 
position, respectively. The amplitude of the $m$=2 harmonic, in each bin, is 
$A^2=(a_2^2+b_2^2)/2$. We define the global amplitudes as 
$A_{\rm global} = \sqrt{\langle A(R)^2 \rangle}$, averaged over radial bins, 
and the central amplitudes as $A_{\rm central} = A(R)$, for stars within 
$R<6R_{\rm D}$.

Figures~\ref{struct-correlations-z0} and~\ref{struct-correlations-z1} show 
the very close relation between the structural evolution of the discs, and 
the evolution of the global tidal force acting on them along their orbits. 
In general, as discs orbit within the group potential well, the relative 
strength of the global tidal force, $\langle \rho_{\rm group} \rangle / \langle \rho_{\rm galaxy} \rangle$,
varies along the orbit, starting from a small tidal force at the virial radius 
of the group, where $\langle \rho_{\rm group} \rangle \approx 0.01 \langle \rho_{\rm galaxy} \rangle$.
As discs approach their first pericentric passage, $\langle \rho_{\rm group} \rangle / \langle \rho_{\rm galaxy} \rangle$
spikes because of the rapid increase in the mean density of the group within 
the disc's orbit. By the time $\langle \rho_{\rm group} \rangle \approx \langle \rho_{\rm galaxy} \rangle$
(or in the case of experiments at ``$z$=1'', $\langle \rho_{\rm group} \rangle \approx 0.3$--$0.5\langle \rho_{\rm galaxy} \rangle$),
the global tidal force is strong enough to trigger significant transformations 
in the structure of discs. In particular for the REF experiments, the 
enhanced mass loss at this point (see Figure~\ref{massbound-evol}) leads to a 
sharp decrease in the scale-lengths of discs. At the same time, the formation 
of central non-axisymmetries (possibly triggered by the generation of tidal 
arms, according to the amplitude of global non-axisymmetries), heats the discs 
vertically, increasing their scale-heights. When
$\langle \rho_{\rm group} \rangle < \langle \rho_{\rm galaxy} \rangle$, the 
global tidal force at pericentres is only able to excite the formation of 
global non-axisymmetries, and of negligible central non-axisymmetries. On the 
other hand, when $\langle \rho_{\rm group} \rangle > \langle \rho_{\rm galaxy} \rangle$, 
the discs experience a series of step-like decreases in their scale-lengths, 
and step-like increases in their scale-heights, at subsequent pericentric 
passages. These step-like changes in the structure of discs highlight the fact 
that most of the evolution suffered by discs takes place at the pericentres.
This is expected since at those points along the orbits of discs, the impulsive 
accelerations on them are the strongest.

Our finding that discs start suffering significant structural transformations 
after $\langle \rho_{\rm group} \rangle \gtrapprox \langle \rho_{\rm galaxy} \rangle$
agrees \emph{only partially} with the recipe for tidal disruption used in the 
semi-analytic model of galaxy evolution by \citet{guo2011}. The following are 
some important differences between this recipe and the results of our simulations. 
First, this criterion is applied only after the galaxy DM halo has been completely 
disrupted. Our simulations show that by the time discs are affected by tidal 
forces, their DM halos still retain a non-negligible fraction of their initial 
mass. Second, their estimation of $\langle \rho_{\rm galaxy} \rangle$ only 
considers baryonic material, while ours takes into account both DM and stellar 
mass. Lastly, \citeauthor{guo2011} assume that the galaxy is completely 
disrupted when the criterion is satisfied. Our simulations show that this 
recipe can underestimate significantly the disruption time of galaxies, since 
they can survive for a much longer time after they start being affected by 
the tidal forces of the global field.

The rest of the experiments at ``$z$=0'' show a behaviour that is qualitatively 
similar to that of the REF experiment. The specific differences between them 
are, firstly, how rapidly along their orbits the discs are dragged into regions 
where $\langle \rho_{\rm group} \rangle \approx \langle \rho_{\rm galaxy} \rangle$; 
and secondly, how much the structure of discs responds to the strong tidal 
forces, based on the initial inclination of their orbits. In the RAD experiment, 
given that the disc has a more eccentric infalling orbit in comparison to the 
disc in the REF experiment, it experiences changes in its structure 
significantly sooner along the orbit. In particular, it reaches the central
region of the group where $\langle \rho_{\rm galaxy} \rangle \approx \langle \rho_{\rm group} \rangle$
during its first pericentric passage. The disc in the RET experiment has a 
structural evolution significantly different from that of the REF experiment, 
despite having a very similar infalling orbit. It shows an almost complete 
absence of both global and central non-axisymmetric features correlated with
pericentric passages. The former can be associated to a slower mass loss rate
(see Figure~\ref{massbound-evol}), and to a slower decrease in the disc scale-length.
The latter can be linked to the formation of a weak central bar that does not
increase the disc scale-height dramatically. The characteristics of the disc 
in the FON experiment lie in between the two extreme behaviours described for 
the REF and RET experiments, thus providing an intermediate case for the 
structural evolution of discs.

Figure~\ref{structure-evol-z0-z1} shows a direct comparison of the evolution of 
the scale-lengths and scale-heights of discs, for all the experiments at ``$z$=0'' 
and ``$z$=1''. We find that the experiments at ``$z$=0'' that evolve the most, 
in terms of scale-lengths, are RAD and MR1:5. In those cases, the initially 
more eccentric orbit and the larger galaxy-to-group mass ratio are linked to a 
more efficient dynamical friction, that drags the galaxies towards the centre
of the group on shorter timescales. In these discs, the scale-lengths decrease 
by $\sim$50 per cent right before reaching the group centre at 
$t\sim$6.5-8.5~Gyr. On the other hand, the discs that evolve the least are found 
in the CIR and MR1:20 experiments, in which the discs spend most of the time at 
the outskirts of the group, away from the denser central region. In those cases, 
the scale-lengths show a decrease of less than 10 per cent over 12~Gyr. The rest 
of the experiments show an intermediate evolution where the scale-lengths 
decrease by $\sim$60 per cent before they reach the group centre at 
$t\sim$11.5-12~Gyr. However, some of these experiments do show some different 
evolution while discs are infalling. The scale-lengths in the FON and RET 
experiments change less over time with respect to REF experiments, which is 
consistent with the different ways these discs generate tidal arms and lose 
their bound mass. On the other hand, the scale-lengths of discs in the BUL 
and COL experiments show practically no differences in comparison to the REF 
experiment.

The evolution of the discs' scale-lengths in the experiments at ``$z$=1'' 
show some interesting differences with respect to those at ``$z$=0''. Discs 
in the RAD and MR1:10, CIR, and REF experiments still exhibit the largest, 
smallest and intermediate evolution in their scale-lengths, respectively.
The scale-lengths in the rest of the experiments show almost no differences
among them, and in general, they present less evolution in comparison to the 
REF experiment. This lack of differentiation between the scale-lengths of most 
of the experiments at ``$z$=1'' could be due to the relatively smaller tidal 
disruption of the discs by a less concentrated group environment in comparison 
to that at ``$z$=0''.

In terms of disc scale-heights, the experiments at ``$z$=0'' that present the 
most evolution are RAD and MR1:5, showing up to $\sim$50 per cent increase by 
the time the discs reach the group centre. This is consistent with what is 
observed for the disc scale-lengths. At the opposite extreme, the CIR experiment 
shows almost no thickening of the disc during 12 Gyr of evolution. Similarly 
to the scale-length evolution, the rest of the experiments, except MR1:20, 
exhibit roughly the same increase in the disc scale-heights, 50-60 per cent, 
right before the time they reach the group centre. Before that time, these 
experiments do show some differences in the amount of disc thickening. Among 
them, the RET experiment appears to be most resistant to thickening, in 
comparison to the REF and COL experiments, followed by discs in the FON and 
BUL experiments.

In the experiments at ``$z$=1'', the disc thickening shows more differences 
between the various cases, in comparison to the scale-length evolution. While 
the largest increase in the scale-heights are in the RAD and MR1:10 experiments, 
with roughly the double mean thickness by $t\sim$6~Gyr, the CIR experiment 
shows no thickening over 8~Gyr of evolution. The rest of the experiments have 
increments in their disc scale-height of $\sim$50-100 per cent between 
$t\sim$8-12~Gyr. Note that the presence of a central bulge does reduce the 
amount of thickening in the disc, with respect to the REF case. Similarly, 
discs in the FON and the RET experiments are also found to retain their 
original thickness for a longer time, compared to the REF experiment.

In summary, the evolution in the radial mass distribution of discs orbiting 
within a group environment is characterised by a steady decline in their 
scale-lengths, caused by the continuous mass loss from the outskirts of the 
discs and the accumulation of mass in their central regions. Interestingly,
the radial mass distributions of discs remain exponential, in spite of this 
evolution. Additionally, disc galaxies also present a significant thickening 
of their vertical structure. This thickening is found to take place only after 
the formation of a central non-axisymmetric feature in the discs. Our 
simulations also show a close relation between the structural evolution of 
discs (in terms of their scale-length, thickness, and the formation of central 
features), and the strength of the global tidal force acting on them. It is 
found that the galaxy-to-group mass ratio, the initial inclination of the 
discs with respect to their orbits, and their initial orbital eccentricities 
are among the most important factors governing the structural evolution of discs.   

\subsection{Disc kinematics}
\label{sec-kine}

\begin{figure*}
\begin{center}
\includegraphics[width=50mm]{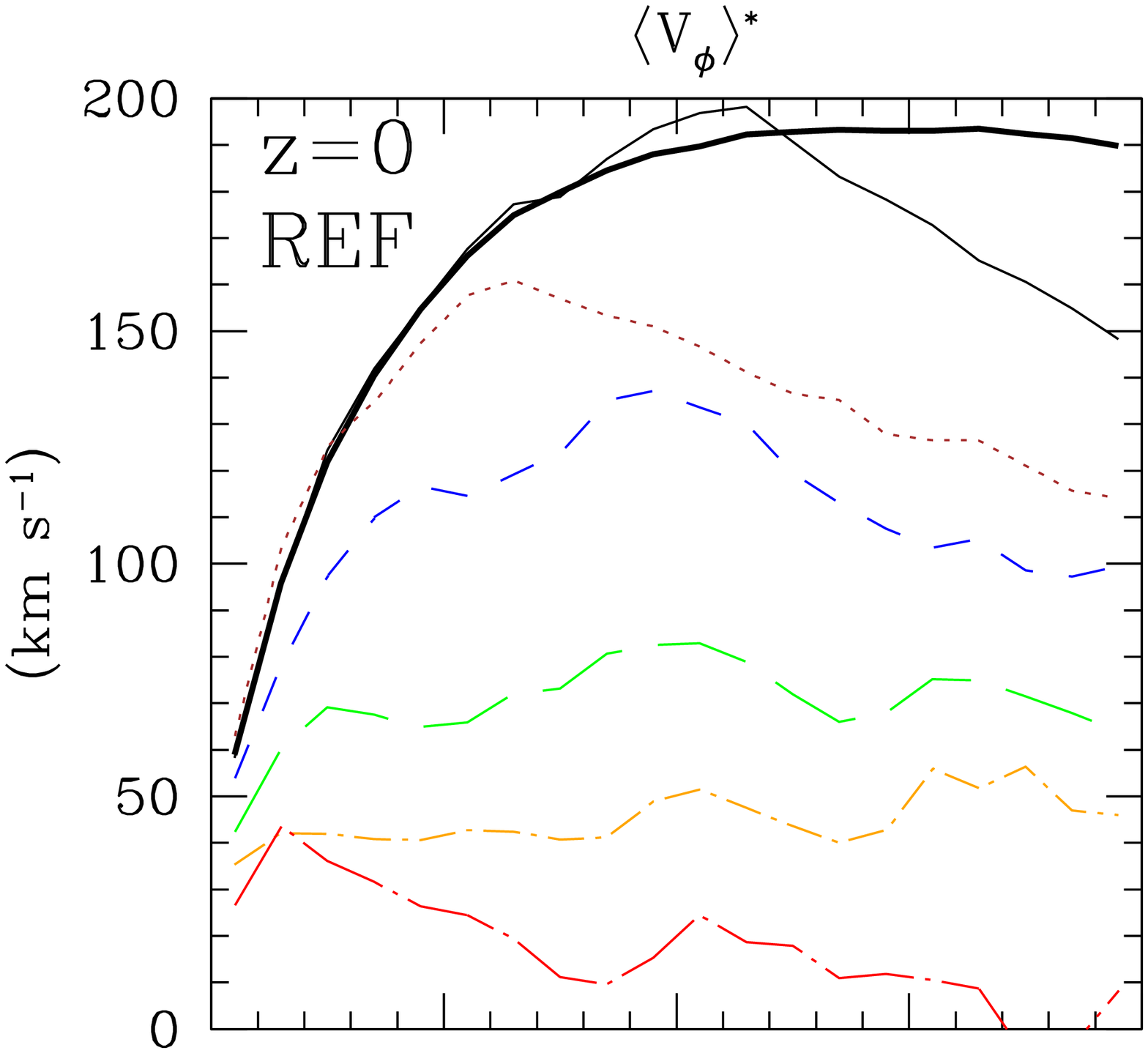}\hspace*{-0.4cm}
\includegraphics[width=50mm]{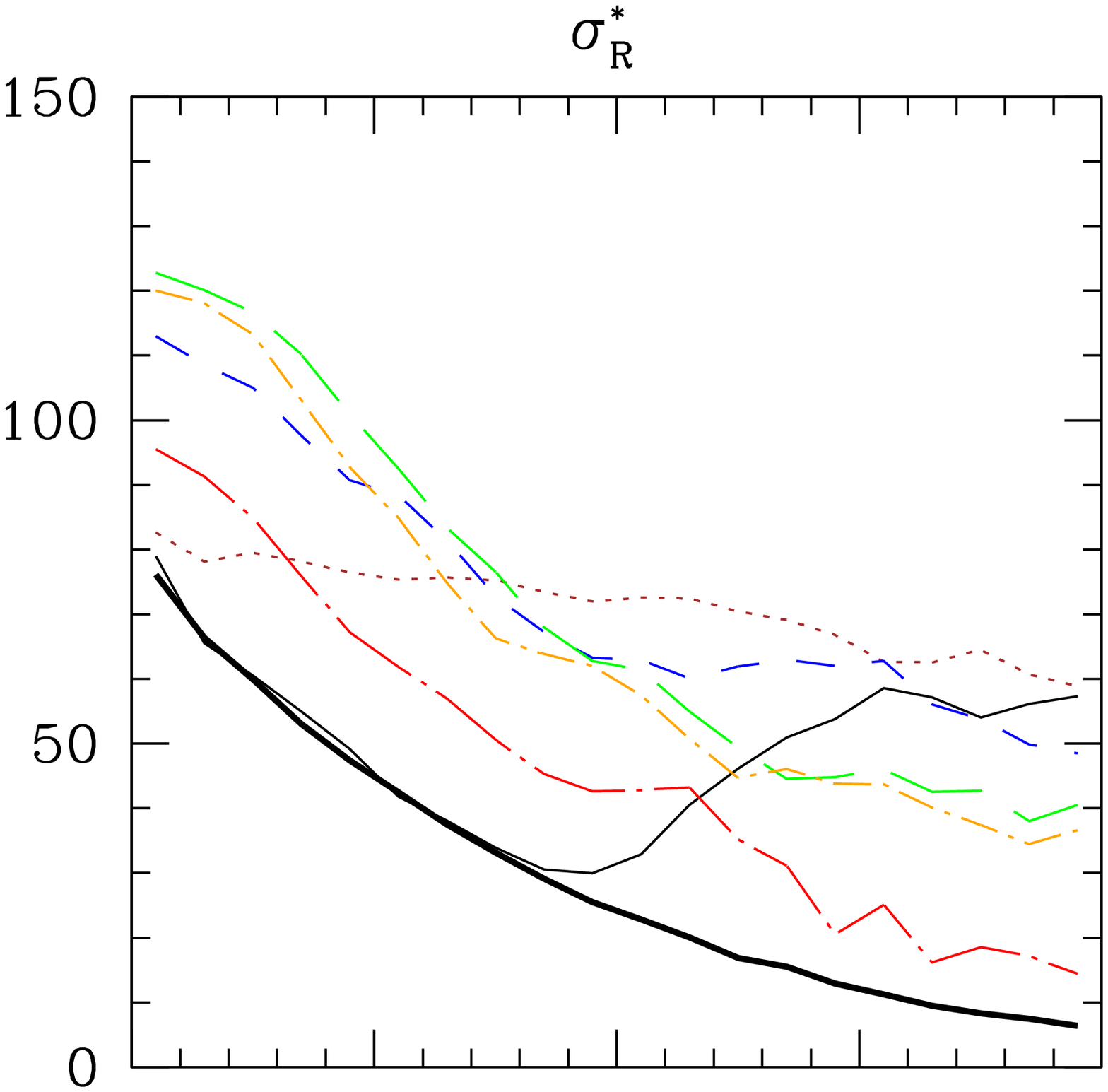}\hspace*{-1.36cm}
\includegraphics[width=50mm]{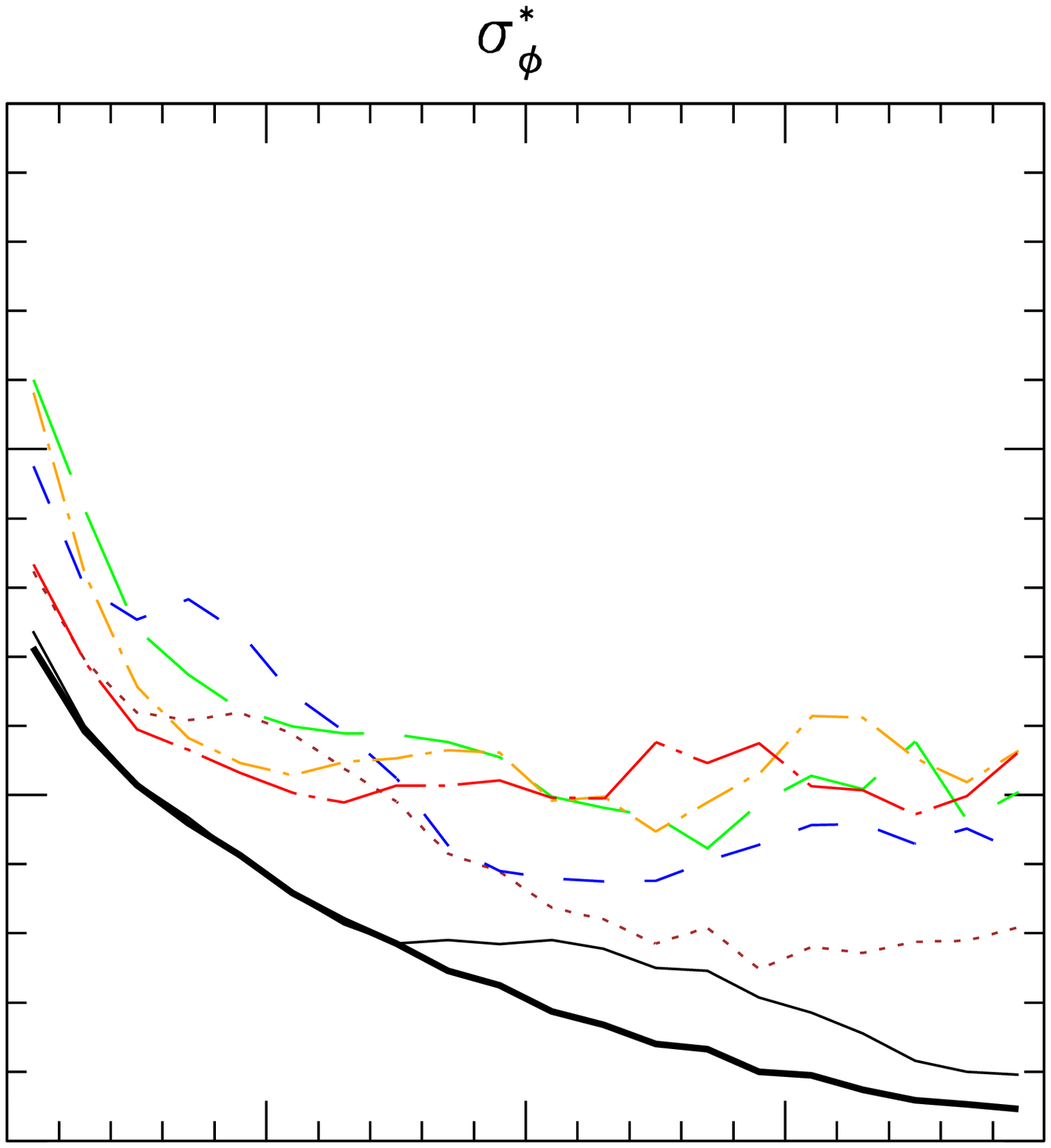}\hspace*{-1.36cm}
\includegraphics[width=50mm]{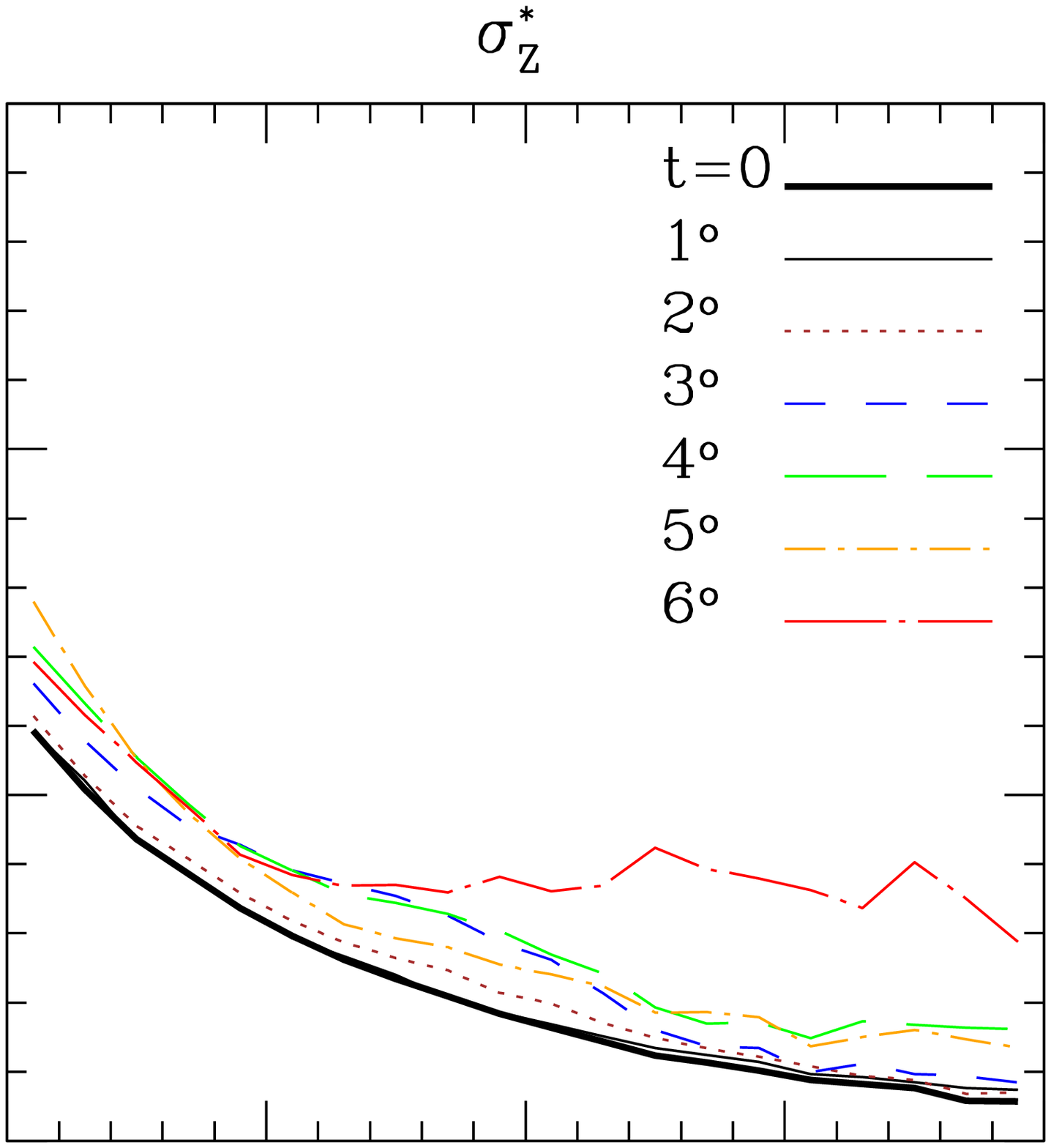}\vspace*{-1.29cm}\\
\includegraphics[width=50mm]{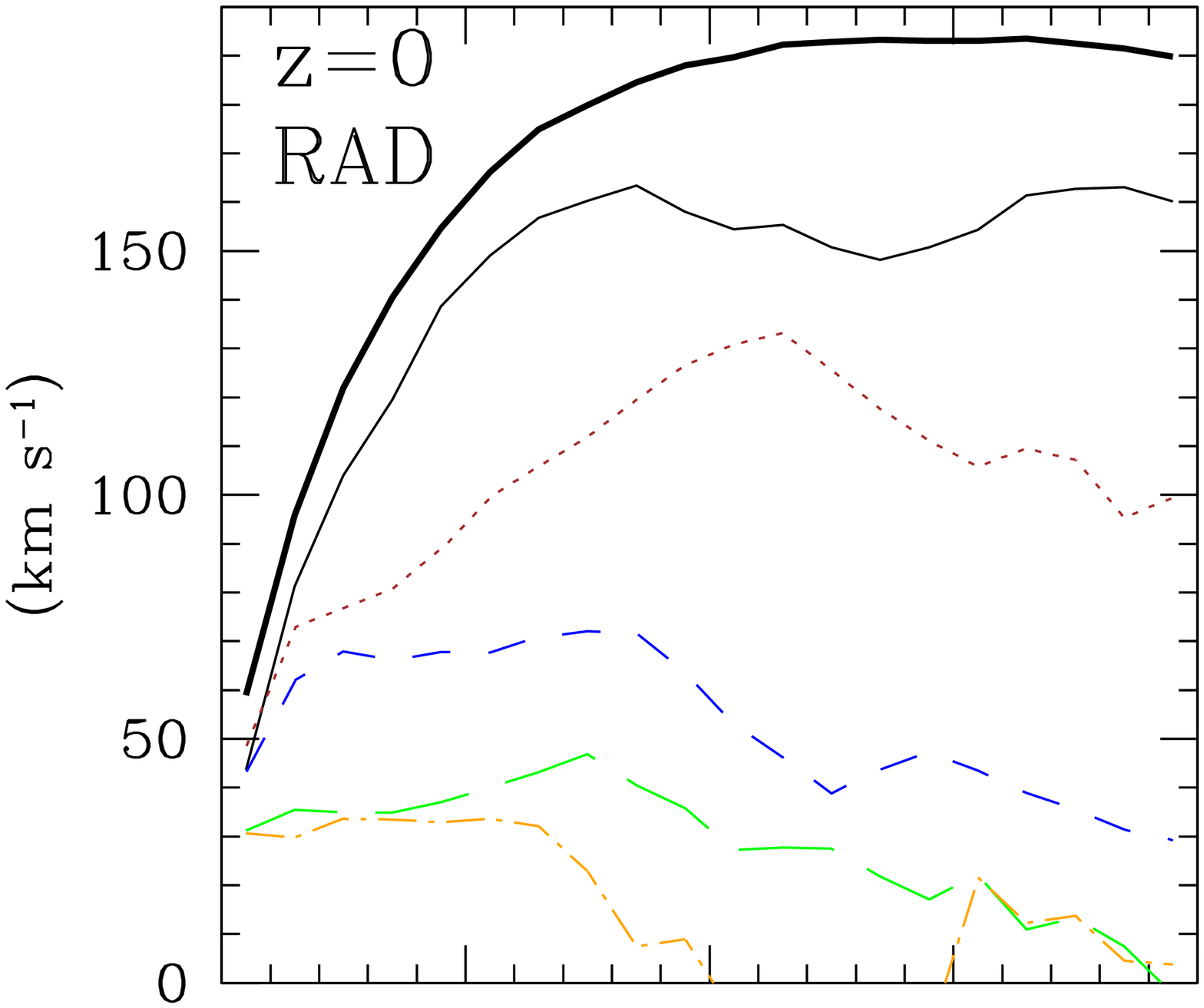}\hspace*{-0.4cm}
\includegraphics[width=50mm]{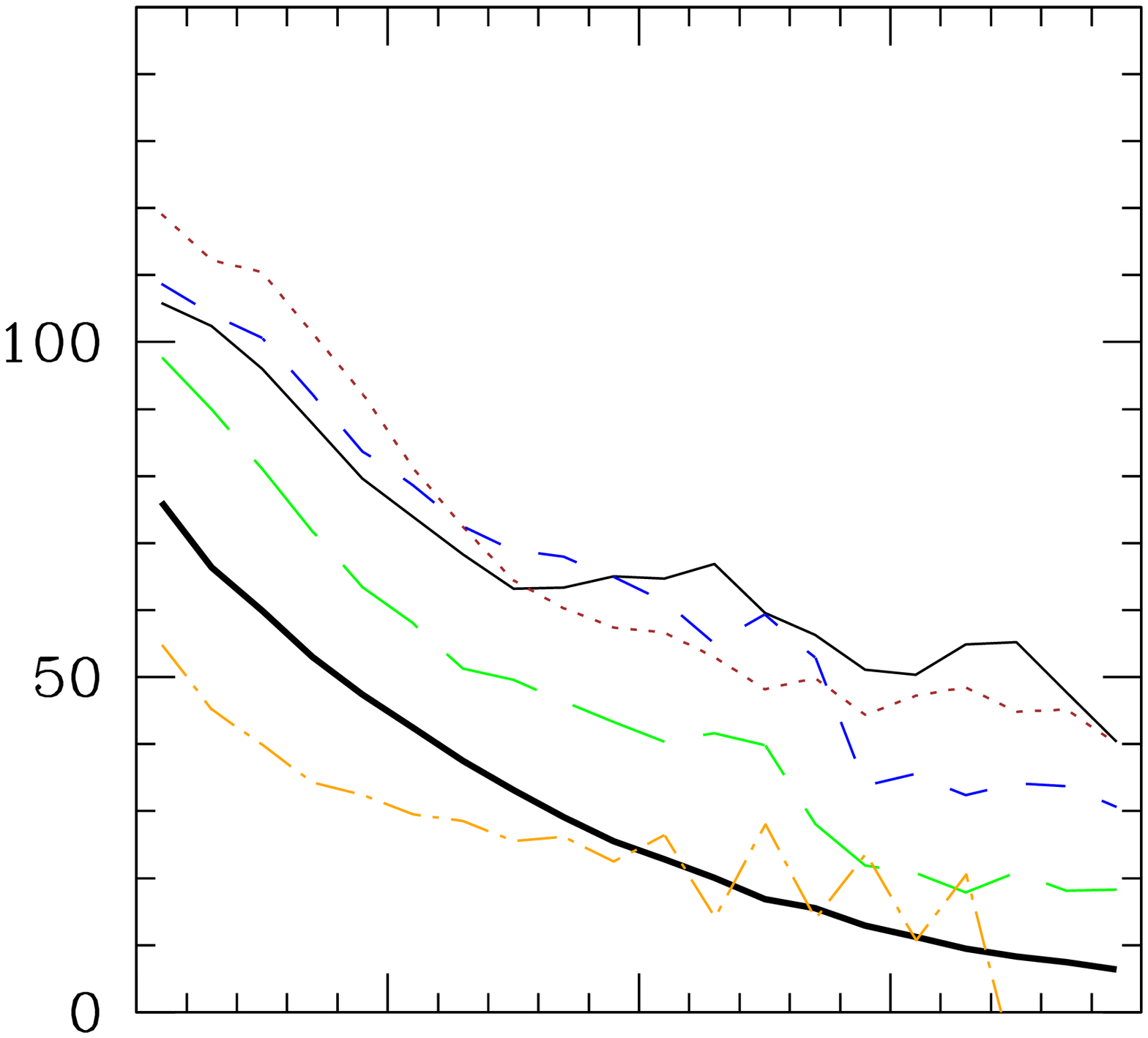}\hspace*{-1.36cm}
\includegraphics[width=50mm]{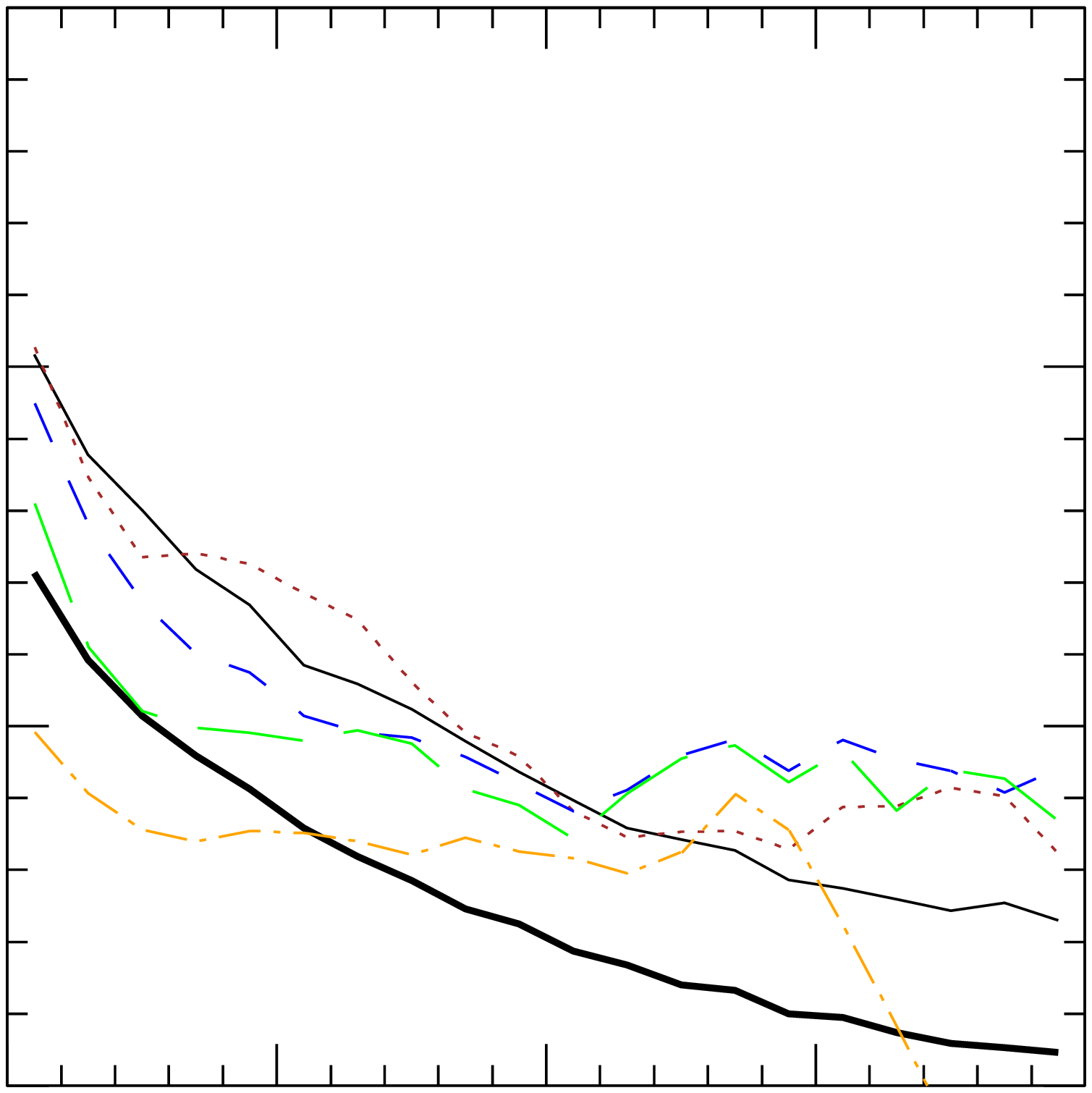}\hspace*{-1.36cm}
\includegraphics[width=50mm]{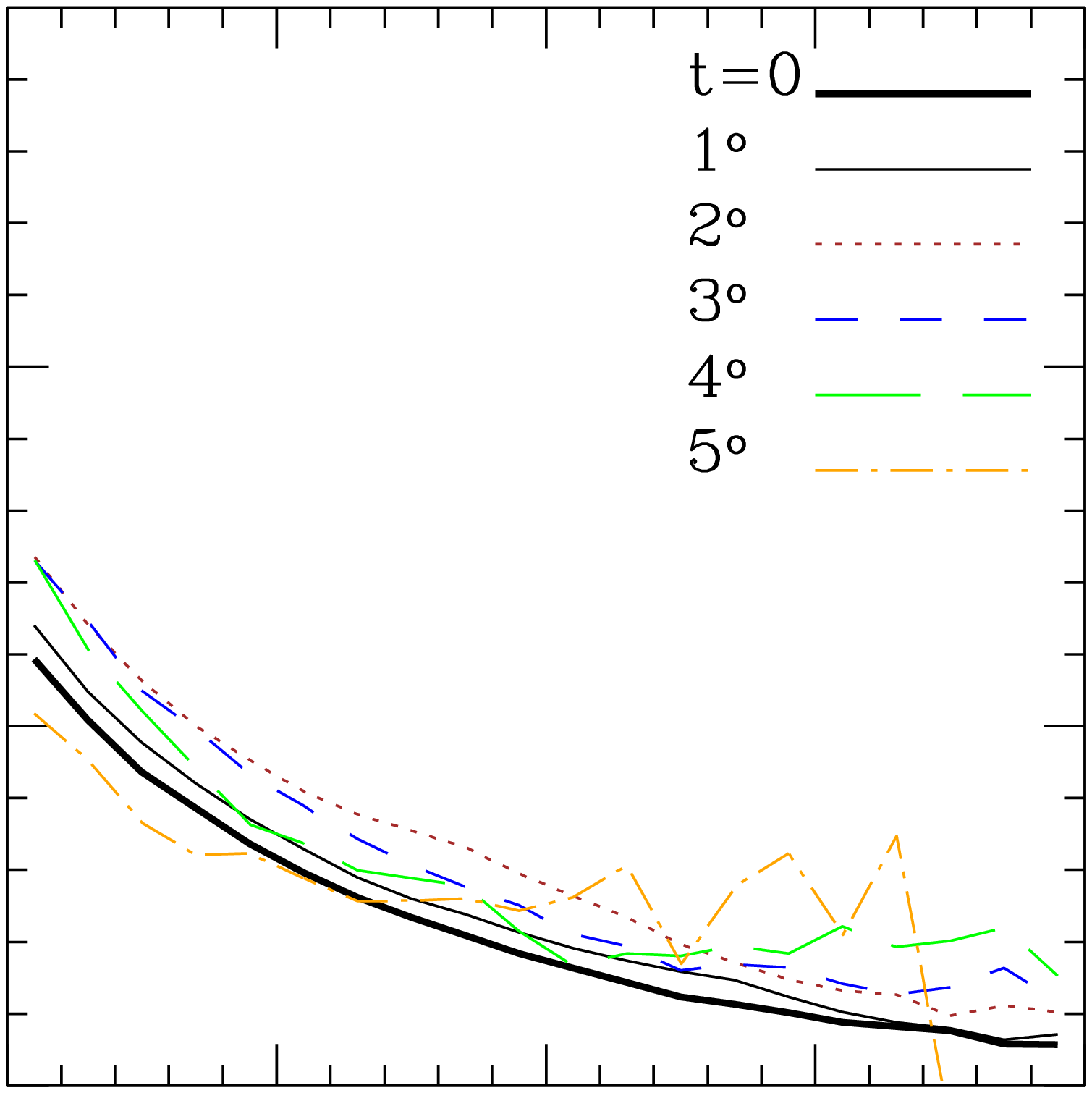}\vspace*{-1.29cm}\\
\includegraphics[width=50mm]{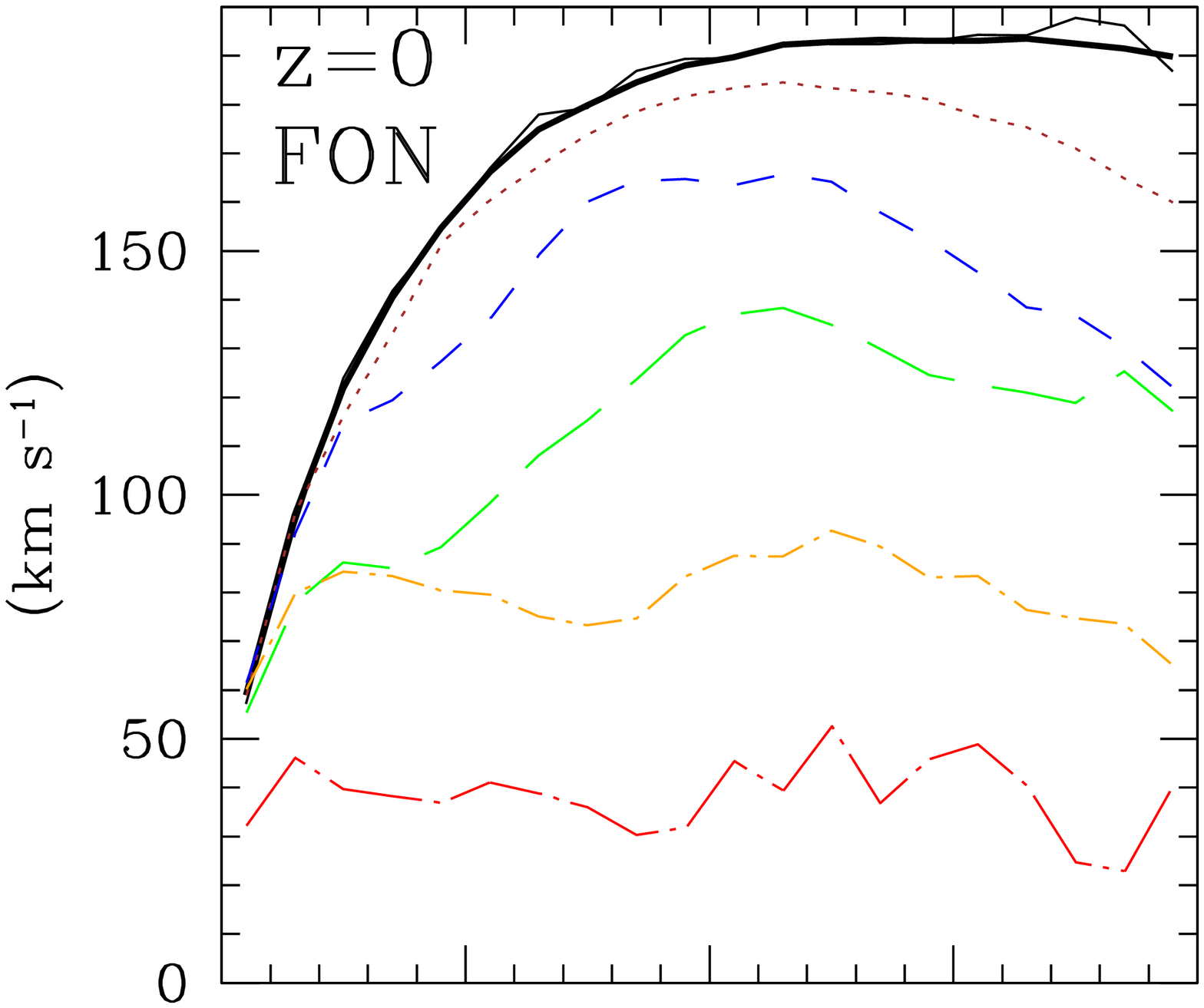}\hspace*{-0.4cm}
\includegraphics[width=50mm]{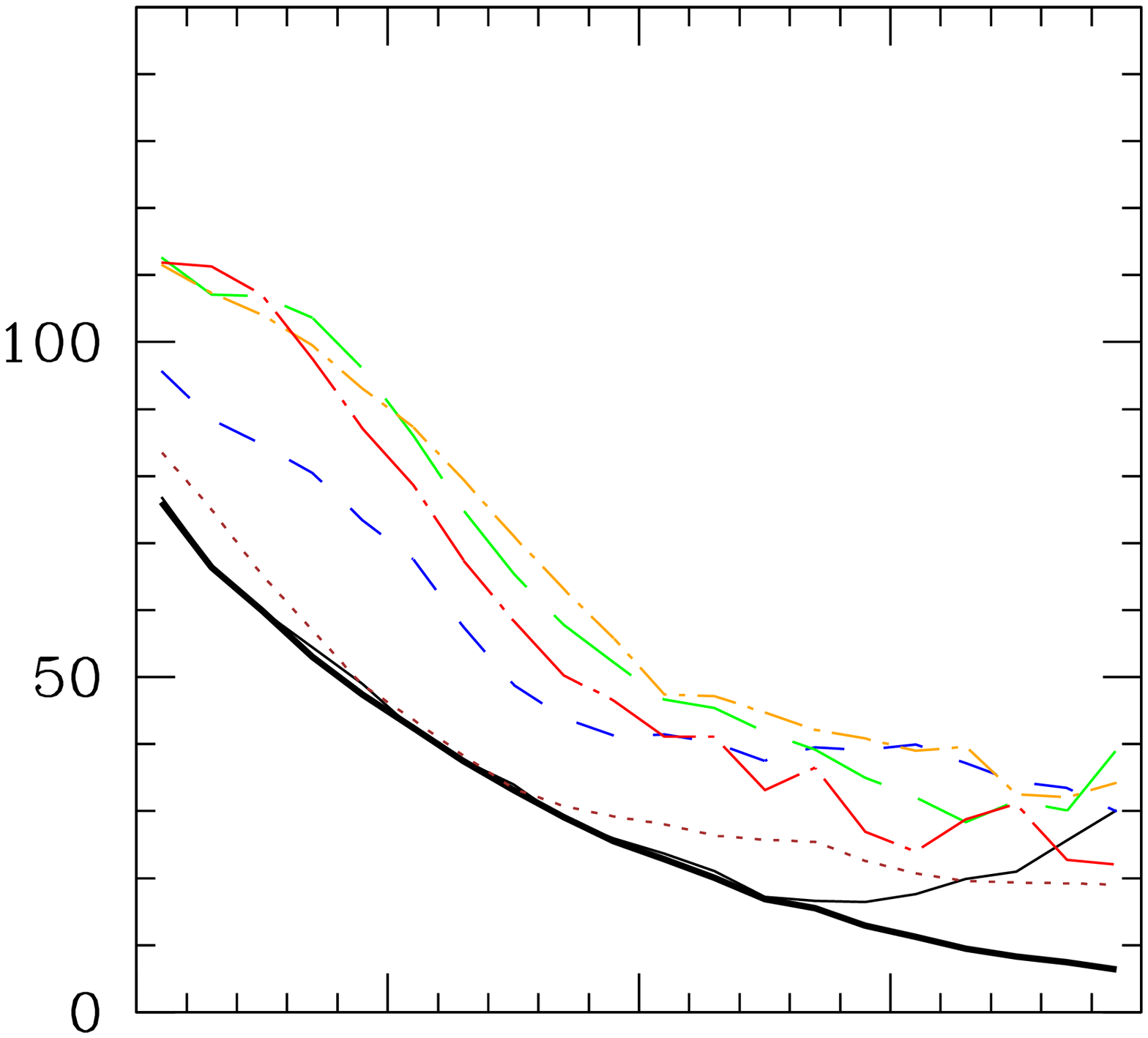}\hspace*{-1.36cm}
\includegraphics[width=50mm]{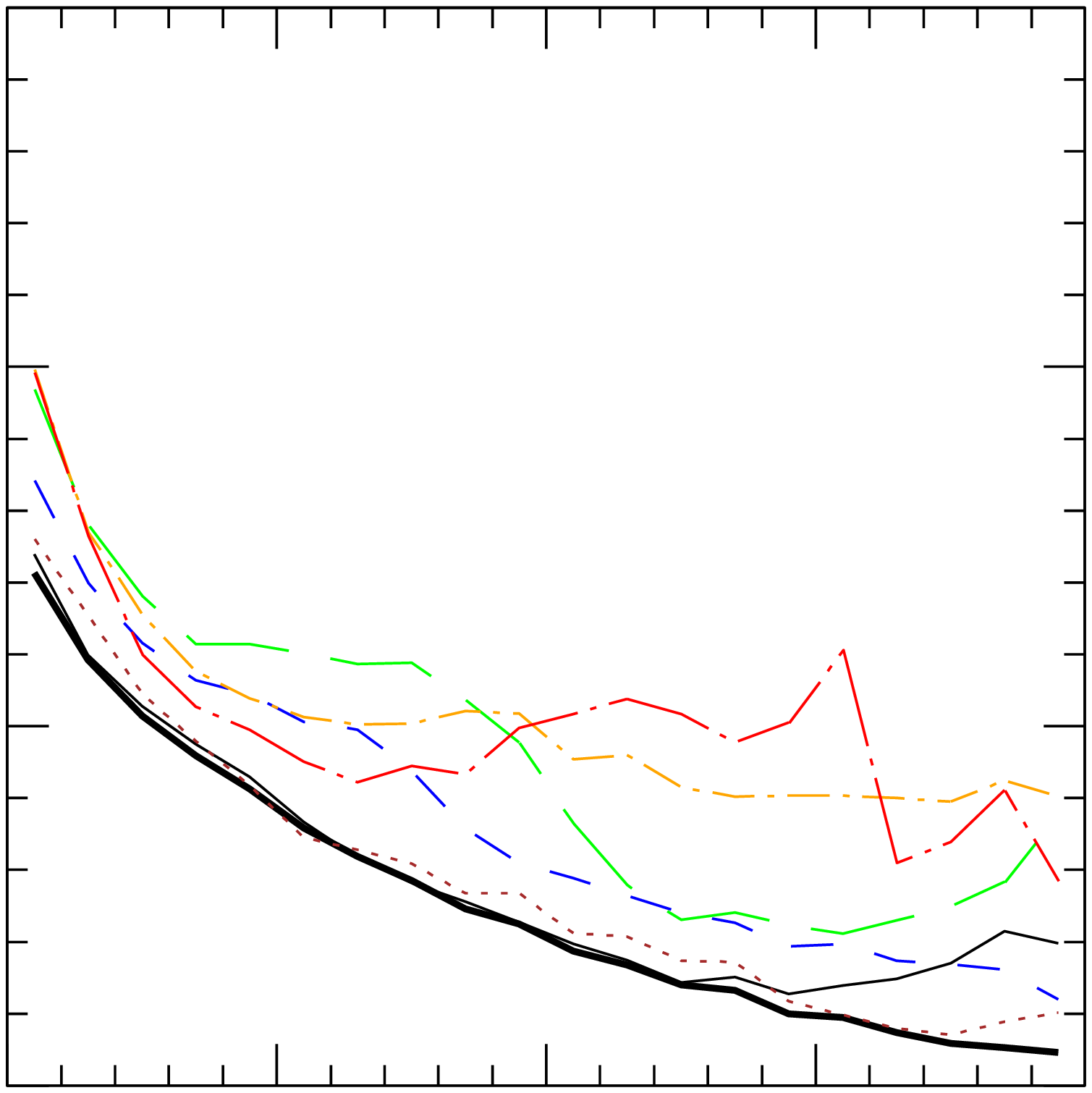}\hspace*{-1.36cm}
\includegraphics[width=50mm]{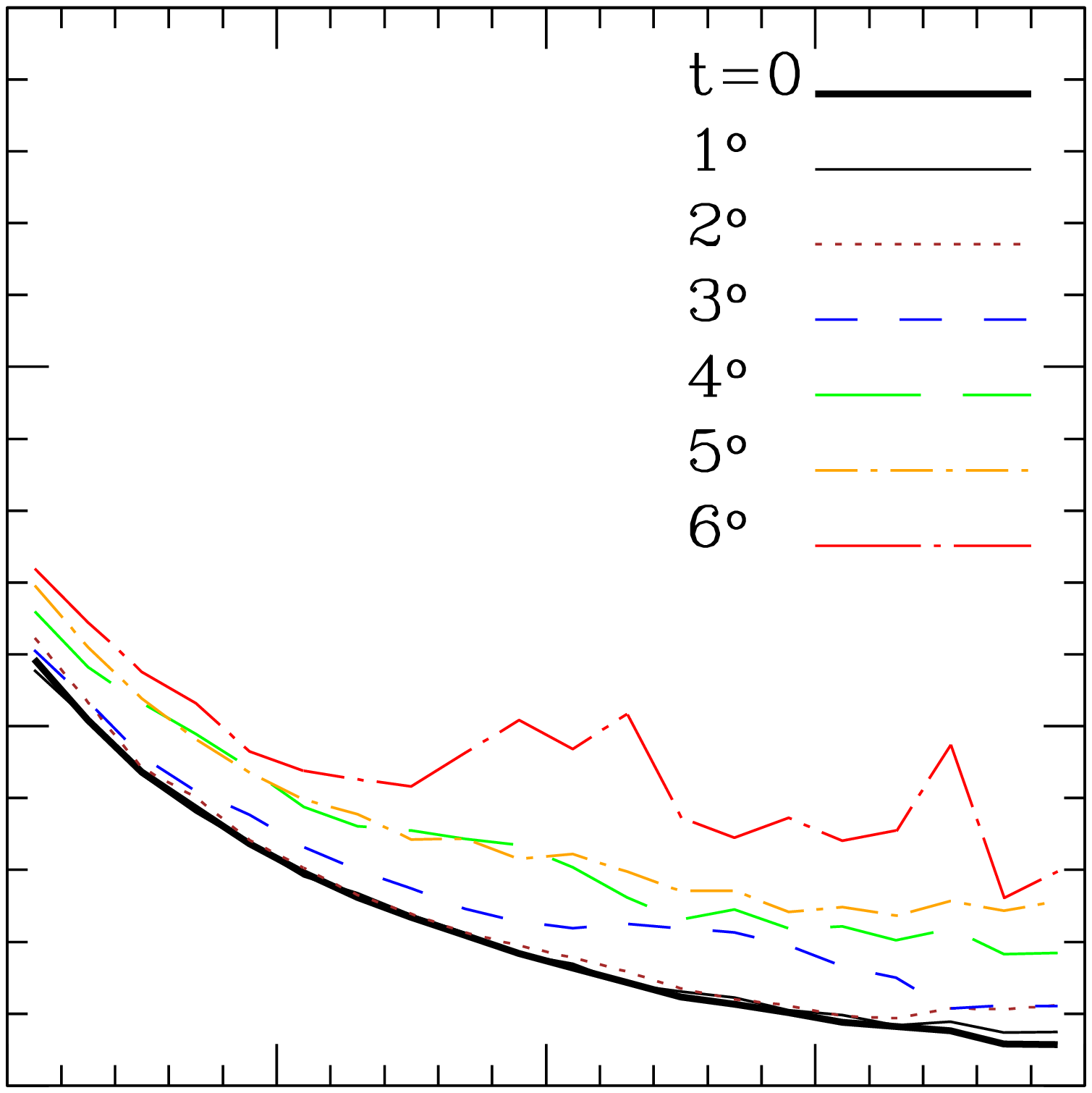}\vspace*{-1.29cm}\\
\includegraphics[width=50mm]{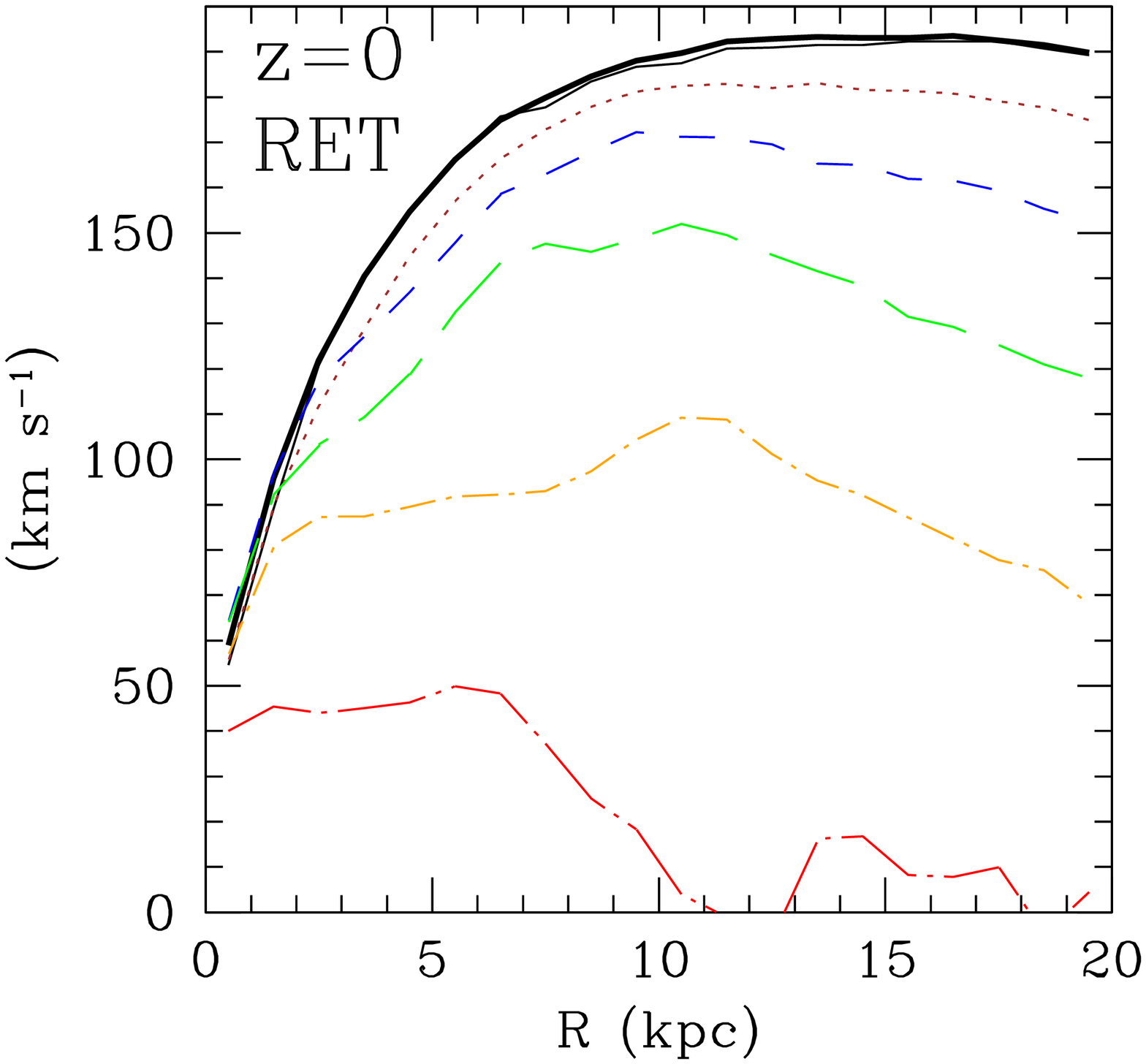}\hspace*{-0.4cm}
\includegraphics[width=50mm]{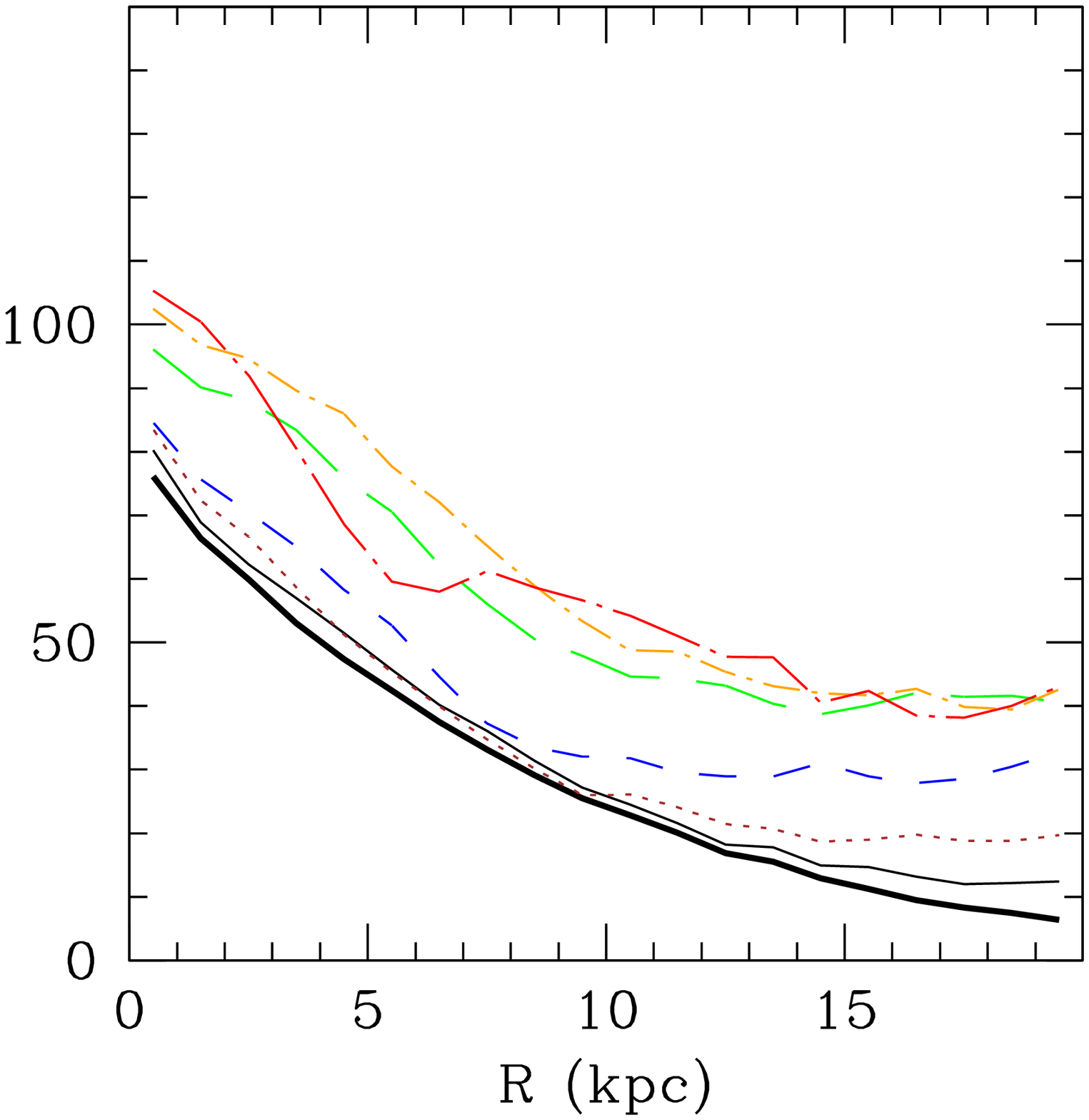}\hspace*{-1.36cm}
\includegraphics[width=50mm]{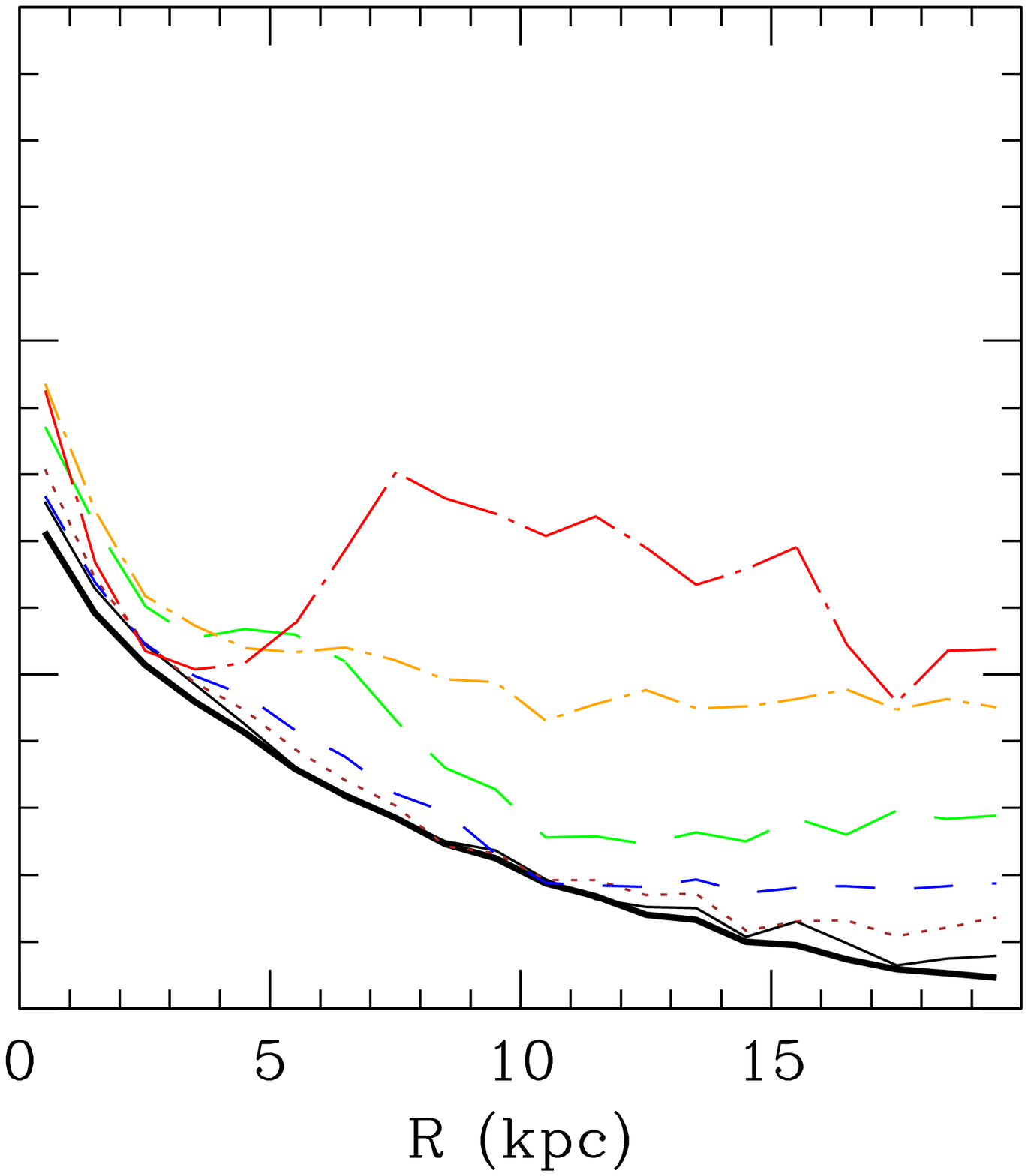}\hspace*{-1.36cm}
\includegraphics[width=50mm]{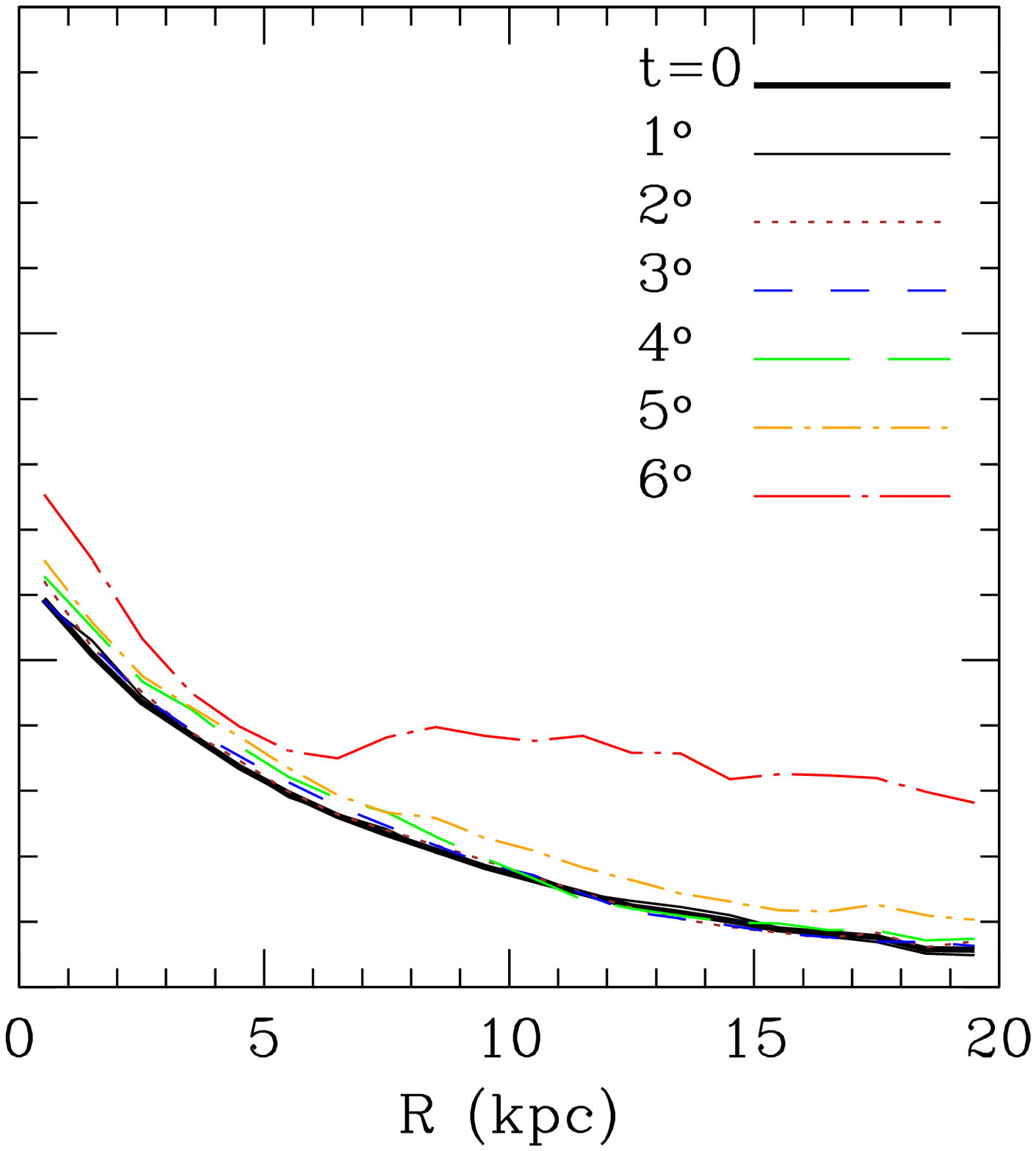}
\end{center}
\caption{
Evolution of the kinematics of disc galaxies in terms of their mean rotation, 
$\langle V_{\phi}\rangle$, and the radial, azimuthal and vertical components
of their velocity dispersions, $\sigma_{\rm R}$, $\sigma_{\phi}$, and 
$\sigma_{Z}$ respectively, after each of their first six pericentric passages.
For clarity, only the REF (reference), RAD (more eccentric infall), FON (face-on 
infall) and RET (retrograde infall) experiments at ``$z$=0'' are shown, while 
the rest of the experiments can be found in the Appendix. The kinematics of 
discs have been computed in concentric rings, 1~kpc wide, considering only 
stars that remain bound and are located within 3~kpc from the midplane, and 
have been corrected by the evolution of the respective disc in isolation.}
\label{kinematics-evol-z0}
\end{figure*}

\begin{figure*}
\begin{center}
\includegraphics[width=50mm]{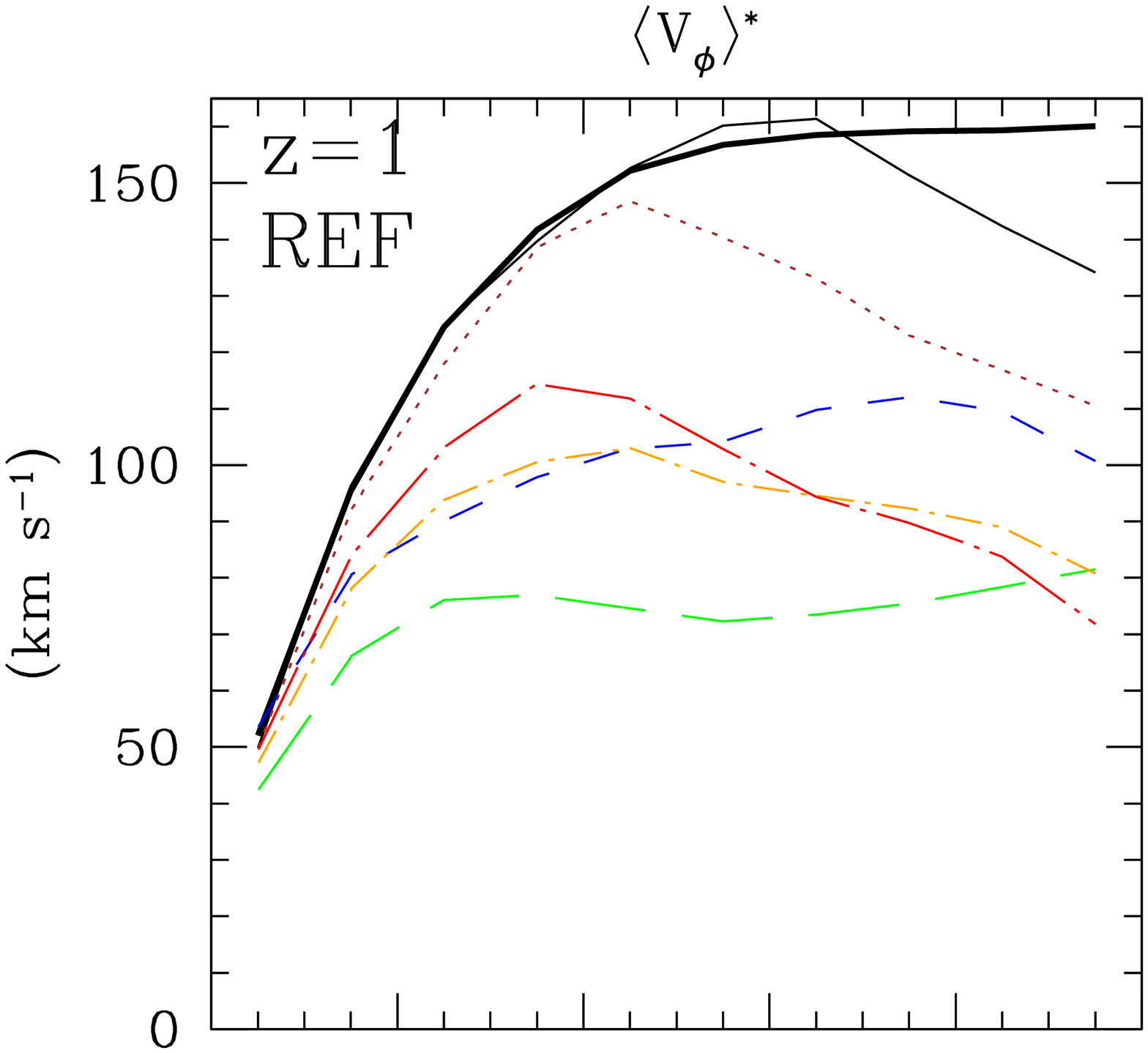}\hspace*{-0.4cm}
\includegraphics[width=50mm]{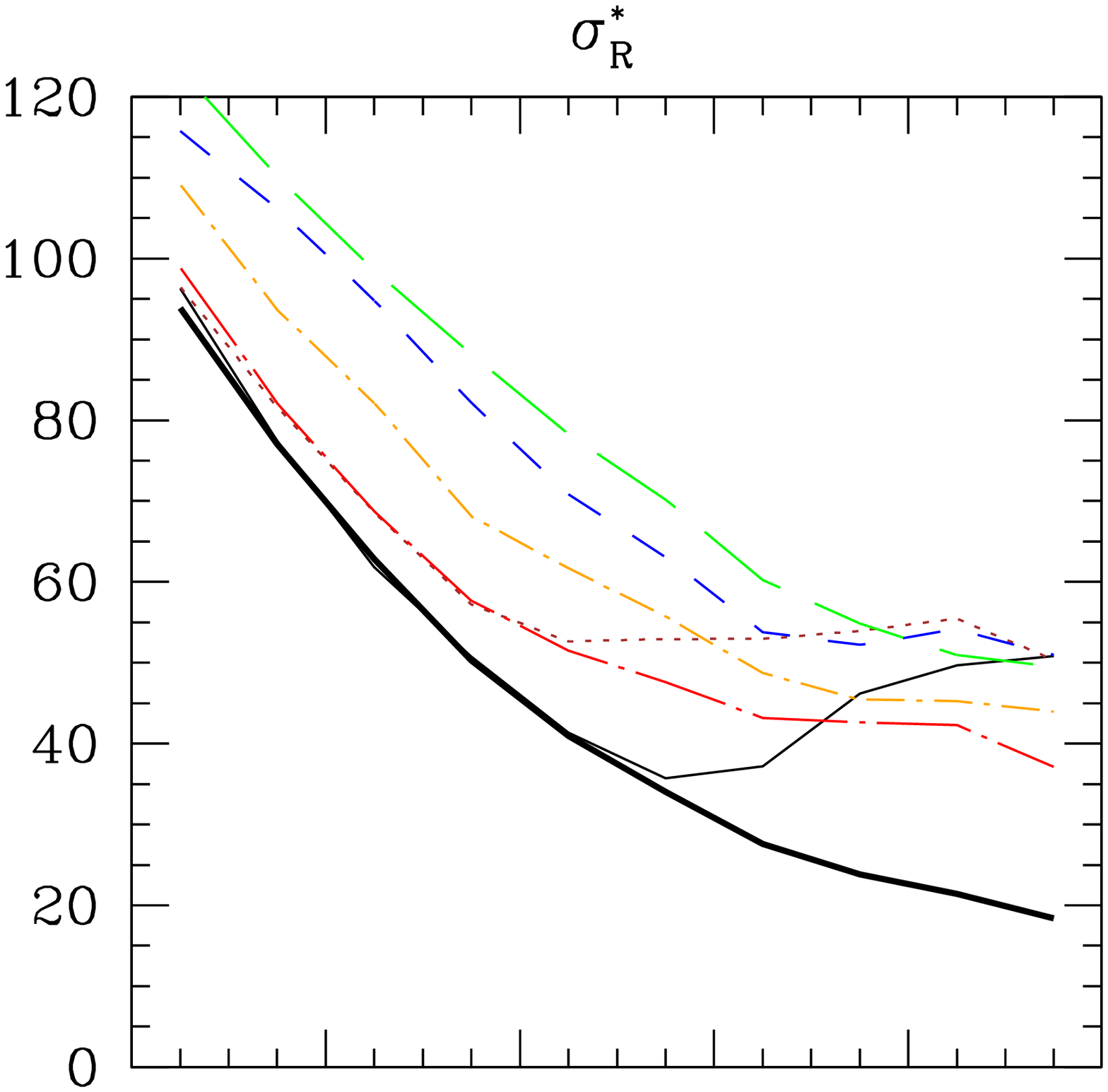}\hspace*{-1.36cm}
\includegraphics[width=50mm]{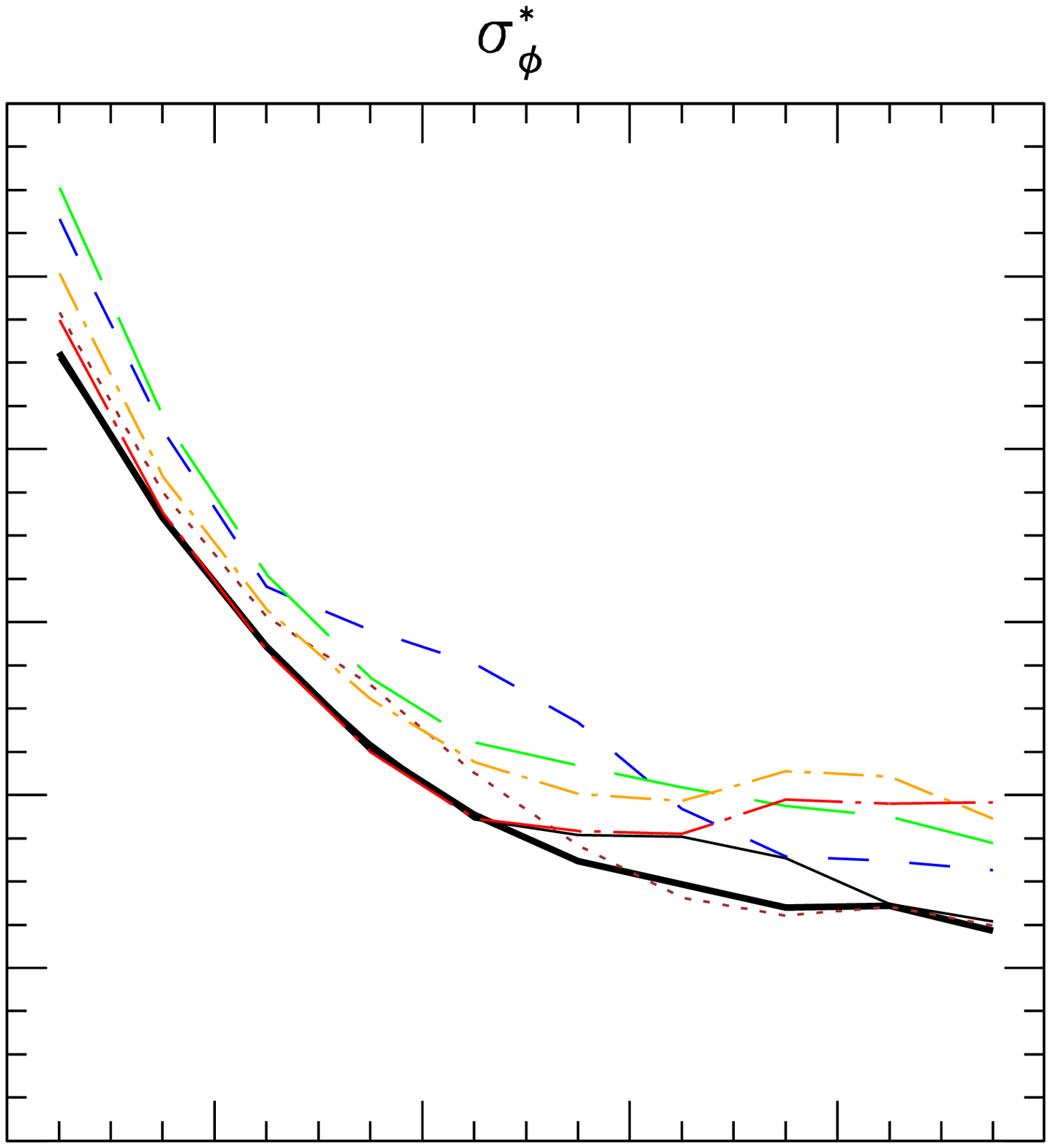}\hspace*{-1.36cm}
\includegraphics[width=50mm]{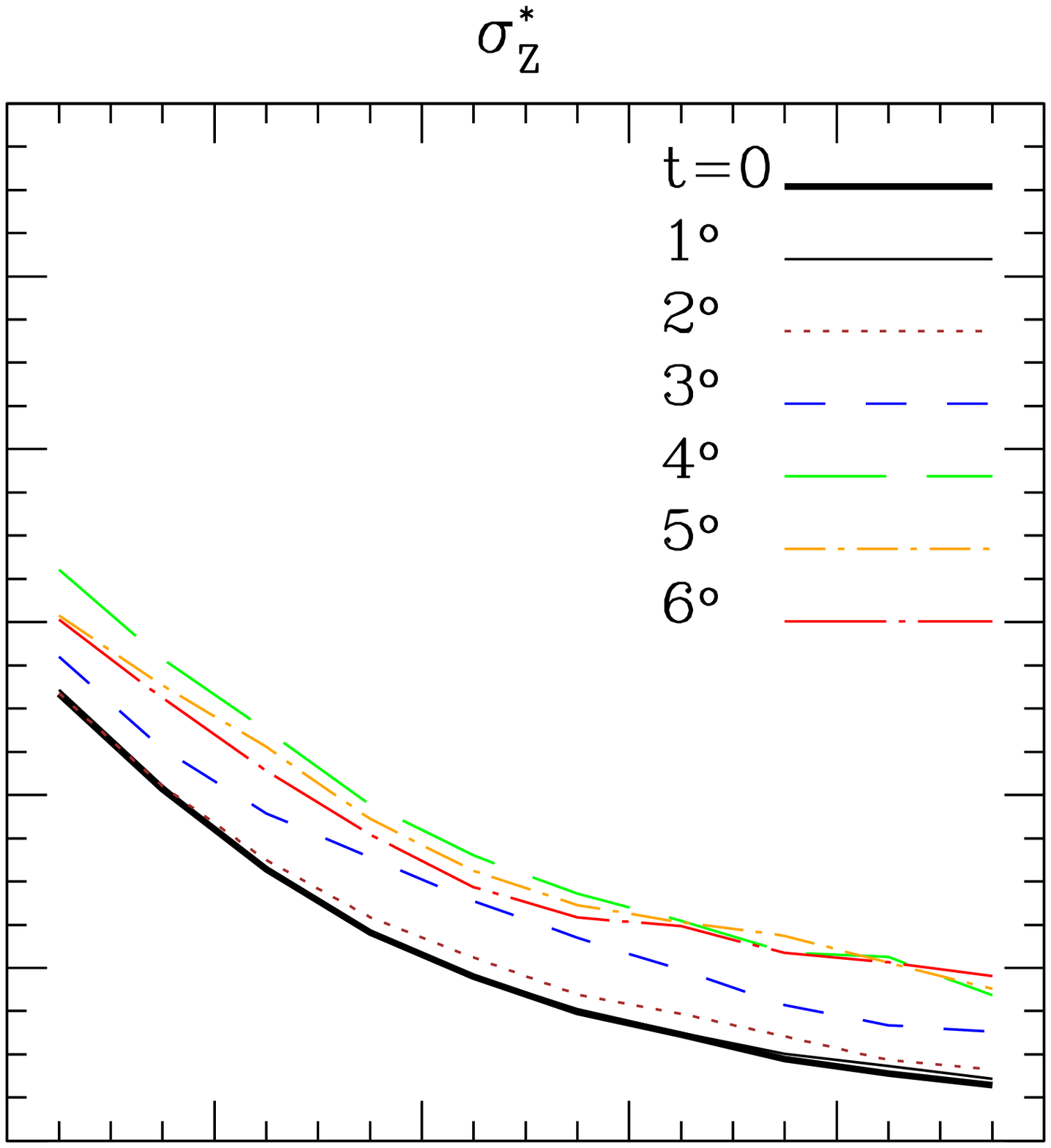}\vspace*{-1.29cm}\\
\includegraphics[width=50mm]{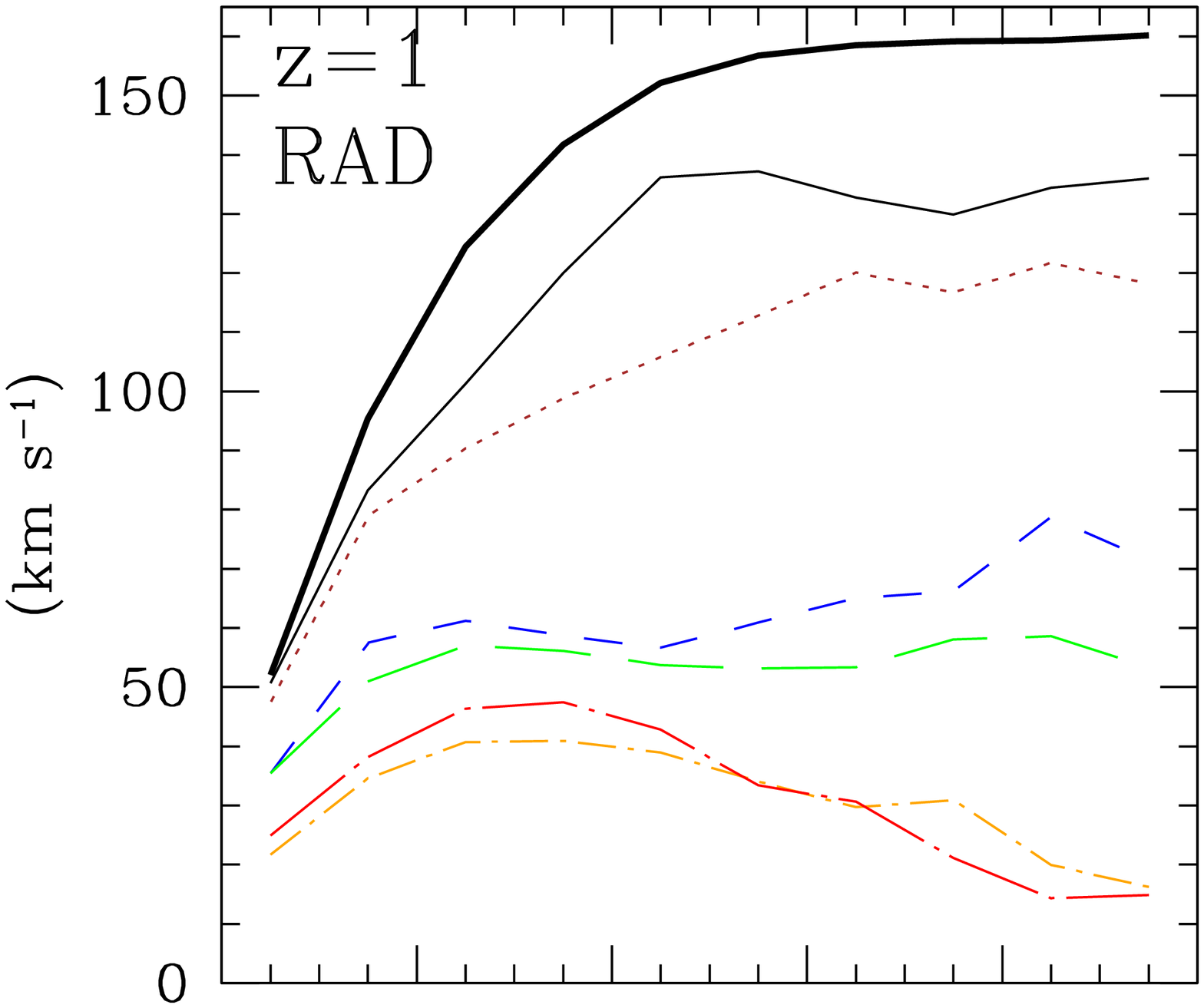}\hspace*{-0.4cm}
\includegraphics[width=50mm]{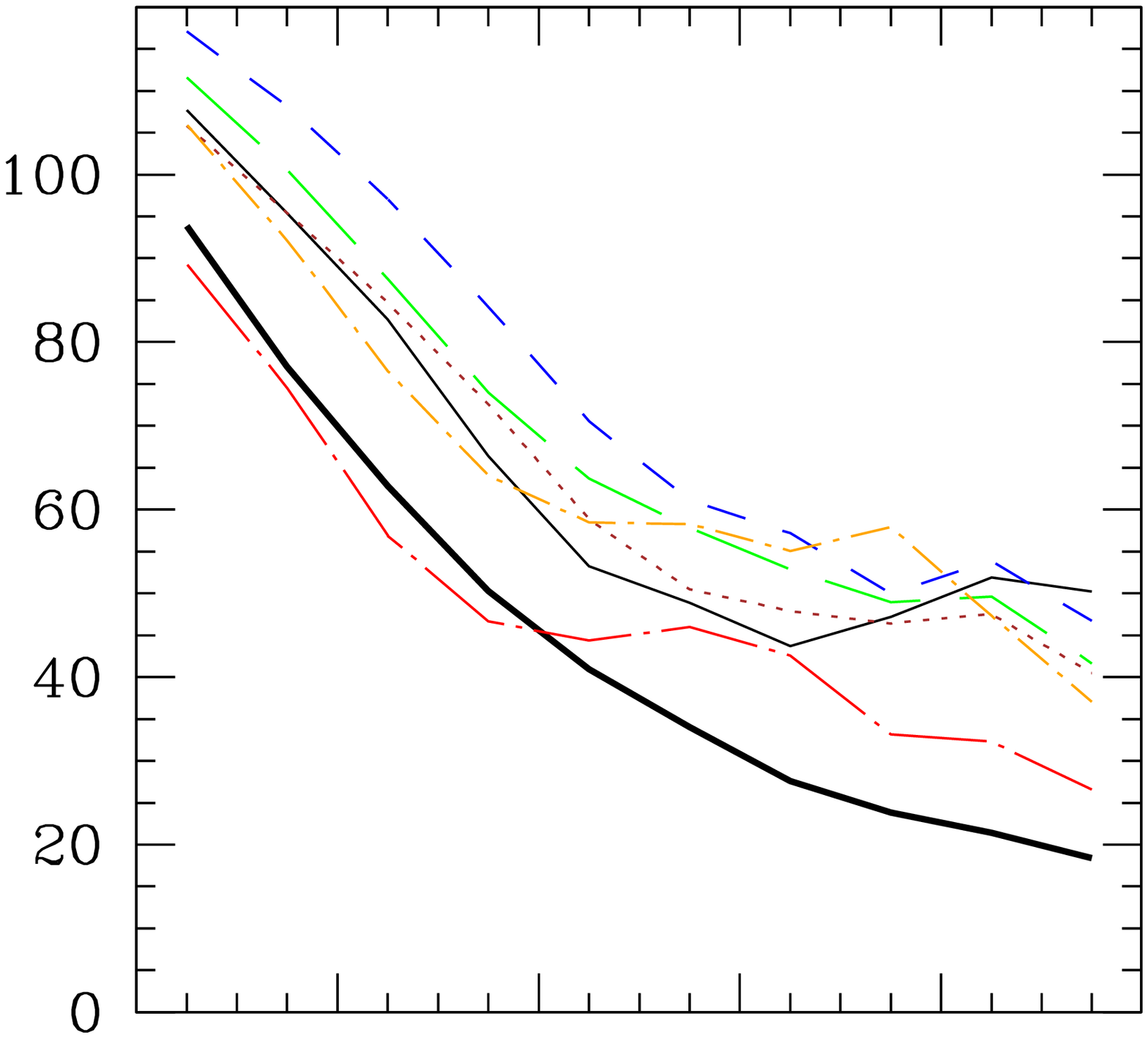}\hspace*{-1.36cm}
\includegraphics[width=50mm]{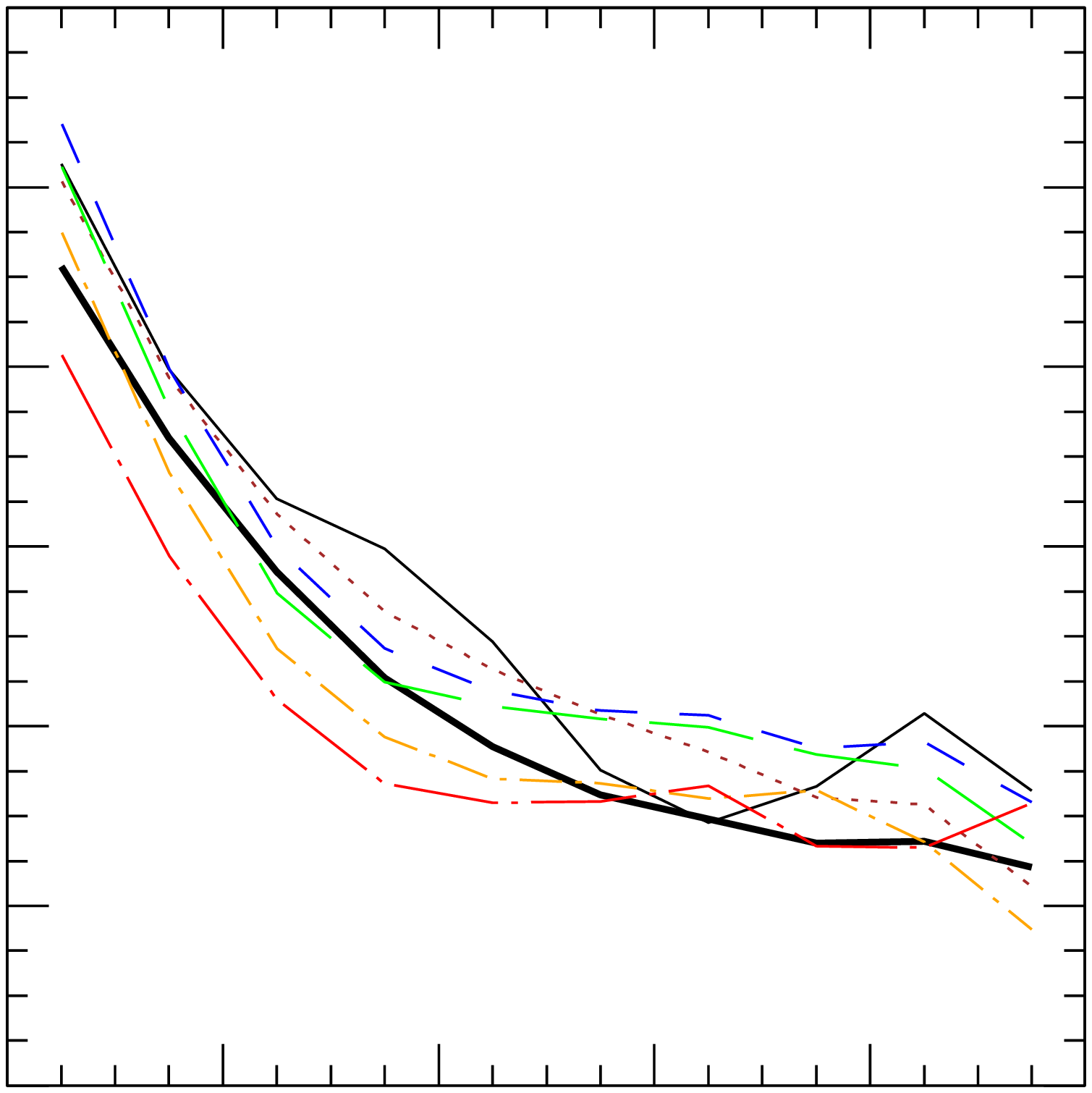}\hspace*{-1.36cm}
\includegraphics[width=50mm]{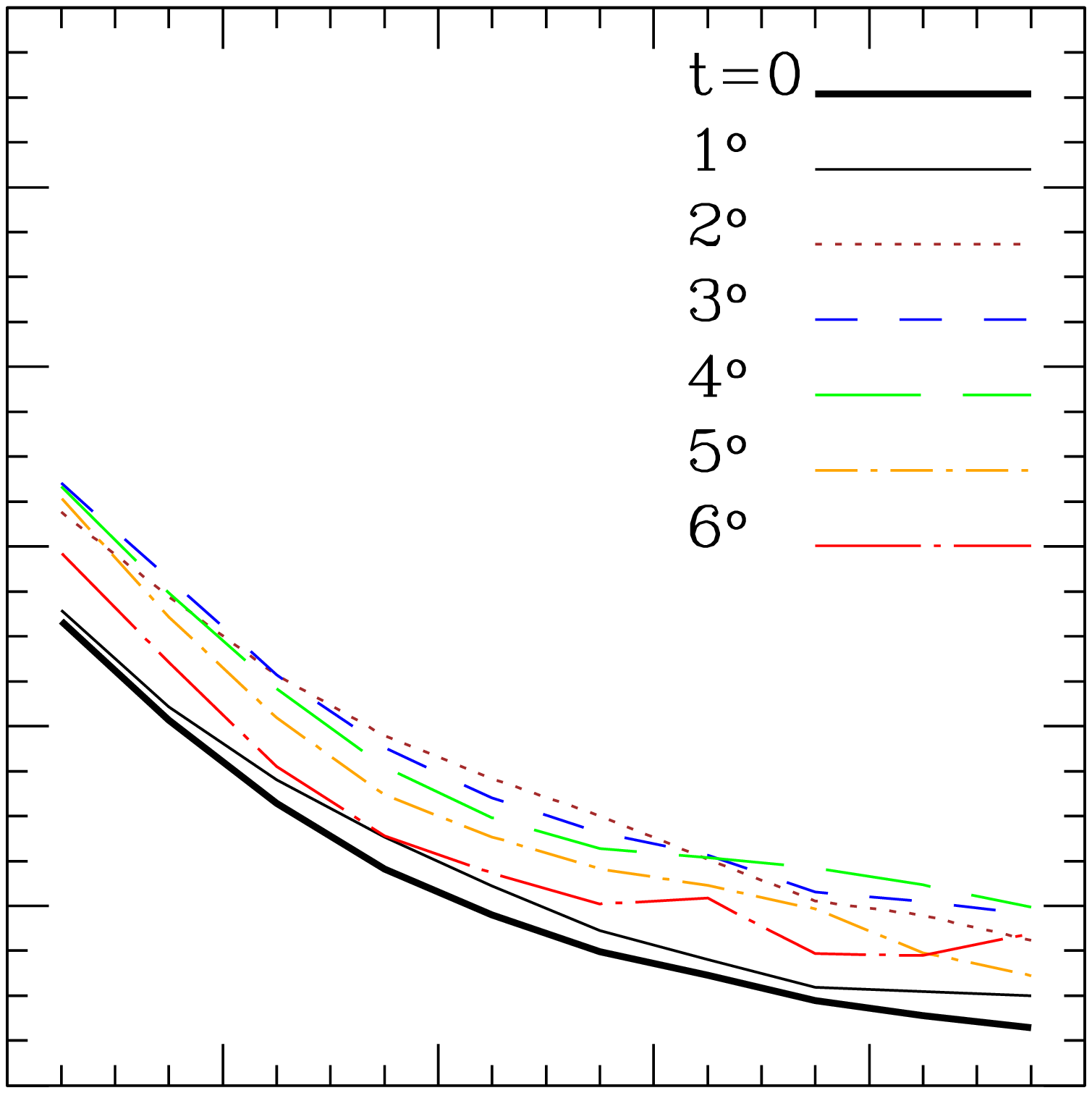}\vspace*{-1.29cm}\\
\includegraphics[width=50mm]{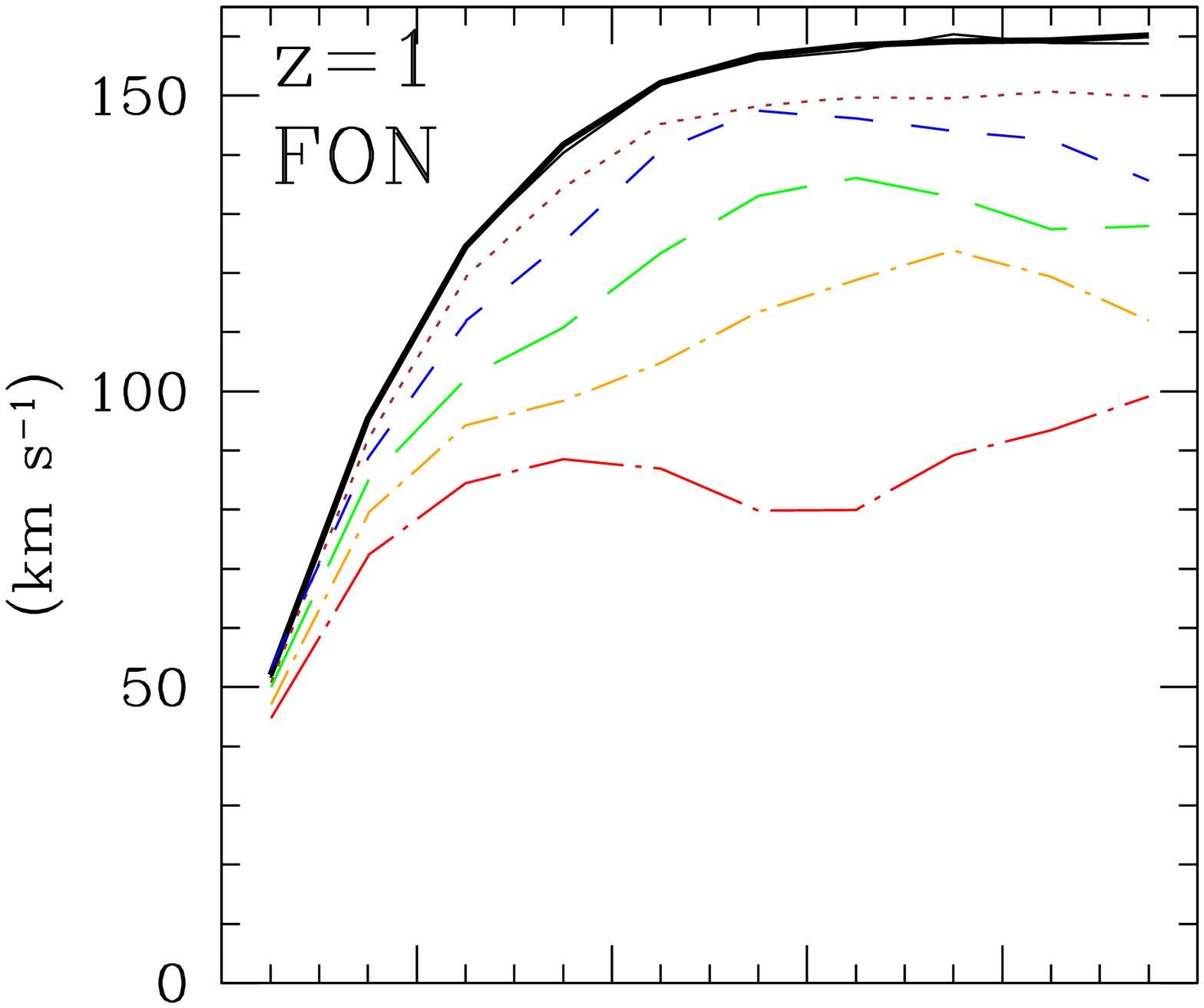}\hspace*{-0.4cm}
\includegraphics[width=50mm]{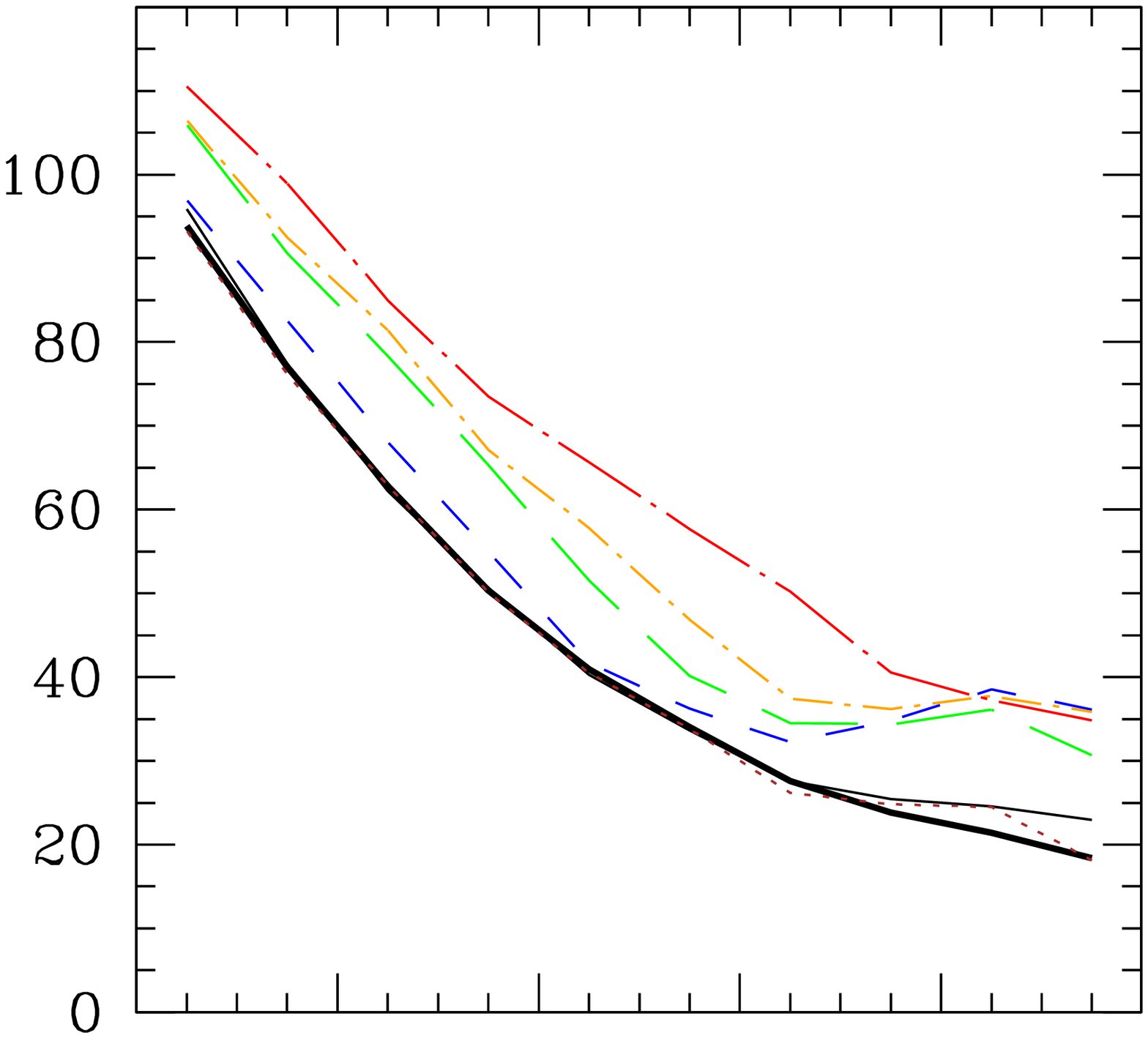}\hspace*{-1.36cm}
\includegraphics[width=50mm]{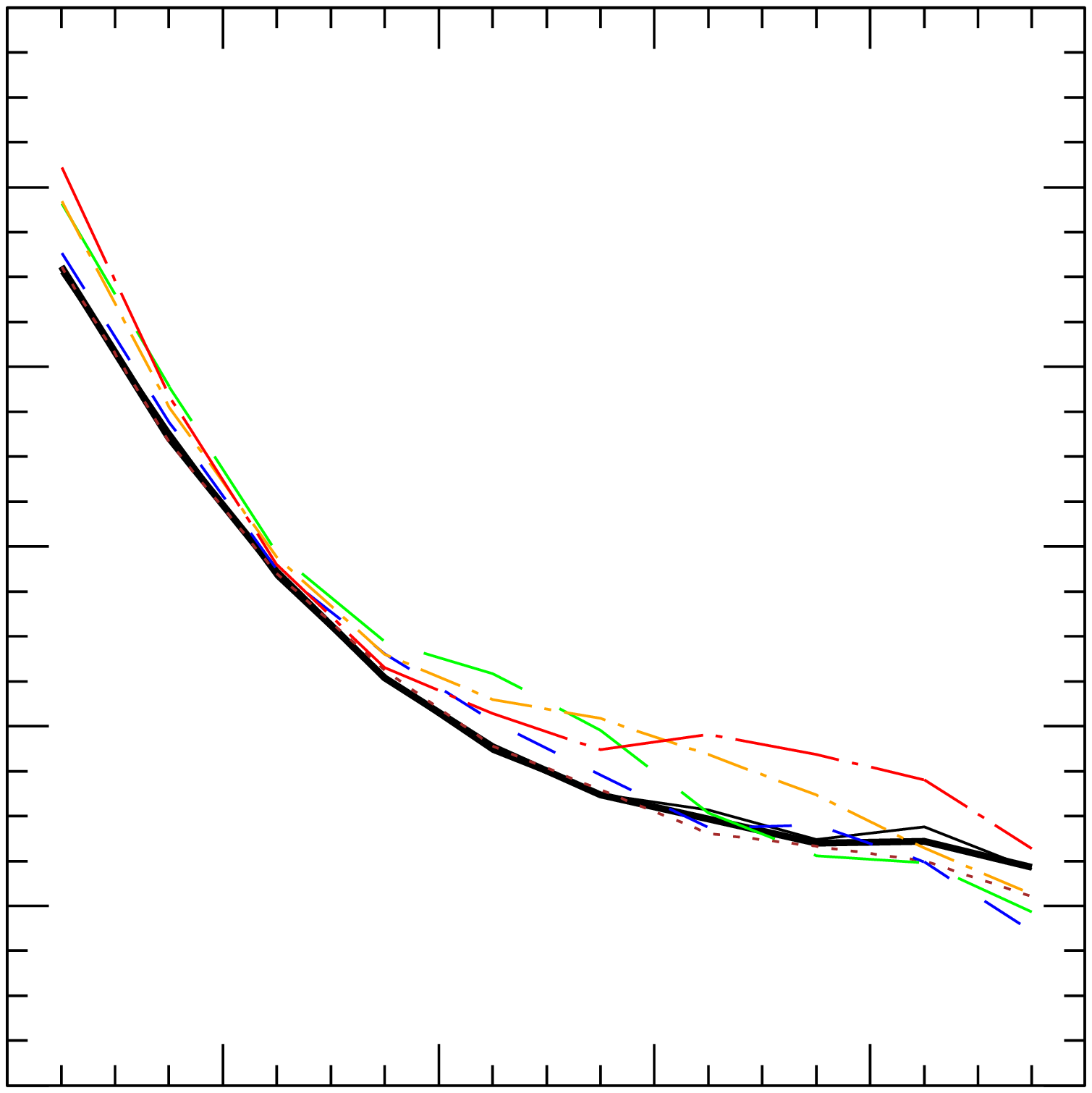}\hspace*{-1.36cm}
\includegraphics[width=50mm]{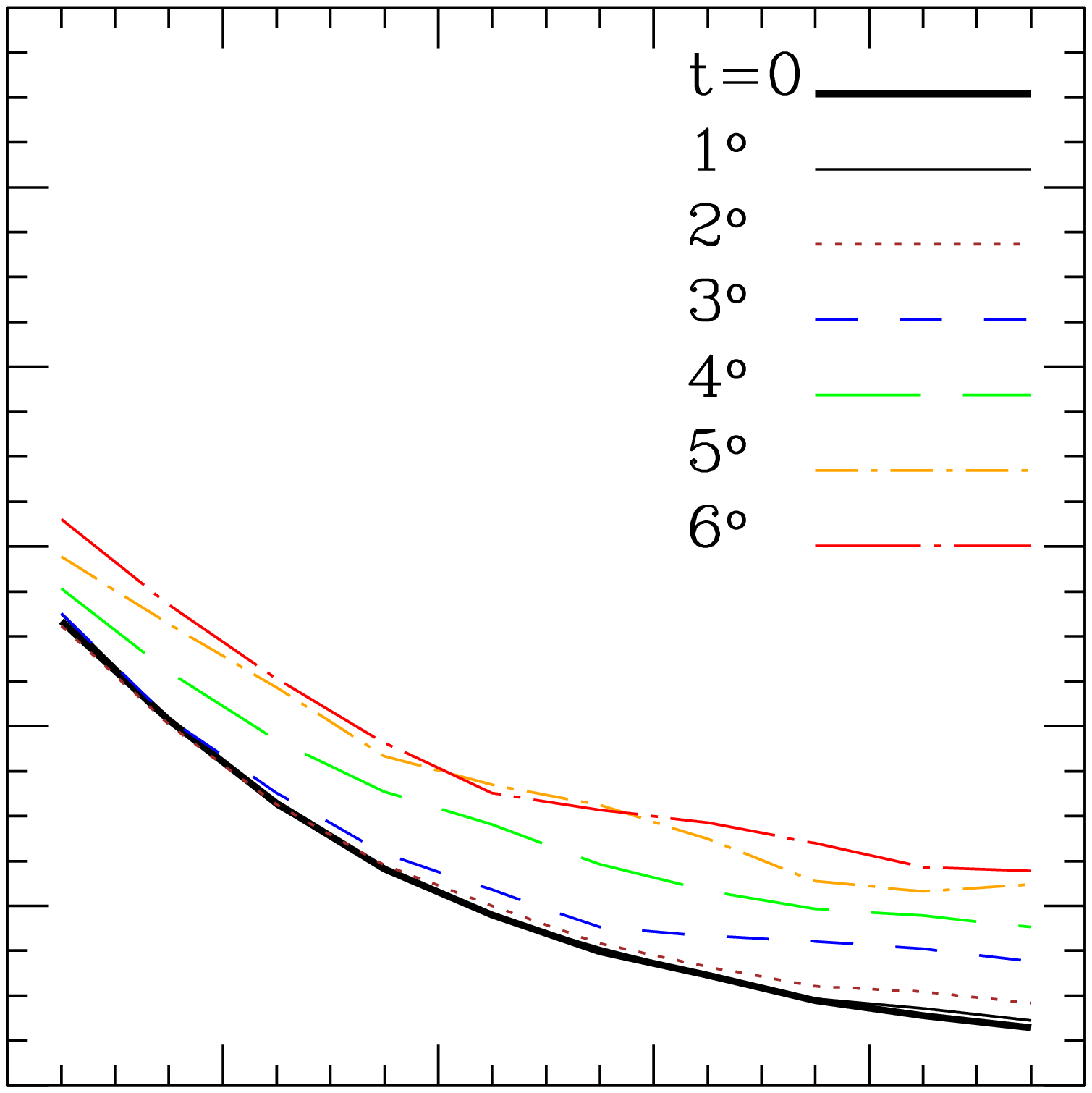}\vspace*{-1.29cm}\\
\includegraphics[width=50mm]{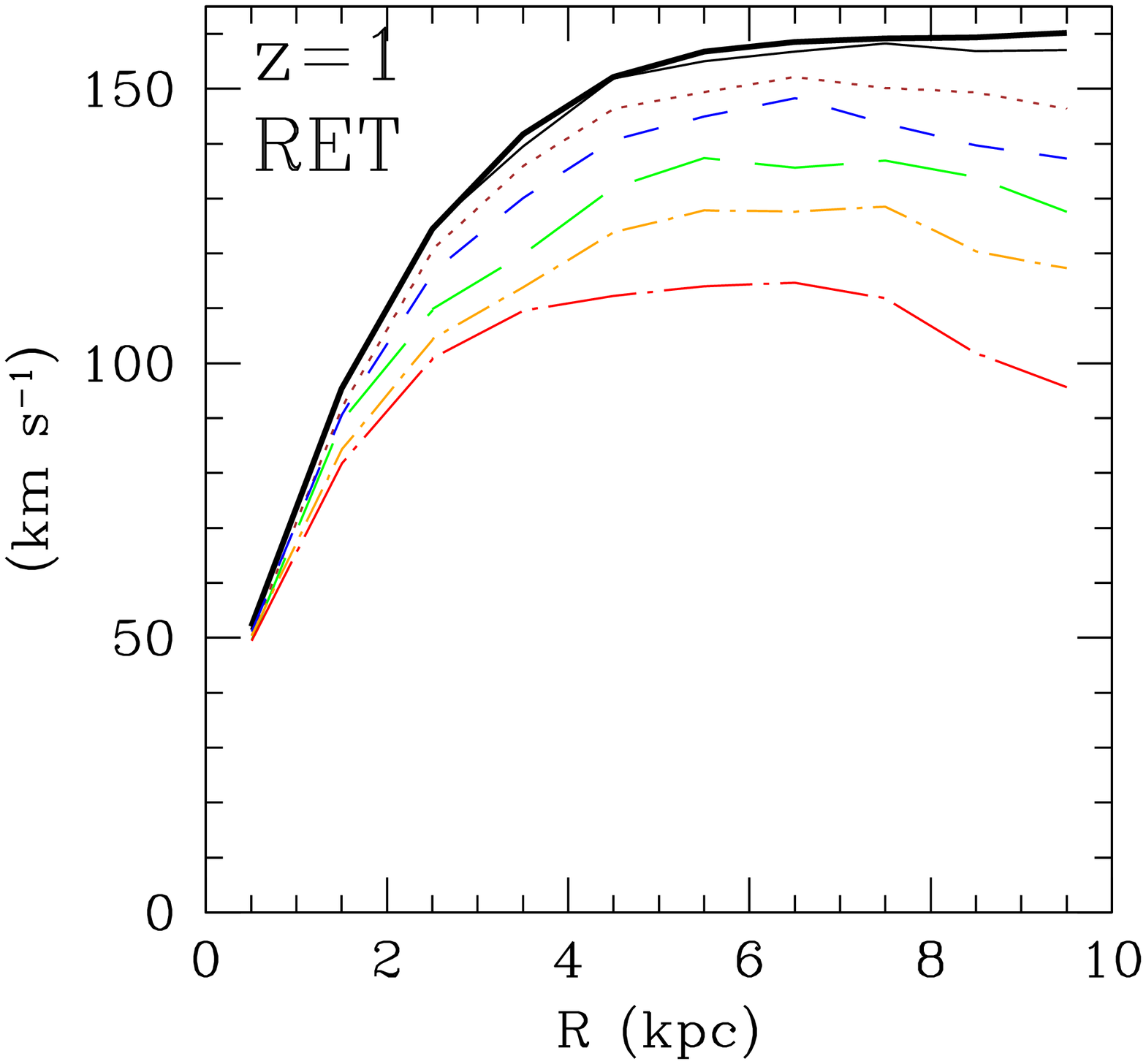}\hspace*{-0.4cm}
\includegraphics[width=50mm]{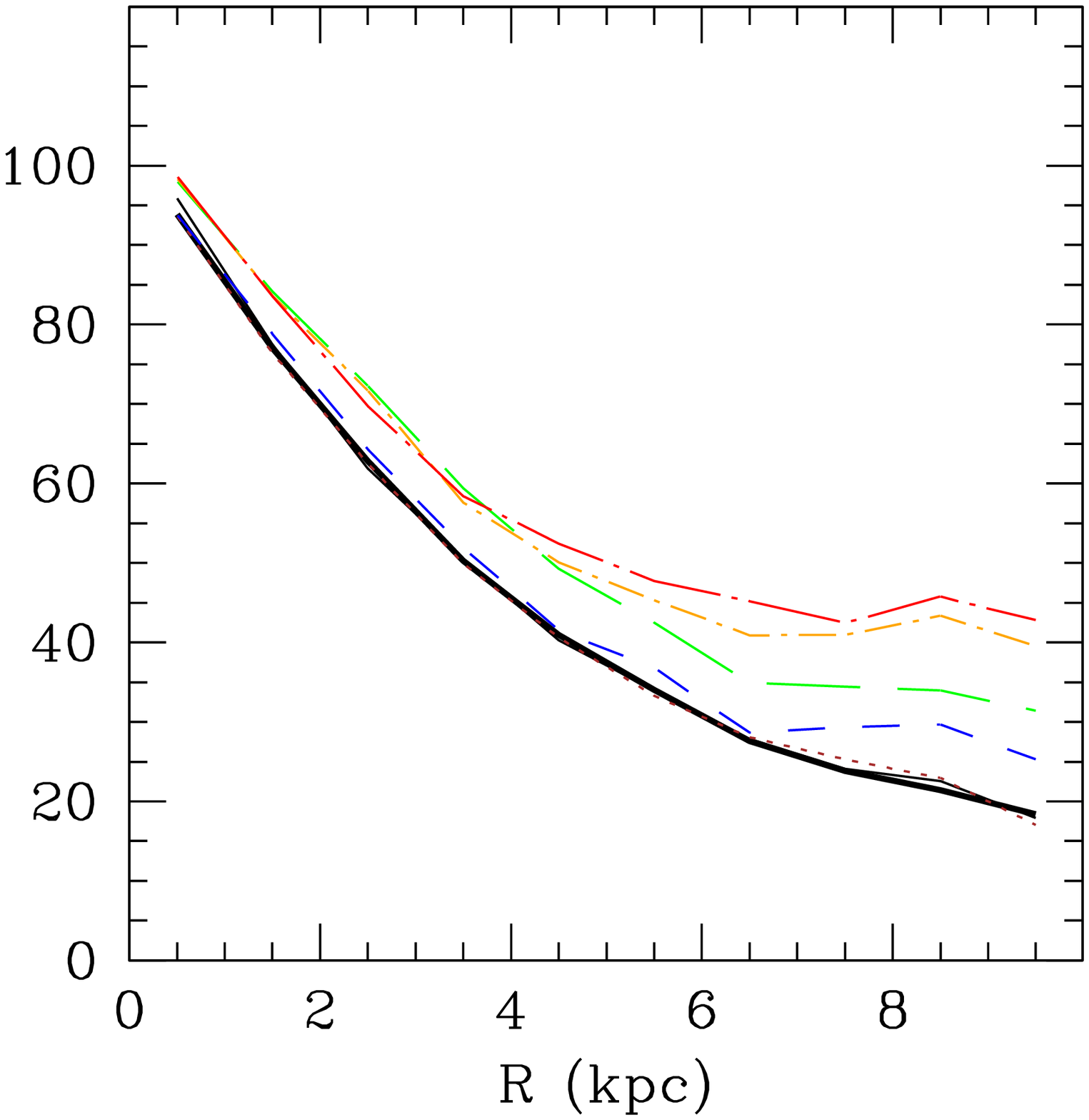}\hspace*{-1.36cm}
\includegraphics[width=50mm]{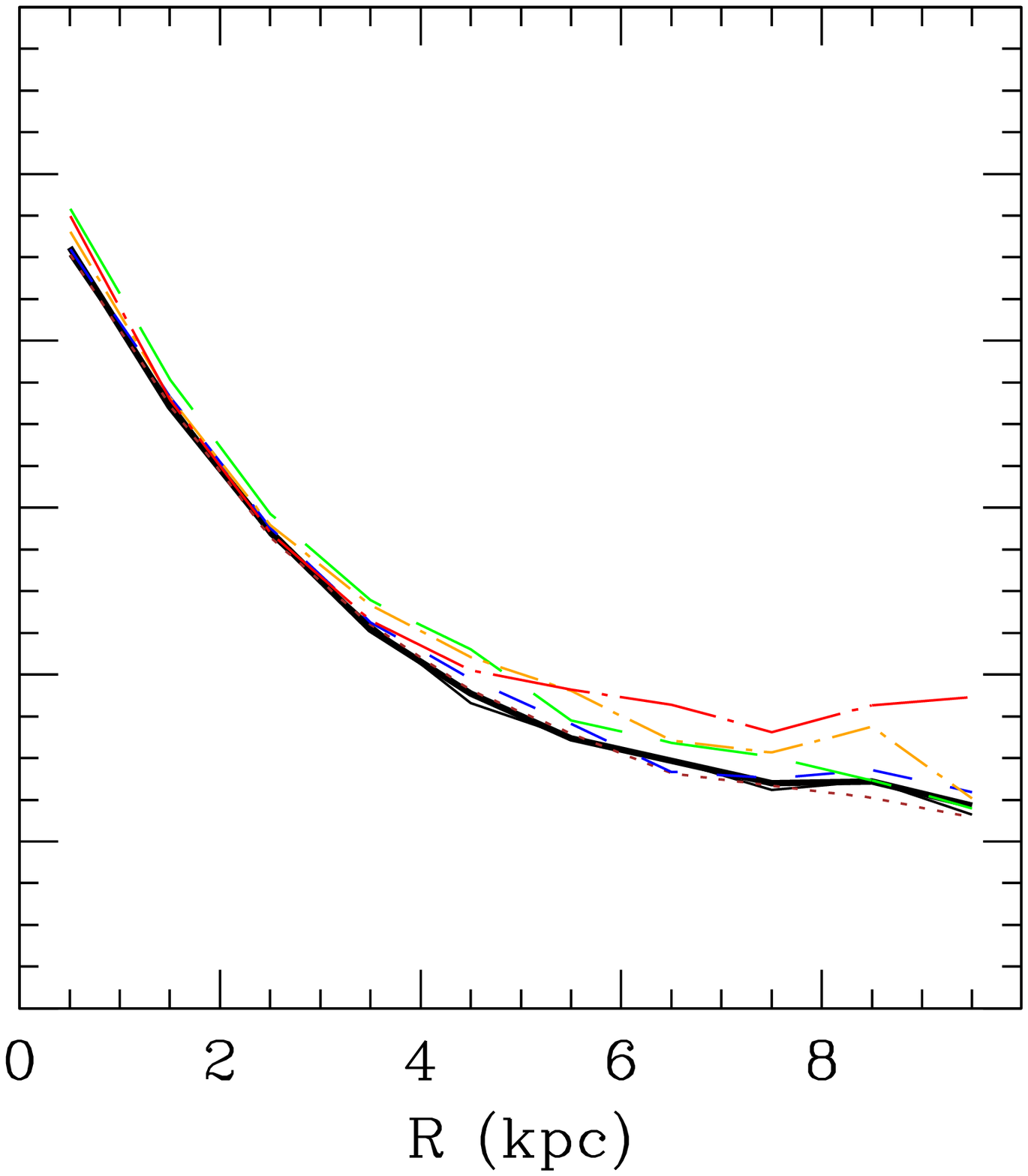}\hspace*{-1.36cm}
\includegraphics[width=50mm]{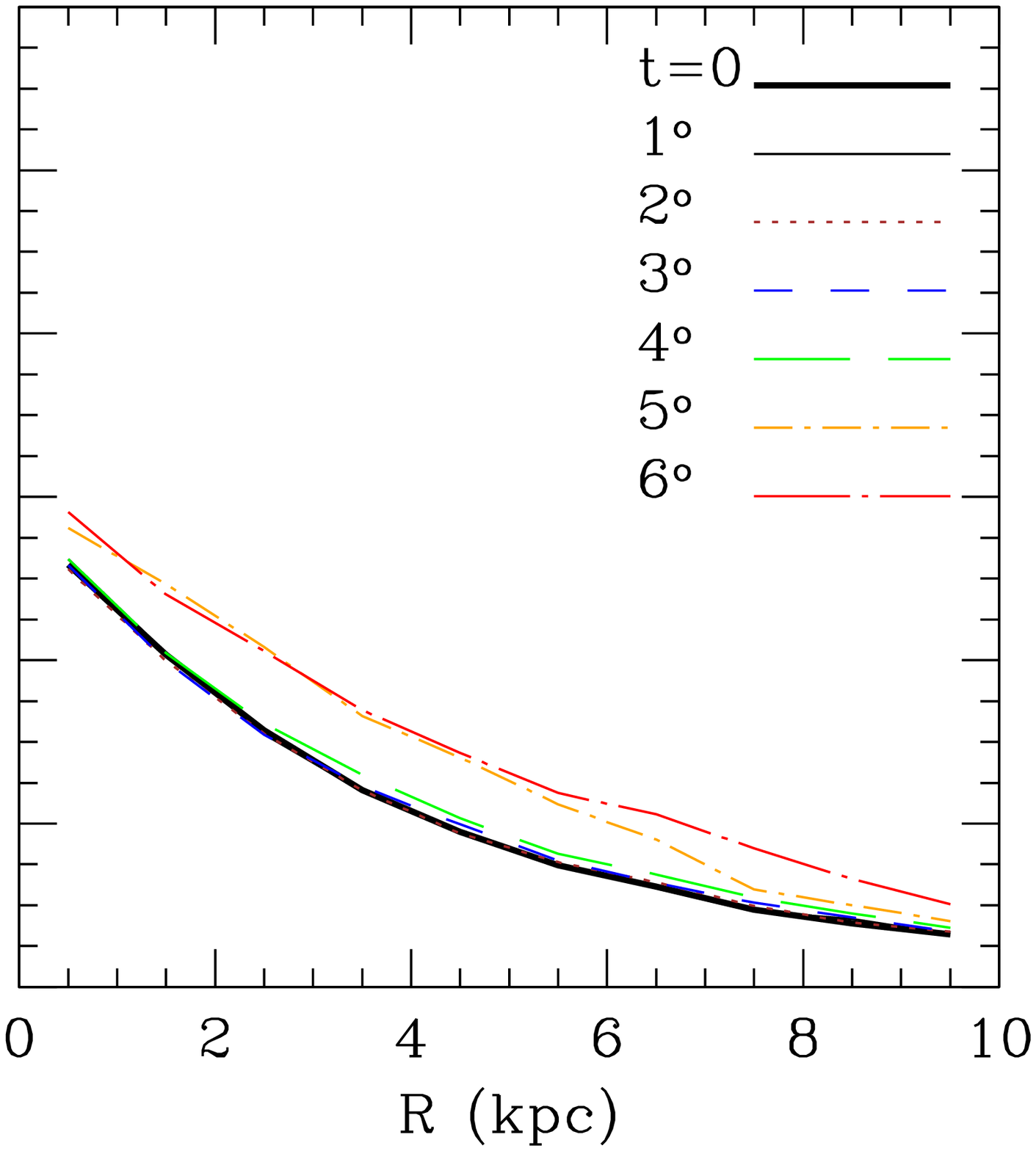}
\end{center}
\caption{Same as Fig.~\ref{kinematics-evol-z0}, but for experiments at ``$z$=1''.
}
\label{kinematics-evol-z1}
\end{figure*}

\begin{figure*}
\begin{center}
\includegraphics[width=50mm]{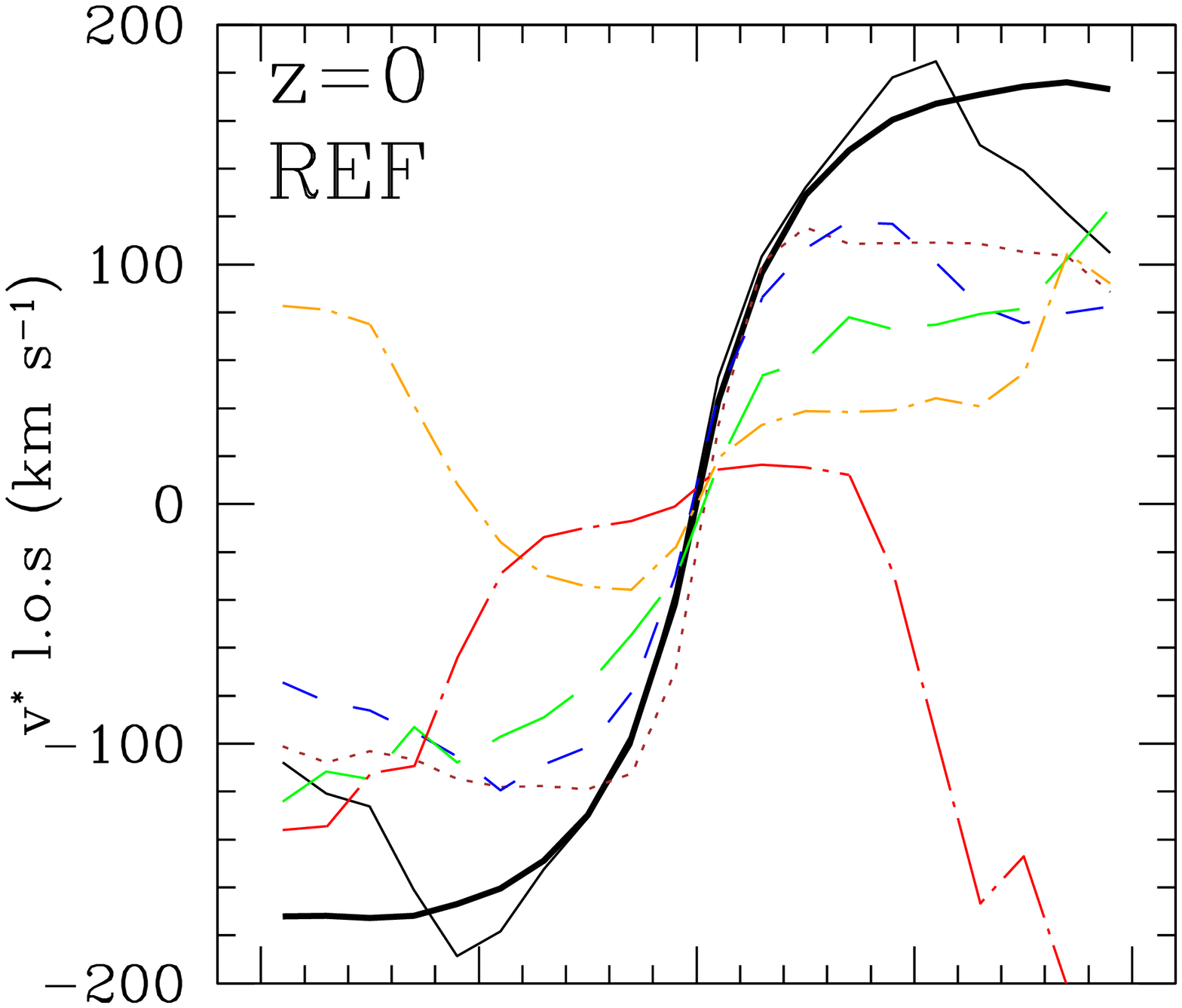}\hspace*{-1.36cm}
\includegraphics[width=50mm]{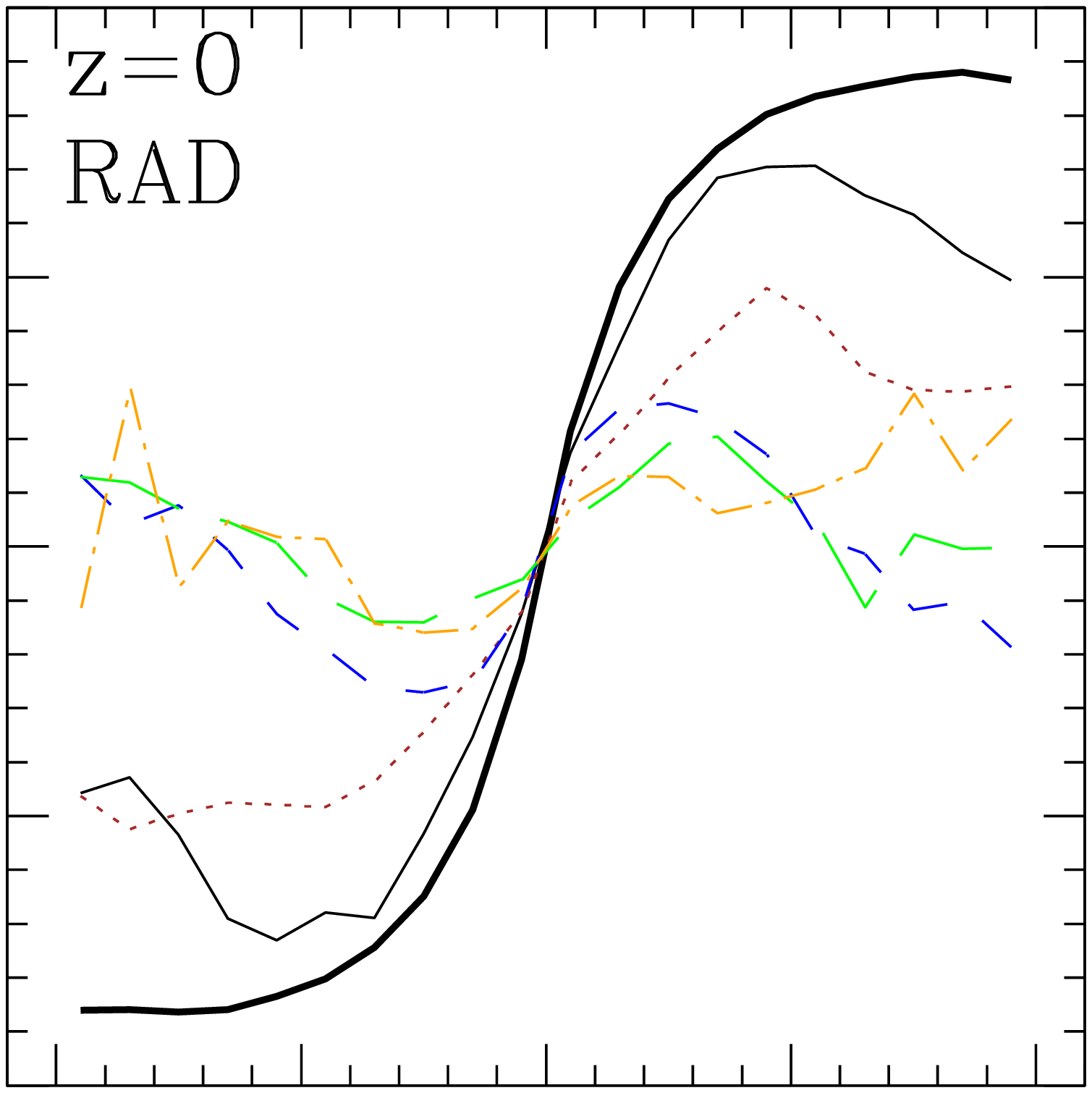}\hspace*{-1.36cm}
\includegraphics[width=50mm]{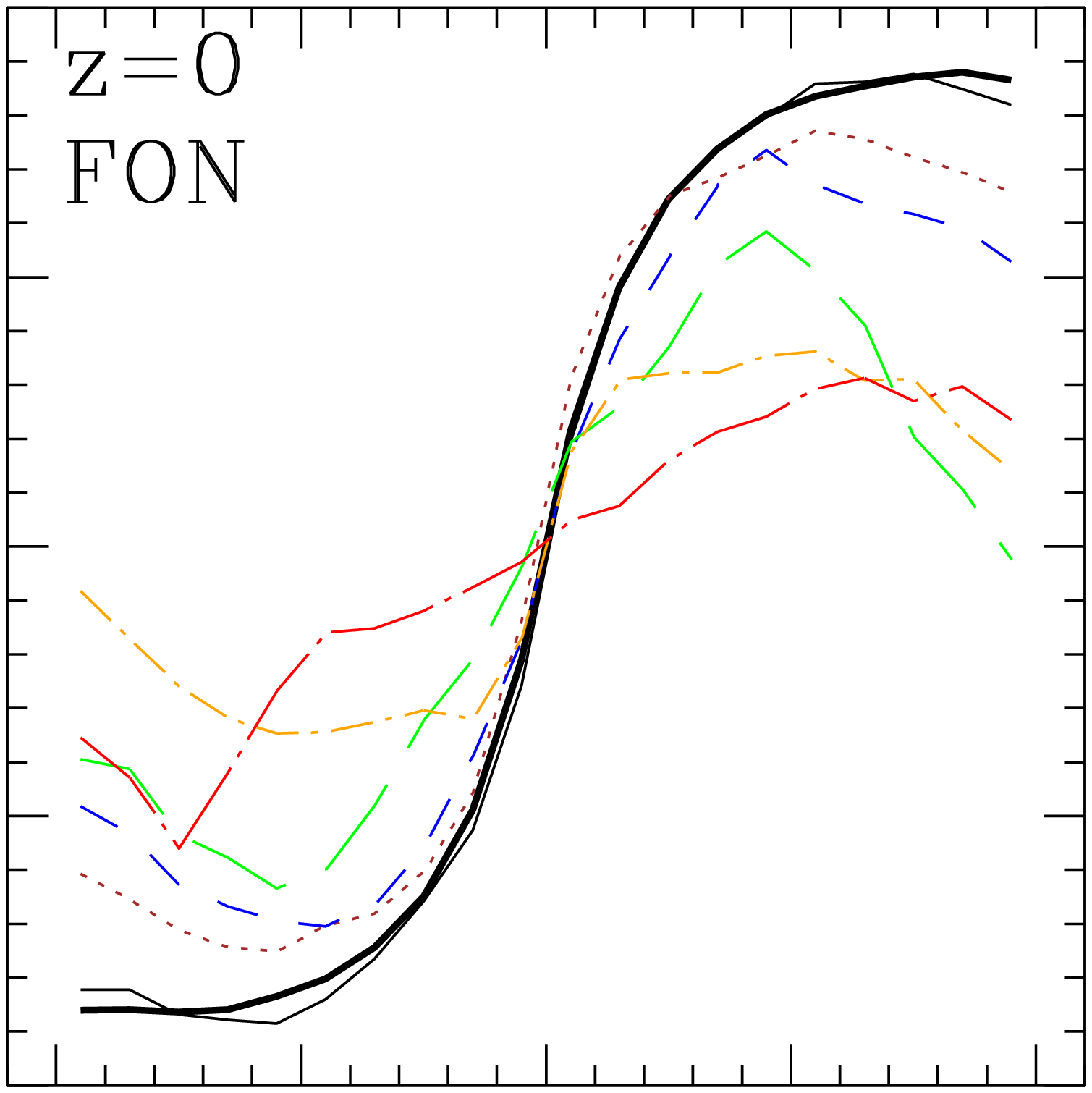}\hspace*{-1.36cm}
\includegraphics[width=50mm]{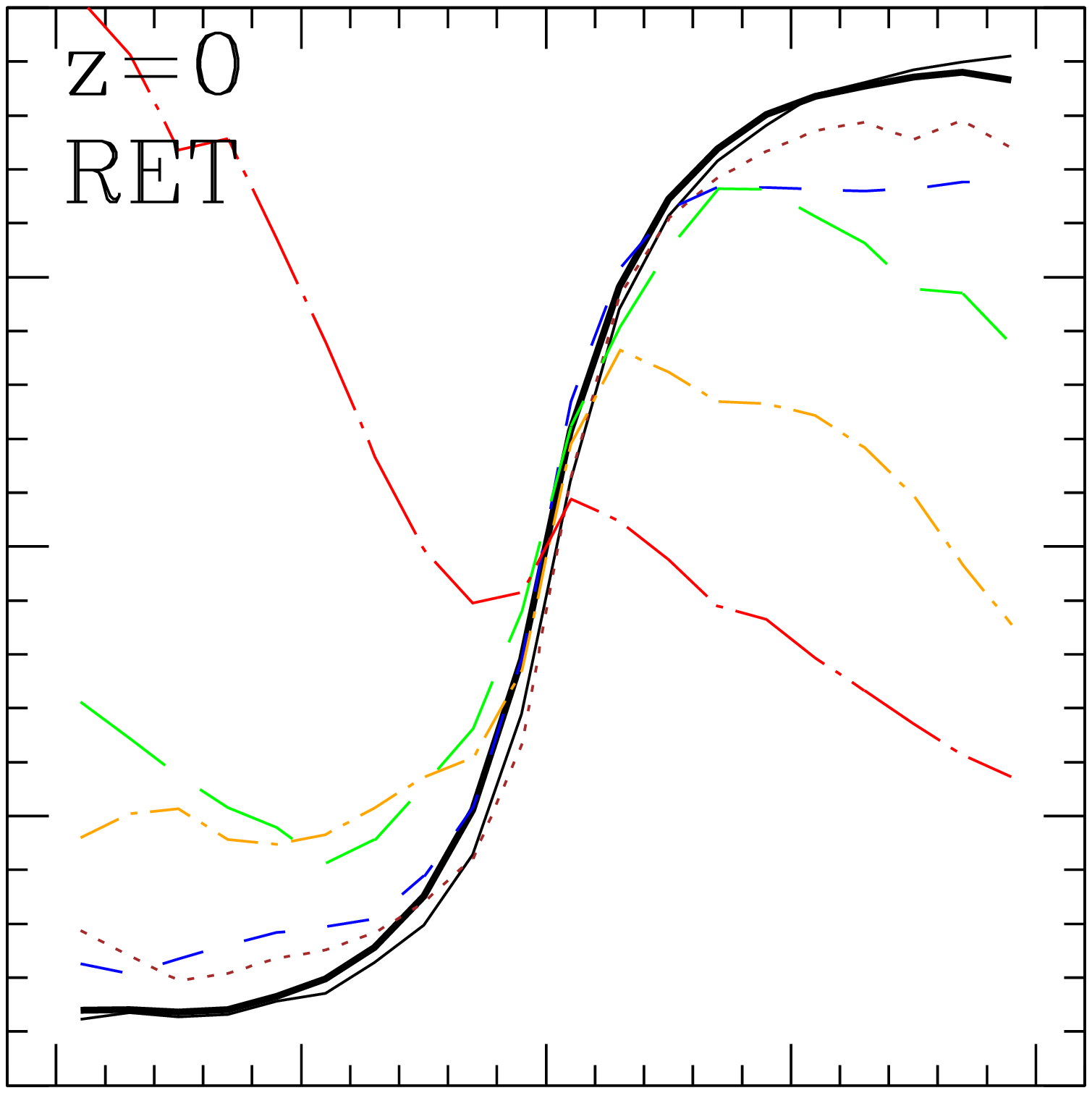}\vspace*{-1.29cm}\\
\includegraphics[width=50mm]{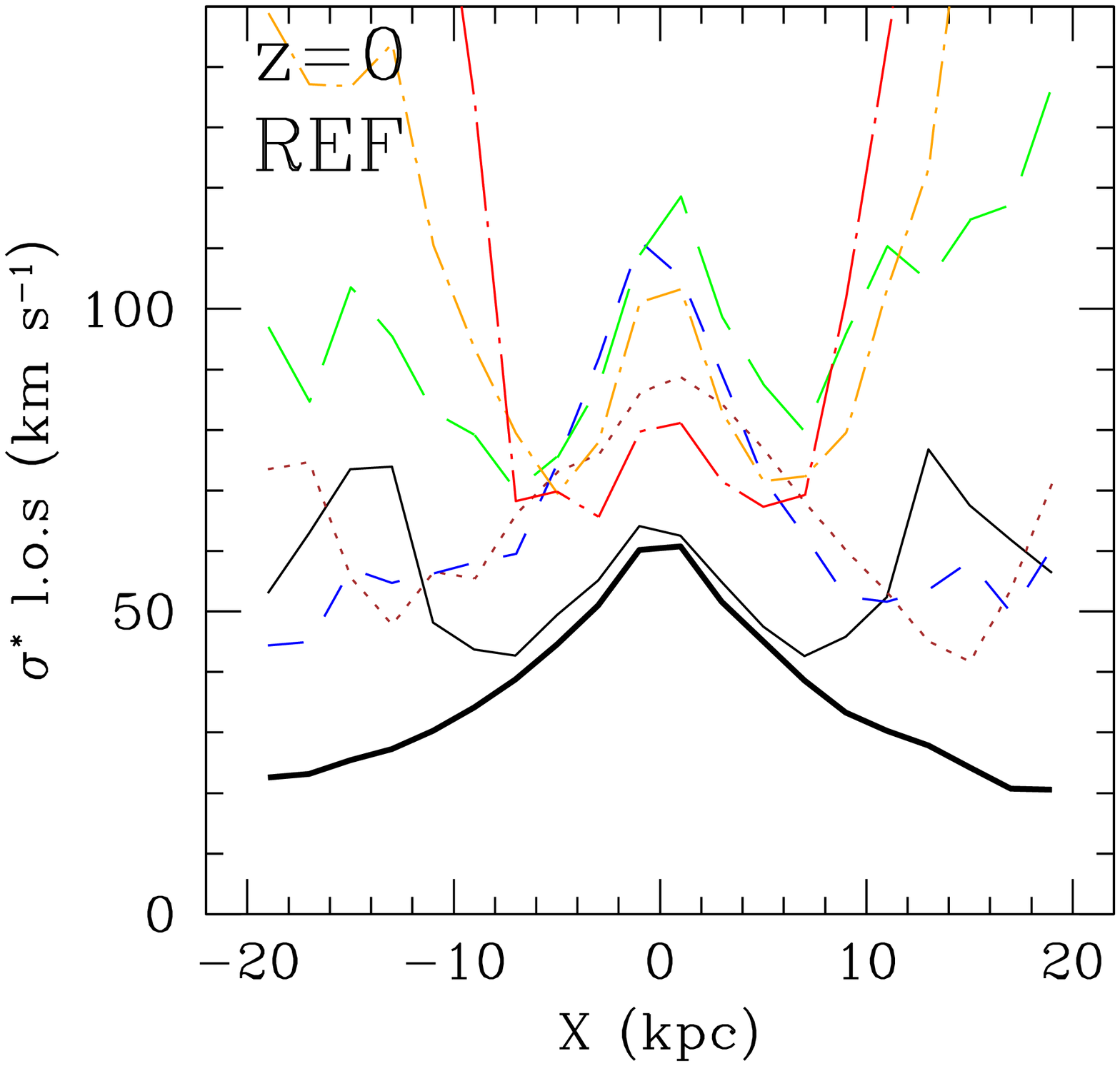}\hspace*{-1.36cm}
\includegraphics[width=50mm]{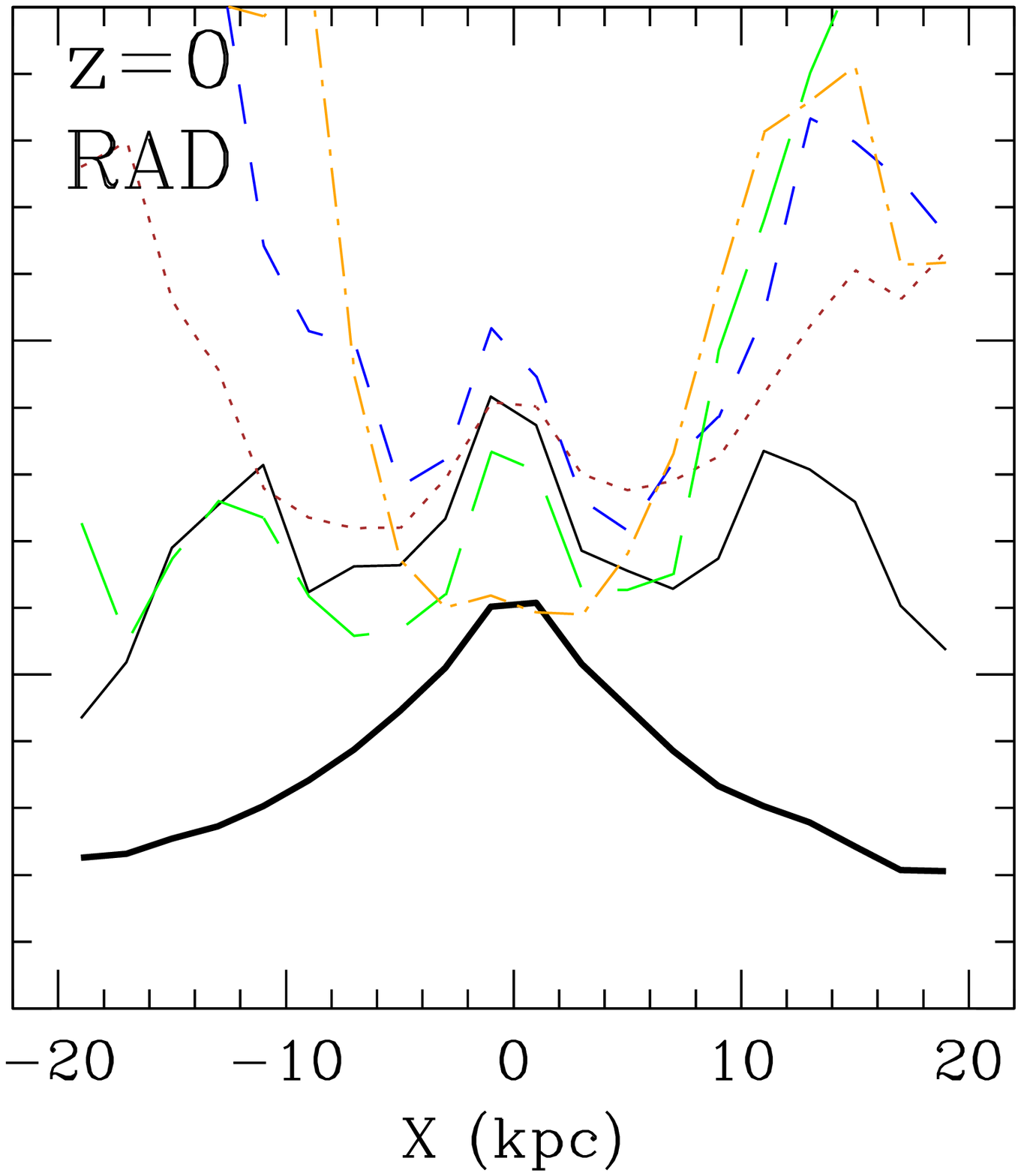}\hspace*{-1.36cm}
\includegraphics[width=50mm]{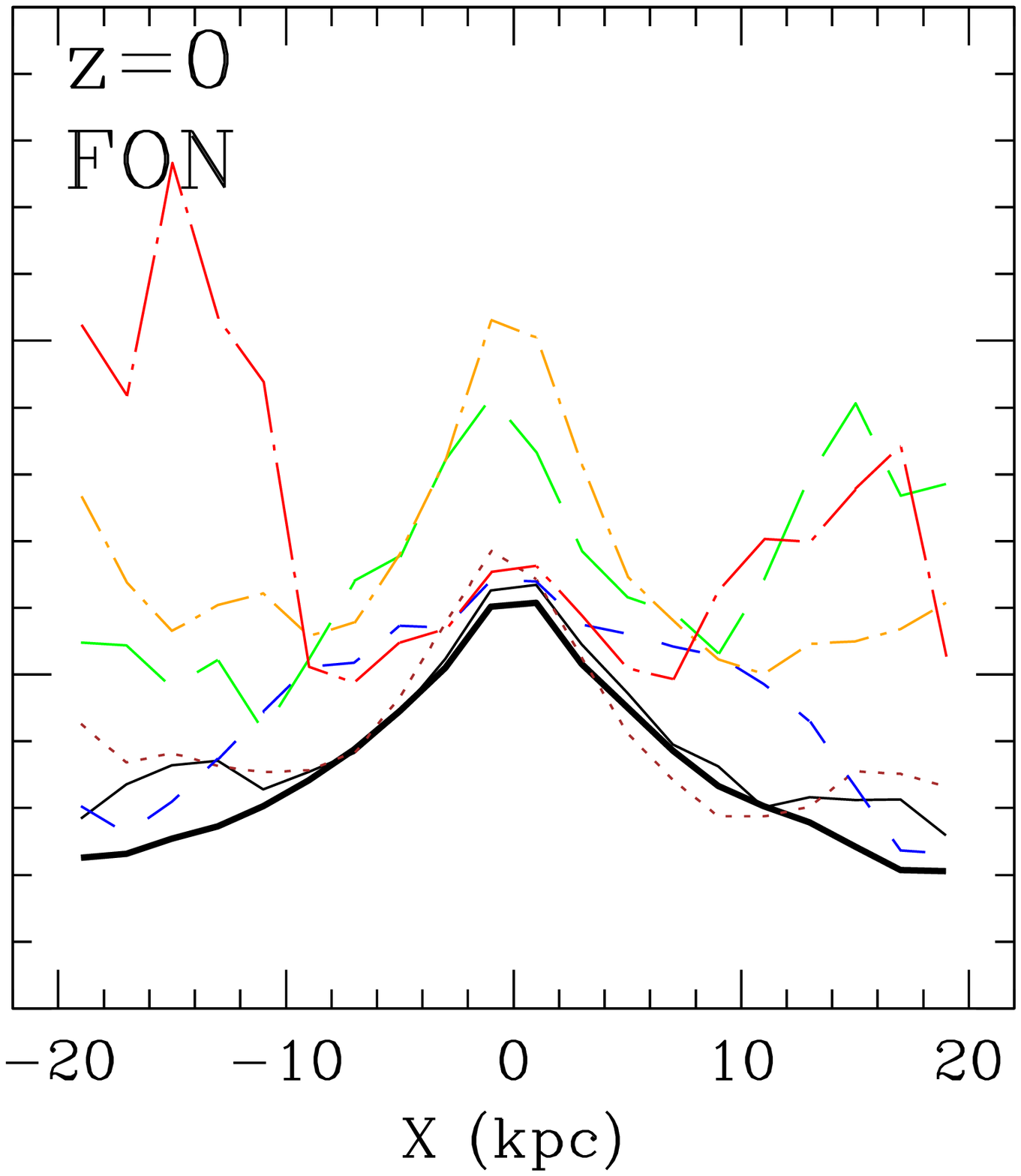}\hspace*{-1.36cm}
\includegraphics[width=50mm]{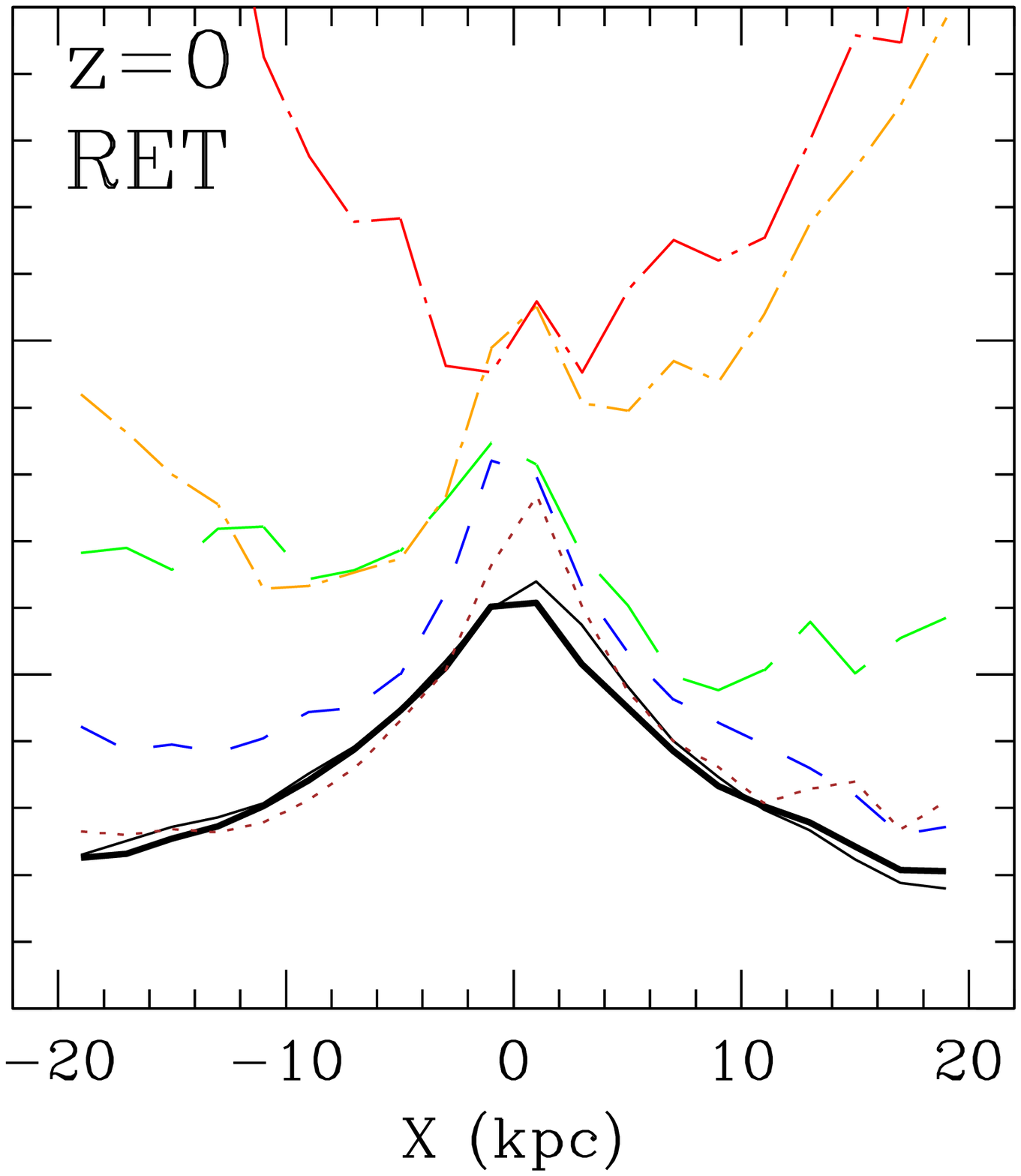}\\[-0.2cm]
\includegraphics[width=50mm]{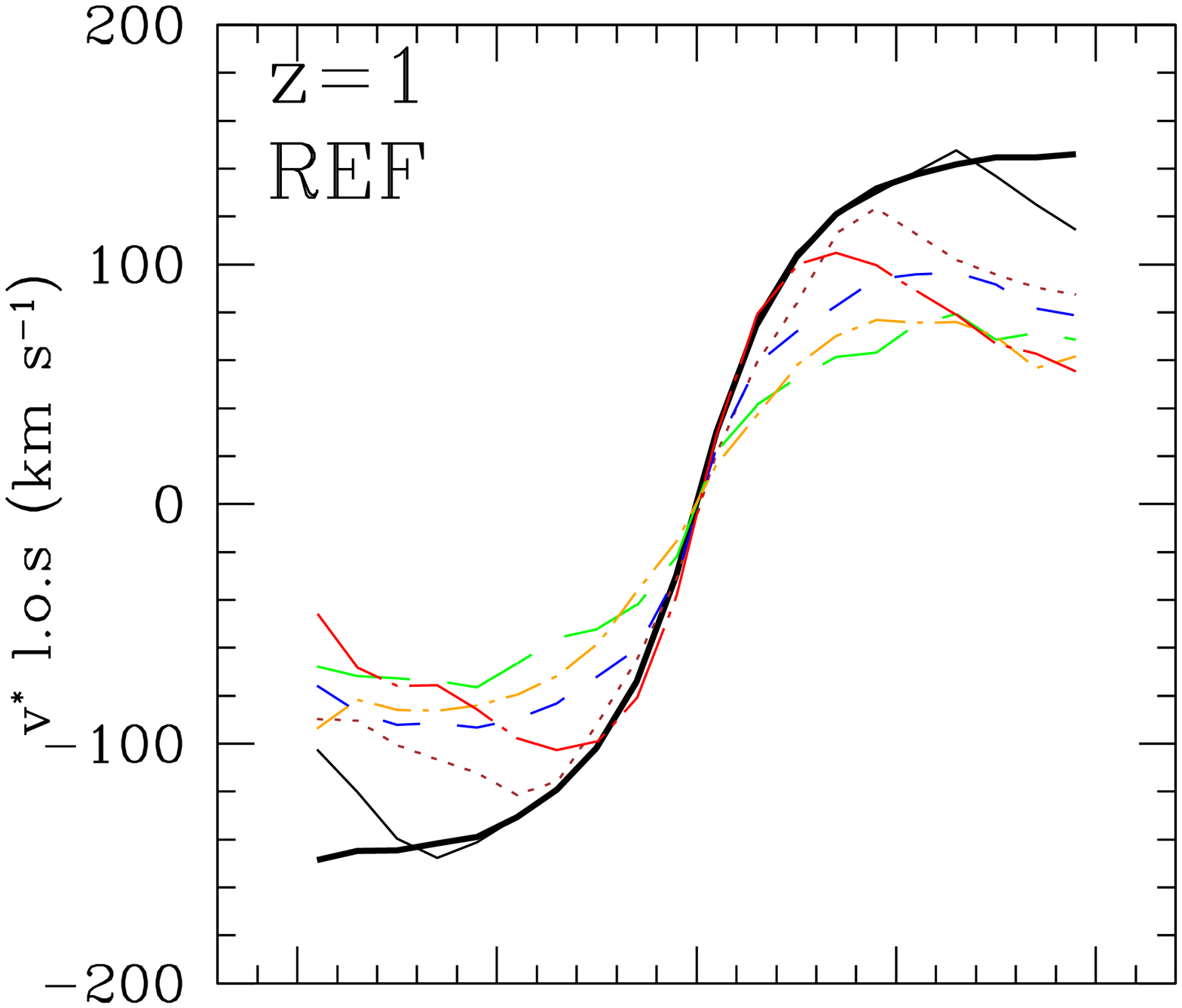}\hspace*{-1.36cm}
\includegraphics[width=50mm]{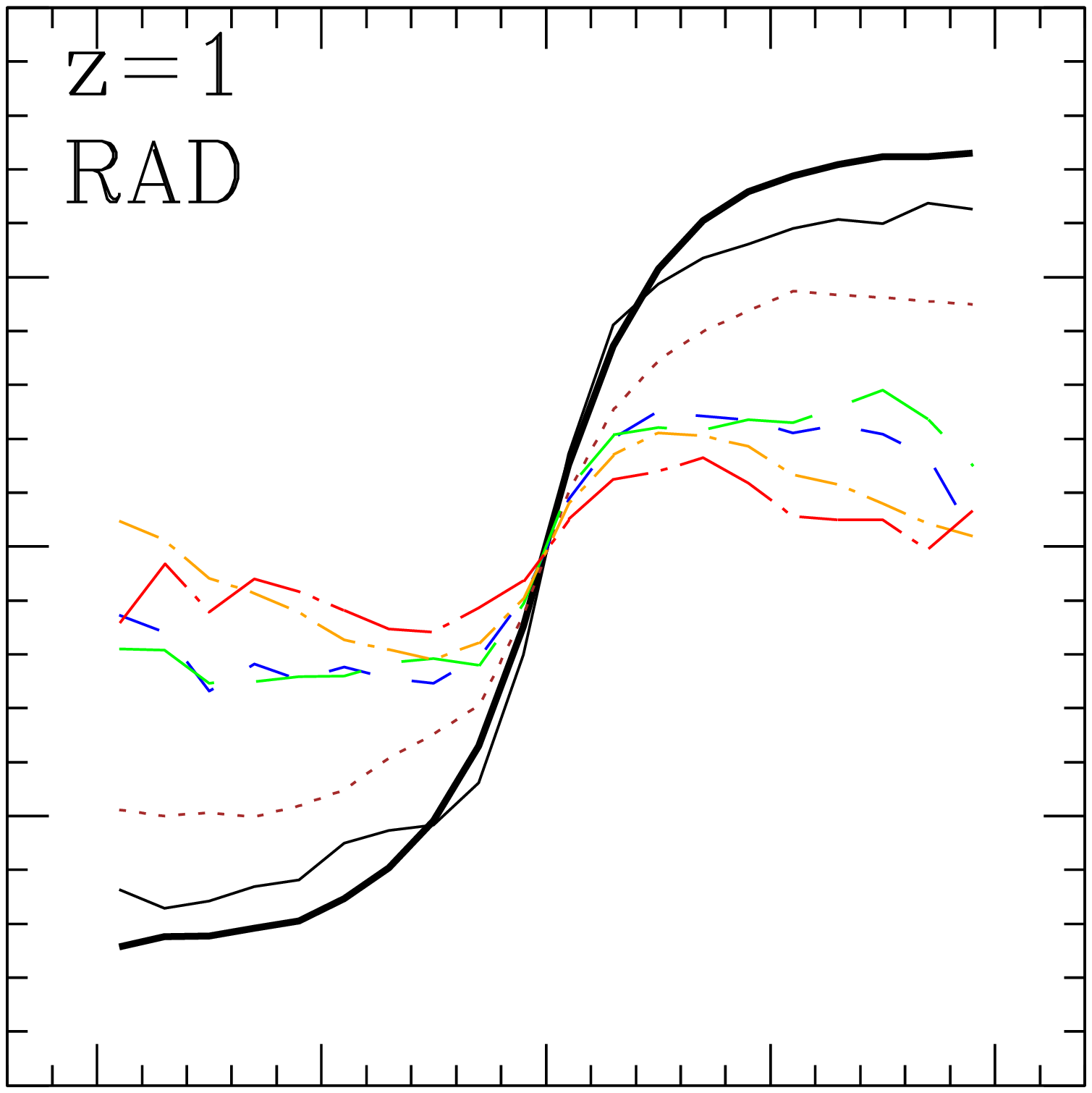}\hspace*{-1.36cm}
\includegraphics[width=50mm]{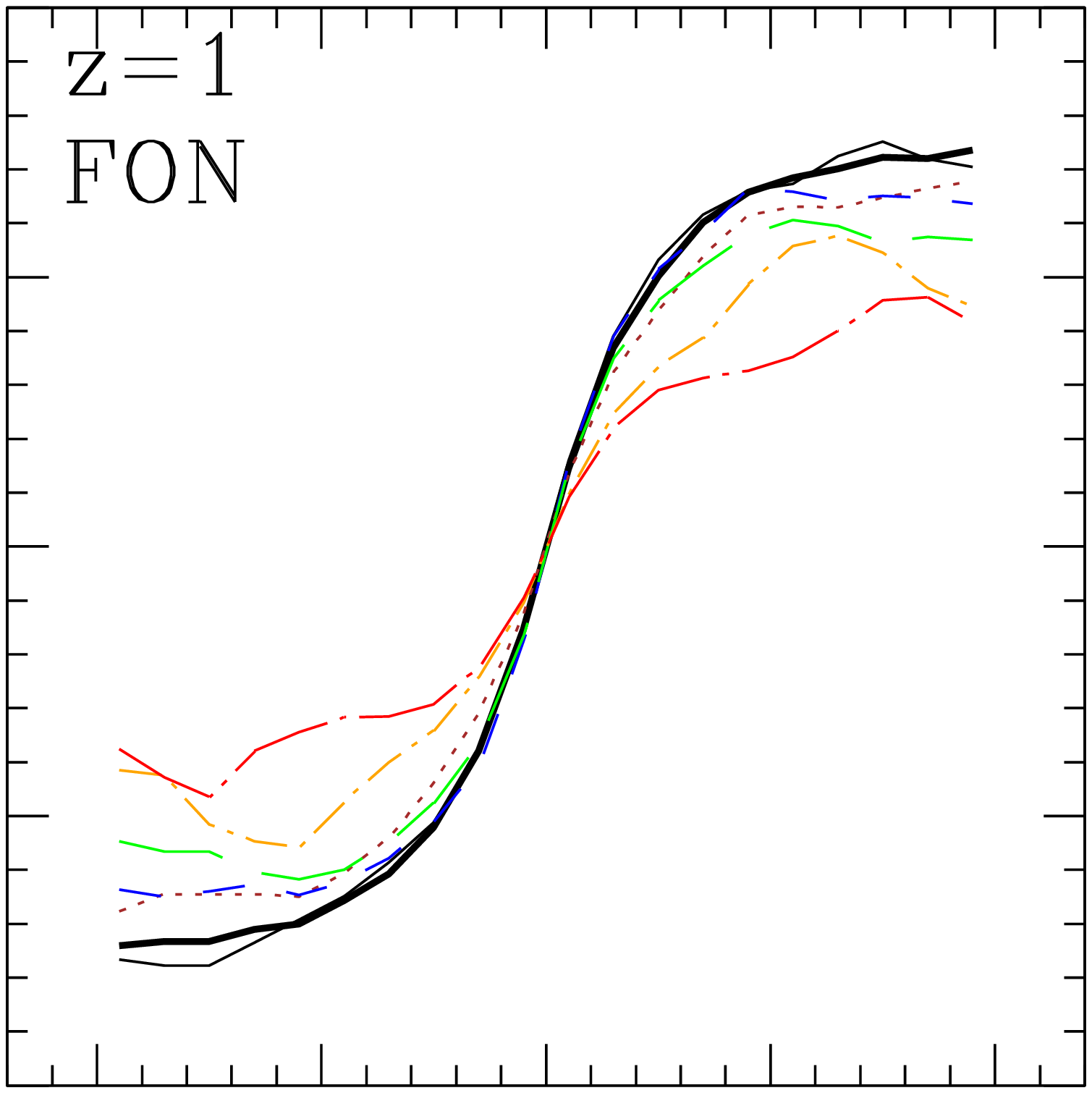}\hspace*{-1.36cm}
\includegraphics[width=50mm]{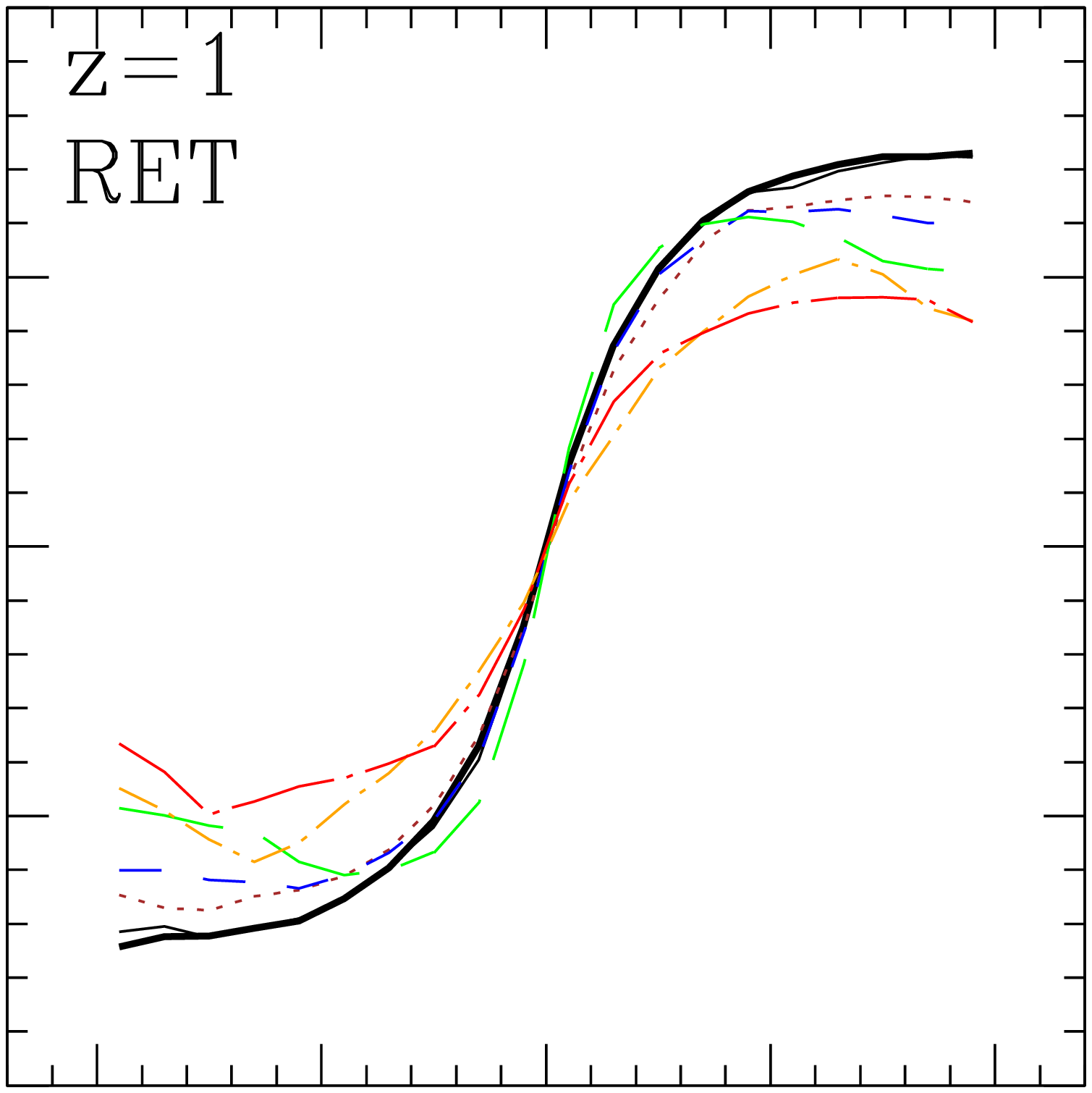}\vspace*{-1.29cm}\\
\includegraphics[width=50mm]{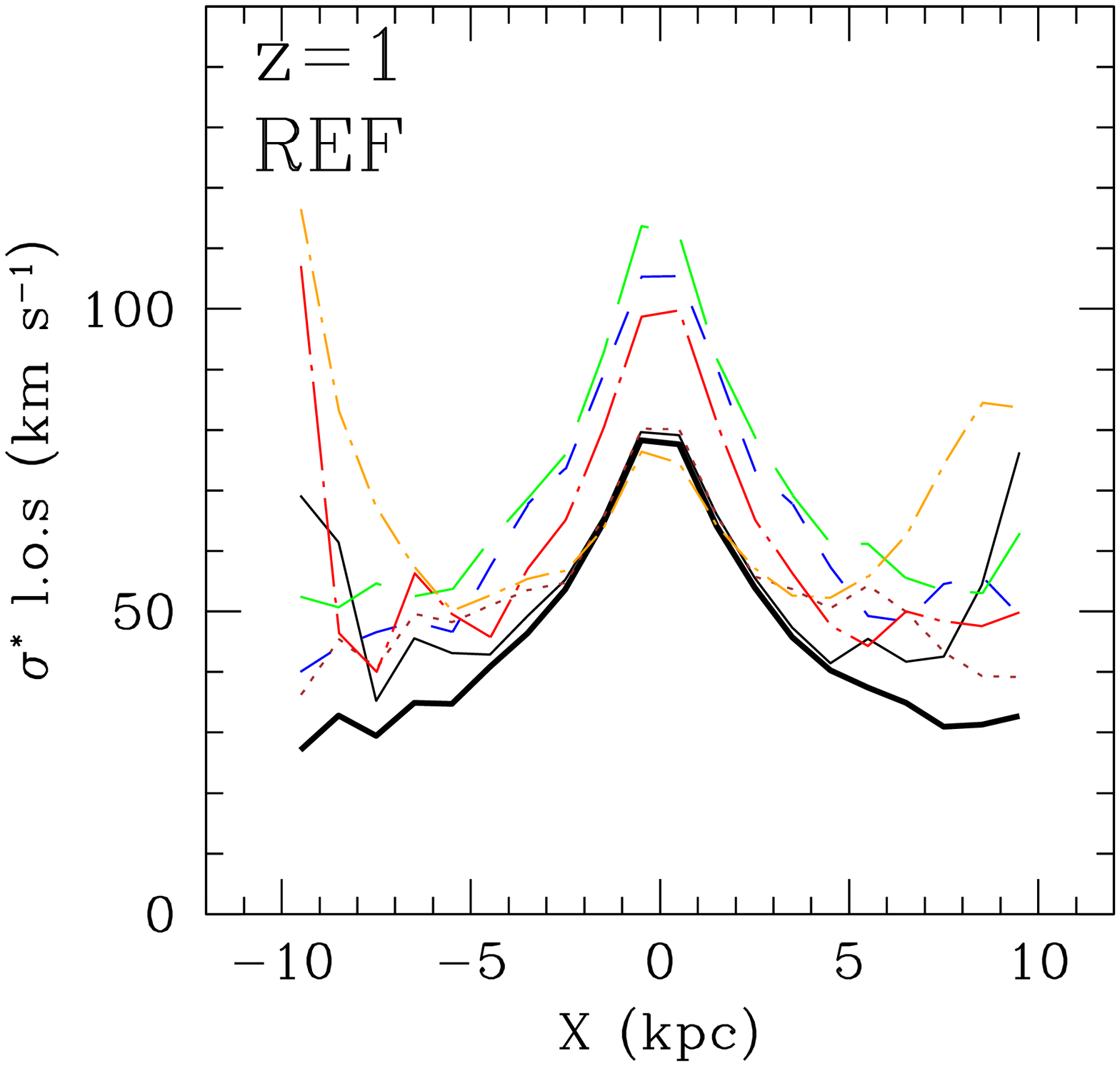}\hspace*{-1.36cm}
\includegraphics[width=50mm]{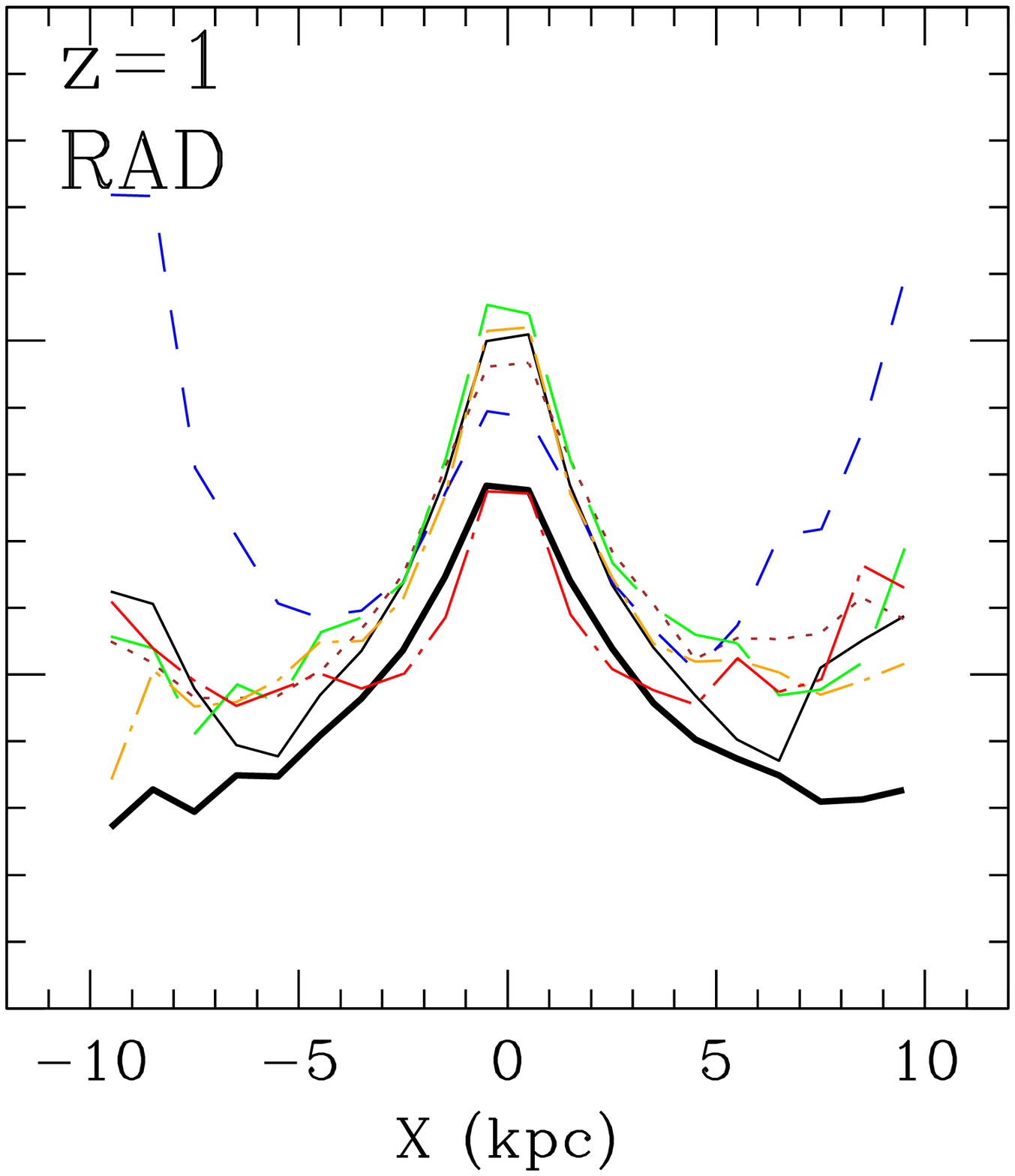}\hspace*{-1.36cm}
\includegraphics[width=50mm]{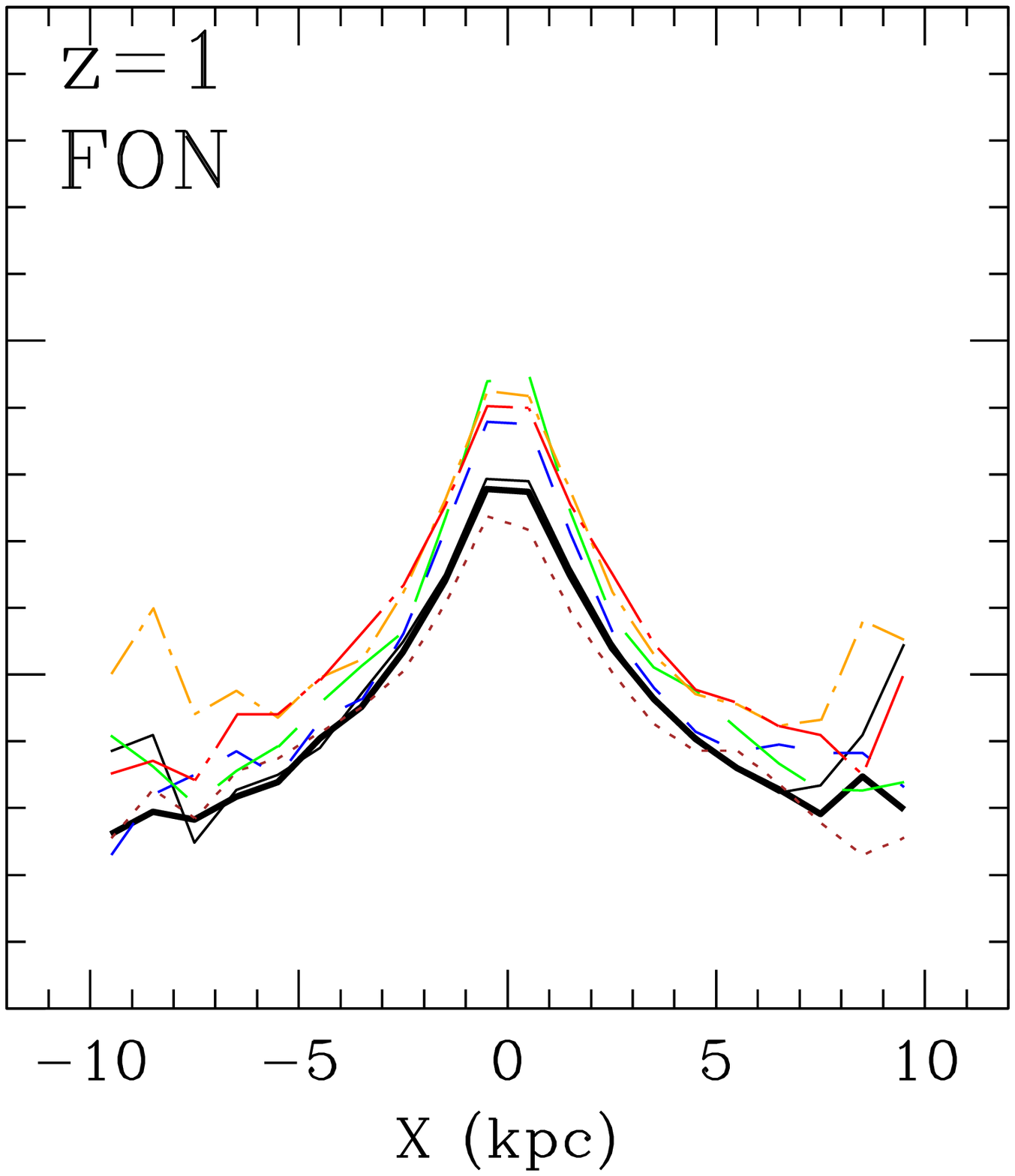}\hspace*{-1.36cm}
\includegraphics[width=50mm]{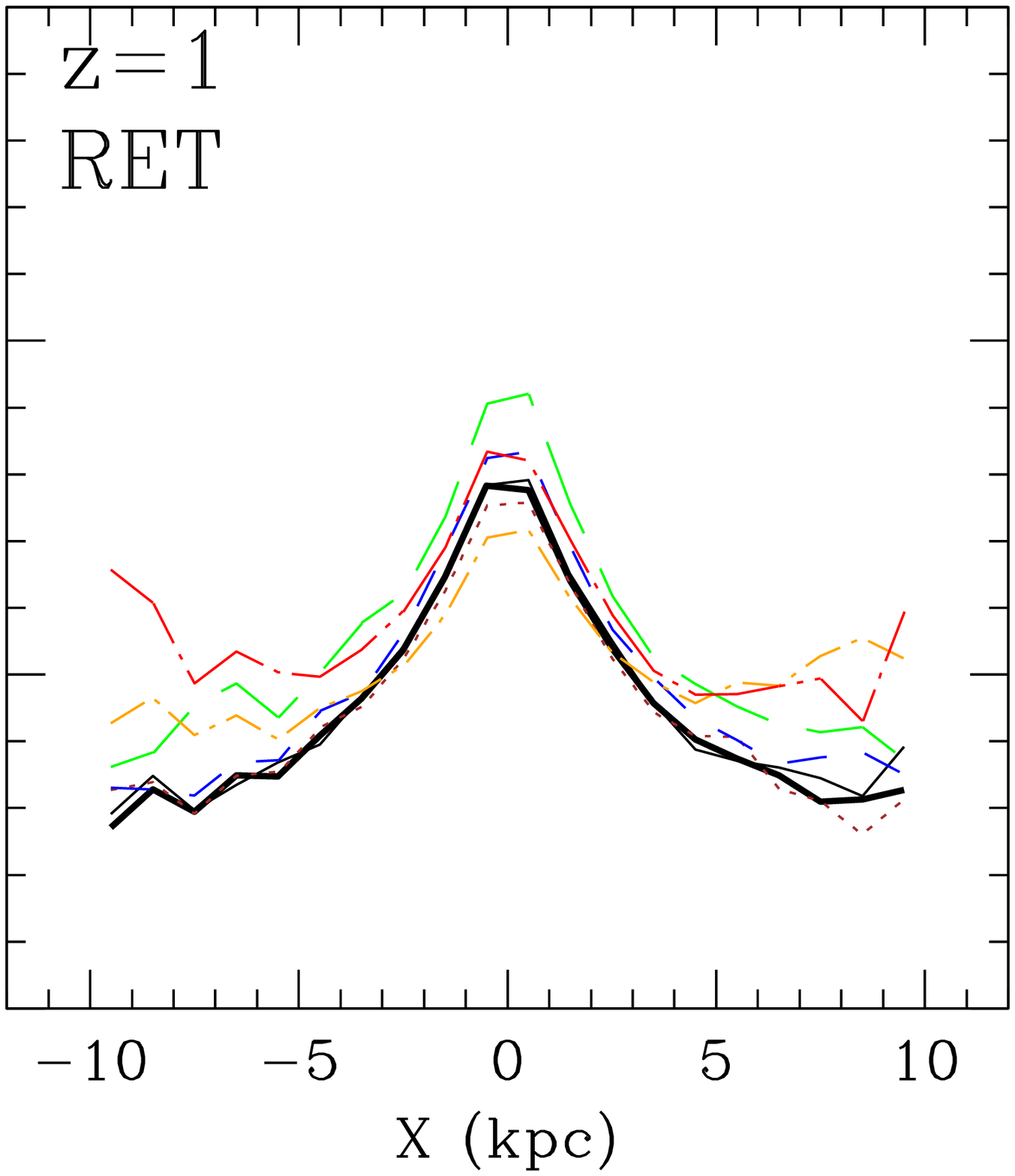}
\end{center}
\caption{
Evolution of the kinematics of disc galaxies in terms of their 
line-of-sight velocities and velocity dispersions, after each of their first 
six pericentric passages. For clarity, only the REF (reference), RAD (more 
eccentric infall), FON (face-on infall) and RET (retrograde infall) experiments 
at ``$z$=0'' and ``$z$=1'' are shown, while the rest of the experiments can 
be found in the Appendix. All discs are placed edge-on, with their midplanes 
along the X-axis. The kinematics have been computed in 1~kpc-wide bins along 
the X-axis, considering \emph{all} disc stars (both bound and unbound) 
located within 3~kpc from the midplane, and have been corrected by the 
evolution of the respective disc in isolation. Both the line and colour codes 
are as in Figures~\ref{kinematics-evol-z0} and~\ref{kinematics-evol-z1}.
}
\label{kinematics-evol-los}
\end{figure*}

In order to study the kinematical evolution of stellar discs as they orbit 
within a group environment, we first align their rotation axes along the 
Z-axis of a Cartesian system centred on the disc centre of mass. Then, the 
disc kinematics are computed in concentric rings 1~kpc-wide, out to 20~kpc 
(10~kpc) in the experiments at ``$z$=0'' (``$z$=1''). We consider only stars
that remain bound to the disc and that are within 3~kpc from the disc midplane. 
Each disc galaxy configuration has been also evolved in isolation (that is, in 
absence of external forces) in order to subtract the self-evolution of the 
discs, attempting to measure  only the evolution caused by the global tidal 
field. The kinematical evolution of discs has been computed as follows:
\begin{equation} \label{kinem-comp}
Q^*(t) \equiv Q(t) - Q_{\rm iso}(t) + Q(t=0) ,
\end{equation}
where $Q \equiv \sigma_{\rm R}, \sigma_{\rm \phi}, \sigma_{\rm Z}$, or 
$\langle V_{\phi} \rangle$. Namely, radial, azimuthal and vertical velocity 
dispersions; or mean rotational velocity.

Figures~\ref{kinematics-evol-z0} and~\ref{kinematics-evol-z1} show the 
kinematical evolution of discs after each of their first pericentric 
passages (up to six), for the REF, RAD, FON and RET experiments at redshift
``$z$=0'' (``$z$=1'').

The kinematic behaviour of discs is closely related to their morphological 
transformation. In general, the kinematical changes in discs are found to 
be mainly driven by the formation of tidal arms, and by the formation of 
central non-axisymmetric features. This, added to the step-like changes in 
both the discs scale-lengths and scale-heights described in 
Section~\ref{sec-struc}, suggest that disc stars are typically subject to 
significant violent relaxation \citep[e.g.,][]{binney-tremaine1987} after 
each pericentric passage. The efficiency of this relaxation appears to be 
linked to that of the dynamical friction in each case.

In general, the REF experiments show that the kinematics of the disc is 
mostly affected by the formation of tidal arms during the first couple of 
pericentric passages, showing a progressive increase in the coplanar velocity 
dispersions (both $\sigma_{\rm R}^*$ and $\sigma_{\phi}^*$) from the outskirts 
of the disc inwards. Note that the formation of tidal arms does not have a 
significant effect on the velocity dispersions at the centre of the disc. This 
increment in the random motions on the plane of the disc is of course 
associated with a steady decrease in the ordered motion of disc stars (i.e., 
mean circular rotation), also from the outskirts inwards. When a central 
non-axisymmetric feature is formed in the disc (typically by the third 
pericentric passage), both $\sigma_{\rm R}^*$ and $\sigma_{\phi}^*$ increase
significantly at the centre of the disc, and this increment in the velocity 
dispersion remains practically unchanged during the rest of the simulation. 
On the other hand, the mean disc rotation is observed to decrease after each 
pericentric passage, featuring a flat mean rotation profile as a function of 
galactocentric radius. In general, the vertical velocity dispersion, 
$\sigma_{\rm Z}^*$, does not show significant variations with respect to the 
initial profile, especially in comparison to the evolution of the velocity 
dispersions on the disc plane. At first sight, this might seem contradictory 
when confronted to the large increase in the thickness of the discs, as 
described in Section~\ref{sec-struc}. According to the isothermal sheet model, 
$\sigma_{\rm Z}^2 \propto \Sigma(R) z_{\rm D}$, where the terms on the 
right-hand side are the mass surface density of the disc and its scale-height, 
respectively. According to this relation, $\sigma_{\rm Z}$ could remain 
roughly constant with time, if an increase in the disc scale-height is 
cancelled out by a decrease in the disc surface density (see 
Figure~\ref{lnsigma-thickness-evol-ref-z0-z1}).

In the RAD experiments, the rapid formation of central non-axisymmetric 
features in discs during the first pericentric passage drives the disc 
kinematical evolution. Therefore, we do not observe a gradual increase in 
both $\sigma_{\rm R}^*$ and $\sigma_{\phi}^*$ from the outskirts inwards 
(related to the formation of tidal arms) but instead a sudden increment at all 
radii, which remains roughly unchanged during several pericentric passages. 
In this context, the mean rotation of the disc appears to decrease faster 
than in the REF experiment, in subsequent pericentric passages.

Discs in the FON and RET experiments show a progressively milder kinematical 
evolution related to tidal arms, in comparison to the REF experiments. This 
is to be expected since the formation of these features is inhibited by the 
initial inclination of the discs, as described in Section~\ref{sec-morph}.
Instead, in these cases the kinematical evolution of discs is mainly driven 
by the slow growth of the central non-axisymmetric feature, reaching a 
maximum increase in the velocity dispersion only after the fourth pericentric 
passage. This is also consistent with the fact that these discs retain for
longer their initial mean rotation compared to the disc in the REF experiment.

The rest of the experiments qualitatively follow a similar kinematical evolution
(see Appendix). Discs with a central stellar bulge (BUL) and that with 
``colder'' kinematics (COL) show a kinematical evolution that is comparable 
to that of the REF experiments. In particular, in the BUL experiment the 
extra stellar mass in the central region allows the disc to retain a higher 
central mean rotation with respect to its outskirts, after each pericentric 
passage. In the COL experiment, the early formation of a central 
non-axisymmetric feature appears correlated to an early increase in the 
coplanar velocity dispersions in the central region of the disc. Even after 
repeated pericentric passages, discs in more circular infalling orbits (CIR) 
show a negligible kinematical evolution due to the lack of interaction with 
the dense central region of the group. Discs in galaxies with a higher (lower) 
galaxy-to-group mass ratio show a rather similar evolution compared to the REF 
experiments. The main difference is that those discs have fewer (more) 
pericentric passages, with respect to the REF experiments, given the efficiency 
of the dynamical friction in each case.

Observationally, it is not easy to obtain three-dimensional measurements 
of the kinematics of disc galaxies. Let alone to distinguish between bound 
and unbound stars in them. Much more accessible are measurements of kinematics
integrated along the line-of-sight. Similarly to Figures~\ref{kinematics-evol-z0} 
and~\ref{kinematics-evol-z1},  Figure~\ref{kinematics-evol-los} shows the 
evolution of the kinematics of discs in our experiments at ``$z$=0'' and 
``$z$=1'', according to measurements of line-of-sight velocities ($v_{\rm los}$) 
and line-of-sight velocity dispersions ($\sigma_{\rm los}$) along their 
edge-on projections. In these measurements we have included \emph{all} disc 
stars (i.e., both bound and unbound) with projected radii $<$20~kpc ($<$10~kpc) 
for our ``$z$=0'' (``$z$''=1) experiments, and within $|z|$$<$3~kpc\footnote{Our 
results do not show significant differences for smaller vertical limits, down 
to one disc scale-height.} from the midplane. These measurements have been 
corrected to account for the evolution of discs in isolation, as described above.
The REF experiment at ``$z$=0'' shows that, after successive pericentric passages, 
$v_{\rm los}$ decreases from the outskirts inwards, denoting that the disc is 
progressively losing rotation in the outer region. As we have seen in the previous 
sections, this behaviour is related to the formation of tidal arms in the external 
regions of the disc. After the last pericentric passage, a low remaining rotation
can still be measured in the central regions of the disc, while at larger radii 
the disc does not exhibit ordered rotation. This is due to the presence of a small 
fraction of stars still bound at the centre of the disc, while the external
region is mostly populated by unbound stars with higher velocity dispersions. 
After a the first pericentric passage, $\sigma_{\rm los}$ increases mostly at 
larger radii, which is consistent with the described decrease of $v_{\rm los}$ 
and the formation of tidal arms. After the following pericentric passages, 
$\sigma_{\rm los}$ increases significantly at smaller projected radii, which as 
described above is related to the formation of a central feature in the disc. 
At larger projected radii, $\sigma_{\rm los}$ also increases dramatically, due 
to the higher fraction of unbound stars with high velocity dispersions. After 
performing similar measurements for different (edge-on) projections of the disc, 
we find that measurements of $v_{\rm los}$ are remarkable consistent.
On the other hand, $\sigma_{\rm los}$ measurements at smaller projected radii 
show a significant dependence on projection, given the non-axisymmetric spatial 
distribution of the central feature formed in the disc. At ``$z$=0'', the RAD 
experiment behaves as the REF experiment, but the disc evolves more violently, 
with larger decreases (increases) of $v_{\rm los}$ ($\sigma_{\rm los}$) after 
each pericentric passage. The FON and RET experiments are also similar to the 
REF experiment, only being more capable of retaining larger amounts of rotation 
after several pericentric passages. The REF, RAD, FON and RET experiments at 
``$z$=1'' are qualitatively similar to those at ``$z$=0''. However, in general, 
discs at ``$z$=1'' retain more of their original rotation. The rest of the 
experiments (see Appendix) show similar behaviours compared to their respective 
REF experiments, while each of them exhibits characteristics that resemble those 
described in the analysis of the three-dimensional disc kinematics.

In summary, the evolution in the kinematics of discs orbiting within a group is 
closely linked to the morphological changes they suffer. The formation of tidal 
arms and of central non-axisymmetric features in the discs increases their 
velocity dispersions, while decreasing their rotational support. The kinematics 
of discs in more eccentric orbits is mostly driven by the formation of central 
features, since their faster orbital decay leaves less time for interactions 
leading to the formation of tidal arms. In a similar vein, as the formation of 
tidal arms is increasingly inhibited in discs with initial inclinations of 
90\degr and 180\degr, their kinematical evolution is also driven by the formation
of central features. Galaxies with a central bulge are able to retain a larger 
fraction of their rotational support, while the evolution of a disc with colder 
kinematics is driven early on by the formation of a central feature. Notably, 
galaxies in less eccentric orbits show a negligible kinematical evolution since 
their structures are less affected by strong global tidal forces. Differences in 
the galaxy-to-group mass ratios of discs are only linked to faster or slower 
kinematical evolution, at timescales dictated by the efficiency of the dynamical 
friction. 

\section{Discussion}
\label{sec-discuss}

\subsection{Comparison to similar studies}

\citet{mastropietro2005} explored the evolution of disc galaxies within a cluster 
environment drawn from a cosmological simulation. In their implementation, they 
randomly chose 20 cluster particles, replacing them with a single $N$-body model 
of a multi-component galaxy, composed by a stellar disc and a spherical DM halo. 
In this way, the authors simulated the evolution of discs accounting for the 
interaction of galaxies with the global potential, galaxy harassment by means
of close encounters among galaxies (and with substructure within the cluster),
as well as the hierarchical growth of the cluster as a function of redshift. 
Their main goal was to study the transformation of disc galaxies into dE/dSph 
galaxies within a cluster environment.

The experiments presented here differ from those of \citeauthor{mastropietro2005} 
in terms of the mass scale of the environment under study, the implementation
of the simulations, the number of interacting galaxies, and the choice of the 
initial orbital parameters of galaxies. However, it is interesting to compare 
how galaxies evolve in the two studies, and to learn how much of the evolution 
of disc galaxies in a cluster environment can already be observed within a 
group environment.

In their study, \citeauthor{mastropietro2005} found that disc galaxies undergo 
major structural transformations, caused by both the interaction with the global 
cluster potential, and the close gravitational encounters with other galaxies 
and substructure. These structural transformations were characterised by the 
formation of spiral patterns and a central bar. Both tidal heating and loss of 
stellar mass ultimately remove the remaining spiral features, and give to the 
bar a rounder shape. In terms of kinematics, in spite of the loss of stellar 
mass, the velocity dispersion is shown to increase at the centre of the disc, 
driven by the formation of the bar. Nonetheless, discs are found to retain a 
significant amount of rotation, which helps them keep some of their original 
disc structure. The authors also showed that discs that orbit mostly in the 
outskirts of the cluster do not lose a significant fraction of their stellar 
mass, while they do experience structural changes, similar to those described 
above.

In our experiments, we observe that disc galaxies orbiting within a group-size 
halo show a similar evolution compared to that within a cluster environment. 
Namely, the discs suffer the formation of spiral patterns and of a central 
bar, along with increasing velocity dispersions. However, it is important to 
highlight some relevant differences, especially when comparing the results by
\citeauthor{mastropietro2005} to our experiments at redshift ``$z$=1''. After 
the formation of a central bar in our discs, the tidal heating provided by the 
global potential is not strong enough to cause a significant mass loss that 
would transform the structure of the galaxies from discs to spheroidals. In our 
experiments, galaxies do show a significant amount of thickening, likely 
related to the vertical heating caused by the formation of the central bar. 
However, we find that the formation of a strong bar, that heats efficiently 
the vertical structure of the disc, is not a generic feature of the interaction 
between a galaxy and a group environment. Instead, it appears to depend strongly 
on the inclination of the disc galaxy with respect to its orbit, at the time of 
infall. Additionally, we observe that global changes in the structure and 
kinematics of discs appear to be closely correlated to the pericentric passages 
experienced by the galaxies during their evolution, in contrast with findings
by \citeauthor{mastropietro2005}.
Finally, we find that galaxies orbiting mostly in the outskirts of a group show 
a negligible loss of stellar mass, like similar galaxies evolving within a 
cluster. However, contrary to \citeauthor{mastropietro2005}, we find 
insignificant evolution in both the structure and kinematics of those discs. 
Our results suggest that the group environment can indeed contribute to 
``pre-process'' galaxies before they are accreted on a cluster. The 
``pre-processing'' appears mild for galaxies sitting at the outskirts of groups.
We stress, however, that our relatively simple experiments do not account for 
close encounters between group members, which could play an important role in 
the evolution of disc galaxies.

Note that since \citeauthor{mastropietro2005} assigned to their galaxies on 
random initial orbits, it is not straightforward to compare their results to 
ours, especially regarding what kind of galaxy evolution would be more likely. 
In addition, since their experiments included the joint effect of global tidal 
forces and harassment, it is not possible to isolate their separate 
contributions to galaxy evolution. 

In a recent study, \citet{kazantzidis2011} examined the formation of dwarf 
spheroidal galaxies as the result of tidal transformations in 
rotationally-supported disc dwarfs that interact with a MW-like galaxy. Such 
interactions were modelled using $N$-body simulations, starting from 
self-consistent models of stellar discs embedded in cosmologically motivated 
DM halos. The authors studied a broad set of configurations for the 
interactions, and their impact on the structure and kinematics of the final 
dwarf galaxies. Even though their study is conceptually different from ours, 
and from that by \citeauthor{mastropietro2005}, we consider instructive to 
draw parallels between disc galaxy evolution across environments of different 
mass scales.

\citeauthor{kazantzidis2011} found that dramatic tidal transformations in 
dwarf disc galaxies were mainly driven by efficient tidally-induced bar 
instabilities, as well as by strong impulsive tidal heating, taking place 
during pericentric passages. These mechanisms ultimately lead to the 
transformation of rotationally-supported discs into pressure-supported 
spheroids. Our results, as well as those found by \citeauthor{mastropietro2005} 
and \citeauthor{kazantzidis2011} confirm that disc galaxies follow a rather
similar evolutionary path, where tidal transformations play a central role, 
independently of the masses of galaxies and the general characteristics of 
their respective environments. This is a consequence of the current 
cosmological paradigm, where encounters between structures, on rather 
eccentric orbits, are of common occurrence.

It is interesting to observe that, contrary to our results, \citeauthor{kazantzidis2011}
did not find a significant variation in the evolution of dwarf discs with 
respect to the inclination of their infalling orbits. This is likely due to 
the fact that they only studied the evolution of disc galaxies on prograde
orbits, with inclinations in the range $0\degr \leq \theta \leq 90\degr$. As
described in previous sections, we find that disc galaxies infalling either 
face-on ($\theta=90\degr$) or on an edge-on retrograde orbit ($\theta=180\degr$) 
are increasingly capable of retaining stars during the infall 
\citep[see also][]{read2006}, and are increasingly resilient against the 
formation of a strong central bar \citep[see also][]{mayer2001}. The 
reduced bar-driven evolution in the discs allows them to maintain their
disc structure for longer times during their interaction with the global 
potential. This result might have potentially relevant implications for the 
morphology-density relation, since it would naturally account for a significant 
fraction of disc galaxies in the inner dense regions of groups. Indeed, 
prograde and retrograde infalls should be equally likely to occur. In the 
context of our experiments, it remains to be seen if such characteristics of 
discs in retrograde orbits remain in place after the addition of close-encounter 
interactions in our models.

Close encounters between galaxies are expected to be frequent in groups, 
given the number of group members in environments of this mass scale. 
\citet{yang2009} have estimated the number of satellite galaxies as up to 
$\sim$~10 in groups of $\sim 10^{13} M_{\sun}$, using a large catalogue of 
galaxy groups in the redshift range $0.01<z<0.2$, compiled from the Sloan 
Digital Sky Survey Data Release 4 \citep{adelman-mccarthy2006}. Given the 
velocity dispersions of groups, these encounters are low-velocity interactions,
as opposed to high-velocity interactions typical of larger environments, known 
as harassment \citep[e.g.,][]{moore1998}. Low-velocity close encounters 
would cause less impulsive but longer lasting asymmetric perturbations in the 
gravitational field surrounding a galaxy. The cumulative effect of these 
frequent perturbations are expected to play an important role in galaxy 
evolution, since they can induce the formation of strong tidal arms and 
central asymmetric features in discs, as well as, episodes of star formation. 
As mentioned in the Introduction, we do not include the effect of close 
encounters in this study. However, in the context of our experiments, we 
expect that these interactions would lead to a more significant morphological
evolution in disc galaxies, inducing the formation of additional tidal arms 
that are not correlated to the pericentric passages of galaxies. This could 
cause earlier formation of (possibly stronger) central asymmetric features,
an even larger increase in the velocity dispersion of discs, and an 
accelerated rate of mass loss. We defer to a future paper of this series a 
detailed study of the effect of close encounters in galaxy groups, and of 
its relative importance with respect to the interactions with the global 
potential.

\subsection{On the formation of S0 galaxies in groups}

Galaxies classified as S0 are characterised by having little gas content, 
practically no signs of spiral arms, and often a prominent central bulge. 
Their formation process is currently unknown, although observational evidence 
suggests that S0 galaxies are formed from disc galaxies that experience structural, 
kinematical and chemical transformations, mostly induced by their environments
\citep[e.g.,][]{dressler1980,vandokkum1998}. It is during those transformations 
that their central bulges should grow significantly, in comparison to those of 
spiral galaxies. In recent years, there has been mounting observational evidence 
suggesting that group environments are more efficient at forming S0 galaxies 
in comparison to cluster environments \citep{wilman2009,just2010}. Our 
simulations show that the mass surface density profiles of disc galaxies
remain exponential out to $\sim$6$R_{\rm D}$, during the evolution within a 
group (see Figure~\ref{lnsigma-thickness-evol-ref-z0-z1}). At face-value, 
this result could have relevant implications for the formation of S0 galaxies 
in groups. Our simulations show that disc galaxies neither form central stellar 
bulges nor enhance pre-existing ones, during their interaction with the global 
tidal field. This implies that if S0 galaxies were formed within galaxy groups, 
then their bulges could not be produced by the effect of group tidal forces 
alone (e.g., via bar instabilities) on old disc stars that were already in place 
when the galaxy was accreted by the group. Thus, if S0 galaxies are mostly formed 
in groups, our simulations would support formation models in which their bulges 
are currently composed by a significant majority of relatively young stars 
\citep[e.g., see][]{bekki2011}. Note, however, that our experiment exploring a 
lower value of the galaxy \emph{Q} parameter (COL experiment) does show a 
relatively small accumulation of stars in the central region of discs, compared 
to the reference experiment. This behaviour is expected since discs that are 
kinematically colder (i.e., have lower \emph{Q} parameter) are less resilient 
against the formation of central non-axisymmetries. A detailed analysis of the 
dependency of the size of the central non-axisymmetric features as a function 
of the \emph{Q} parameter of disc galaxies is, however, beyond the scope of 
this paper.

\subsection{On models of tidal disruption/stripping}
\label{sec-disrup}

All recent semi-analytic models of galaxy formation (SAMs) exhibit an 
excess of faint and passive satellites with respect to observational measurements
\citep[see e.g.,][]{wang2007,fontanot2009}. In a recent paper, 
\citet{weinmann2011} have shown that the semi-analytic models predict a higher 
ratio of dwarf-to-giant galaxies within clusters compared to observations. In 
addition, they also show that SAMs do not exhibit a decrease in the dwarf-to-giant 
ratio, as observed in the central regions of clusters
\citep[e.g.,][]{sanchez-janssen2008,barkhouse2009}. \citeauthor{weinmann2011}
argue that the most likely solution to this problem is the adoption of a more 
efficient tidal disruption for faint galaxies, particularly at the centre of 
clusters.

\citet{diemand2004} analysed a statistical sample of dark matter substructures 
in a set of simulated clusters, and ascribed the lack of `slow' subhalos to 
the fact that tidal stripping and tidal disruption is very efficient for these 
systems. In this scenario, faint galaxies that are also slow would be disrupted 
more easily by the strong tidal field of the dense central regions of clusters. 
In addition, faint galaxies are also less dense and have a shallower potential 
well, so that they are less resilient to the influence of the tidal field of 
the parent halo. Such a mechanism would certainly lower the overall 
dwarf-to-giant ratio within a cluster, especially at its centre where the tidal 
field is stronger.  It would also explain the more extended spatial distribution 
of dwarf galaxies observed in clusters, as well as their broader velocity 
distributions. Our simulations, however, do not support this interpretation. 
Figure~\ref{massbound-evol} shows that galaxy disruption has a much stronger 
dependency on both the mass ratio between the systems and on the orbital 
eccentricity of the infalling orbit, than on the orbital energy of the galaxy. 
Note that our disc galaxies in both the CIR and RAD experiments initially have 
practically the same orbital energy (see Table~\ref{list-exper}), however, they
are disrupted in significantly different ways. Thus, our simulations offer 
indirect support to alternative explanations for the lower central concentration 
of dwarf galaxies and their higher velocity dispersion in clusters. Namely, either 
observed dwarf galaxies are actually an infalling population mostly located in 
the outer regions of cluster/groups \citep[e.g.,][]{conselice2001}, or their 
characteristic radial distribution within a group/cluster is a natural 
consequence of energy equipartition \citep[e.g.,][]{white1976}.

On a related subject, as pointed out by \citet{weinmann2011}, currently there 
are contradictory implementations in SAMs for the tidal stripping suffered by 
galaxies in clusters.  On the one hand, it is modelled as proportional to 
$M/m$ \citep[e.g.,][]{kim2009}, where $M$ is the mass of the primary system 
and m is that of the galaxy. Even though there is not a detailed physical 
mechanism behind this scaling, usually it is based on the assumption that 
stars in smaller galaxies can be more easily stripped, since these galaxies 
have smaller tidal radii. On the other hand, in other studies the efficiency 
of tidal stripping is modelled as proportional to $m/M$ \citep{wetzel2010}. 
This scaling is based on the assumption that dynamical friction is more efficient 
for larger galaxies, causing them to reach more rapidly the inner regions of 
a cluster/group where tidal disruption/stripping is more intense. According 
to our simulations (compare MR1:20 vs. REF vs. MR1:5 in 
Figure~\ref{massbound-evol}), implementations of tidal stripping scaling as 
$m/M$, i.e. more efficient for larger infalling galaxies, offer a better 
description of the evolution of galaxies within a group.

\section{Conclusions}
\label{sec-conclusion}

In this work, we studied the evolution of disc galaxies orbiting within a 
group-size system, using $N$-body simulations. We focus specifically on the 
changes induced by global tidal forces on the infalling galaxies. Both the group 
and the disc galaxies are modelled as ``live'' multi-component systems, composed 
by both a DM halo and a stellar component. The stellar component embedded in the 
group, that accounts for its central galaxy, is modelled with a spherical mass 
distribution. The stellar disc of galaxies follows an exponential density profile.
Our strategy is to examine the evolution of a single galaxy at a time, in order to 
isolate the effect of the global tidal forces on the disc, avoiding that of close 
encounters with other galaxies. The galaxy is released on an infalling orbit from 
the virial radius of the group, and sinks towards the inner region of the group 
under the effect the dynamical friction. This set-up is repeated to probe an 
ample parameter space that covers a number of relevant aspects of the galaxy-group 
interaction. Specifically, we study the different outcomes obtained from: prograde 
and retrograde infalls (with respect to the disc rotation), different orbital 
eccentricities (consistent with distributions of orbital parameters from 
cosmological studies), different disc inclinations, the presence of a central 
stellar bulge in the disc, different internal kinematics of discs, and finally, 
different galaxy-to-group mass ratios. For each of these aspects, our simulations 
are ran at two redshift epochs, ``$z$=0'' and ``$z$=1''. At ``$z$=0'', our fiducial 
disc galaxy resembles a bulge-less, slightly less massive Milky Way.

Our results can be summarised as follows: 

\begin{itemize}
\item Our simulations confirm that the evolution of disc galaxies due to a 
group-size tidal field is consistent with the evolution observed in previous 
studies across different environment mass scales (clusters and Milky Way-like 
environments). The general evolution of disc galaxies is characterised by the 
formation of tidal arms and of central non-axisymmetric features. This 
consistency across different environment mass scales is a natural consequence 
of the current cosmological paradigm, where interactions between structures 
along eccentric orbits are common.
\item We address the question of when, along their orbits, disc galaxies start 
presenting significant structural transformations due to the group tidal field. 
Our simulations show that they take place not before the mean density of the 
group, within the orbit of the galaxy, is $\sim$0.3--1 times the central mean 
density of the galaxies. We find that this agrees only partially with 
recipes of tidal disruption currently used in some semi-analytic models of 
galaxy evolution. 
\item We find that all disc galaxies on prograde orbits (i.e, where a disc 
rotates in the same sense as its infalling orbit) show a very similar 
morphological evolution. This is found to be independent of the initial 
orbital parameters of discs, initial galaxy-to-group mass ratio, presence of a 
central bulge, and internal disc kinematics. We also find that the only significant
difference among these experiments is the timescale at which the morphological 
changes take place. This is essentially connected to how efficient the 
dynamical friction affecting the galaxies is in each experiment. For instance, 
a more efficient dynamical friction and a faster evolution in discs are observed 
in experiments with larger galaxy-to-group mass ratios and more eccentric orbits.
\item Our simulations show that both the morphological and structural evolution 
of discs are strongly dependent on the initial inclination of discs with respect 
to their orbits. For instance, discs infalling face-on (i.e. with their midplanes 
perpendicular to the orbital plane) and on retrograde orbits (i.e. discs rotating 
contrary to the sense of their orbits), are found to be increasingly resilient 
against the formation of tidal arms and of strong central non-axisymmetries, with 
respect to discs in prograde orbits. This makes both face-on and retrograde discs 
more capable to retain their original disc structure and kinematics for longer 
times. This would explain naturally the presence of a possibly significant 
fraction of disc galaxies in the inner region of groups (and maybe clusters). 
However, such resilience of discs could be altered by the effect of close 
encounters with other galaxies, which are not included in this study.
\item The evolution in the structure of the disc galaxies during their infall is 
characterised by a steady decrease in the scale-length of their mass distributions, 
and by significant vertical thickening, that is more pronounced towards the outskirts 
of discs. Faster variations in the structure of discs are observed during pericentric 
passages, where discs are affected the most by strong tidal impulses from the group 
potential. The kinematical evolution of discs appears closely related to both the 
morphological and structural changes they suffer. In general, it is mainly driven 
by the formation of tidal arms and non-axisymmetric central features, that typically 
increase the velocity dispersions in the discs from the outskirts inwards. Disc 
galaxies evolve towards being less rotationally-supported systems, with hotter 
(although not significantly flatter) velocity dispersion profiles.
\item Discs evolving within a group tidal field neither form significant central 
bulges, nor enhance the pre-existing ones. Then, under the assumption that most 
S0 galaxies (which often have larger bulges than found in spiral galaxies) 
are formed in group environments, these cannot be formed by the effect of the
global tidal field on ``old'' stars originally in the discs. Instead, our 
simulations would favour formation models of S0 galaxies where their central 
bulges are composed by a significant fraction of young stars (i.e. formed after the
accretion of a galaxy onto a group).
\item Our simulations do not support the concept that tidal disruption of galaxies 
depends mostly on their orbital energy. Such premise has been proposed in the 
literature to invoke a more efficient tidal disruption of slow, faint galaxies in
semi-analytic models of galaxy evolution. This being aimed to improve the agreement 
with observations of the dwarf-to-giant ratio of galaxies in clusters, as well as 
its radial distribution. Our results show instead a stronger dependency of tidal 
disruption on both the galaxy-to-group mass ratio and the initial orbital 
eccentricity of discs.
\item Finally, our results contradict some of the current models of tidal stripping 
of galaxies in clusters, used in semi-analytic models, where tidal stripping is 
more efficient for less massive galaxies.
\end{itemize}

The present work can be expanded along several lines, in order to improve our 
knowledge on how group-scale environments shape the main properties of disc 
galaxies. For instance, the addition of gas components to disc galaxies would 
allow the study of more complex physical processes, such as, strangulation 
and ram-pressure. In a future study we plan to also include the effect of 
close encounters on the evolution of disc galaxies, aiming to quantify the 
relative importance of such encounters against the effect of the global tidal 
field.

\section*{Acknowledgements}

We thank A. Biviano, F. Fontanot, A. Macci\`o, S. Weinmann, and D. Wilman for 
useful discussions and suggestions. \'AV and GDL acknowledge financial support 
from the European Research Council under the European Community's Seventh Framework
Programme (FP7/2007-2013)/ERC grant agreement n. 202781. This work has been 
partially supported by the PRIN-INAF 2009 Grant ``Towards an Italian Network 
for Computational Cosmology'', by the European Commission's Framework Programme 7, 
through the Marie Curie Initial Training Network CosmoComp (PITN-GA-2009-238356), 
and by the PD51 INFN Grant. The simulations were carried out at the ``Centro 
Interuniversitario del Nord-Est per il Calcolo Elettronico'' (CINECA, Bologna), 
with CPU time assigned under INAF/CINECA and ISCRA-B grants.

\bibliographystyle{mn.bst}
\bibliography{mn-jour,biblio}

\appendix
\section{Figures}

The following Figures are included here for completeness and are the continuation 
of Figures~\ref{struct-correlations-z0}, \ref{struct-correlations-z1}, 
\ref{kinematics-evol-z0}, and \ref{kinematics-evol-z1}. They illustrate the 
structural and kinematical evolution of disc galaxies for the remaining BUL, CIR, 
COL, MR1:5, MR1:20, MR1:10, and MR1:40 experiments, at the respective redshift
epoch ``$z$=0'' and ``$z$=1''.

\begin{figure*}
\begin{center}
\includegraphics[width=88mm]{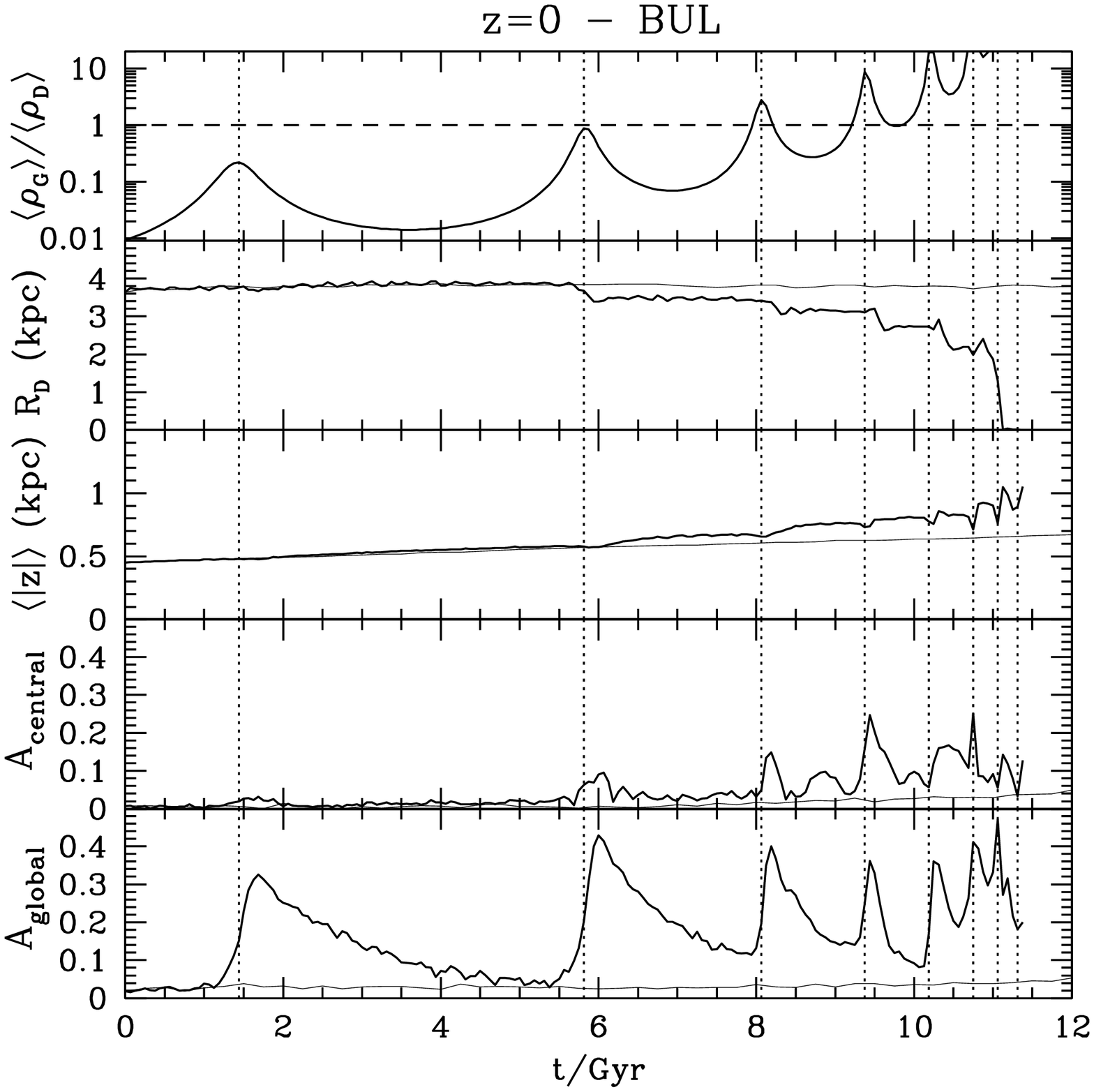}
\includegraphics[width=88mm]{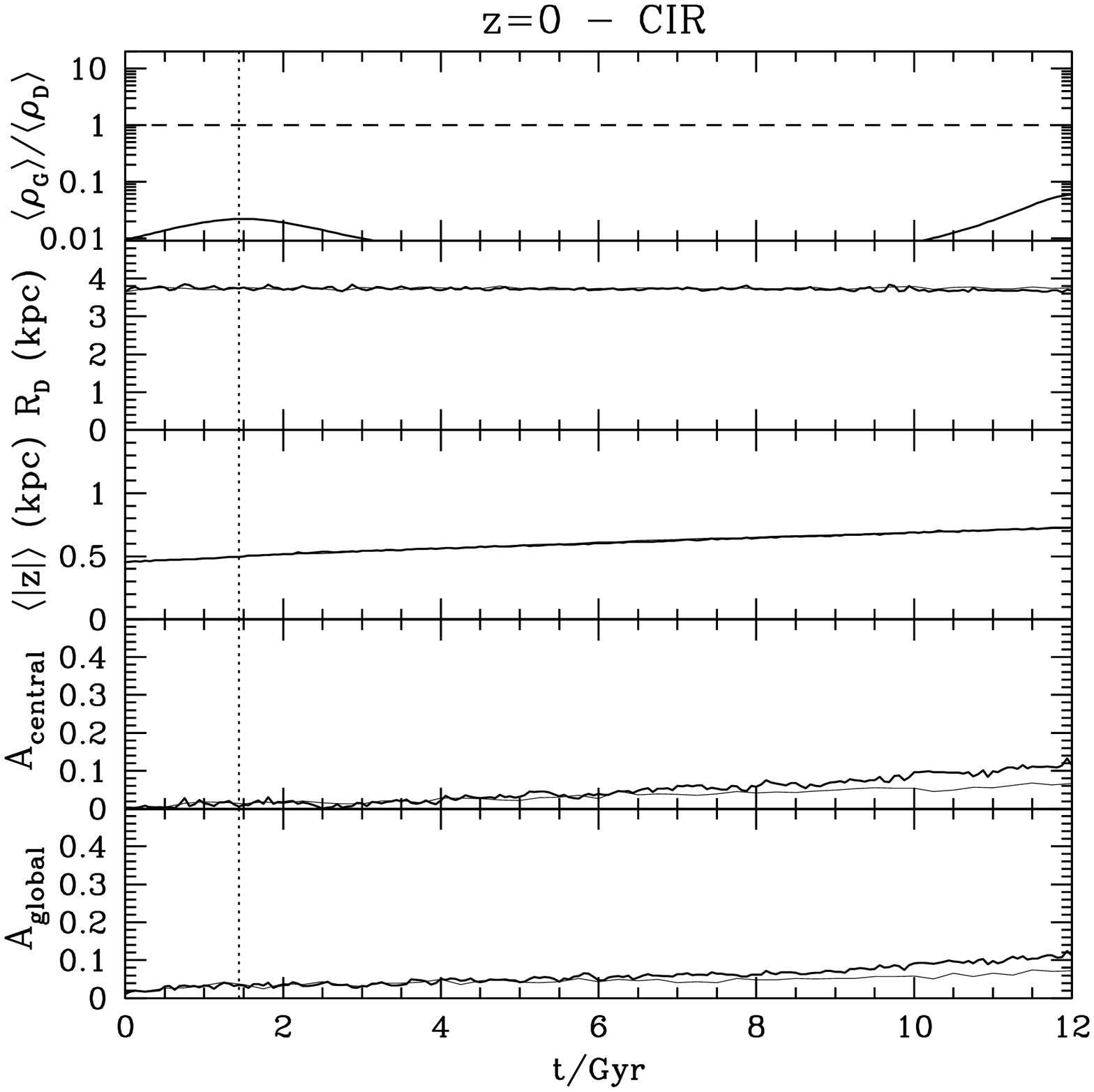}
\includegraphics[width=88mm]{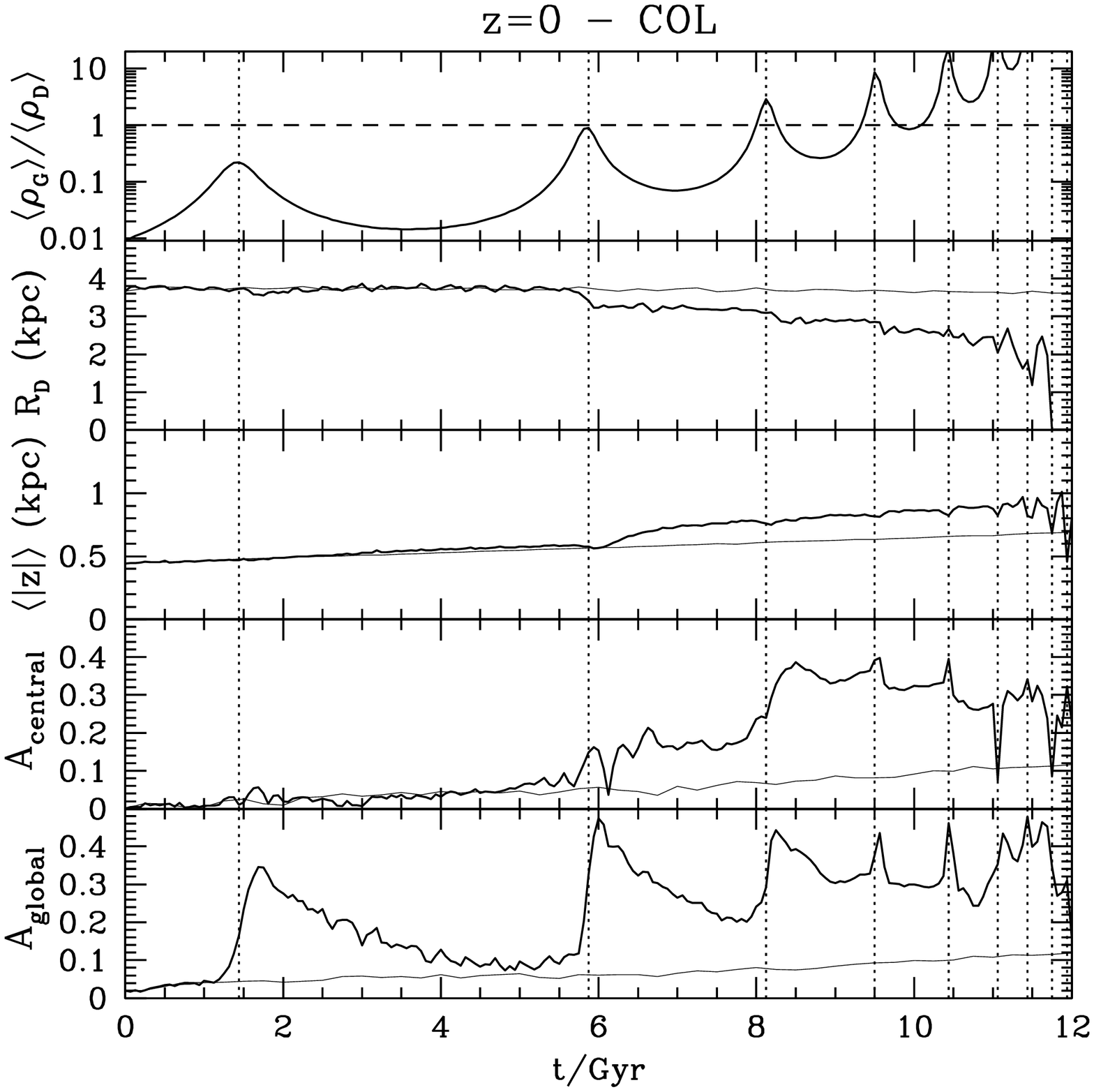}
\includegraphics[width=88mm]{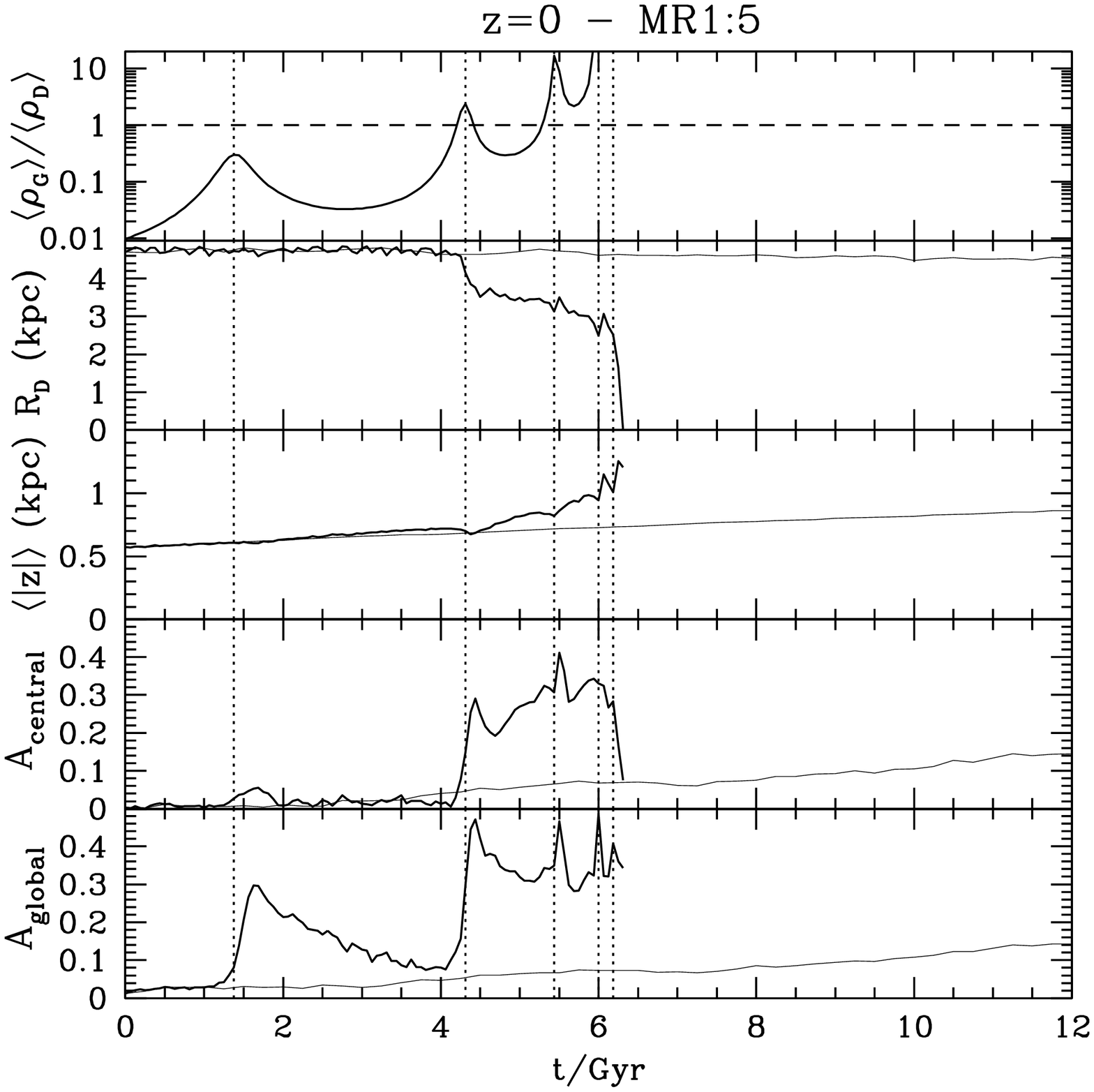}
\end{center}
\caption{Continuation of Figure~\ref{struct-correlations-z0}. 
Evolution of the structure of disc galaxies in terms of their scale-lengths, 
$R_{\rm D}$, mean thickness, $\langle|z|\rangle$, and both central and global 
$m$=2 Fourier amplitudes, $A_{\rm central}$ and $A_{\rm global}$. The evolution 
of discs is shown along with the strength of the global tidal force acting upon 
them along their orbits, estimated as the group-to-galaxy mean density ratio,
$\langle \rho_{\rm group} \rangle / \langle \rho_{\rm galaxy} \rangle$ (see 
details in the text). For comparison, the evolution of the structure of isolated
disc galaxies (thin lines), and the times of pericentric passages (vertical 
dotted lines), are also included.}
\label{app-f1}
\end{figure*}

\begin{figure*}
\begin{center}
\includegraphics[width=88mm]{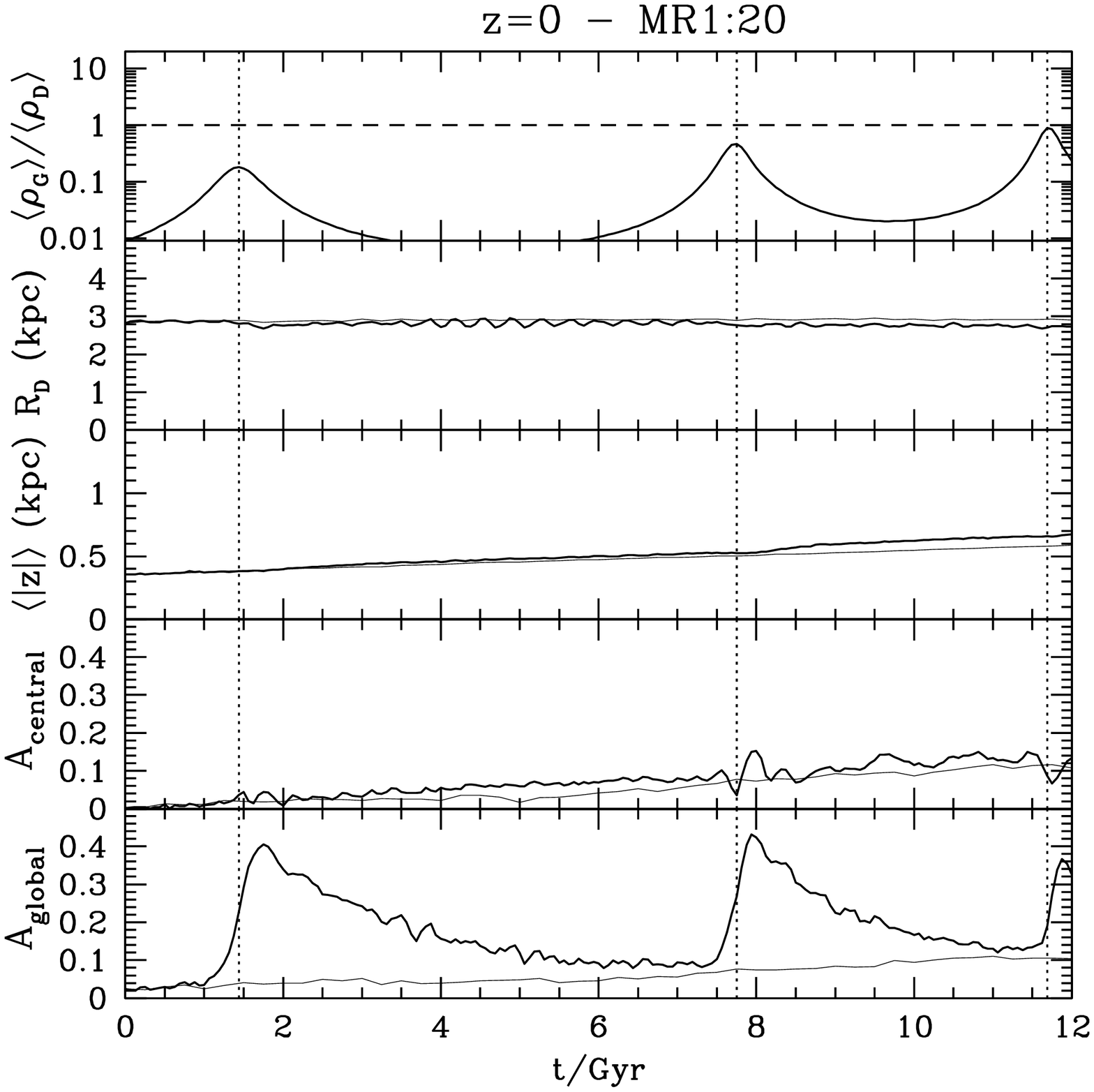}
\end{center}
\caption{Same as Figure~\ref{app-f1}}.
\label{app-f2}
\end{figure*}

\begin{figure*}
\begin{center}
\includegraphics[width=88mm]{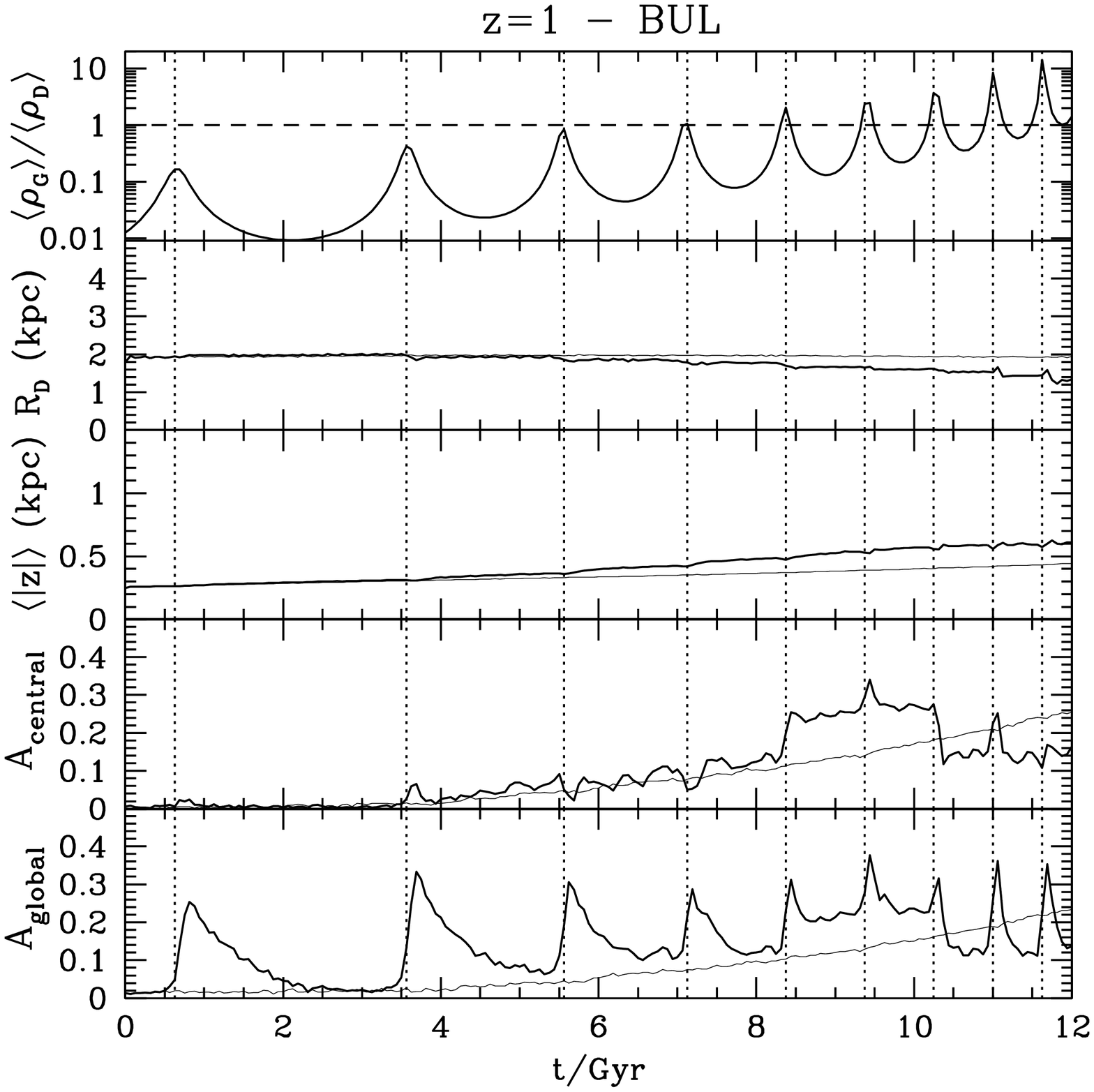}
\includegraphics[width=88mm]{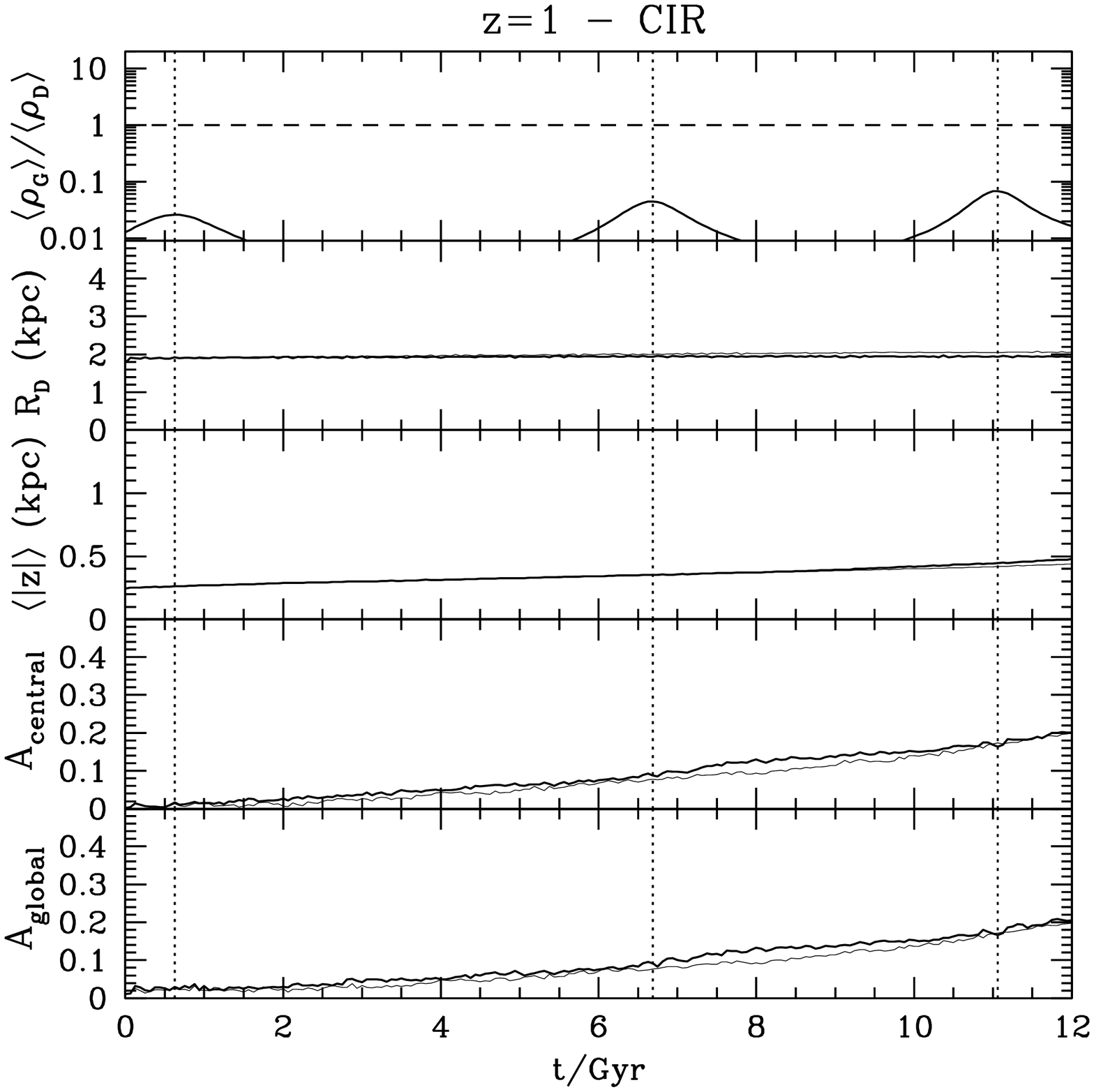}
\includegraphics[width=88mm]{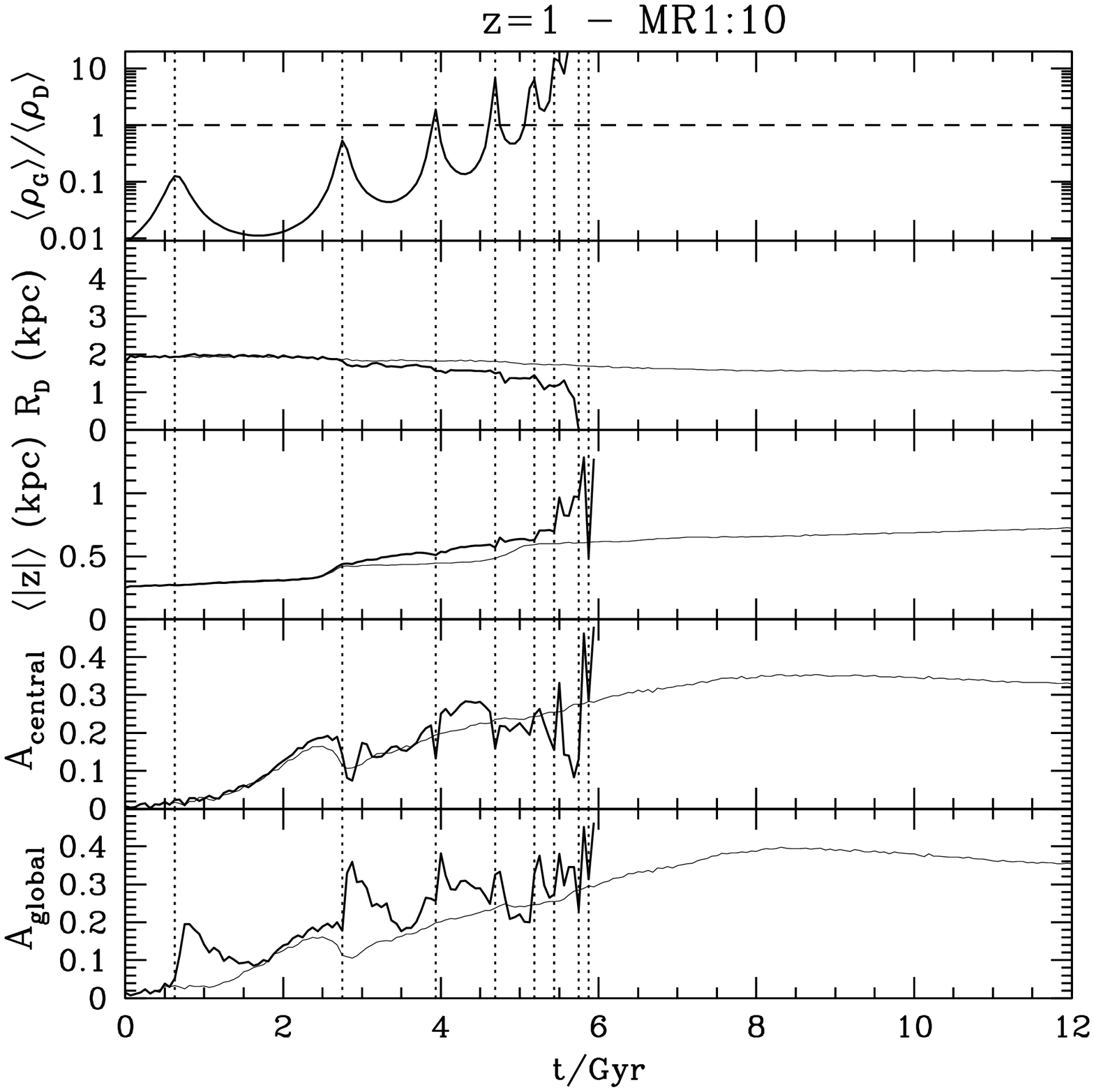}
\includegraphics[width=88mm]{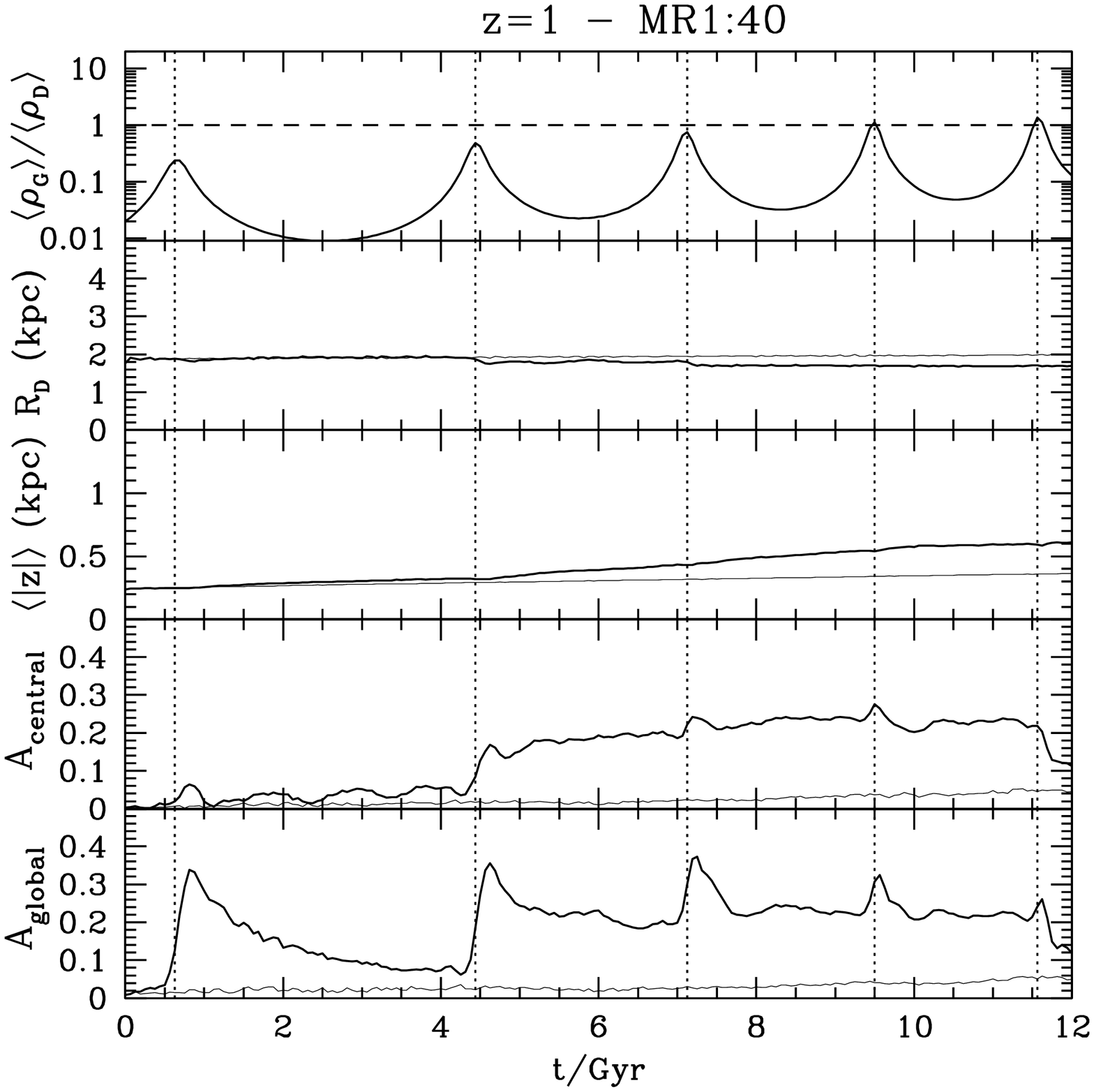}
\end{center}
\caption{Continuation of Figure~\ref{struct-correlations-z1}. Same description 
as Figure~\ref{app-f1}.}
\label{app-f3}
\end{figure*}

\begin{figure*}
\begin{center}
\includegraphics[width=50mm]{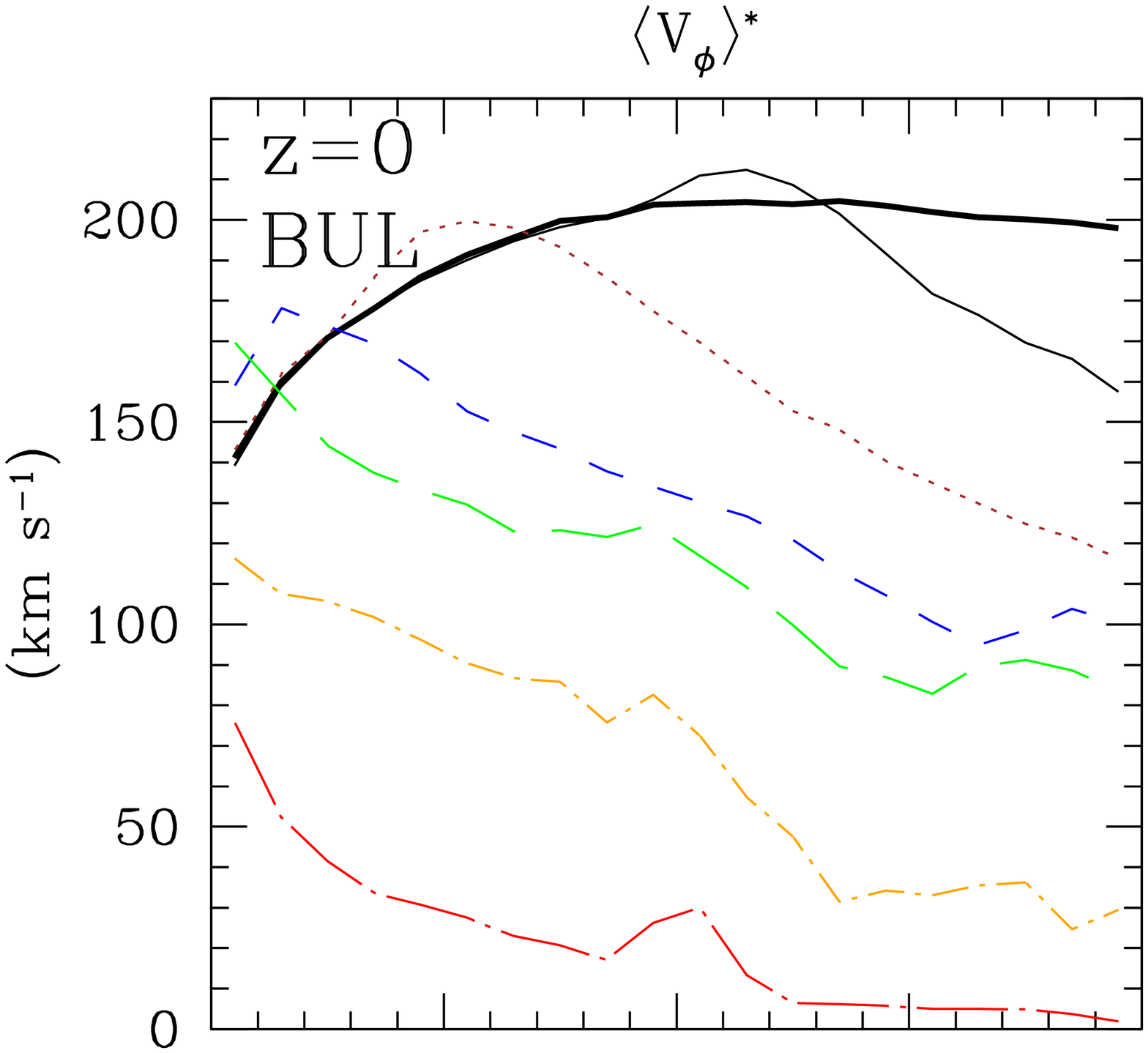}\hspace*{-0.4cm}
\includegraphics[width=50mm]{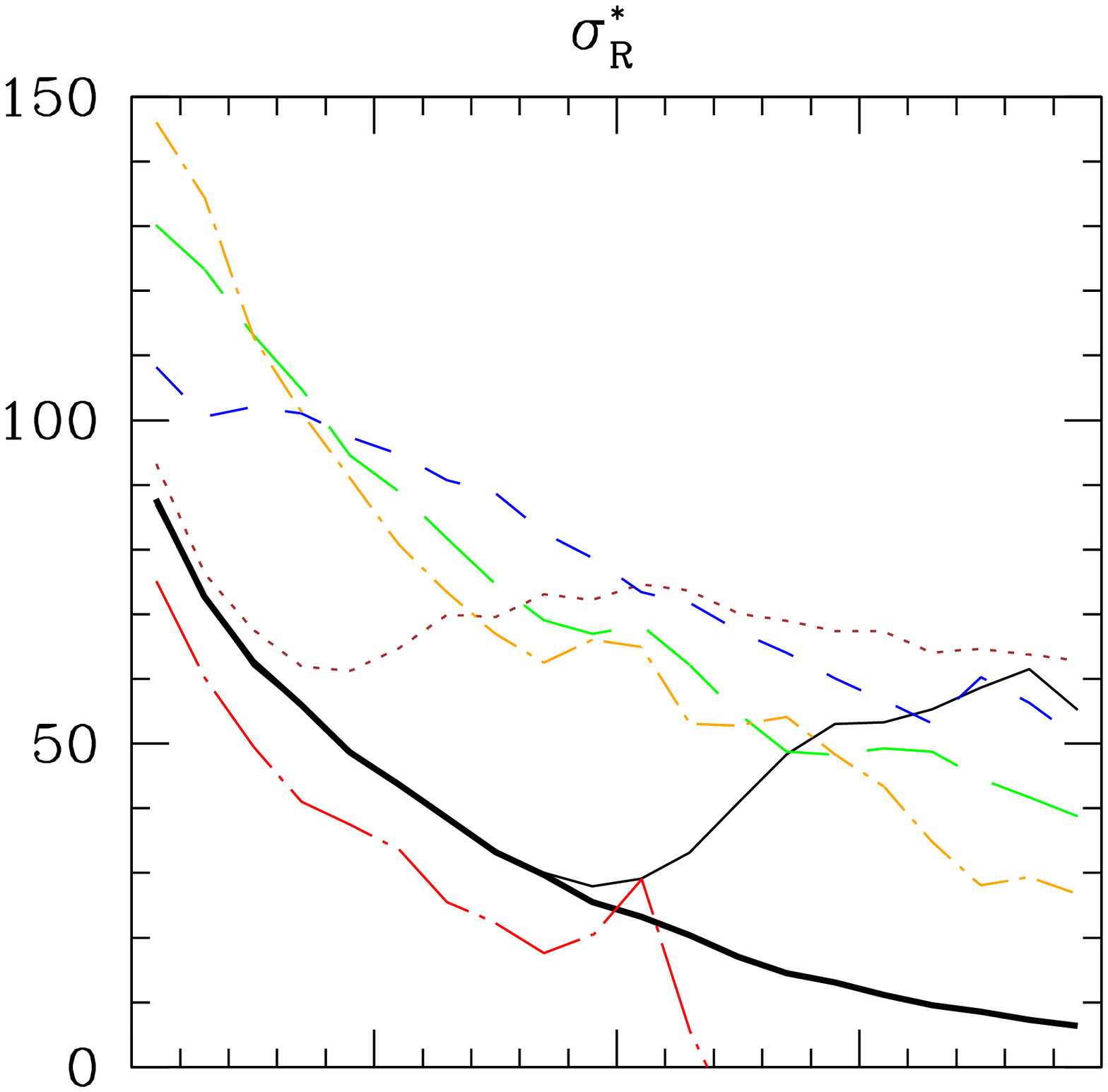}\hspace*{-1.36cm}
\includegraphics[width=50mm]{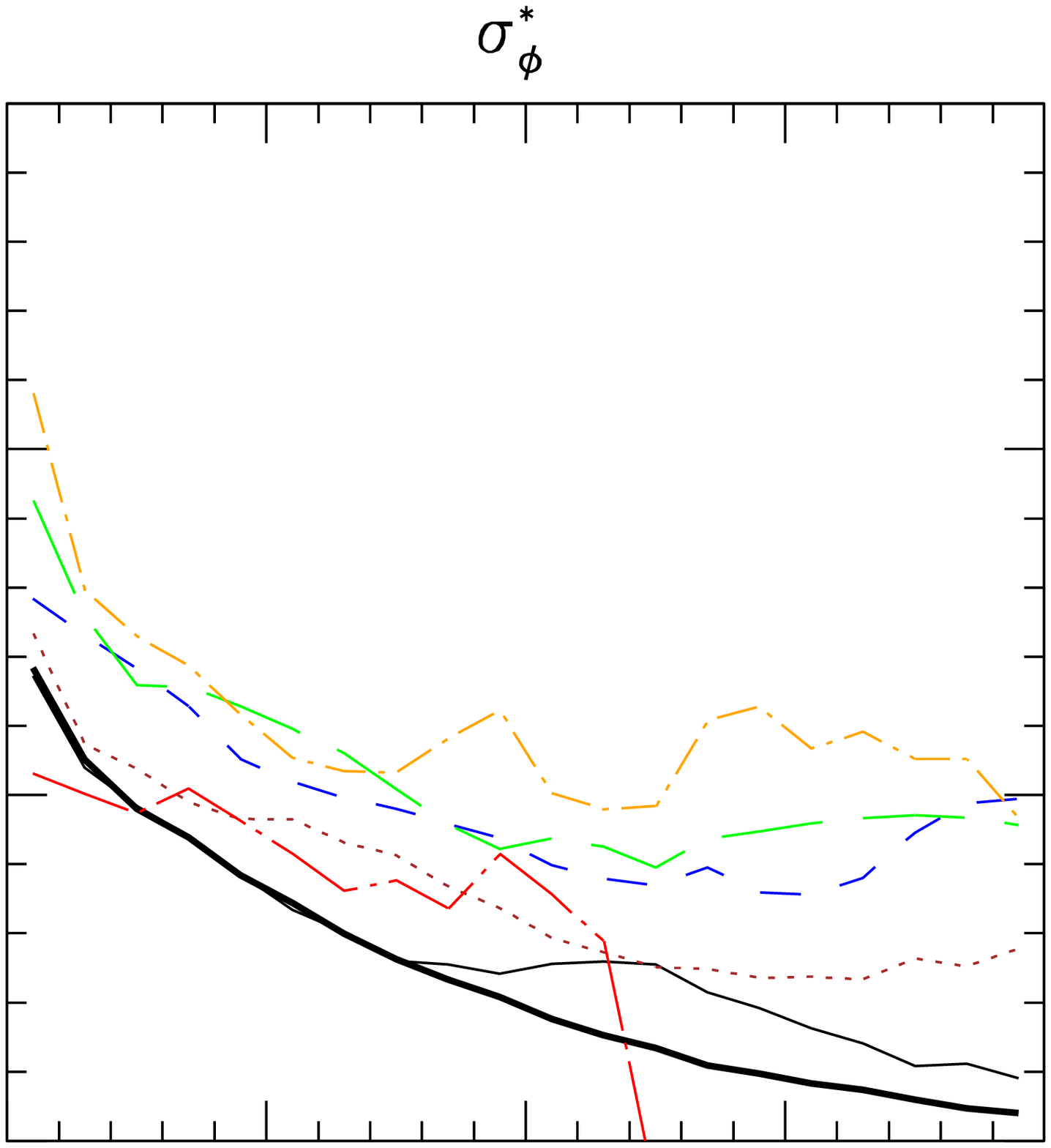}\hspace*{-1.36cm}
\includegraphics[width=50mm]{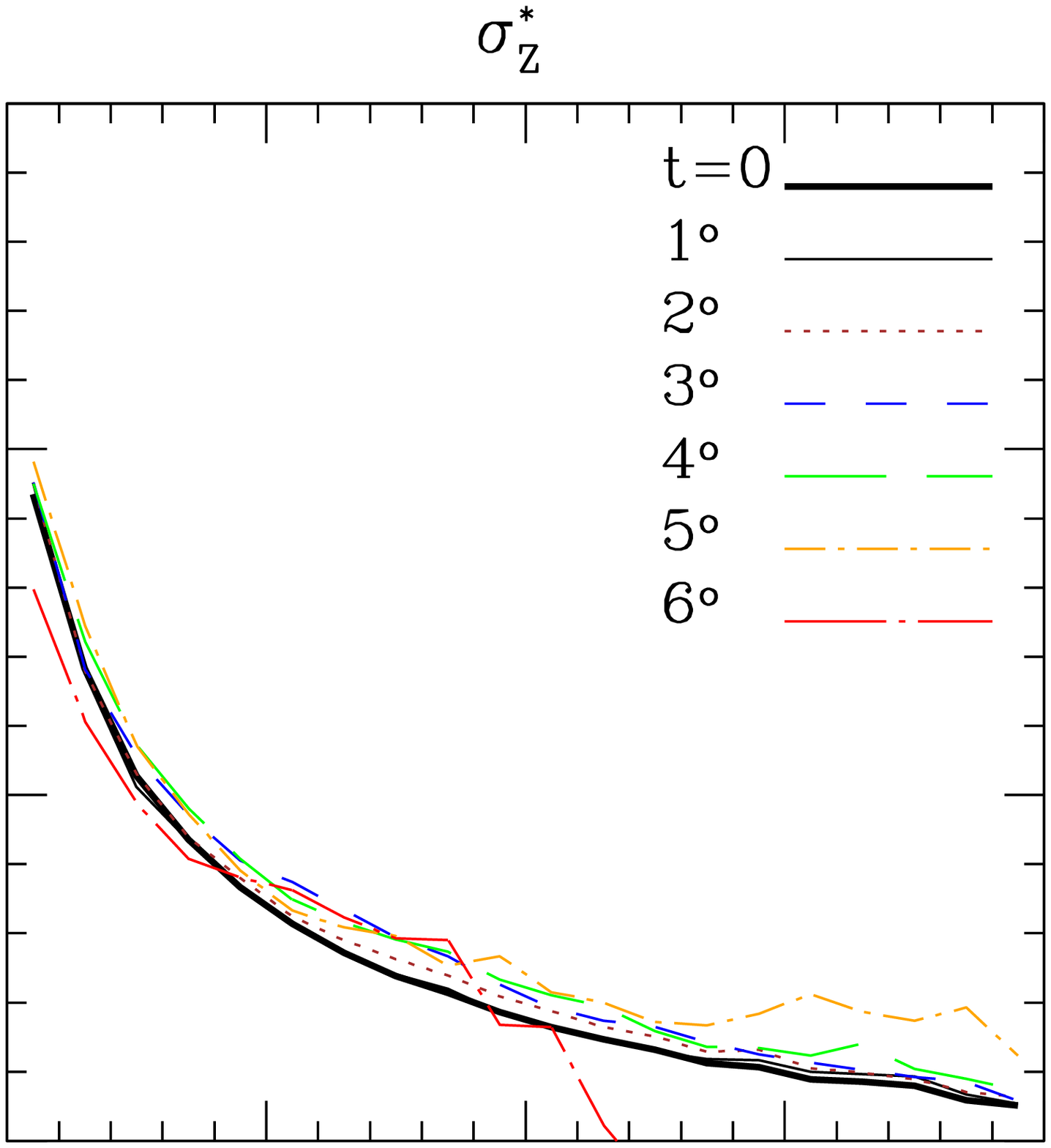}\vspace*{-1.29cm}\\
\includegraphics[width=50mm]{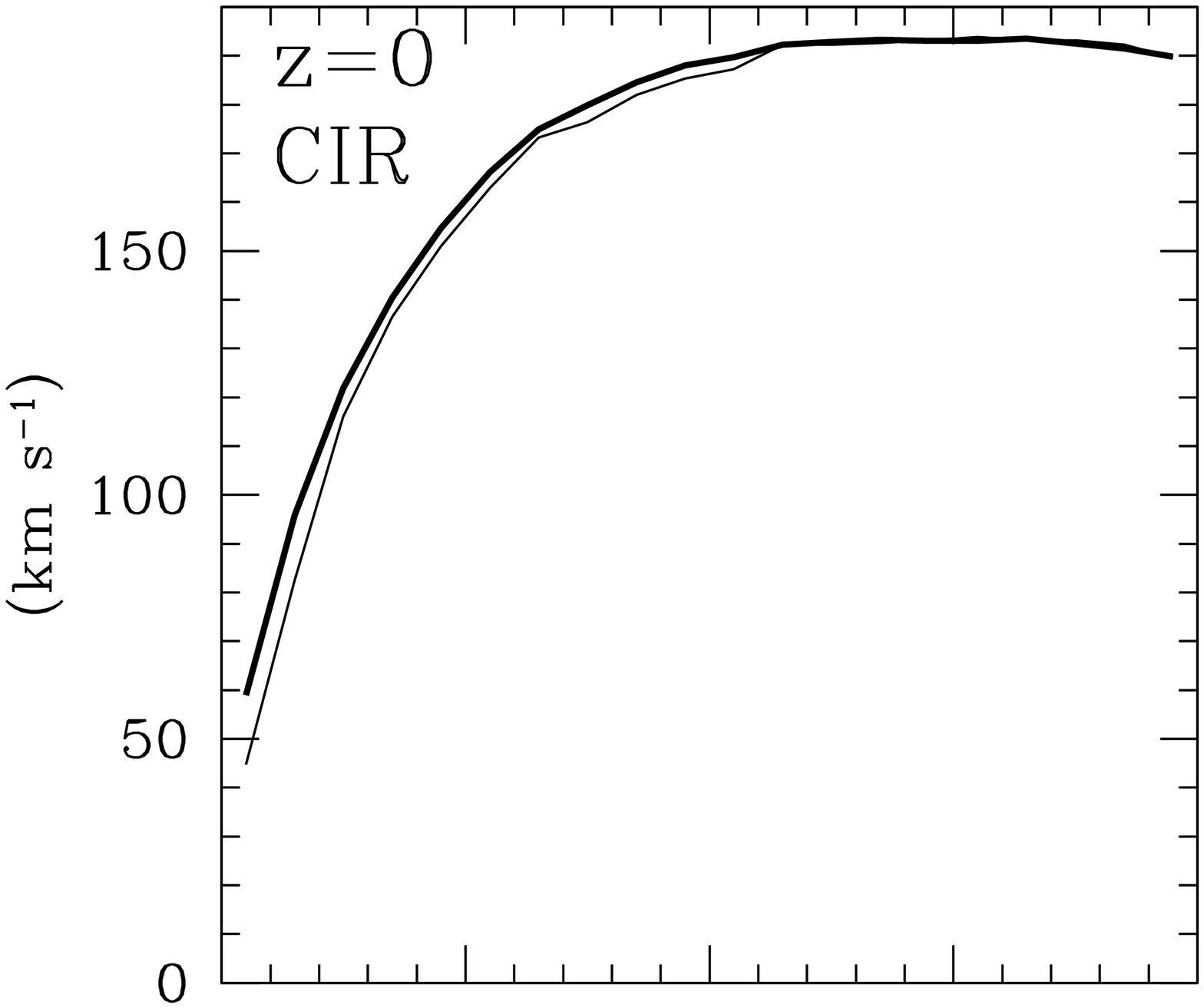}\hspace*{-0.4cm}
\includegraphics[width=50mm]{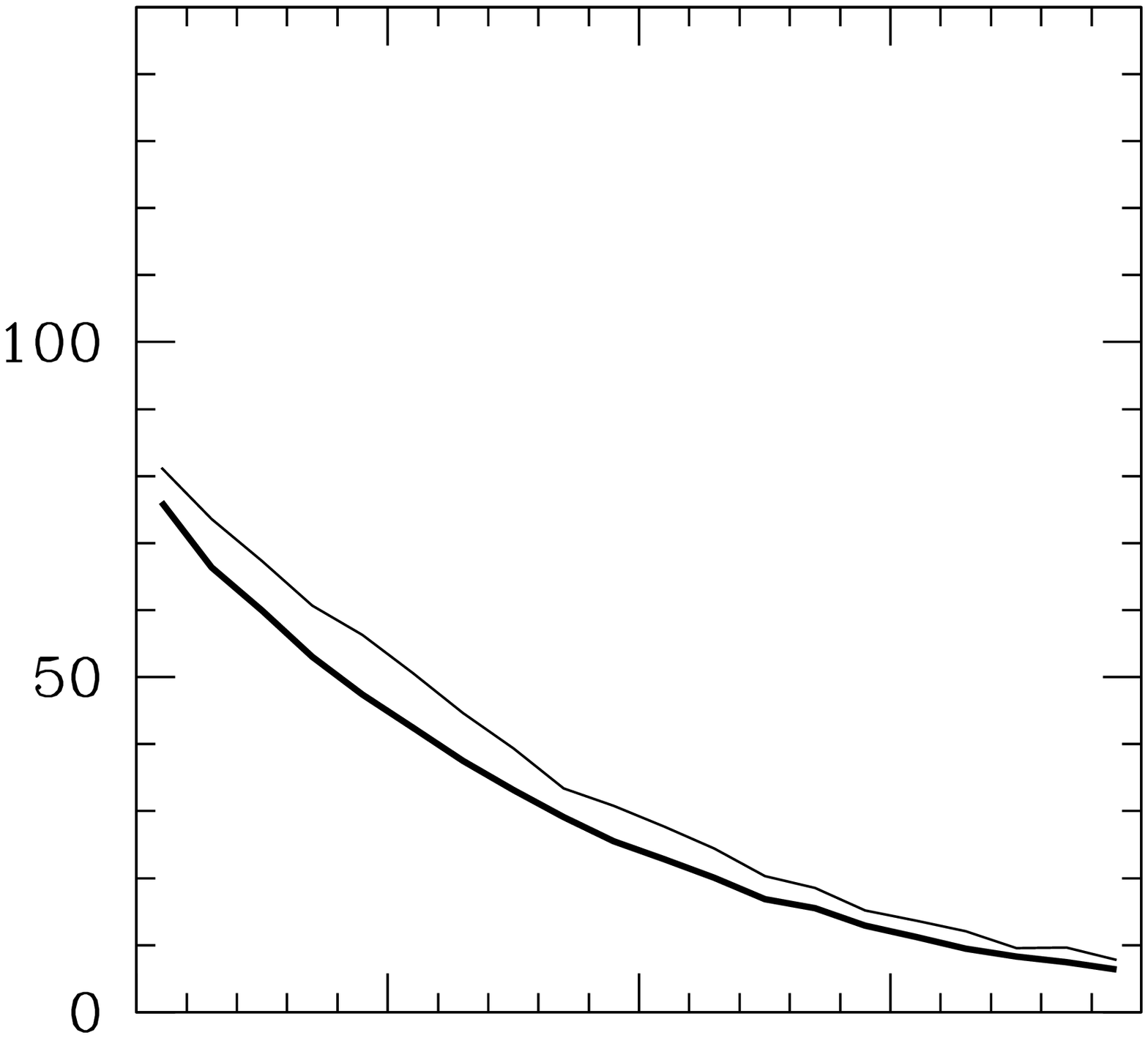}\hspace*{-1.36cm}
\includegraphics[width=50mm]{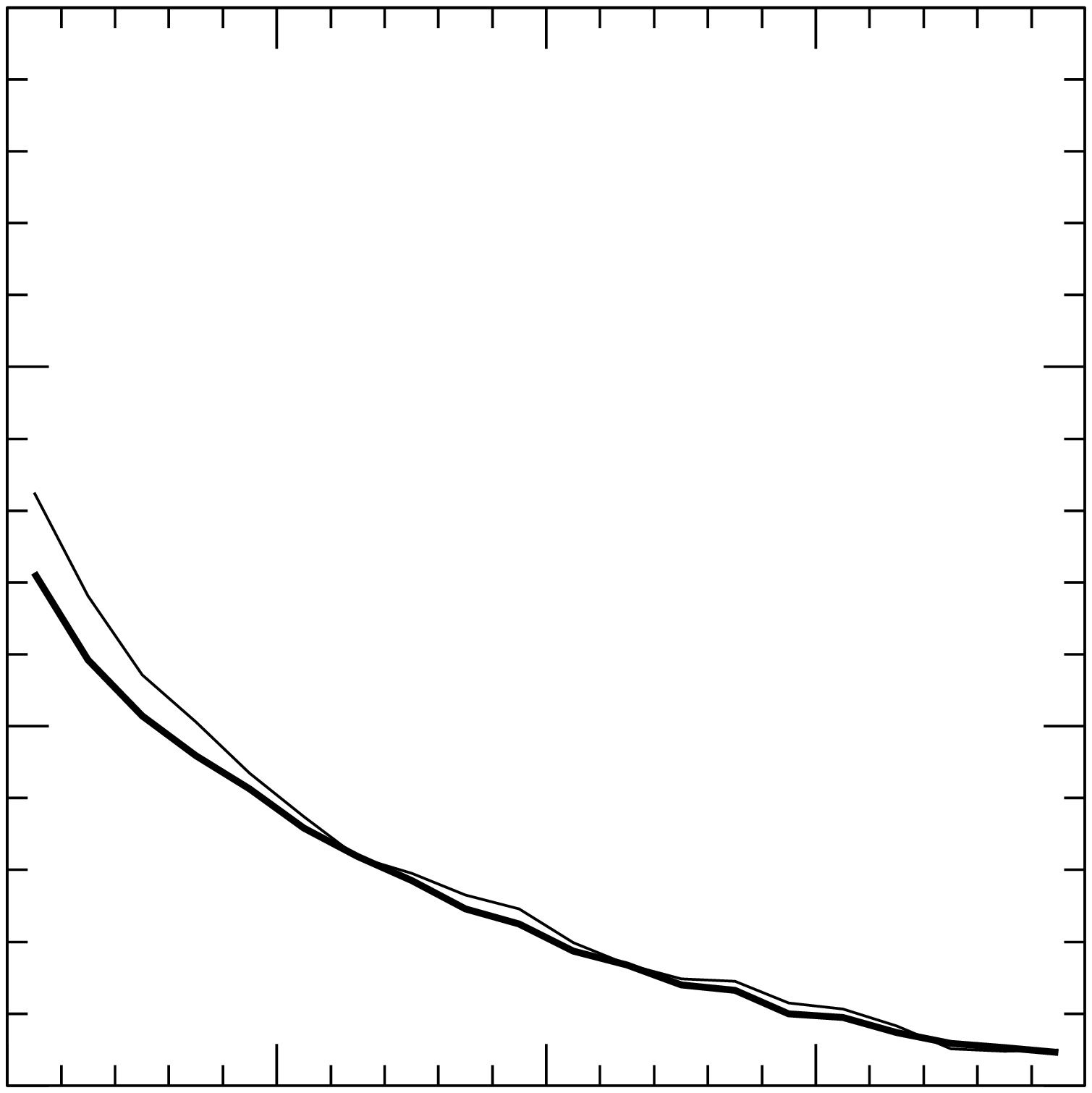}\hspace*{-1.36cm}
\includegraphics[width=50mm]{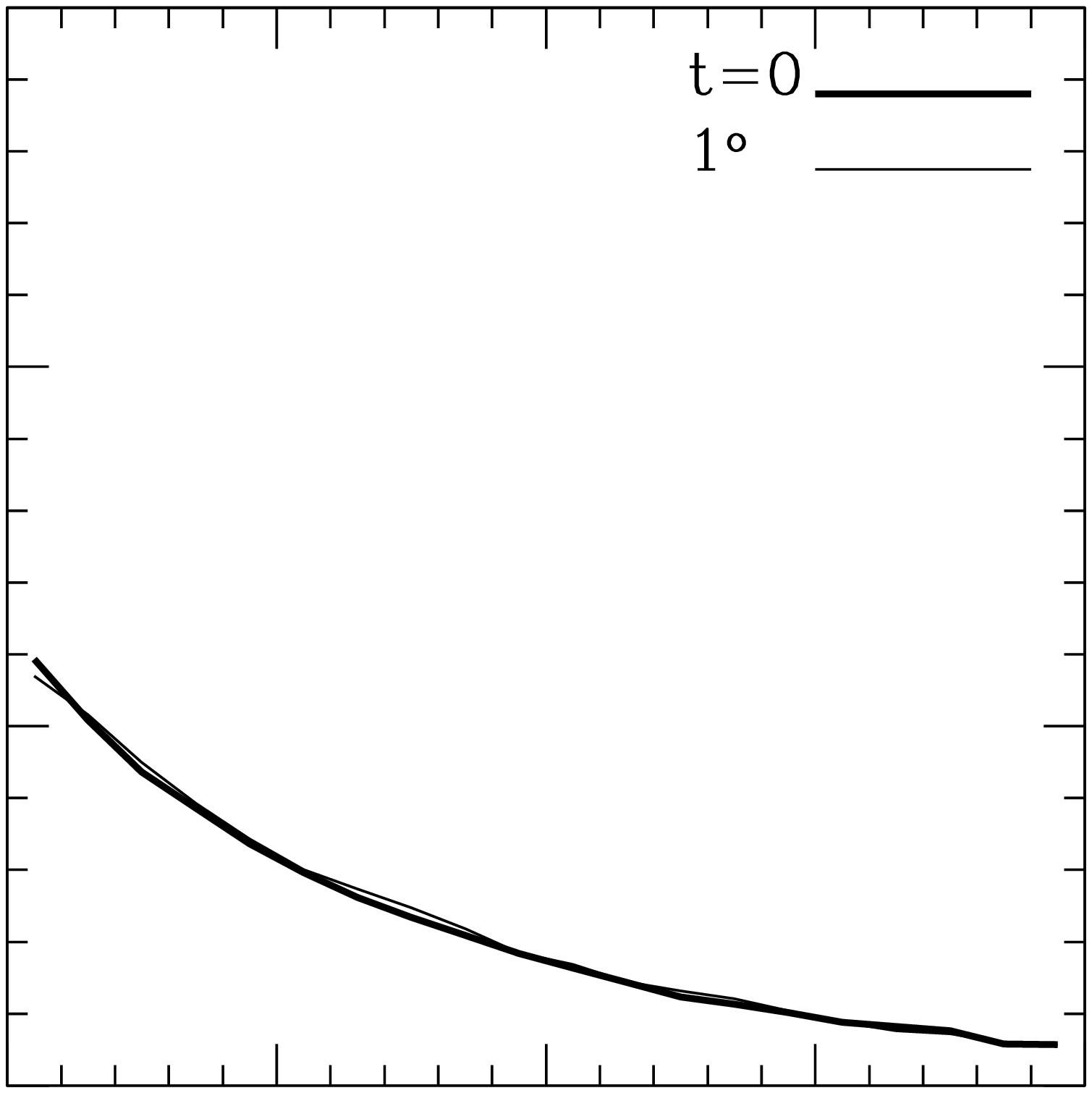}\vspace*{-1.29cm}\\
\includegraphics[width=50mm]{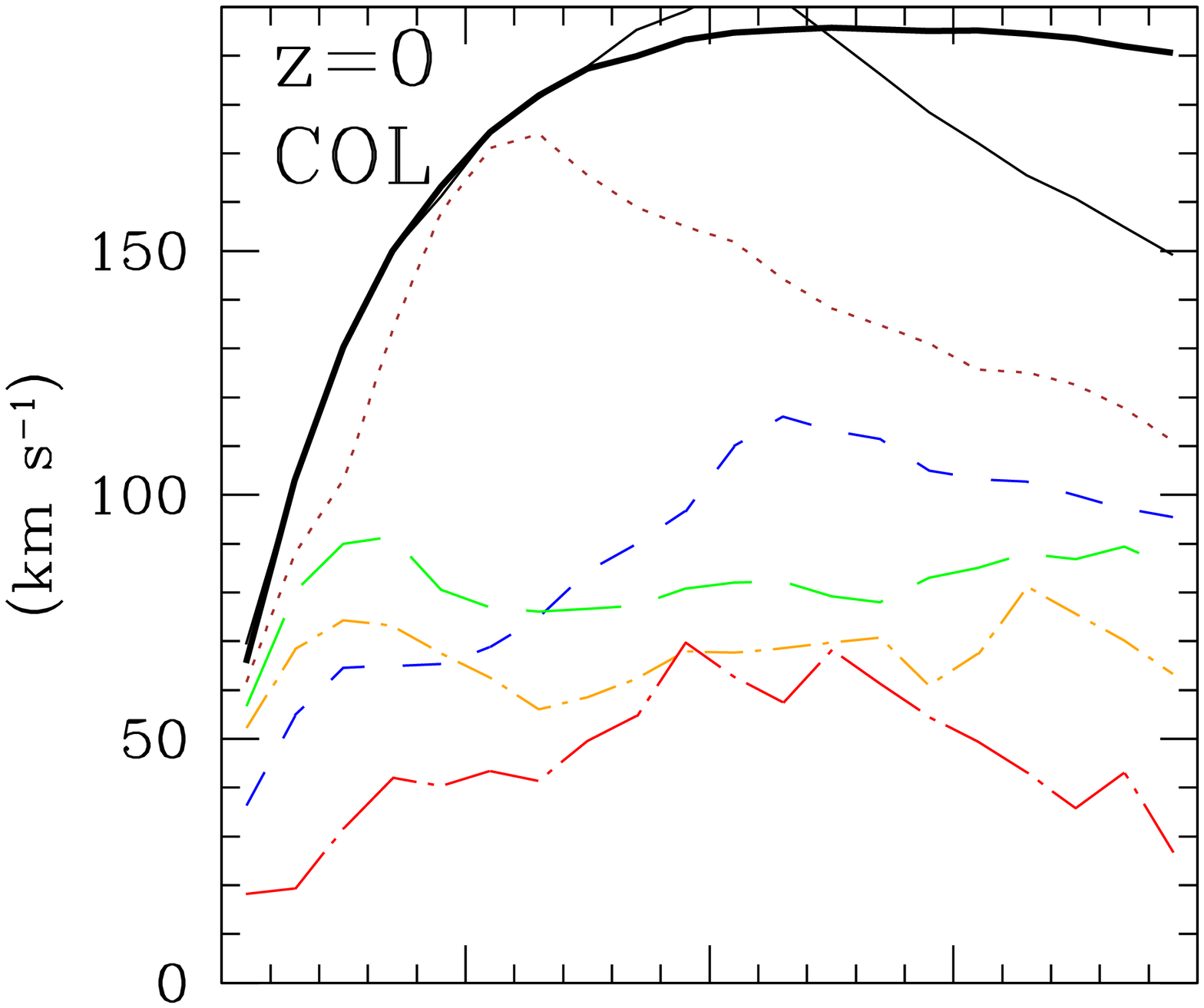}\hspace*{-0.4cm}
\includegraphics[width=50mm]{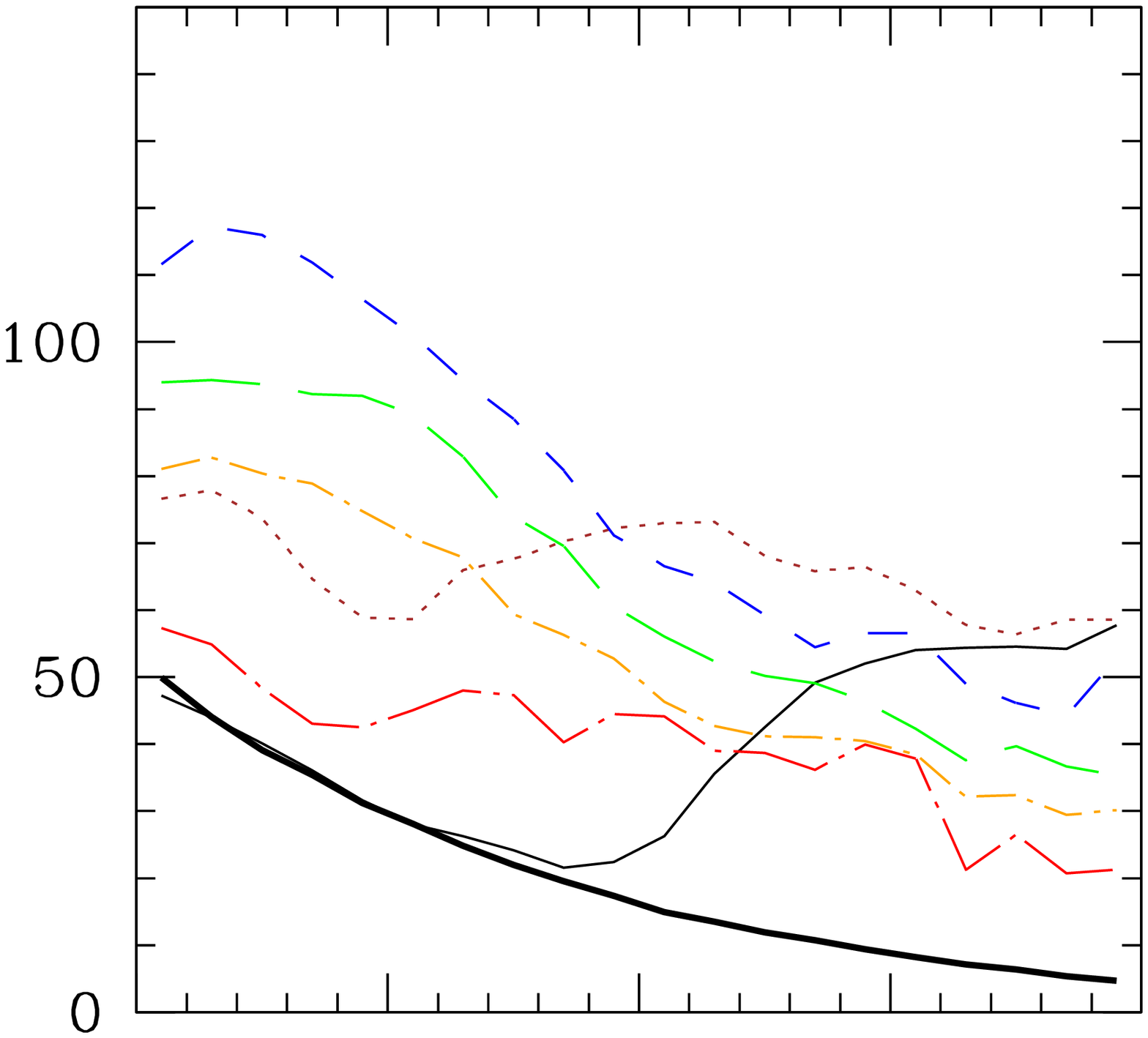}\hspace*{-1.36cm}
\includegraphics[width=50mm]{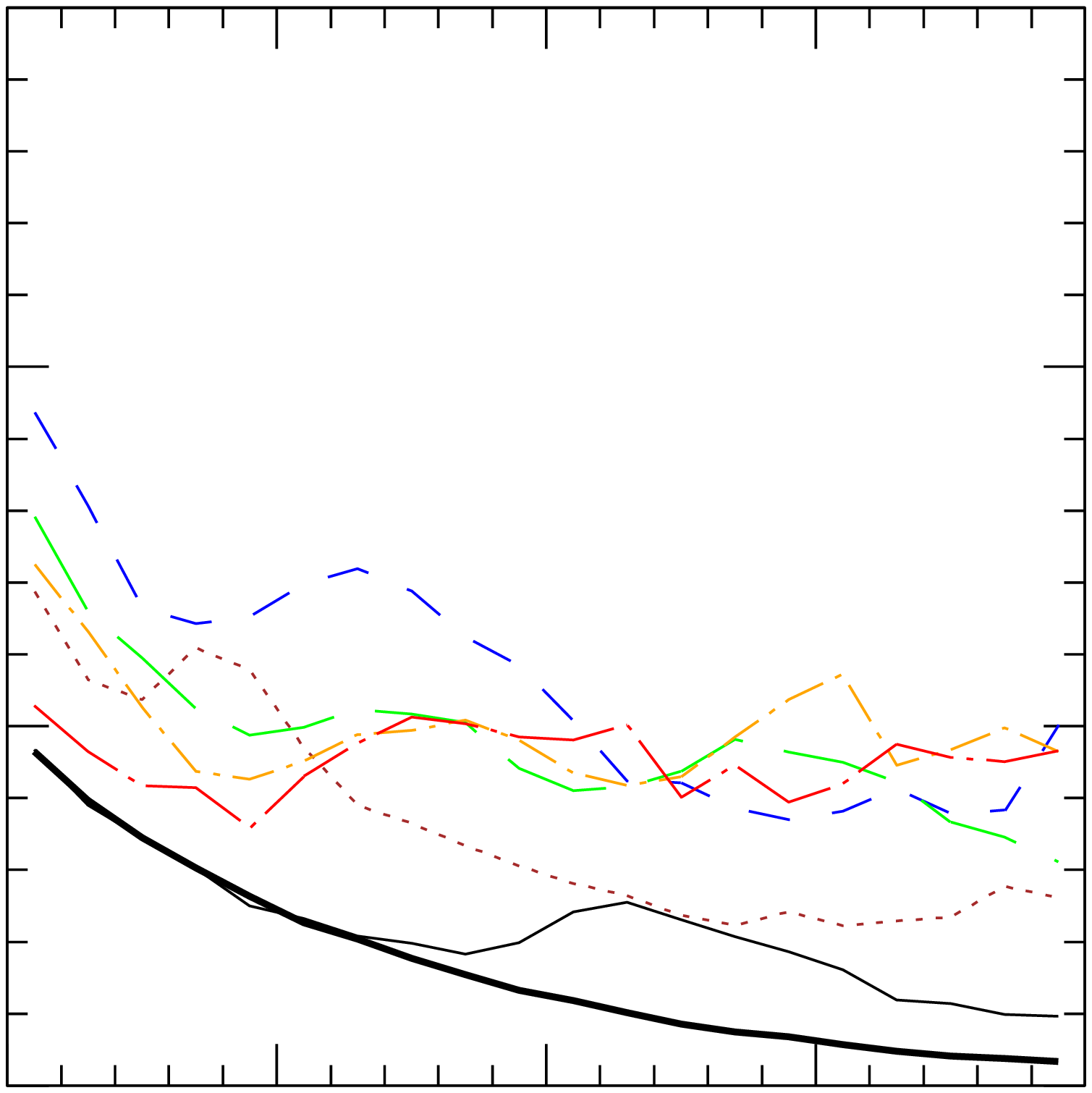}\hspace*{-1.36cm}
\includegraphics[width=50mm]{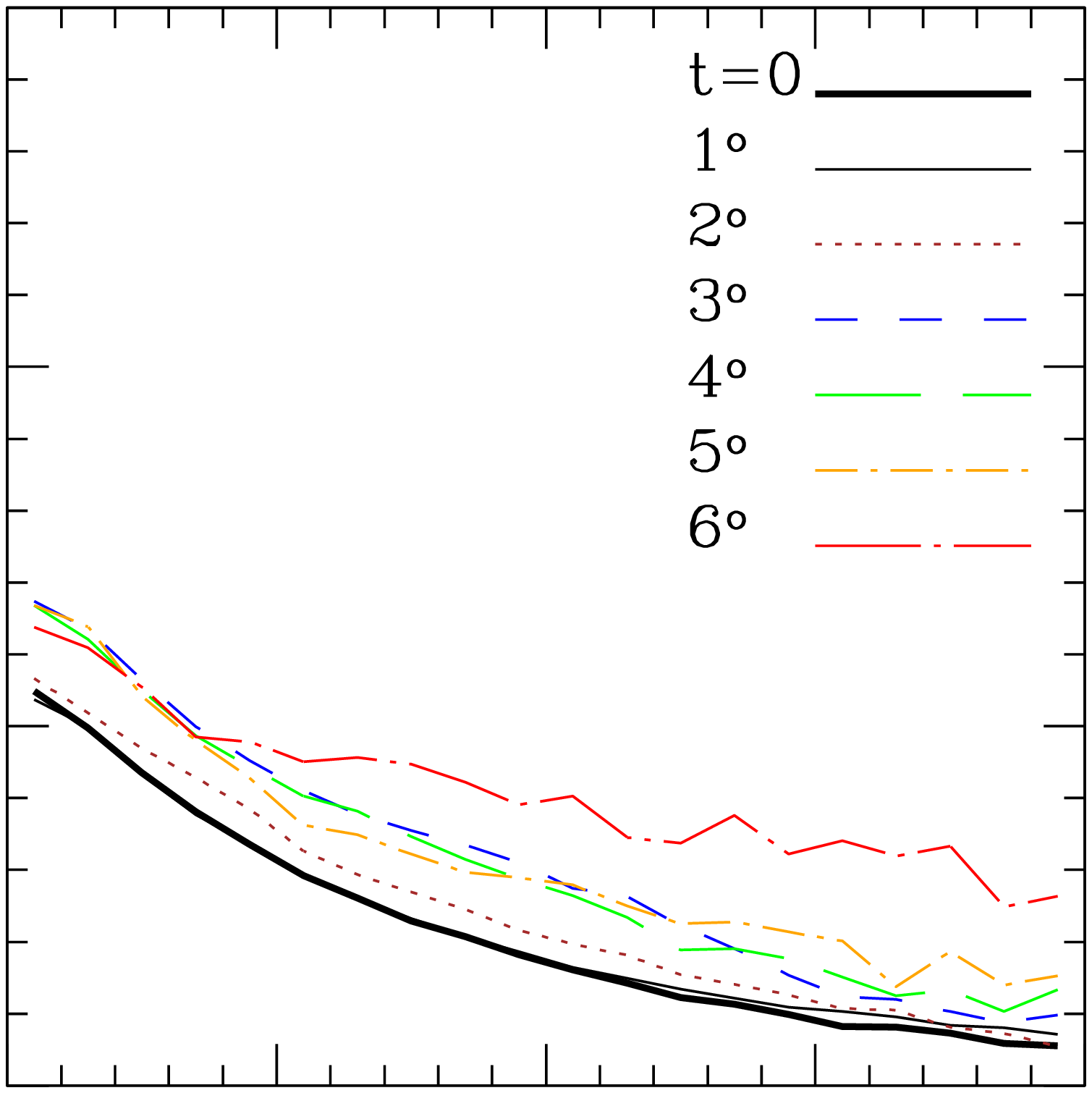}\vspace*{-1.29cm}\\
\includegraphics[width=50mm]{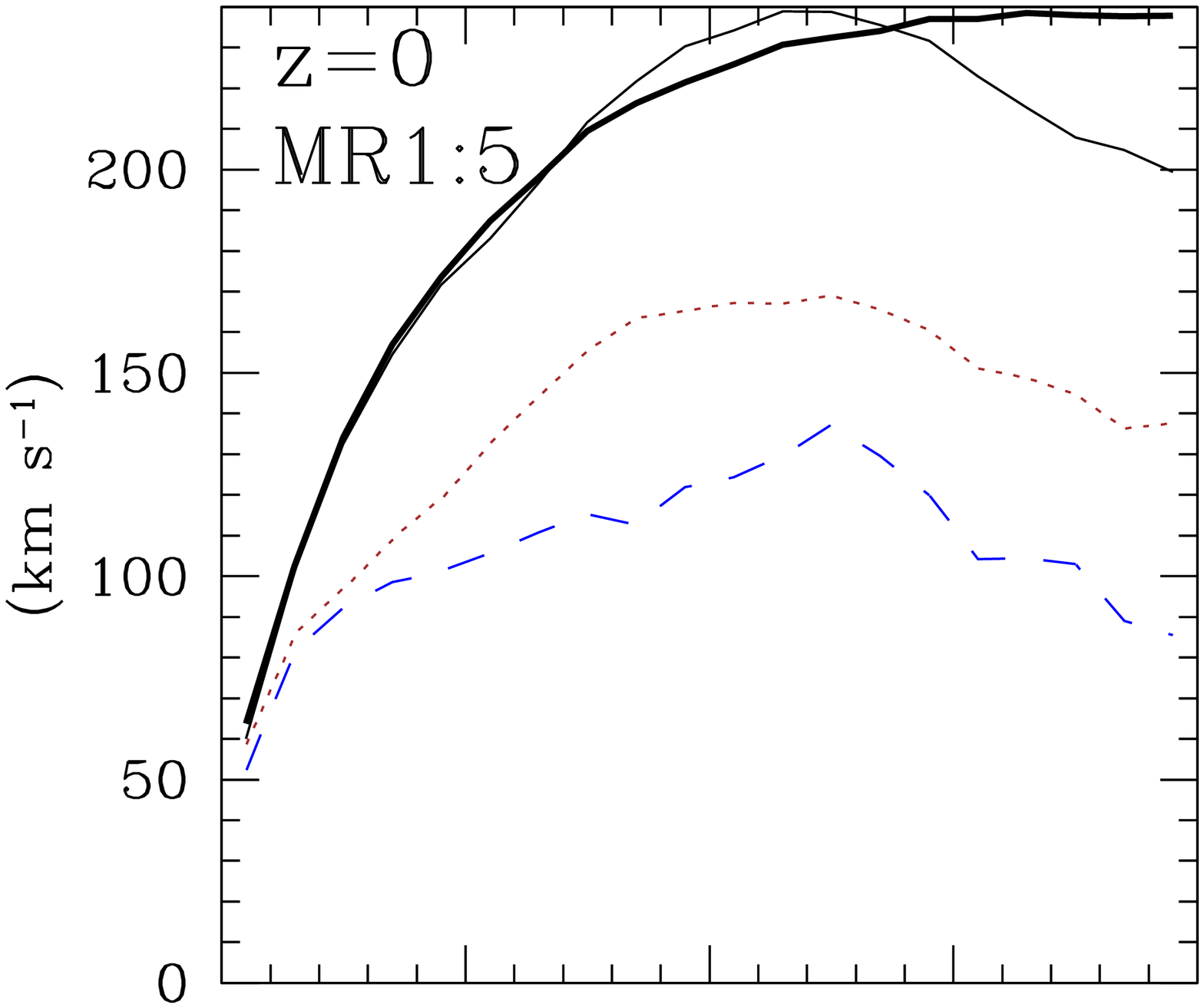}\hspace*{-0.4cm}
\includegraphics[width=50mm]{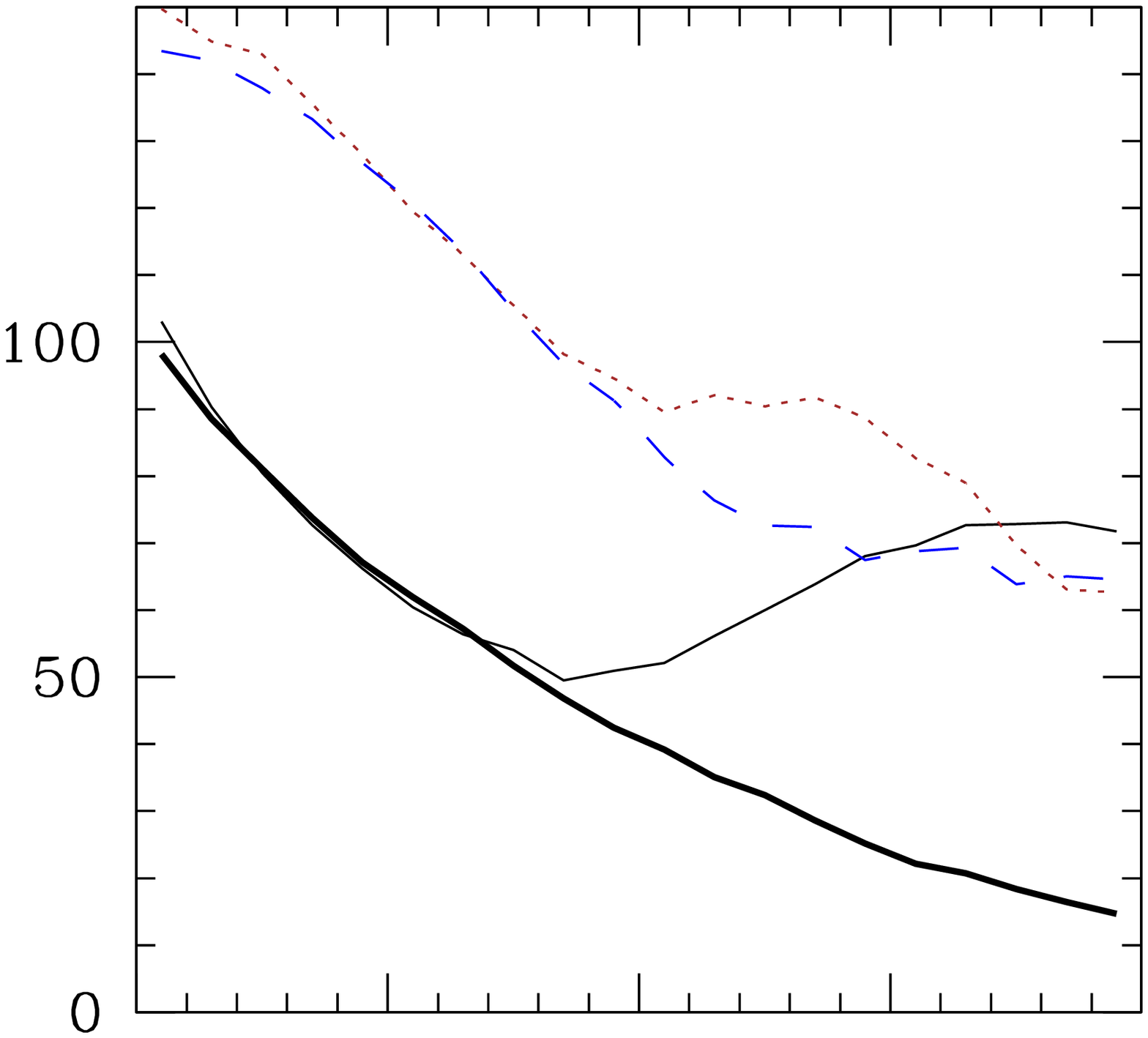}\hspace*{-1.36cm}
\includegraphics[width=50mm]{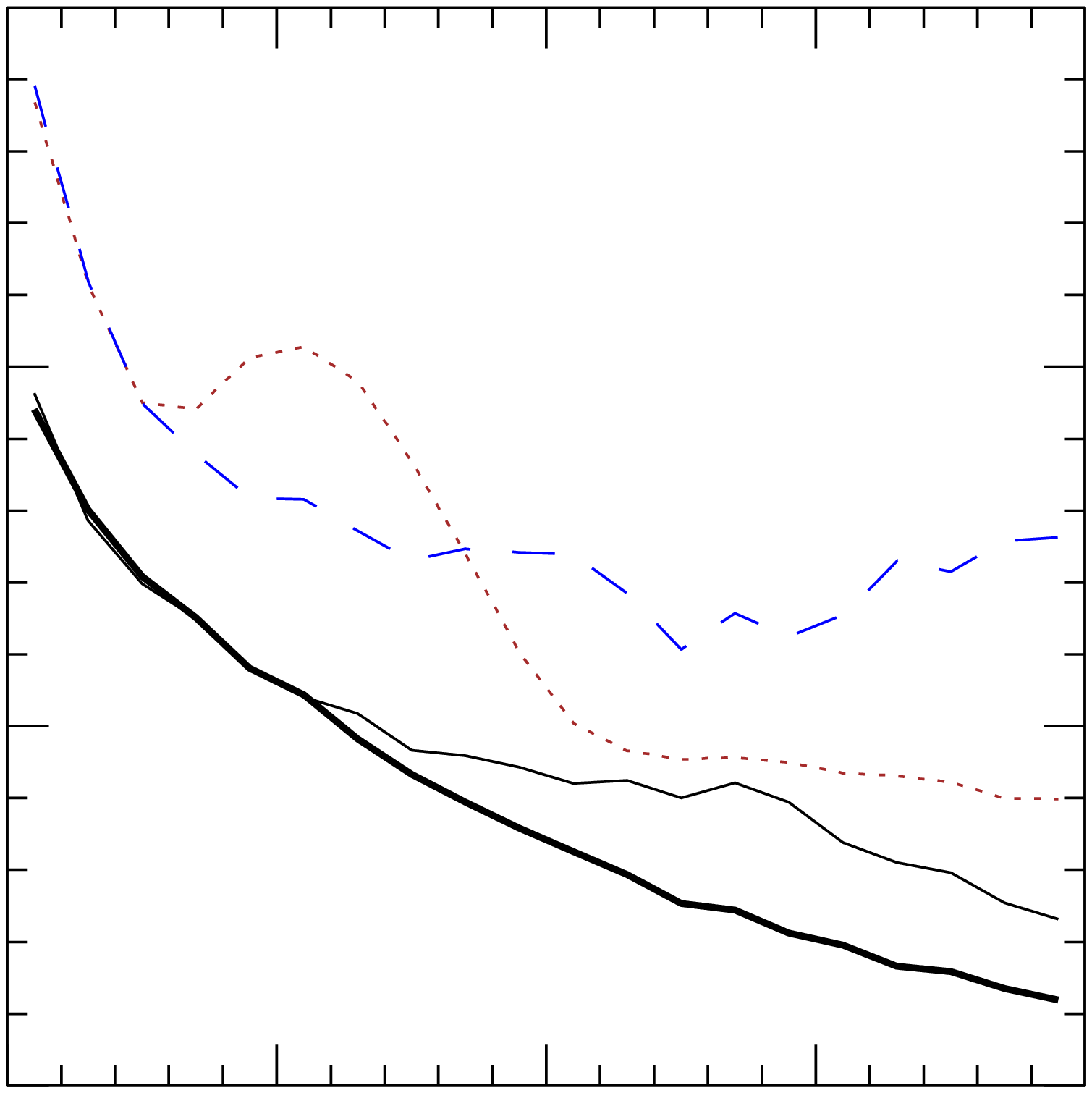}\hspace*{-1.36cm}
\includegraphics[width=50mm]{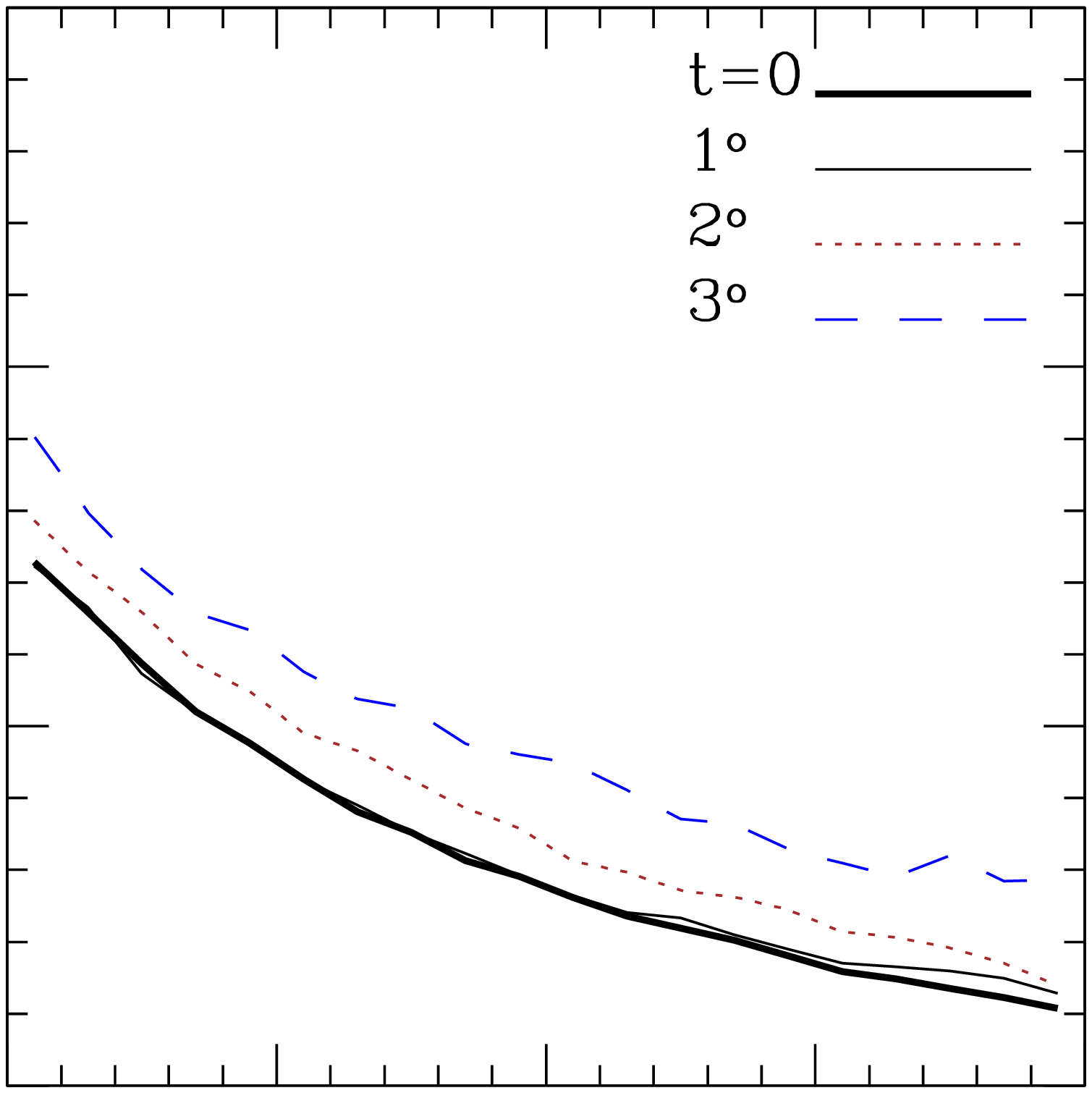}\vspace*{-1.29cm}\\
\includegraphics[width=50mm]{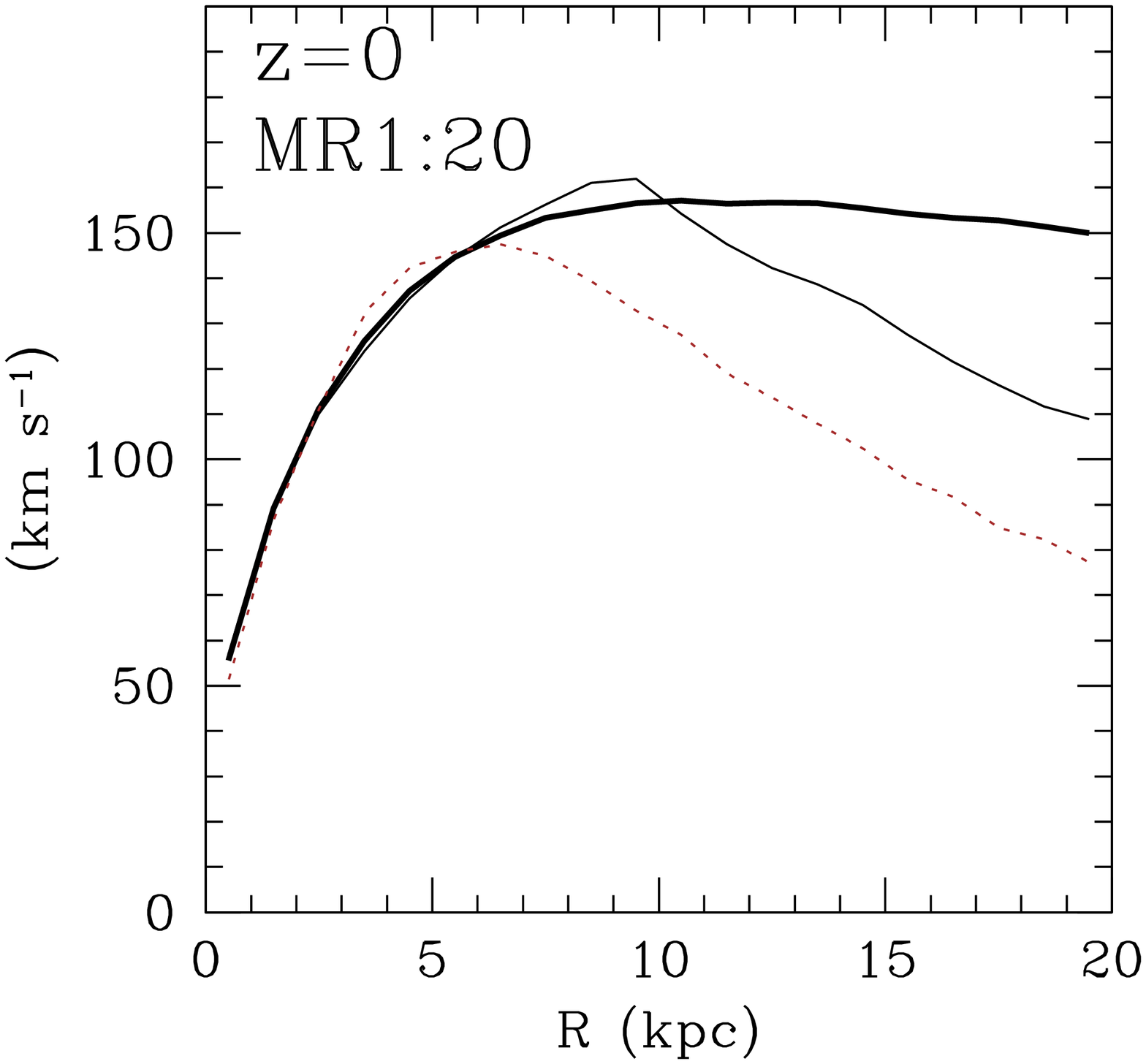}\hspace*{-0.4cm}
\includegraphics[width=50mm]{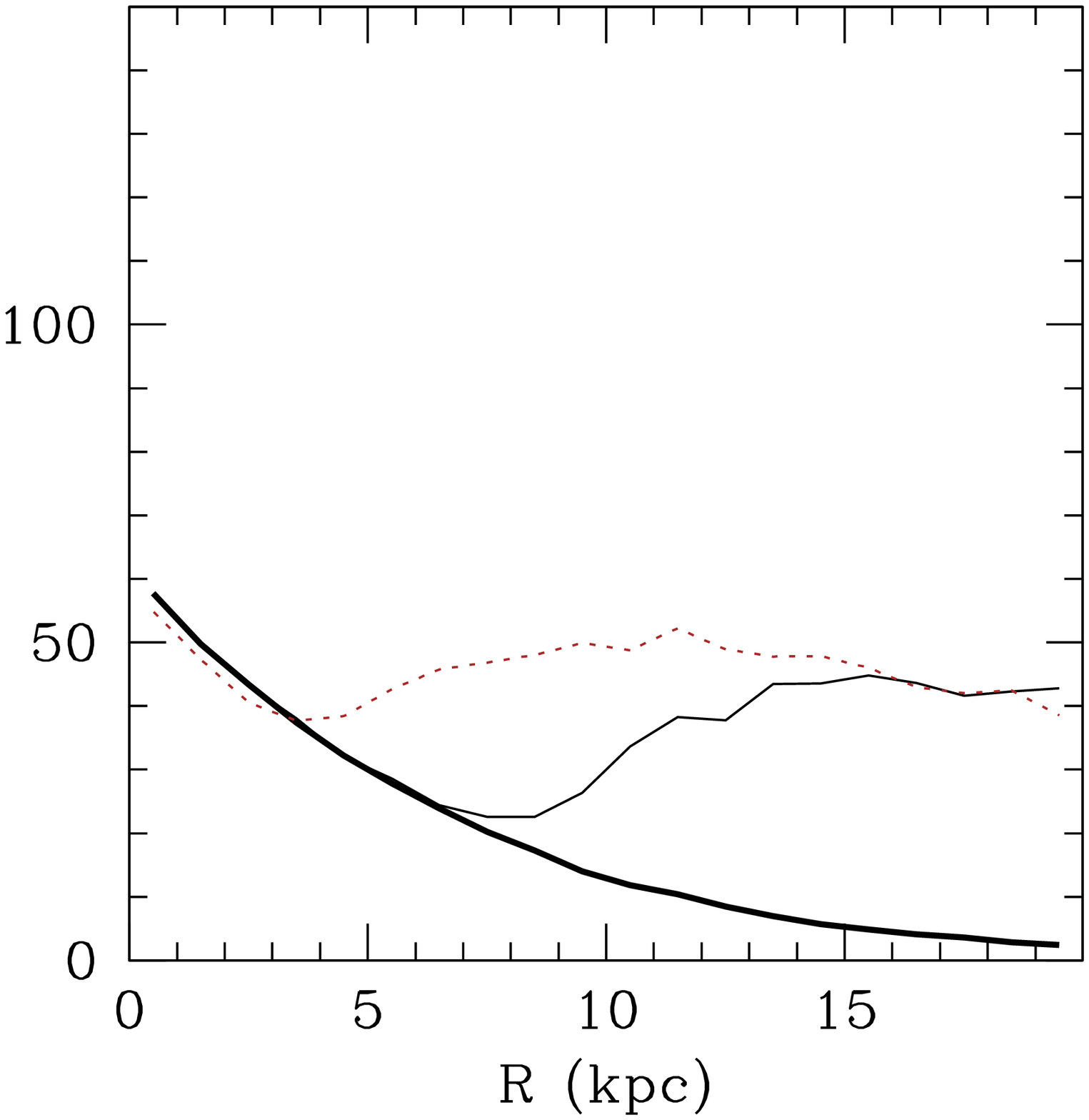}\hspace*{-1.36cm}
\includegraphics[width=50mm]{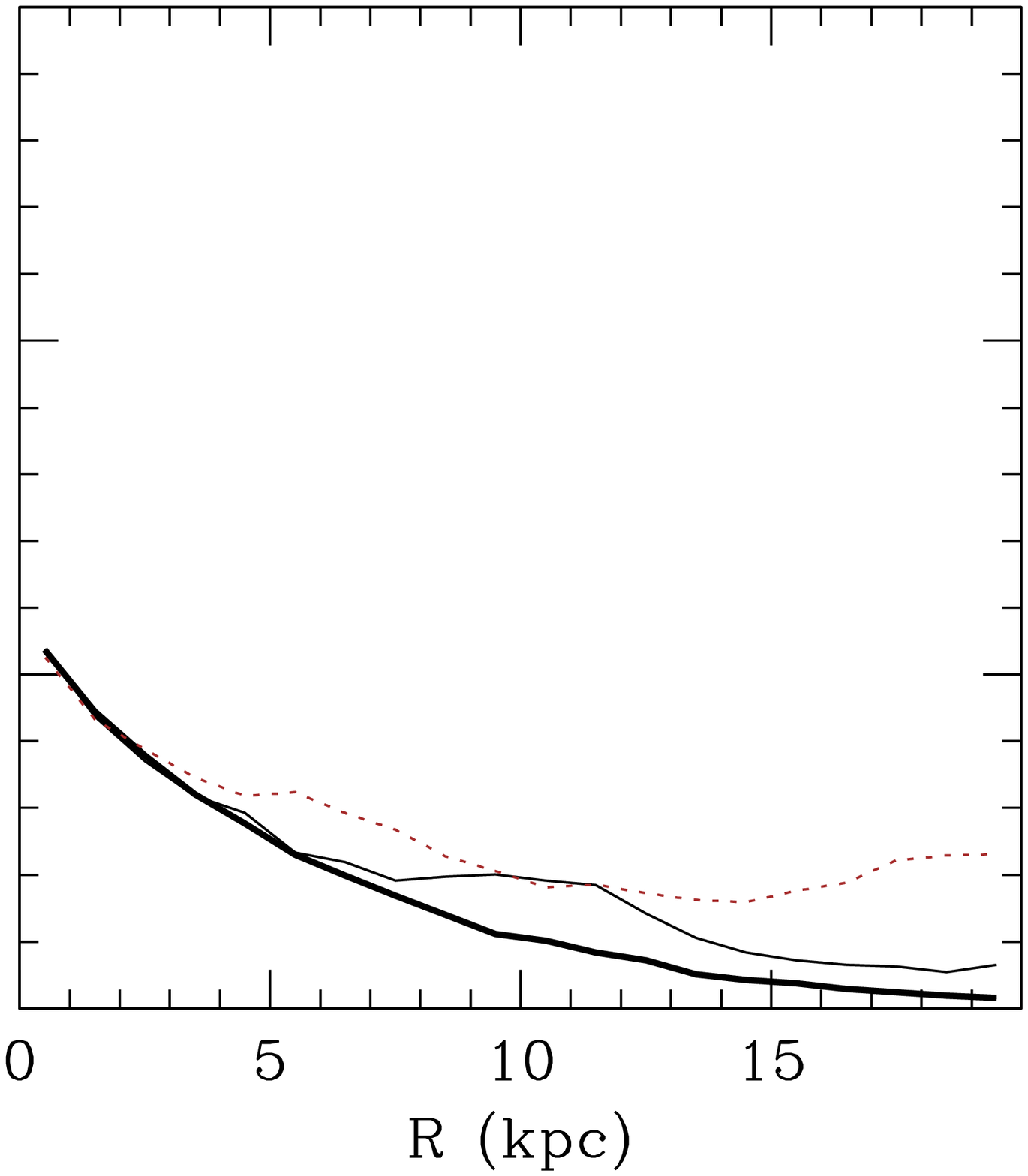}\hspace*{-1.36cm}
\includegraphics[width=50mm]{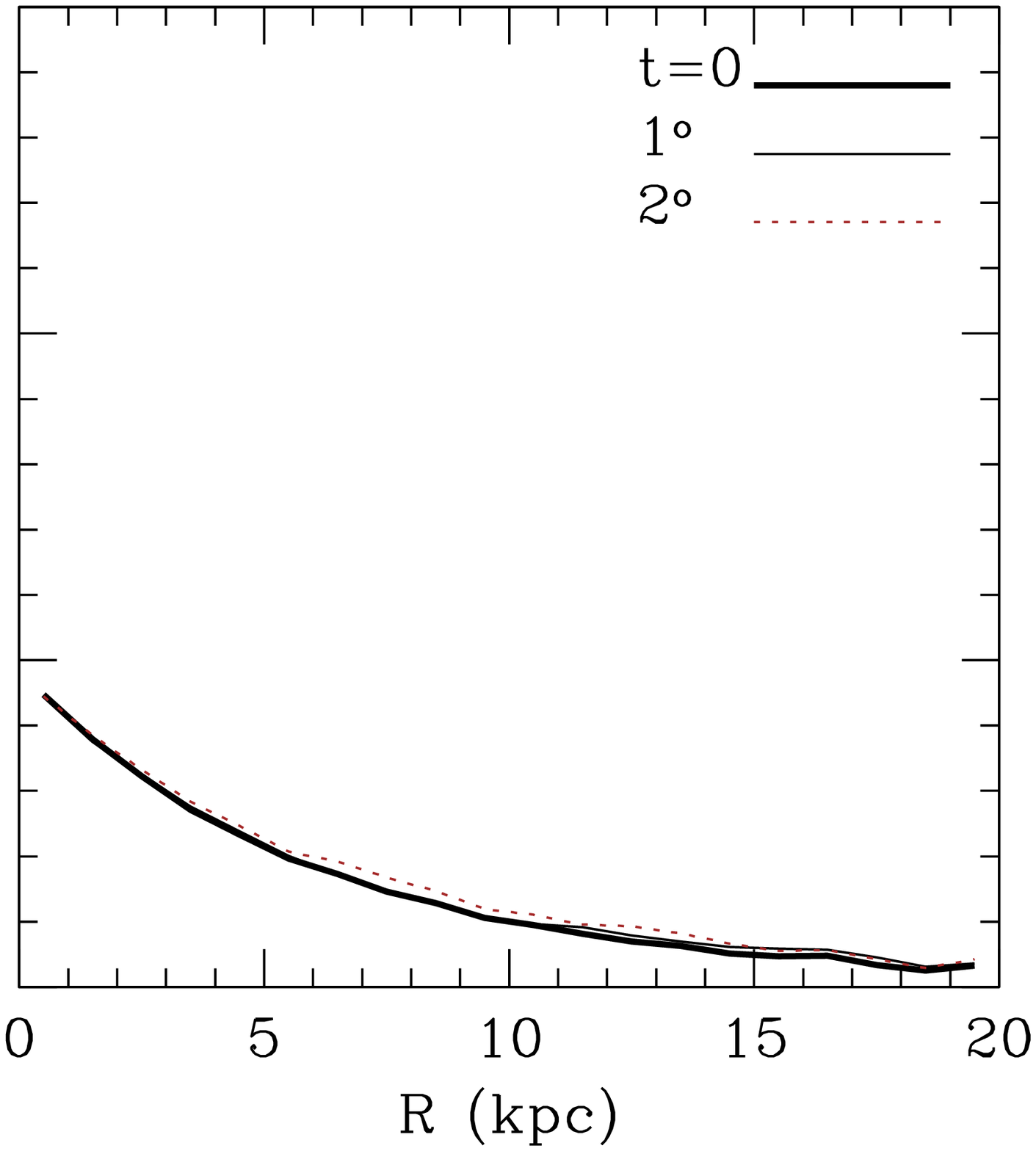}
\end{center}
\caption{
Continuation of Figure~\ref{kinematics-evol-z0}. 
Evolution of the kinematics of disc galaxies in terms of their mean rotation,
$\langle V_{\phi}\rangle$, and the radial, azimuthal and vertical components
of their velocity dispersions, $\sigma_{\rm R}$, $\sigma_{\phi}$, and 
$\sigma_{Z}$ respectively, after each of their first six pericentric passages.
The kinematics of discs have been computed in concentric rings, 1~kpc wide, 
considering only stars that remain bound and are located within 3~kpc from the 
midplane, and have been corrected by the evolution of the respective disc in 
isolation.
}
\label{app-f4}
\end{figure*}

\begin{figure*}
\begin{center}
\includegraphics[width=50mm]{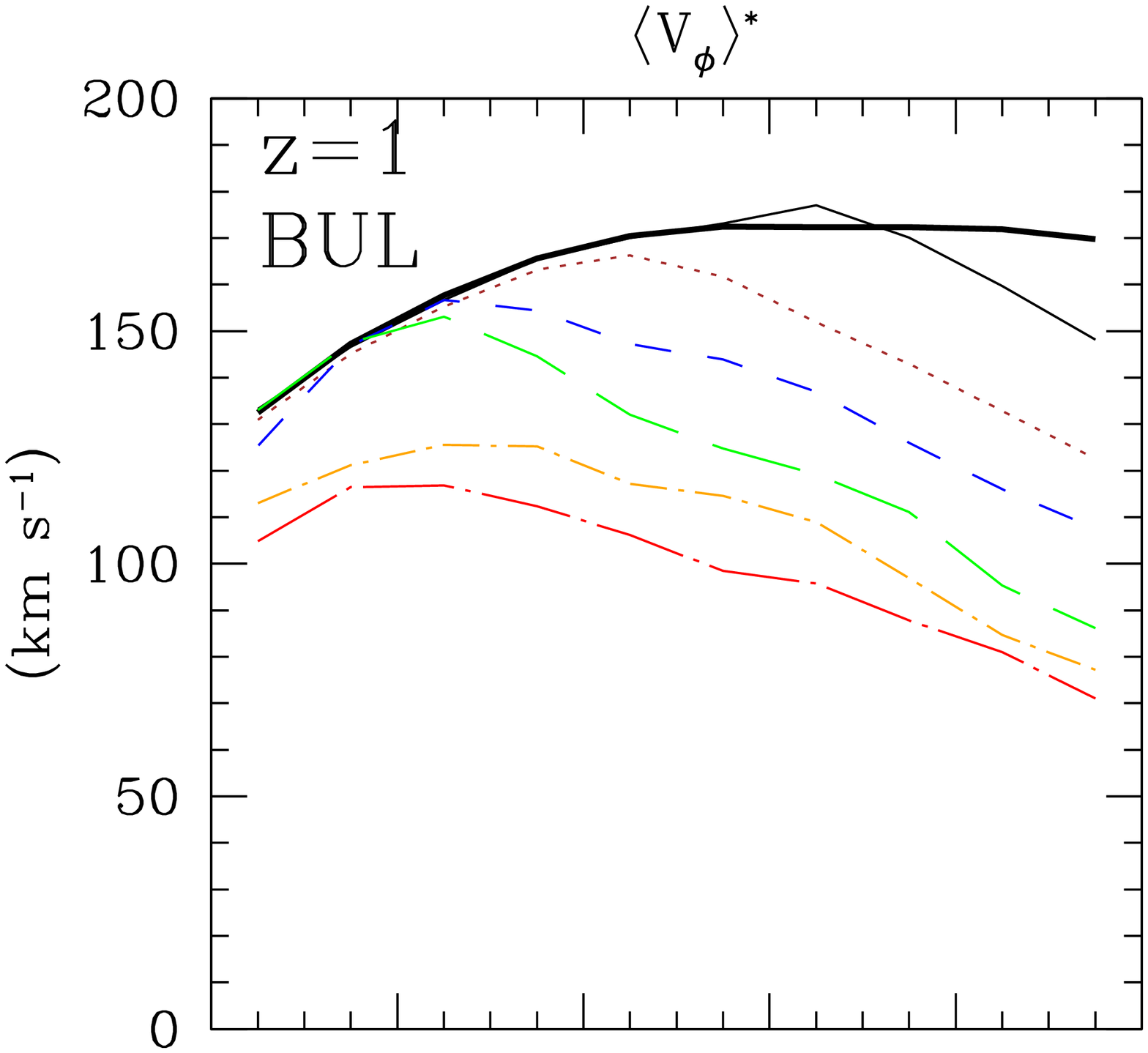}\hspace*{-0.4cm}
\includegraphics[width=50mm]{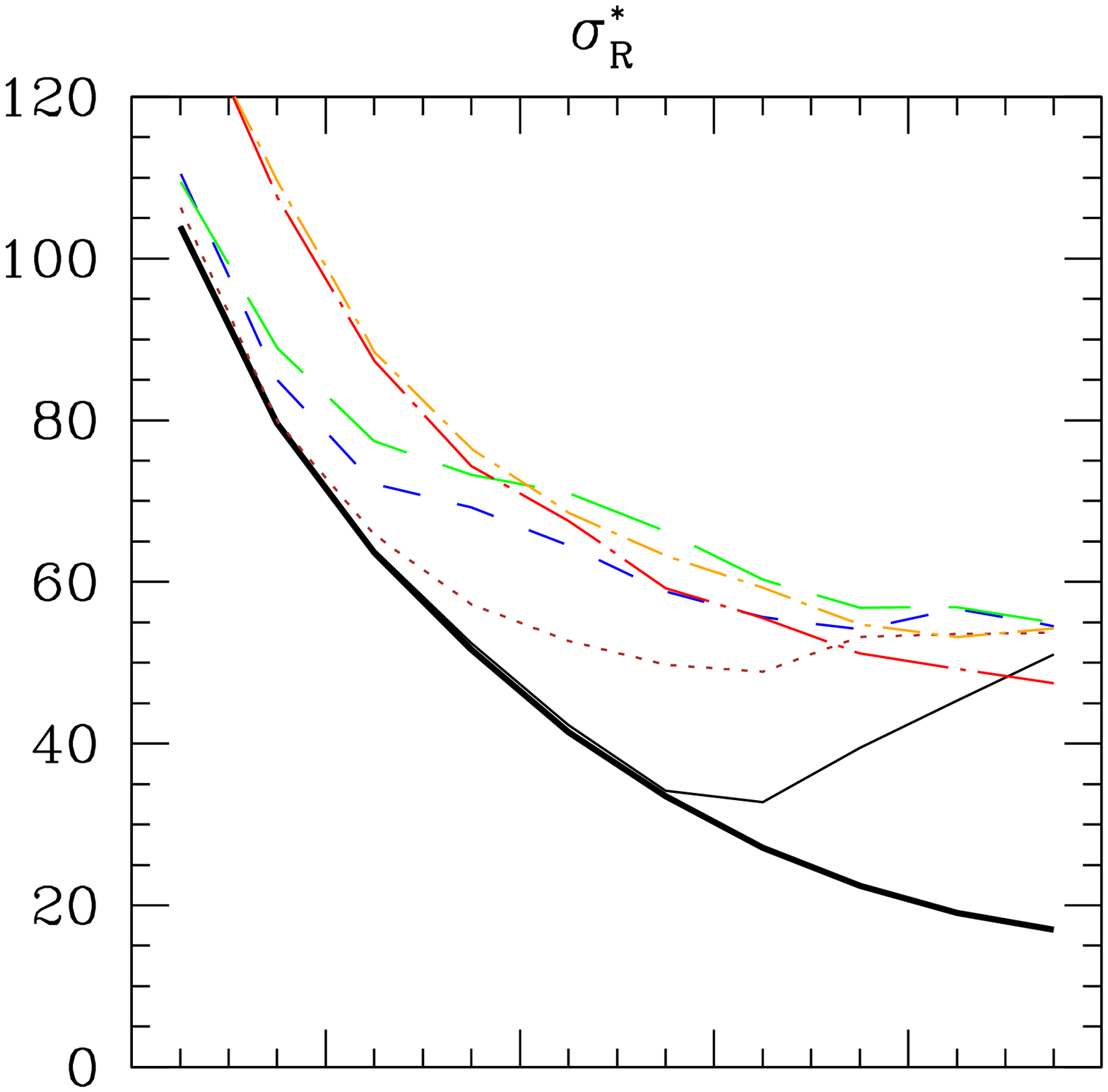}\hspace*{-1.36cm}
\includegraphics[width=50mm]{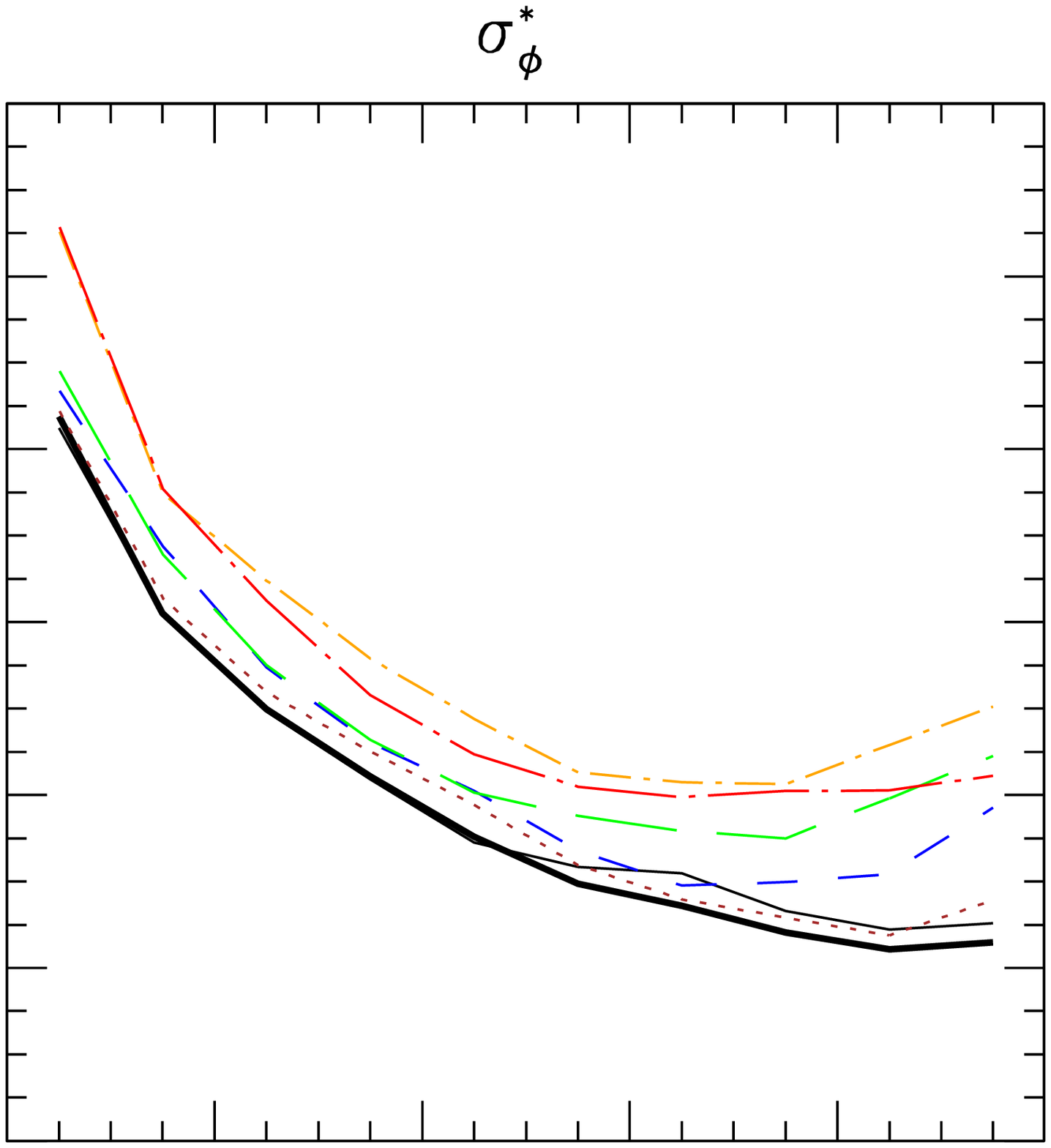}\hspace*{-1.36cm}
\includegraphics[width=50mm]{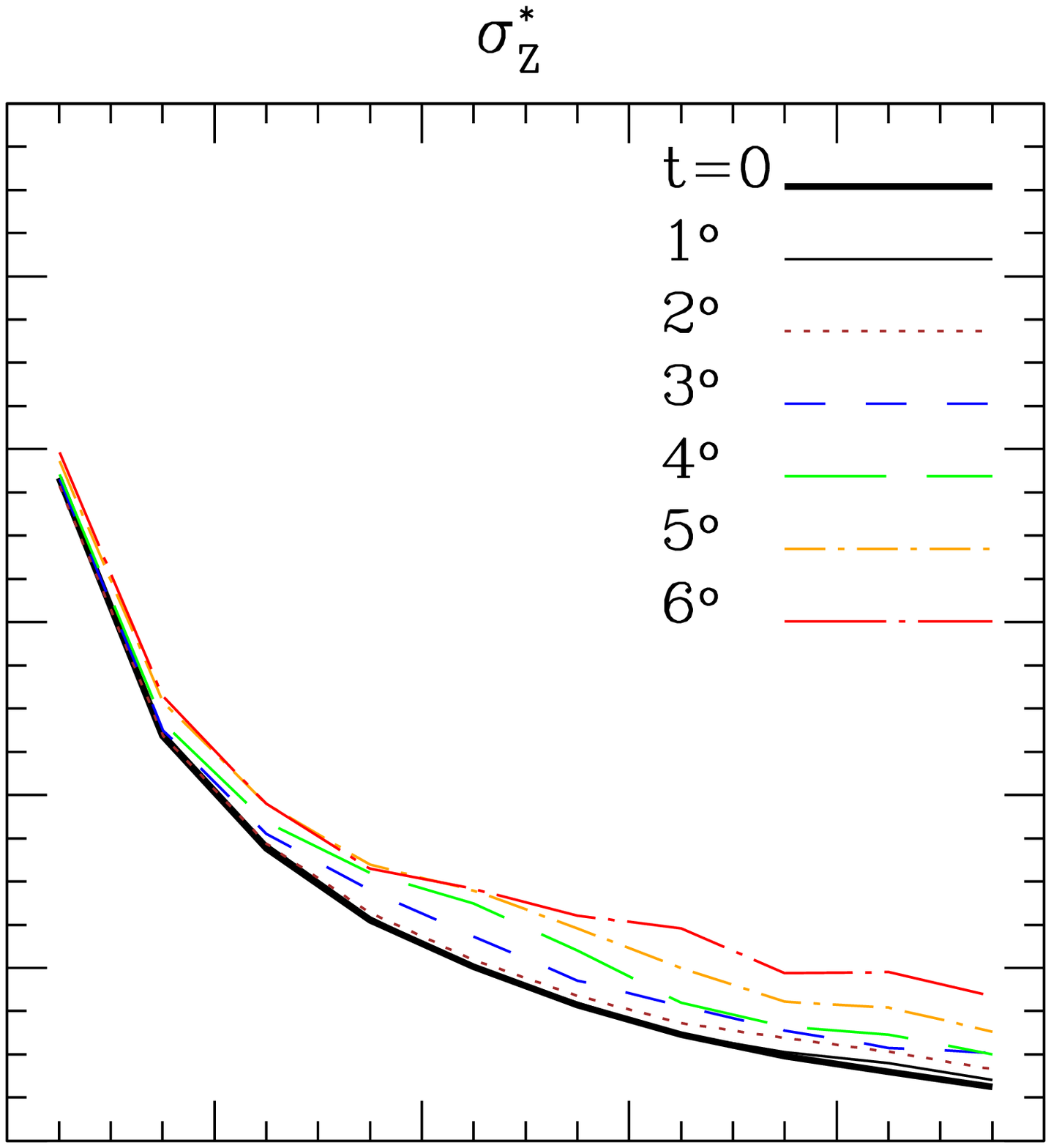}\vspace*{-1.29cm}\\
\includegraphics[width=50mm]{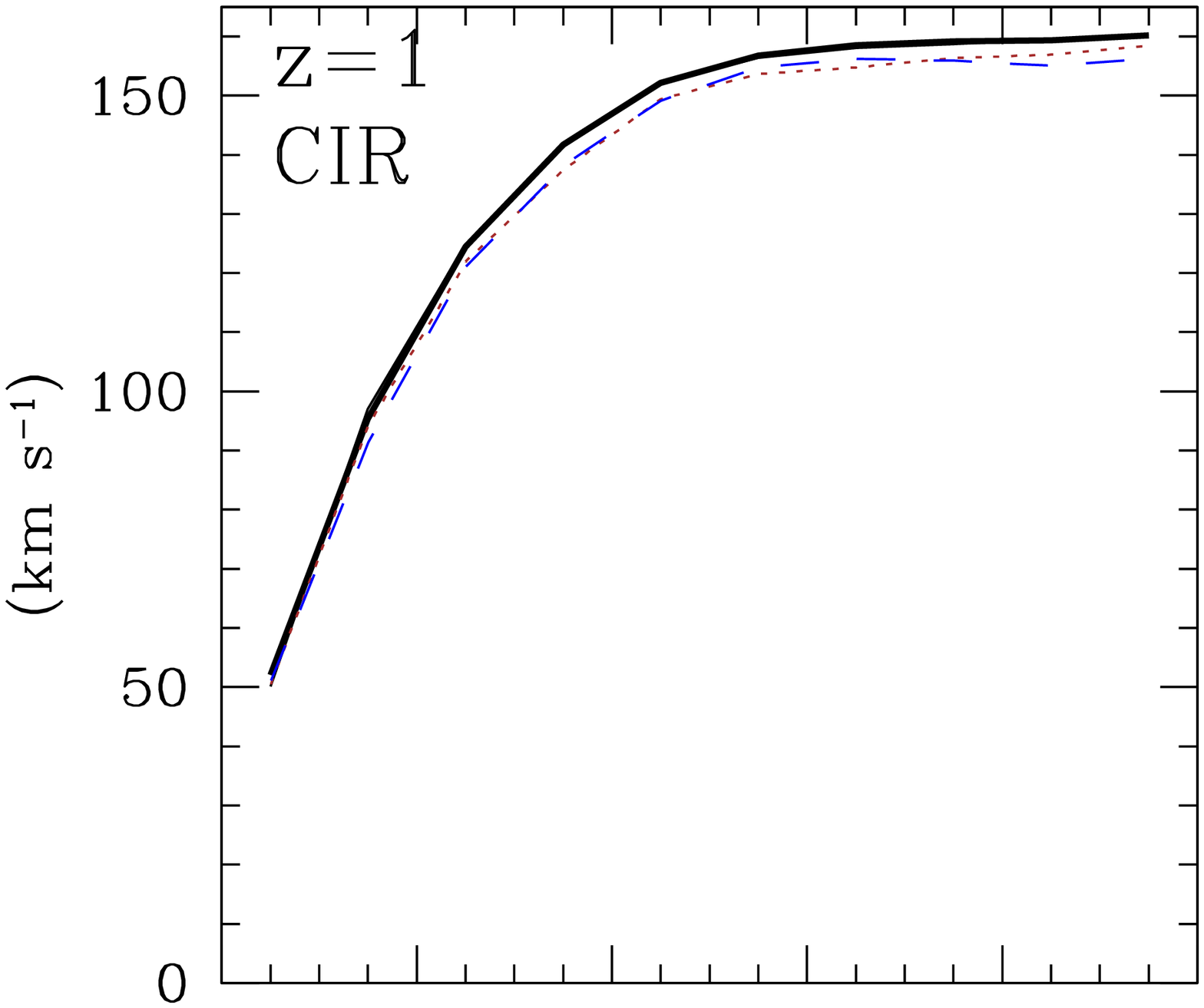}\hspace*{-0.4cm}
\includegraphics[width=50mm]{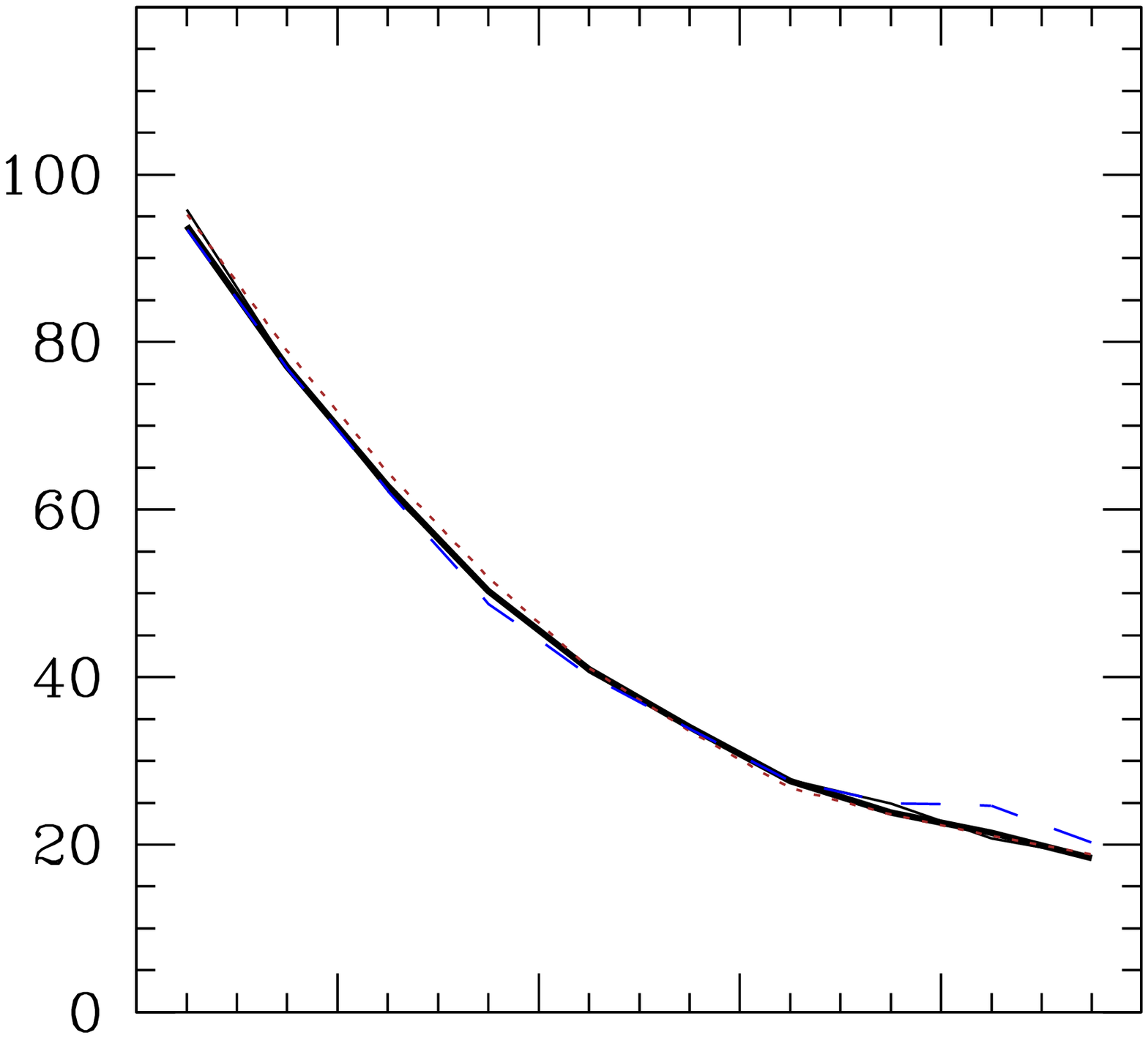}\hspace*{-1.36cm}
\includegraphics[width=50mm]{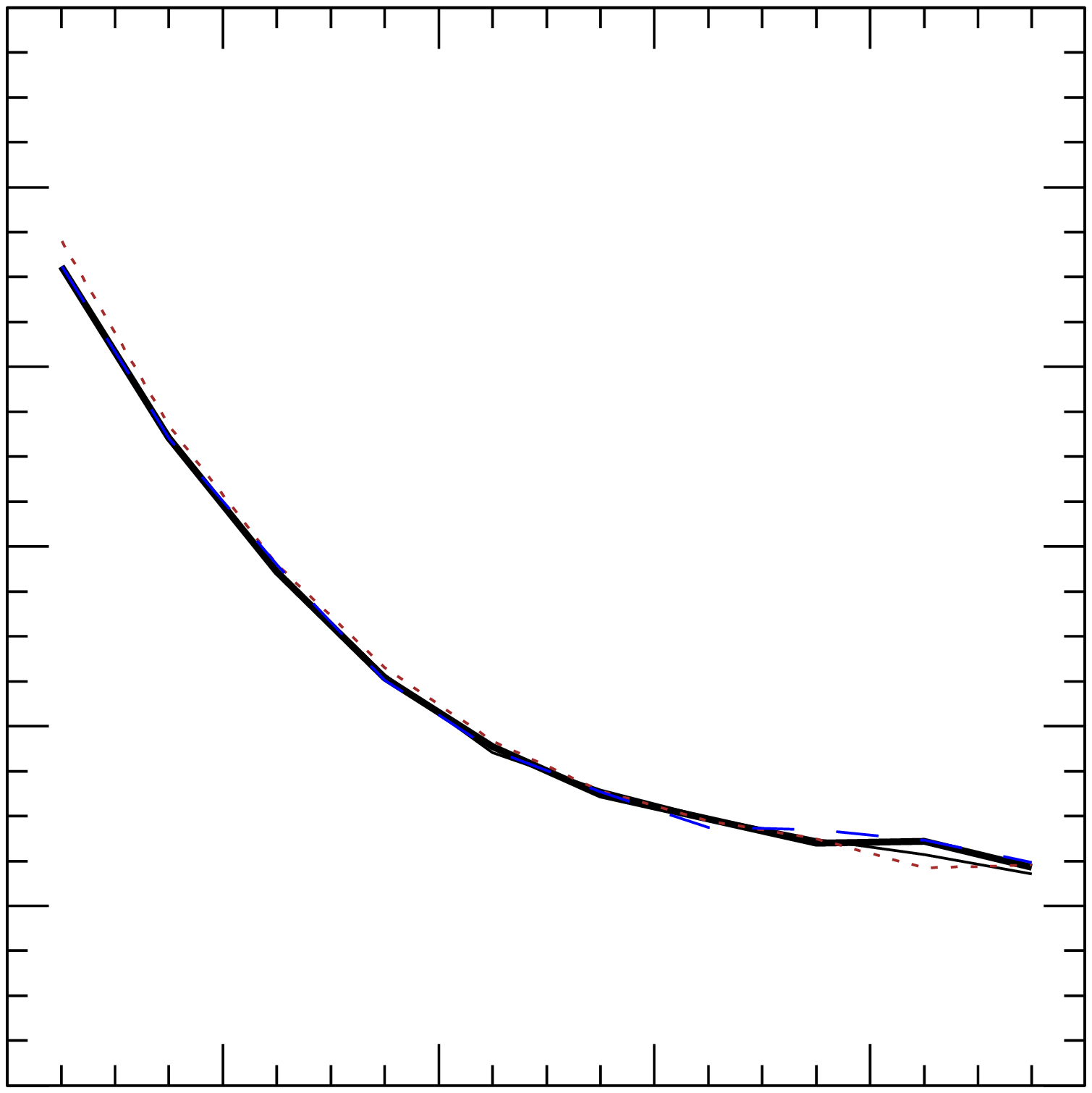}\hspace*{-1.36cm}
\includegraphics[width=50mm]{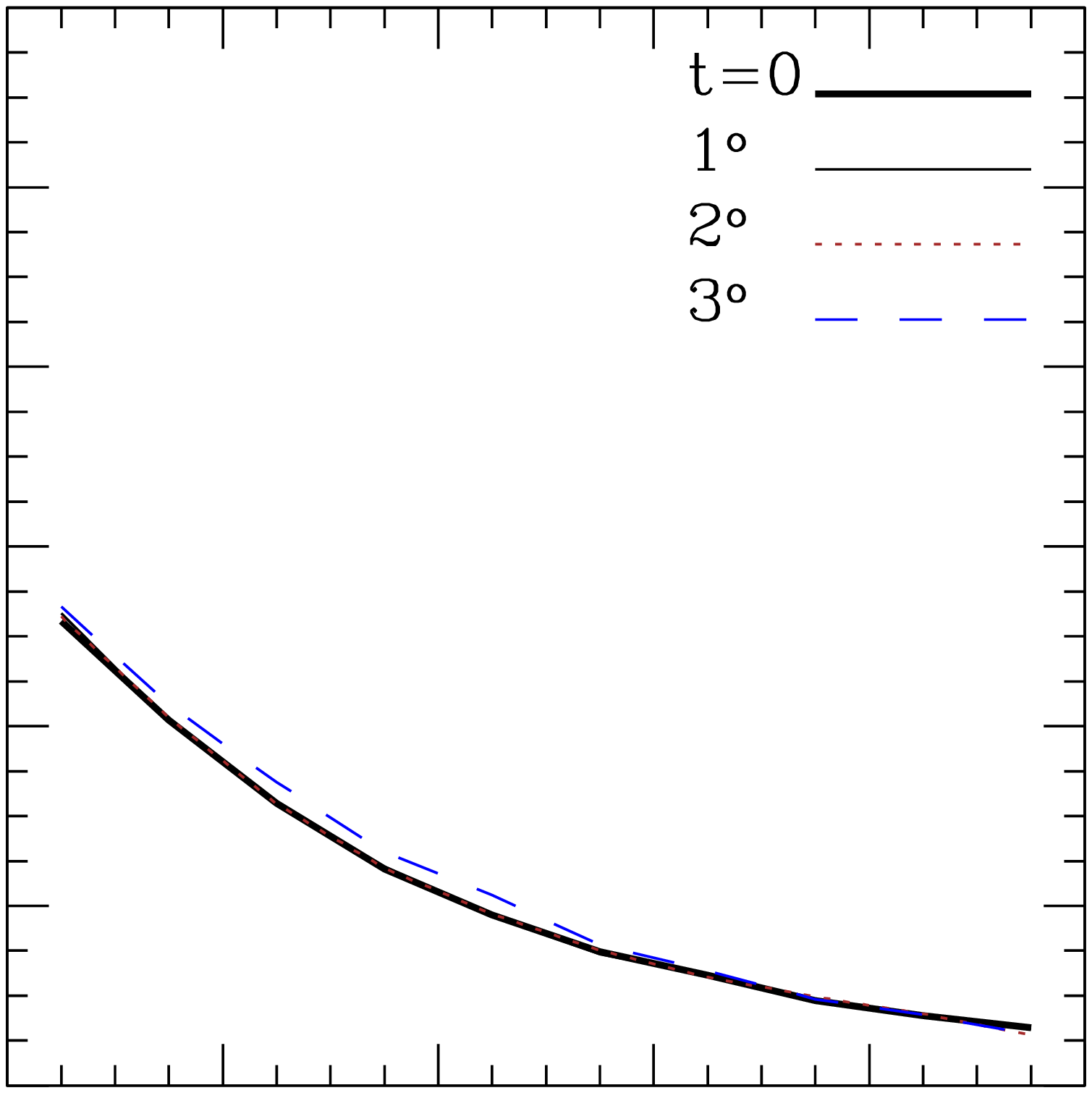}\vspace*{-1.29cm}\\
\includegraphics[width=50mm]{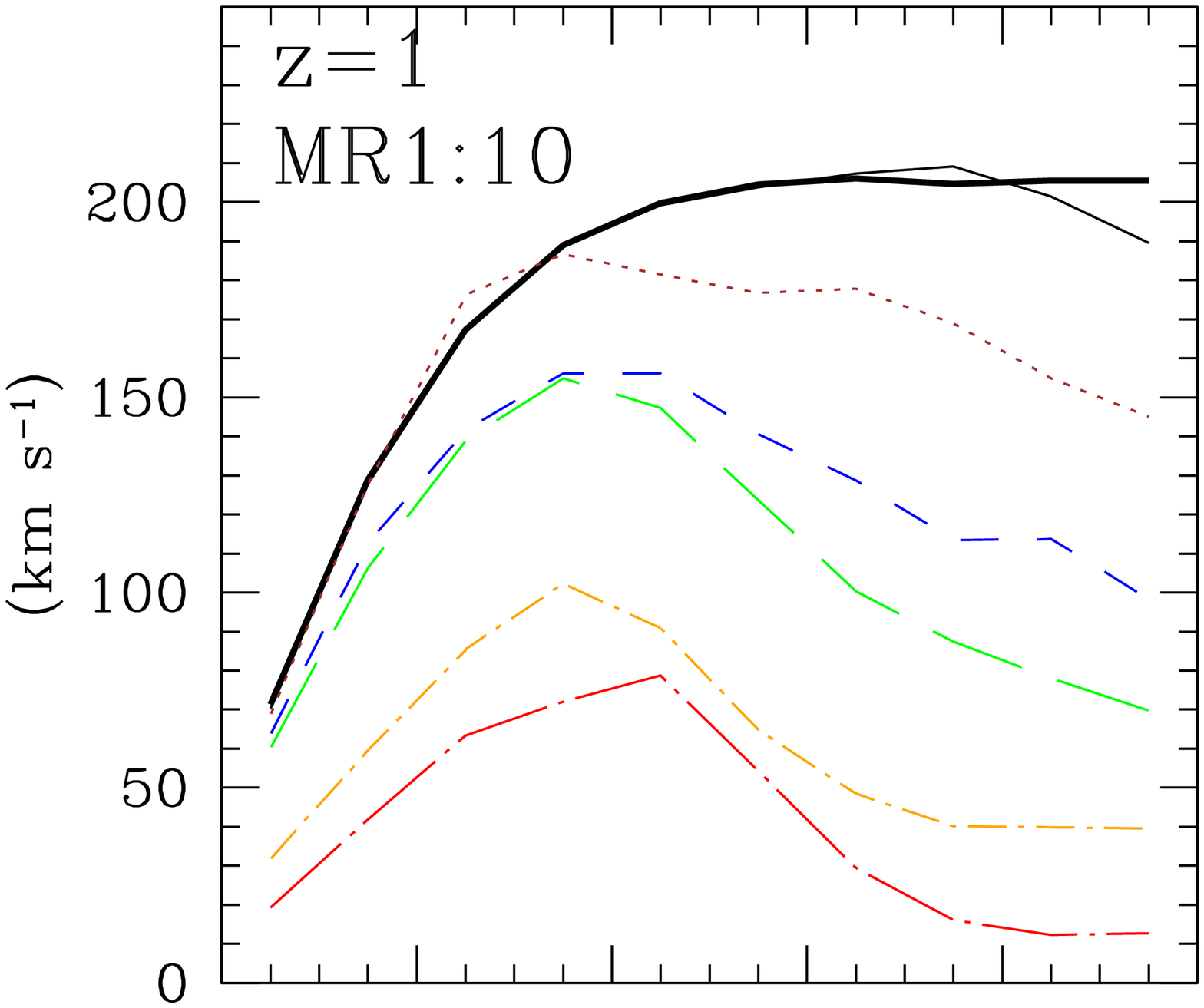}\hspace*{-0.4cm}
\includegraphics[width=50mm]{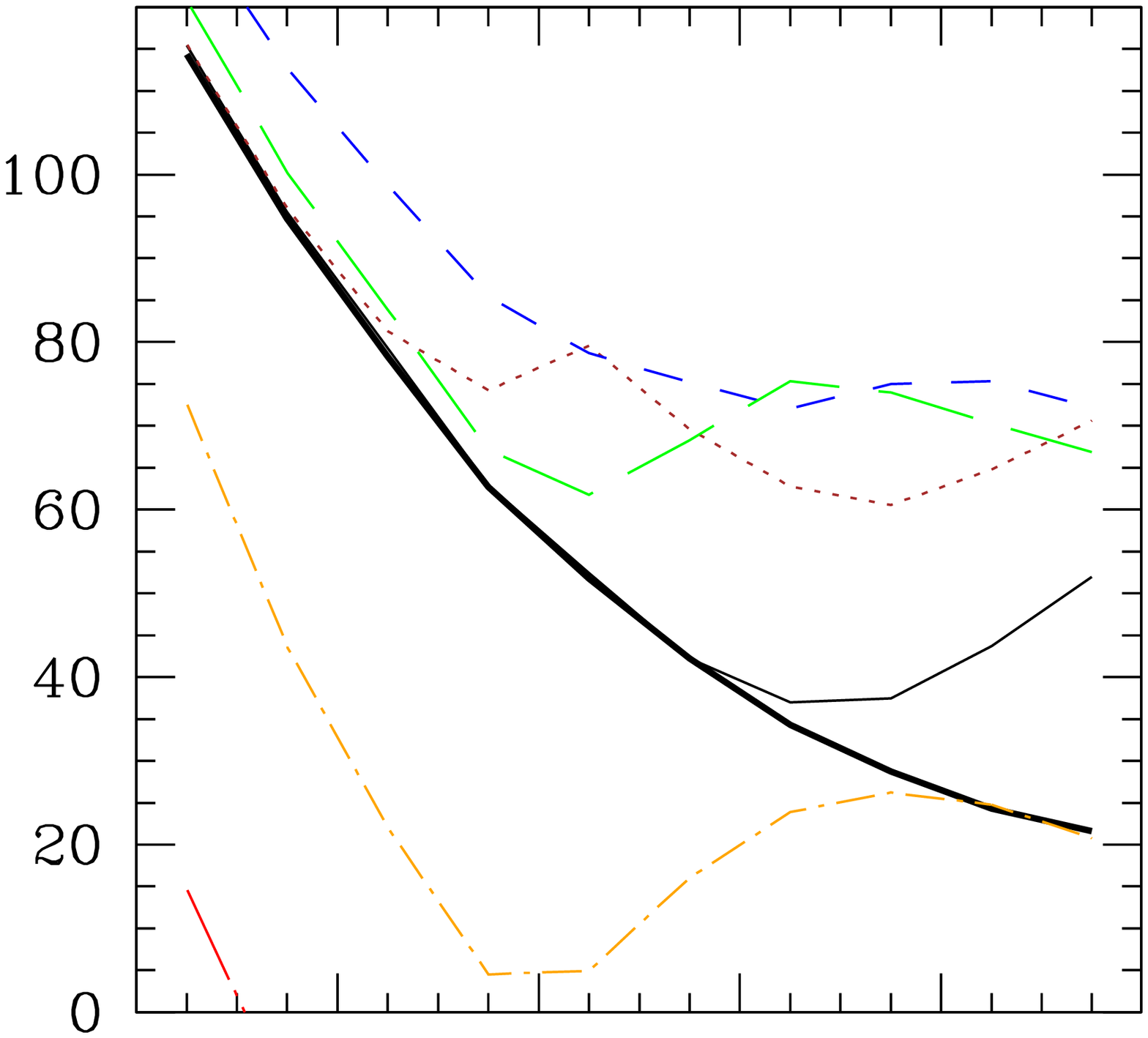}\hspace*{-1.36cm}
\includegraphics[width=50mm]{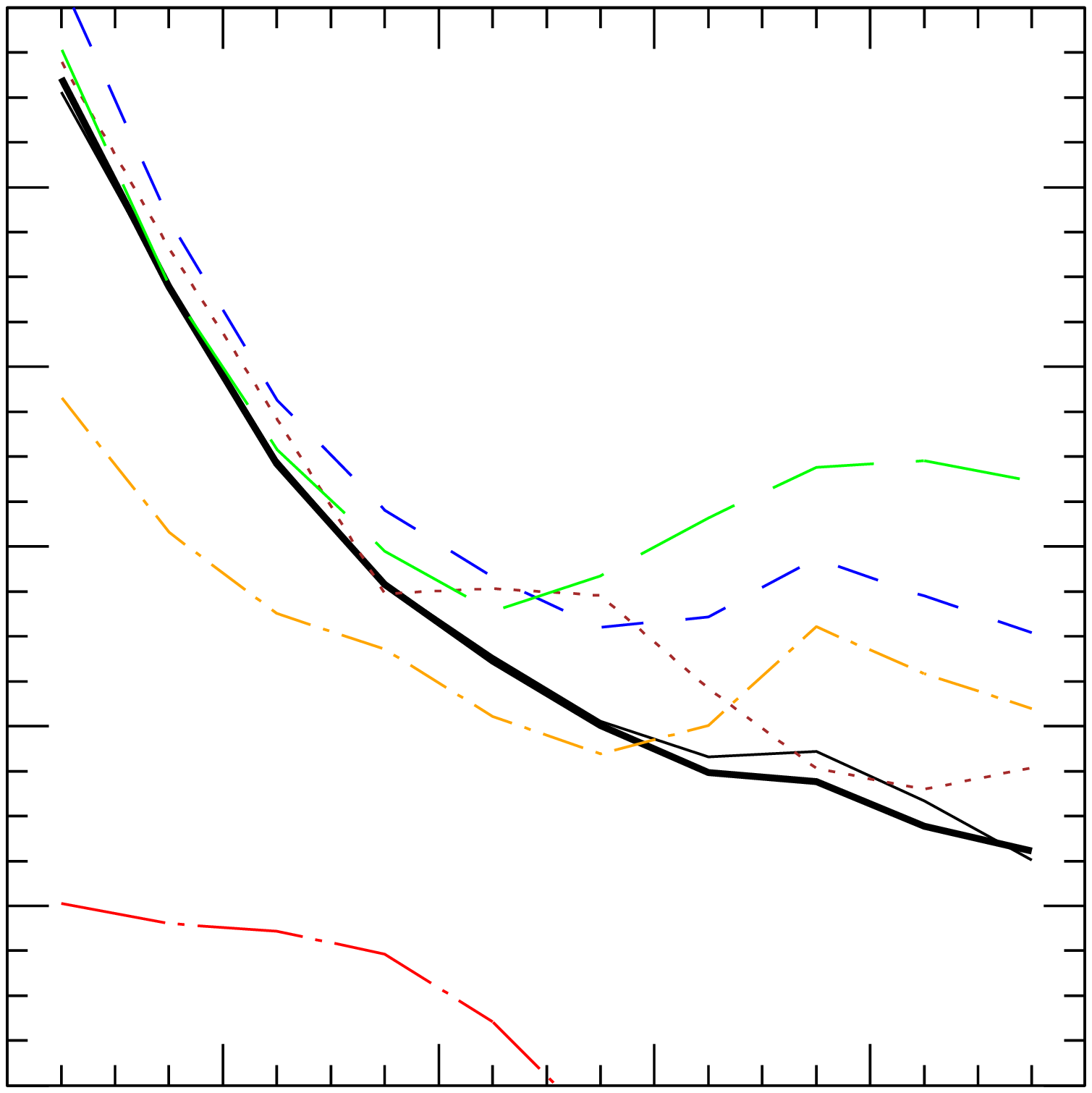}\hspace*{-1.36cm}
\includegraphics[width=50mm]{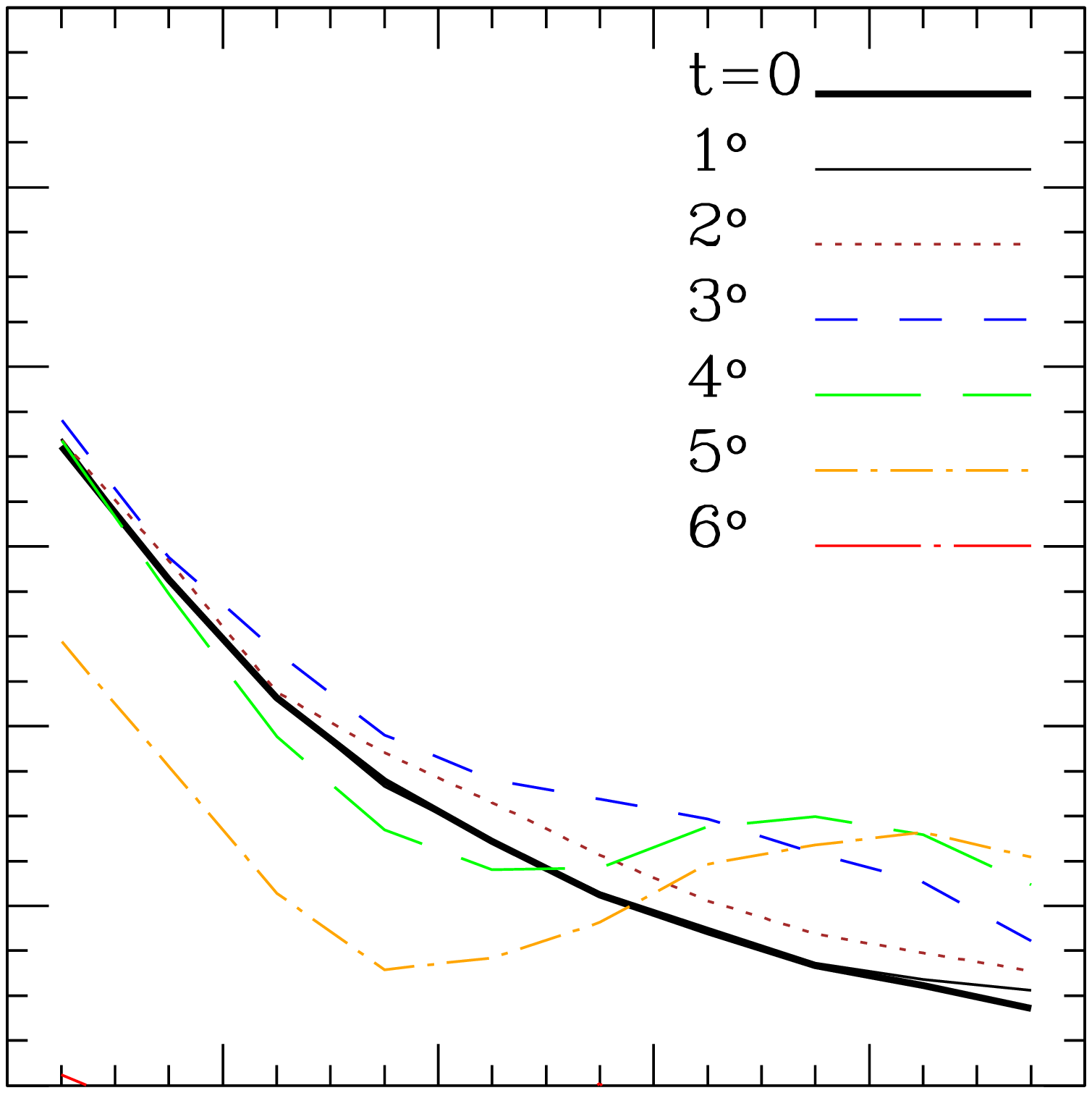}\vspace*{-1.29cm}\\
\includegraphics[width=50mm]{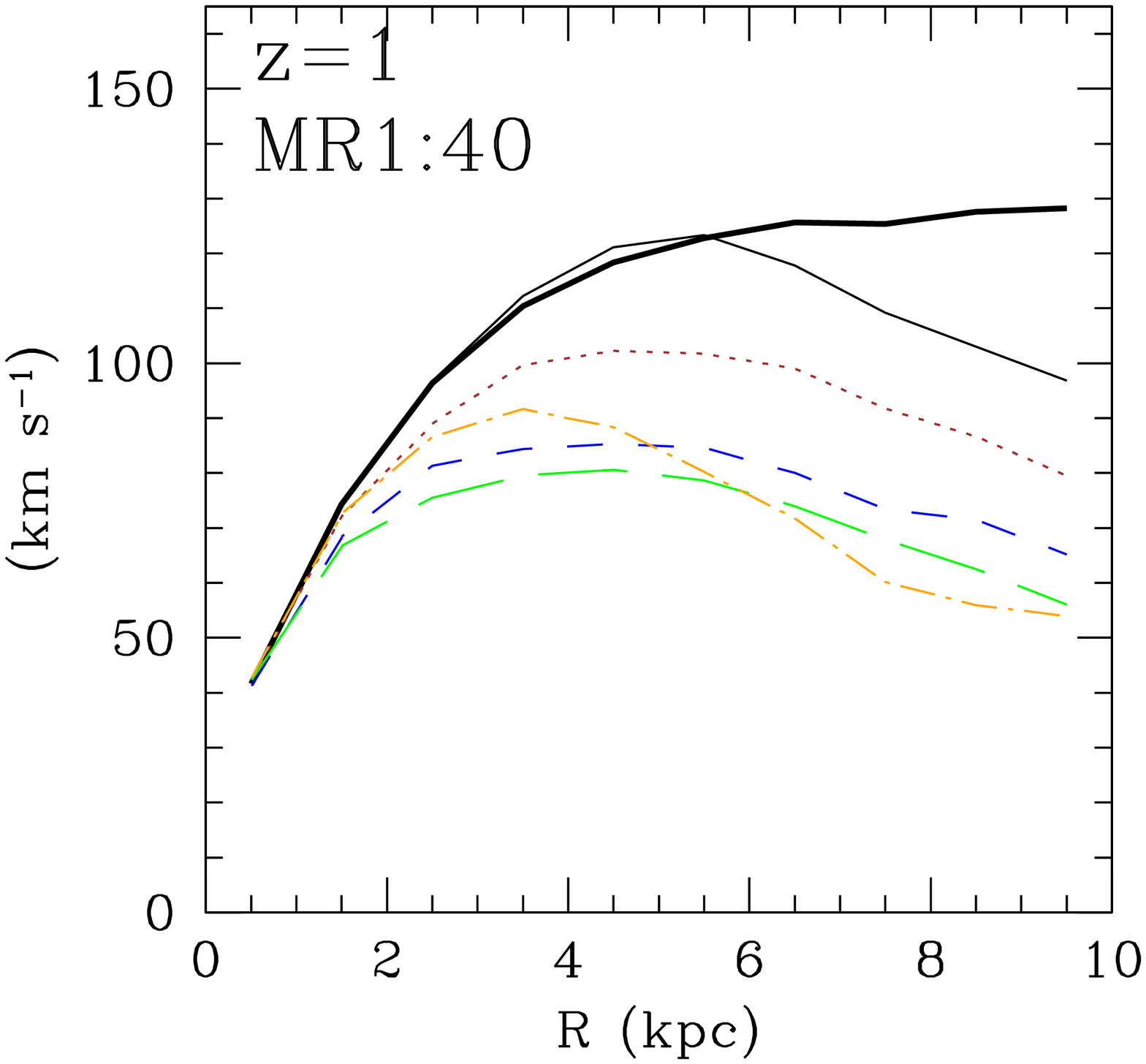}\hspace*{-0.4cm}
\includegraphics[width=50mm]{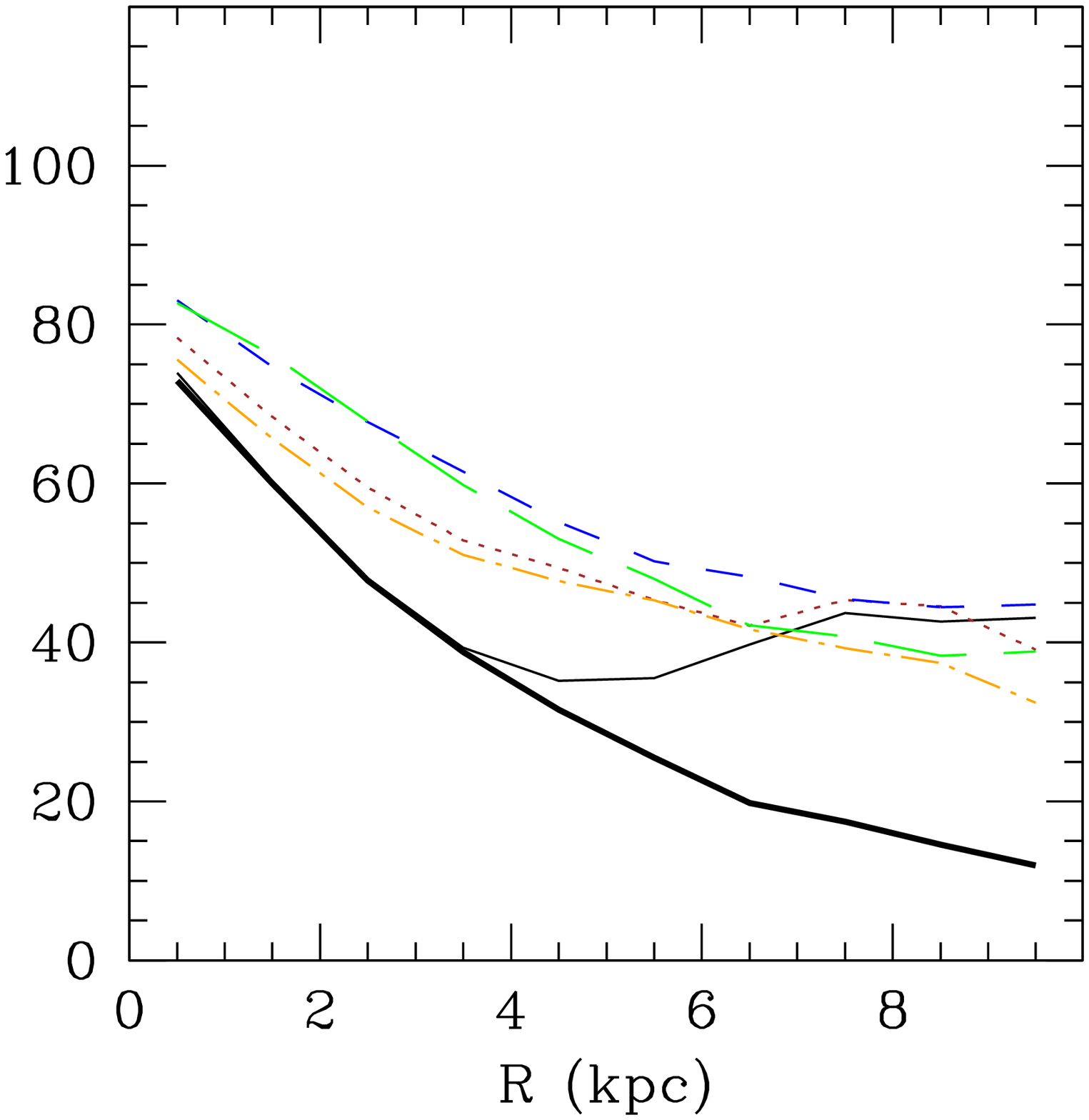}\hspace*{-1.36cm}
\includegraphics[width=50mm]{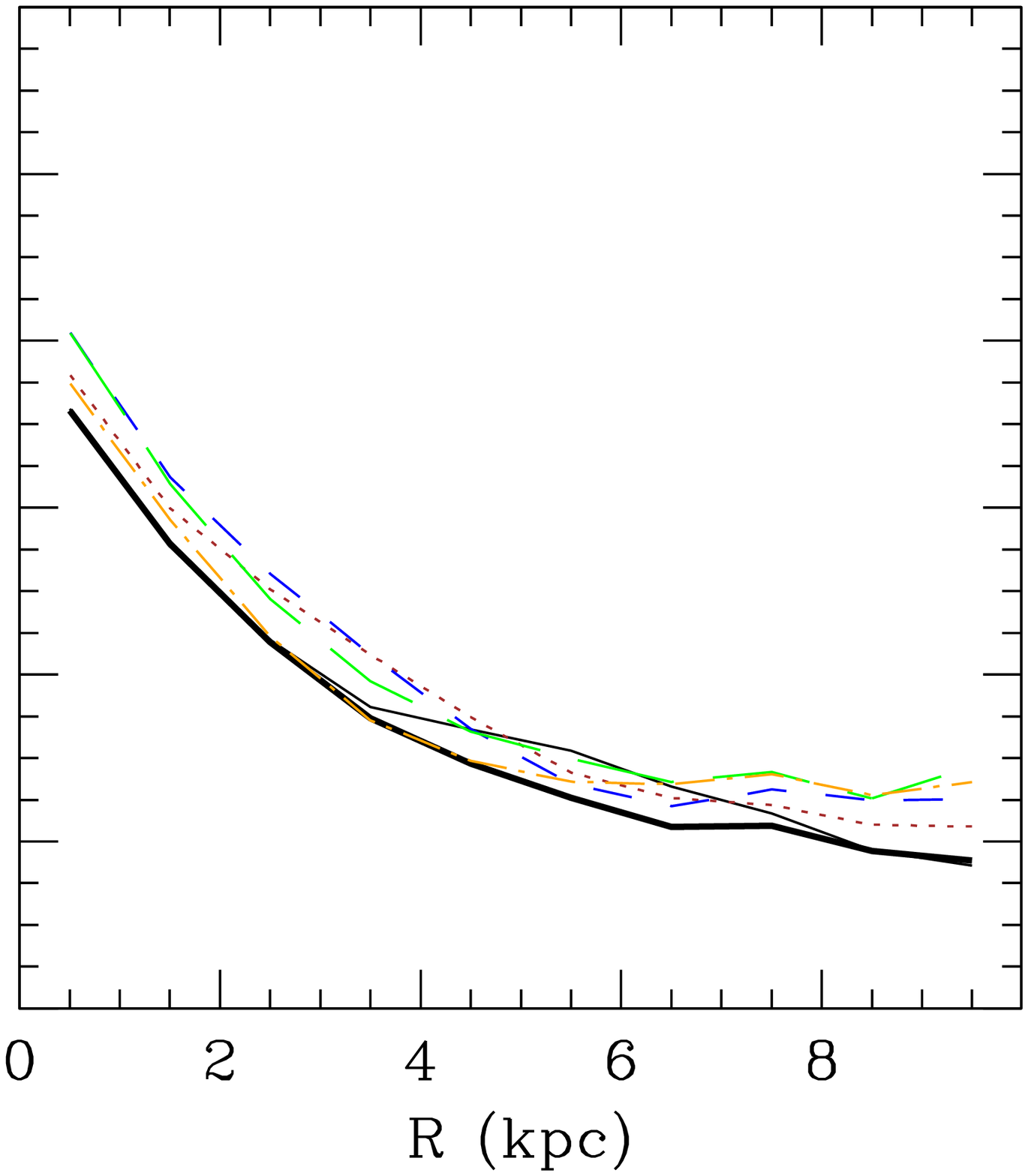}\hspace*{-1.36cm}
\includegraphics[width=50mm]{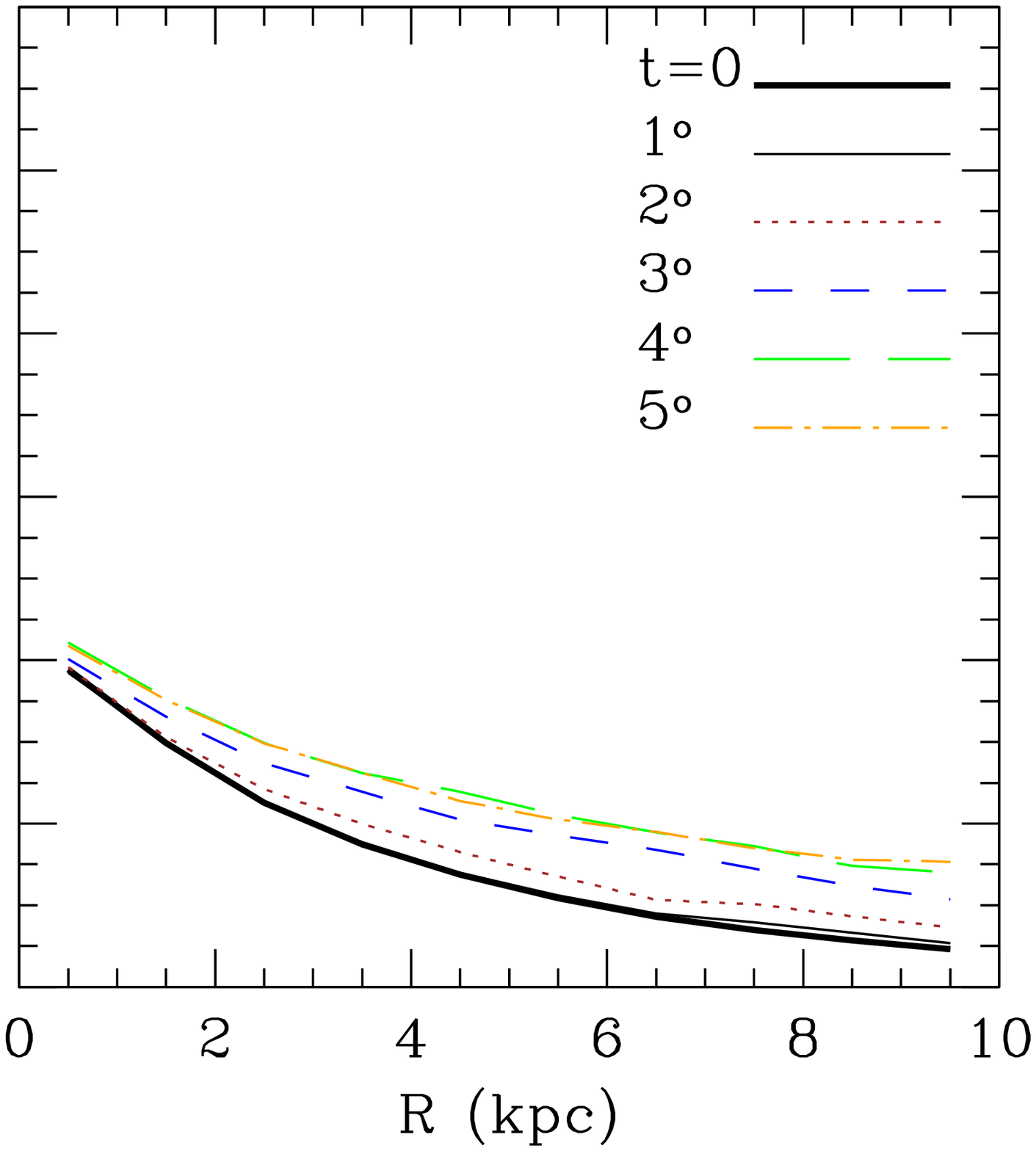}
\end{center}
\caption{Continuation of Figure~\ref{kinematics-evol-z1}. Same description as 
Figure~\ref{app-f4}.}
\label{app-f5}
\end{figure*}

\begin{figure*}
\begin{center}
\includegraphics[width=42mm]{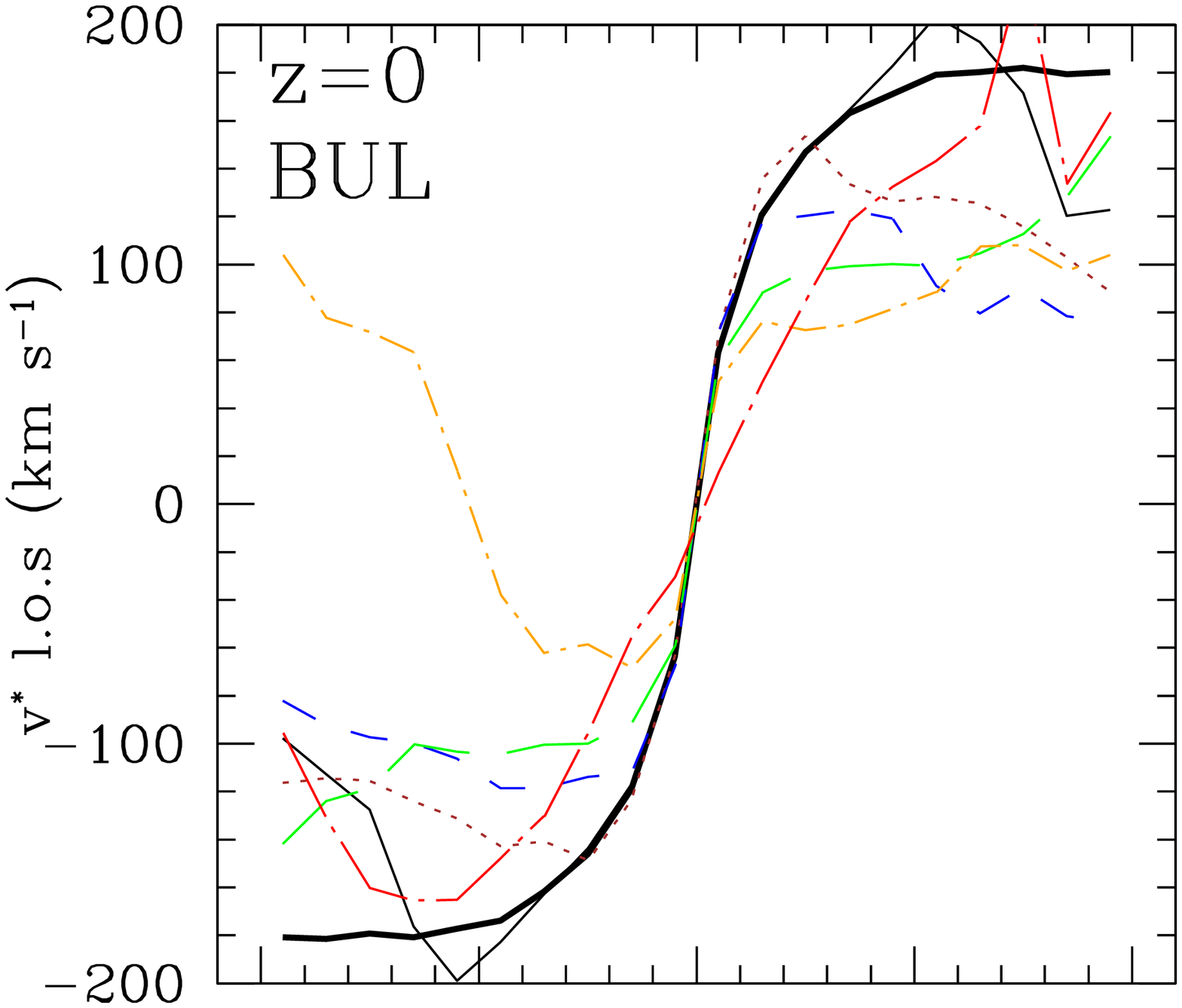}\hspace*{-1.16cm}
\includegraphics[width=42mm]{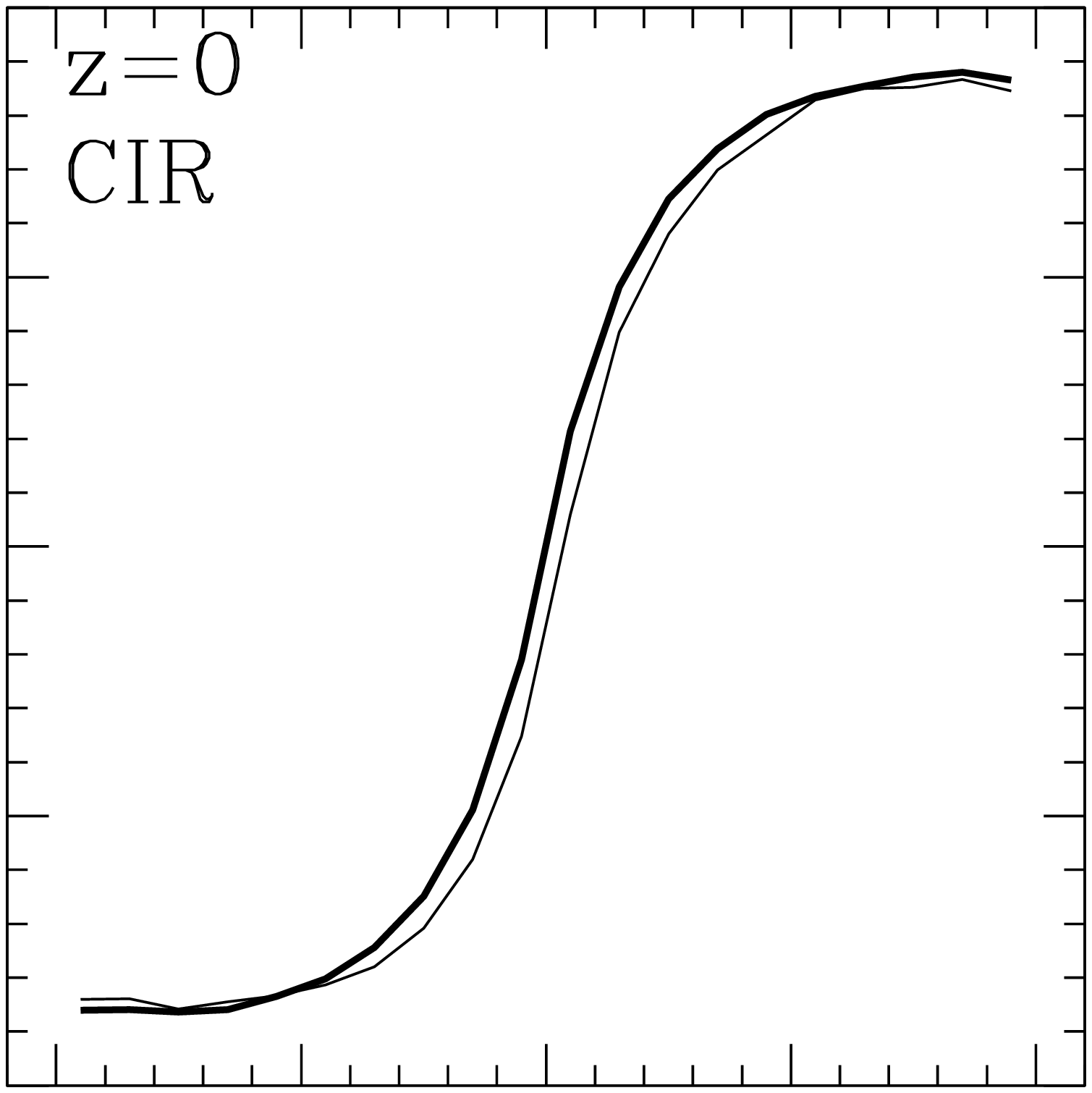}\hspace*{-1.16cm}
\includegraphics[width=42mm]{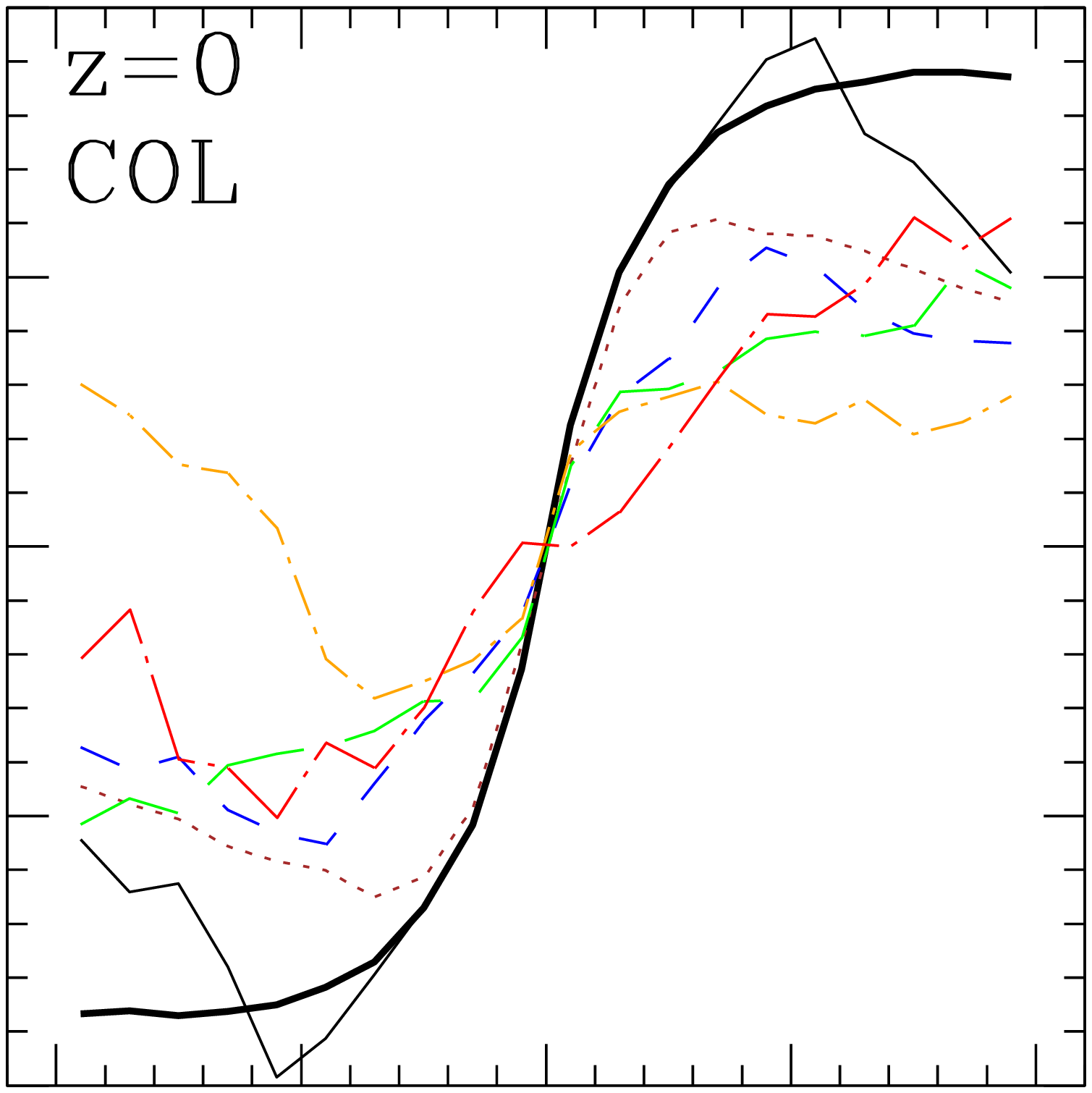}\hspace*{-1.16cm}
\includegraphics[width=42mm]{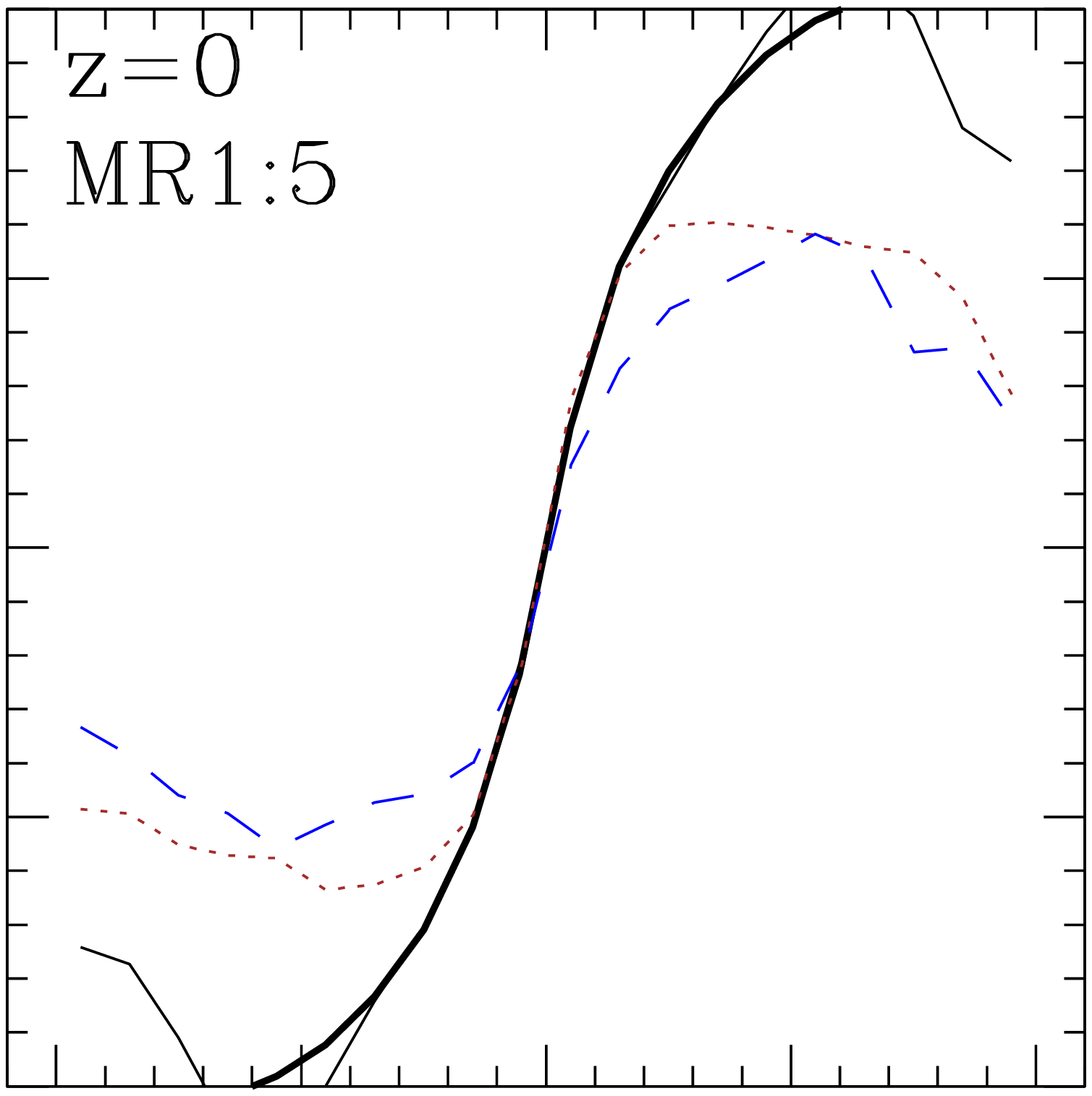}\hspace*{-1.16cm}
\includegraphics[width=42mm]{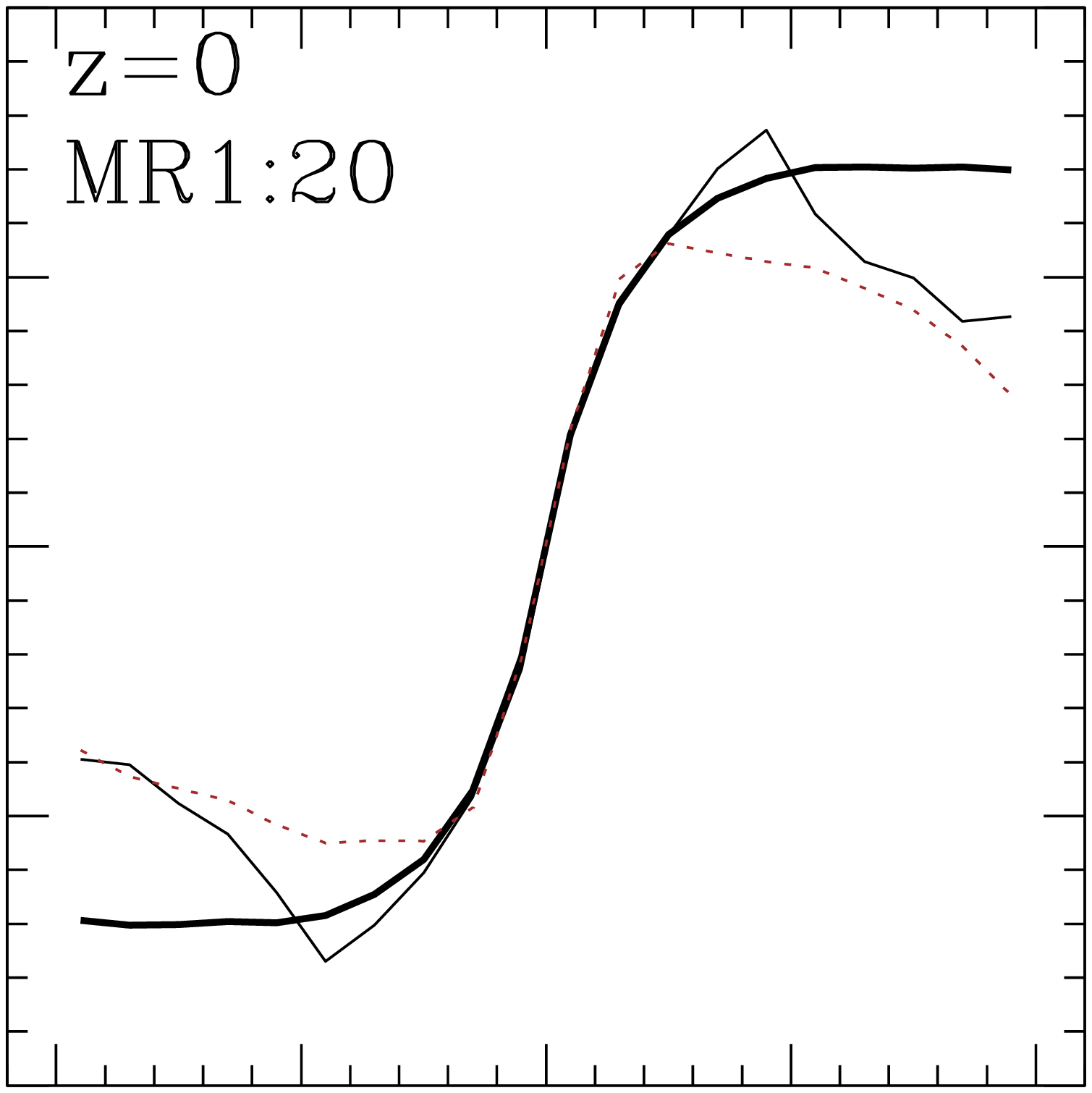}\vspace*{-1.09cm}\\
\includegraphics[width=42mm]{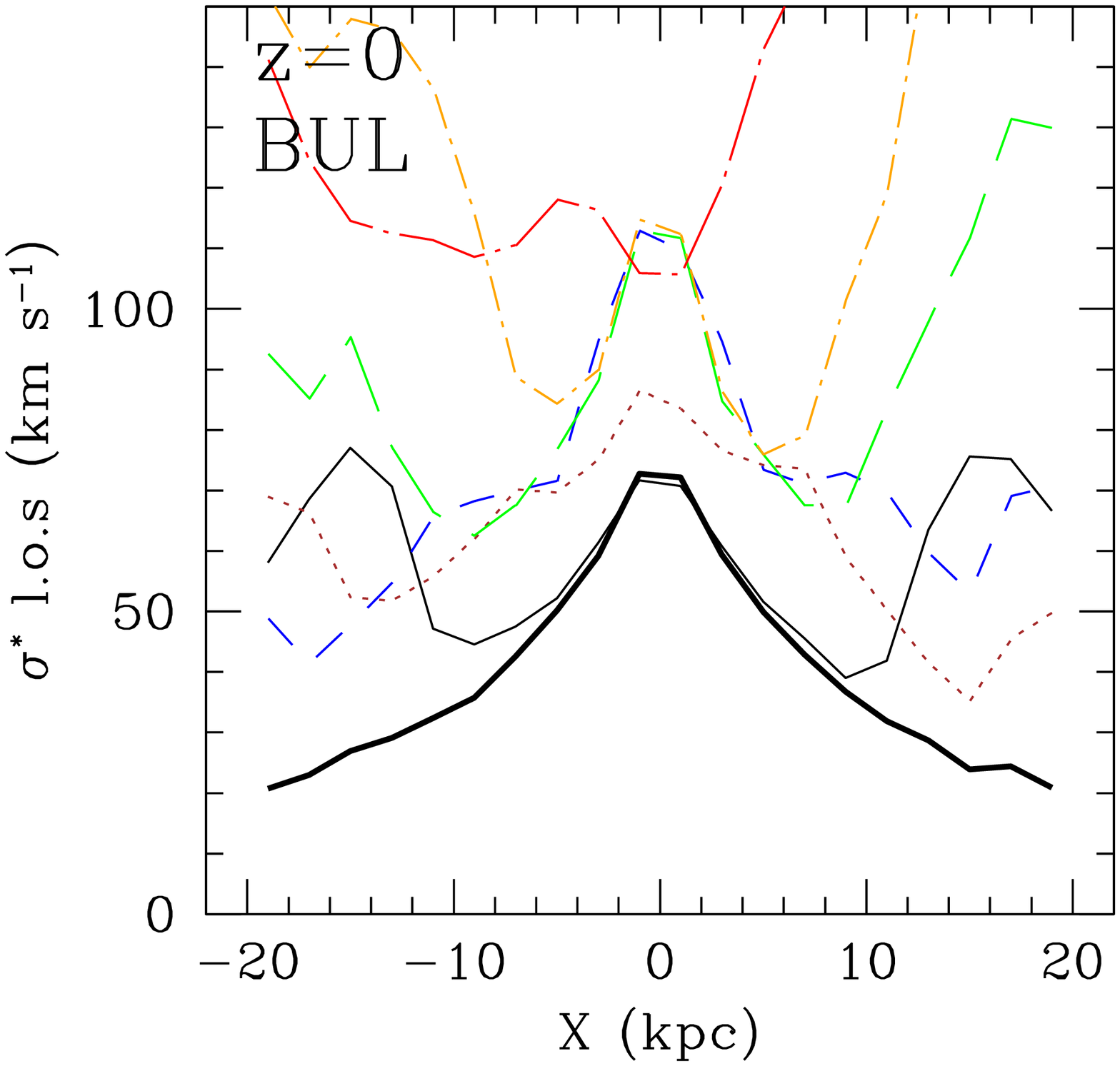}\hspace*{-1.16cm}
\includegraphics[width=42mm]{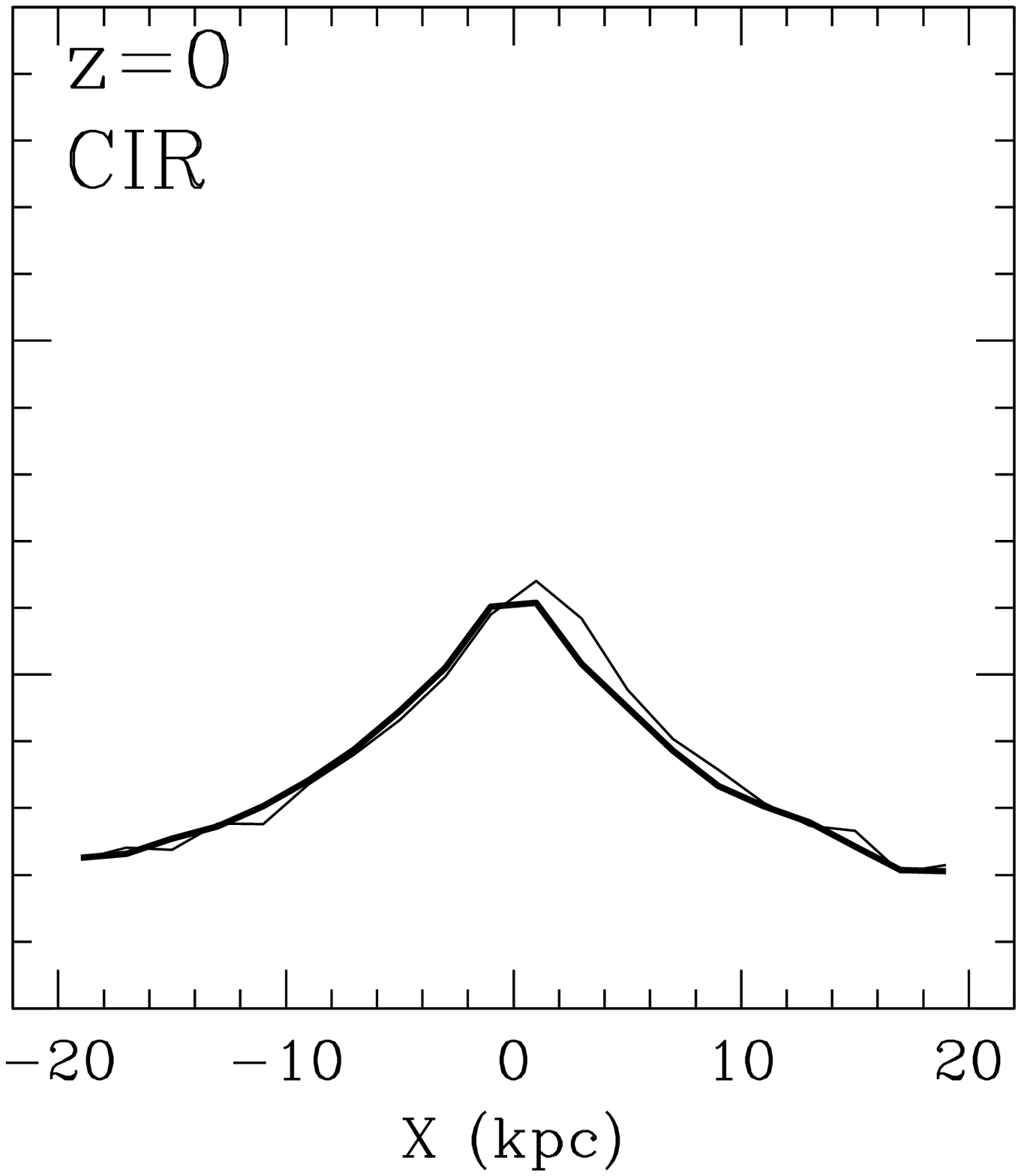}\hspace*{-1.16cm}
\includegraphics[width=42mm]{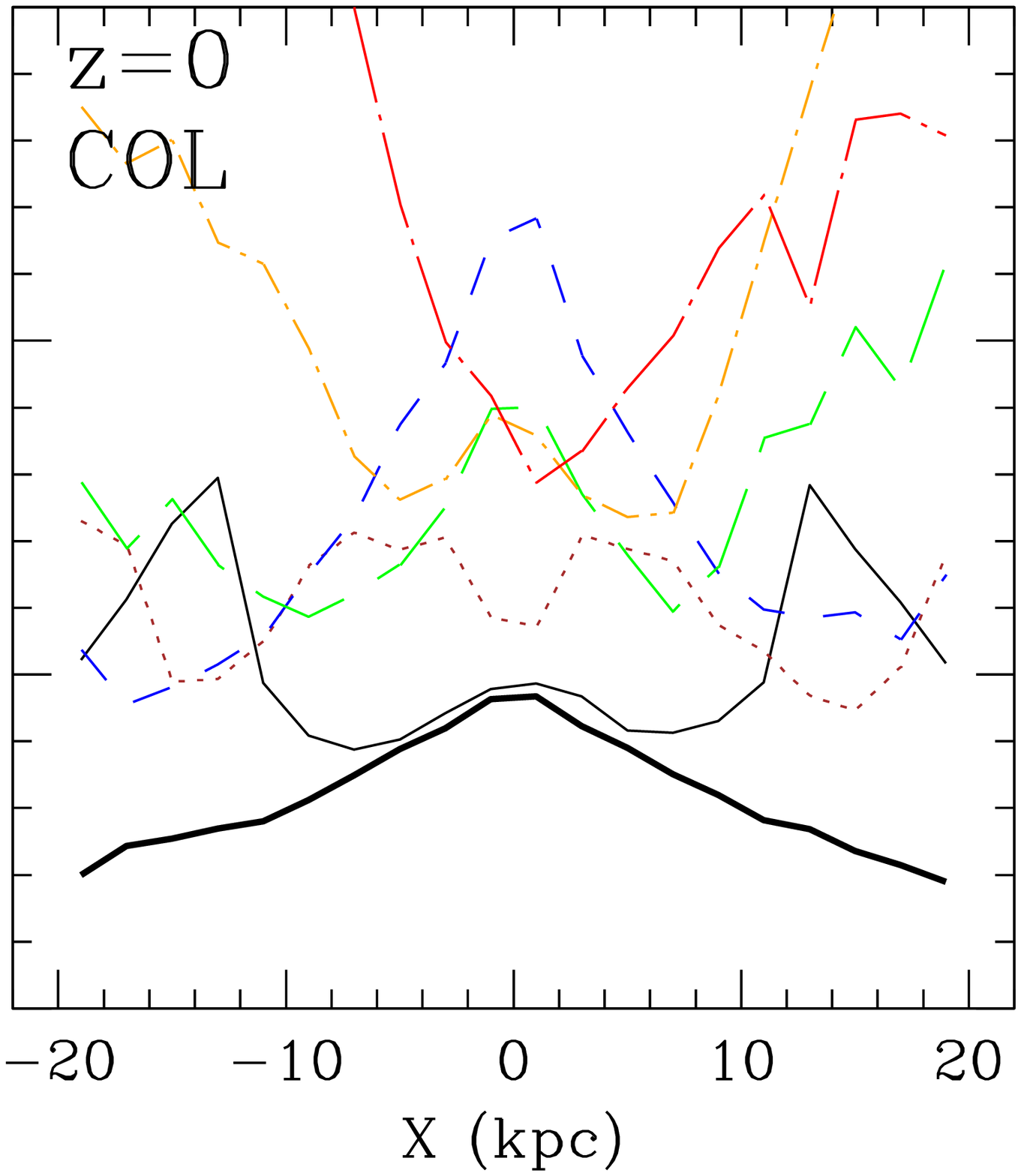}\hspace*{-1.16cm}
\includegraphics[width=42mm]{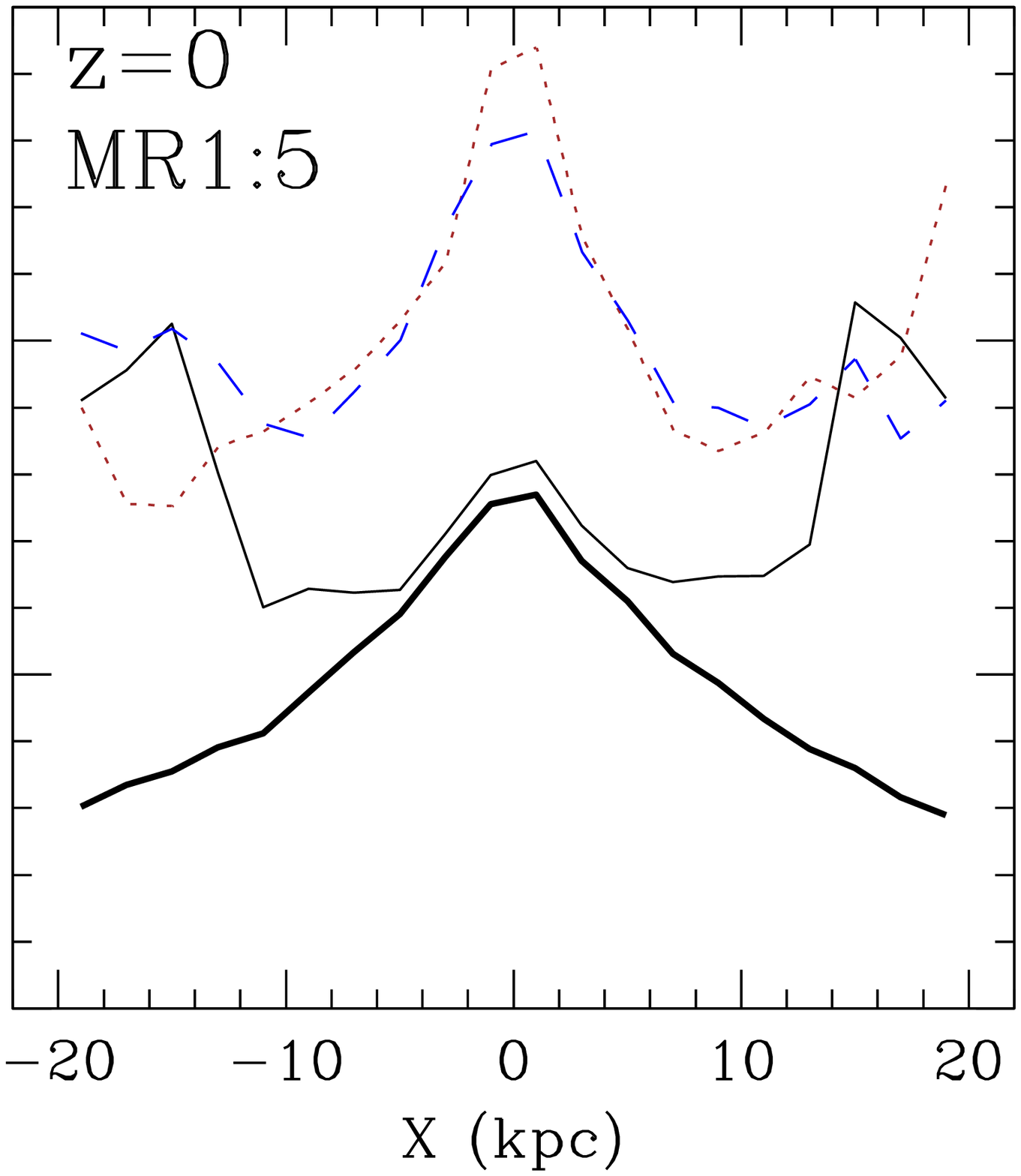}\hspace*{-1.16cm}
\includegraphics[width=42mm]{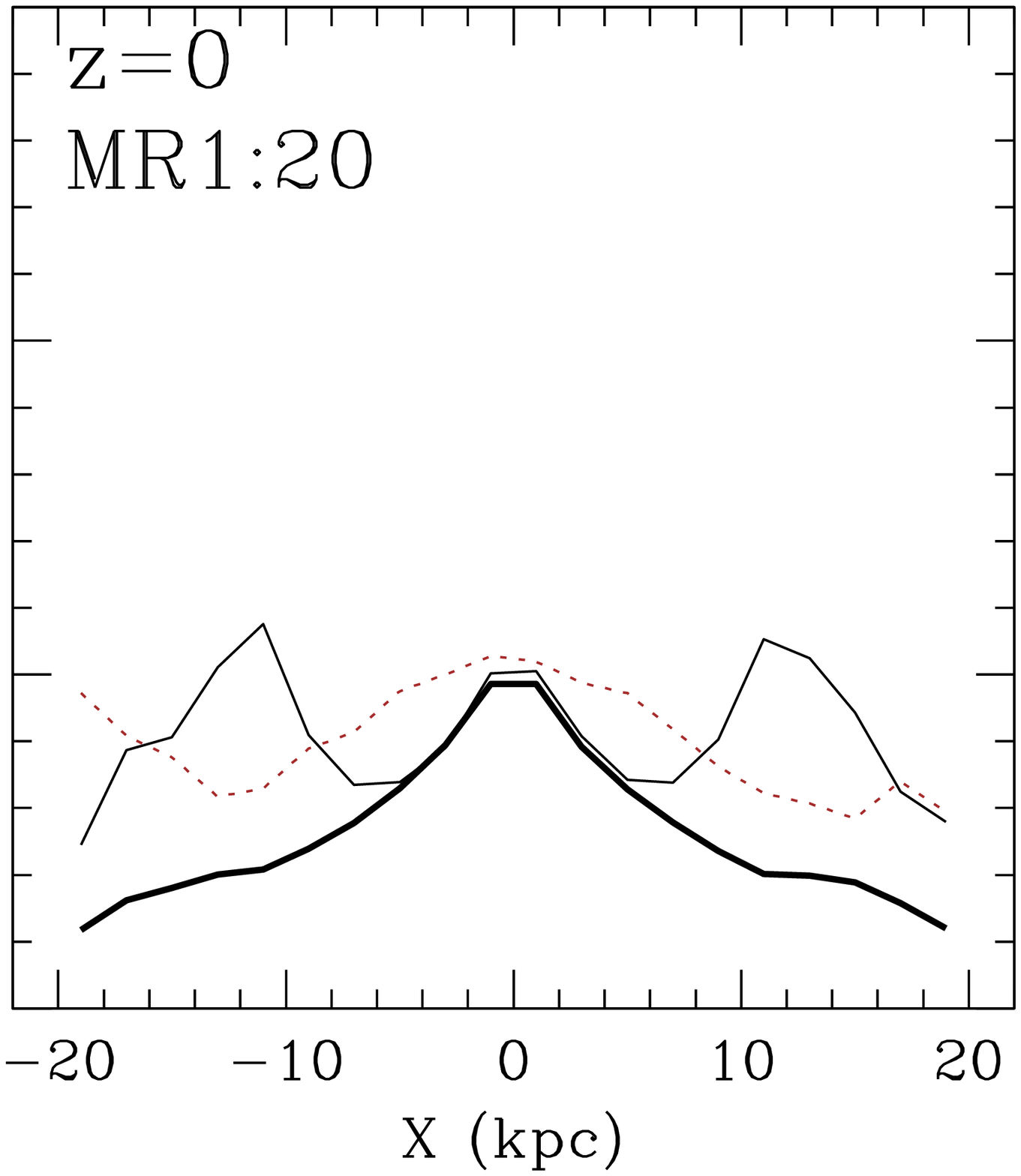}\\[-0.2cm]
\includegraphics[width=50mm]{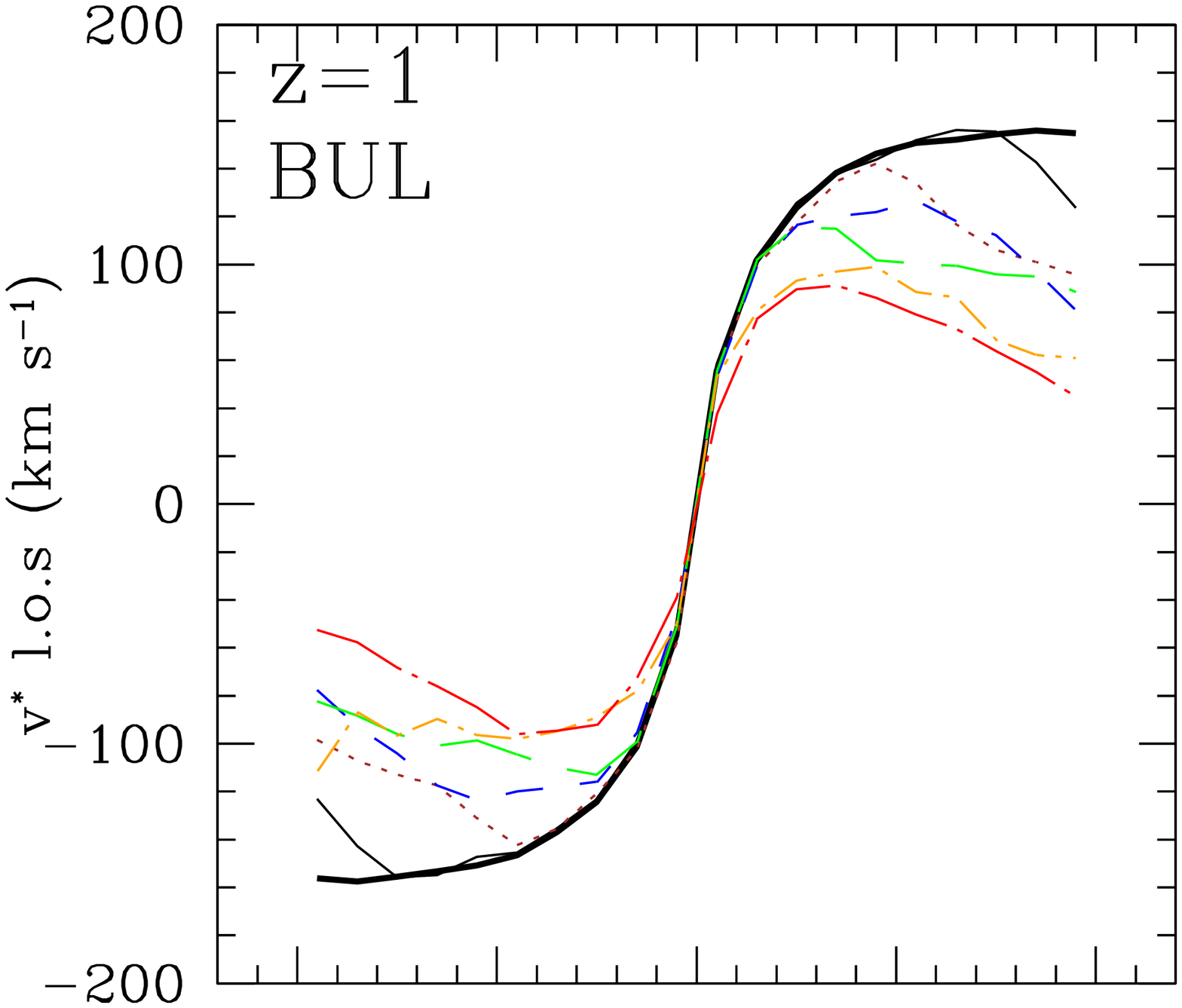}\hspace*{-1.36cm}
\includegraphics[width=50mm]{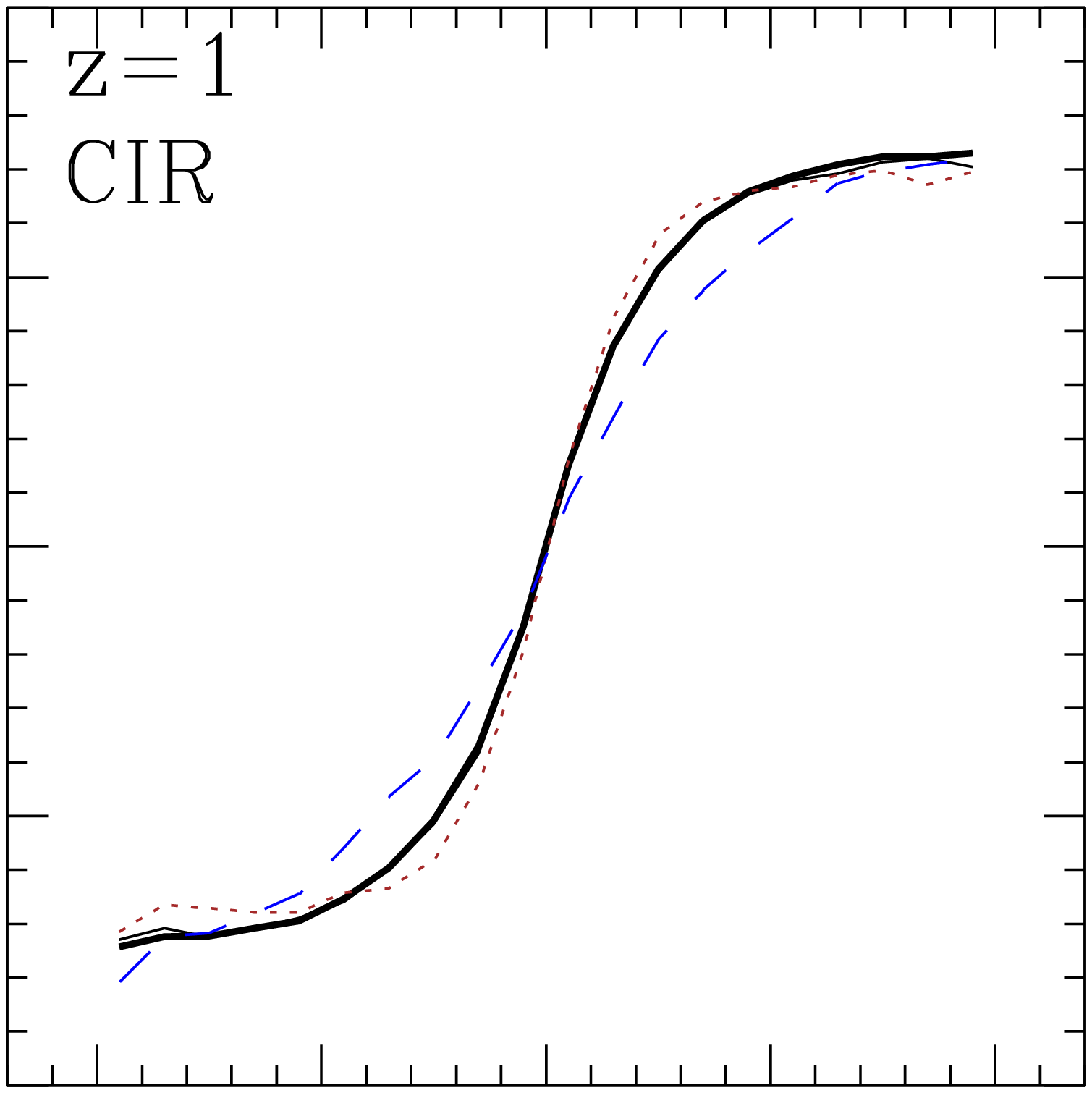}\hspace*{-1.36cm}
\includegraphics[width=50mm]{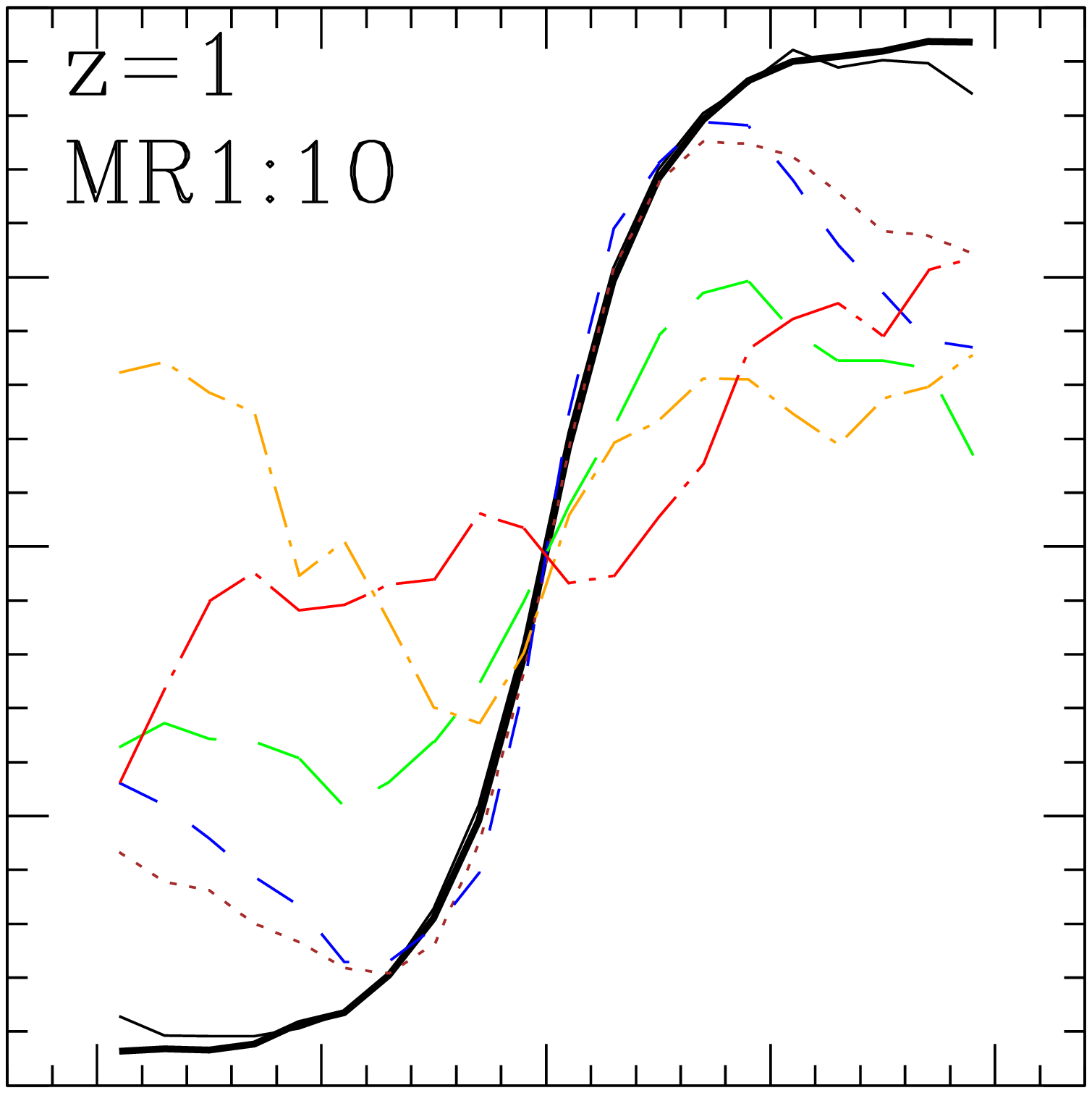}\hspace*{-1.36cm}
\includegraphics[width=50mm]{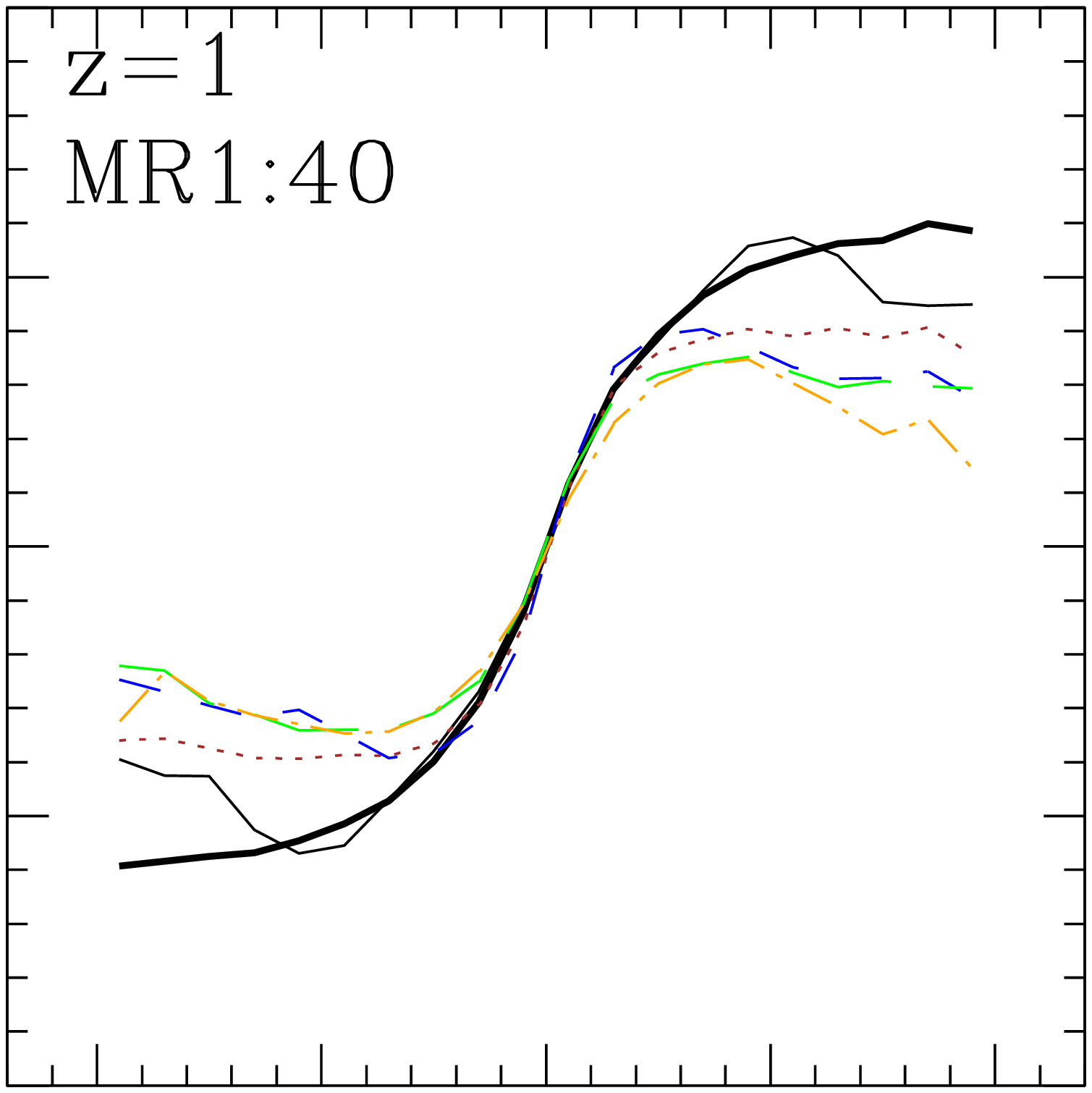}\vspace*{-1.29cm}\\
\includegraphics[width=50mm]{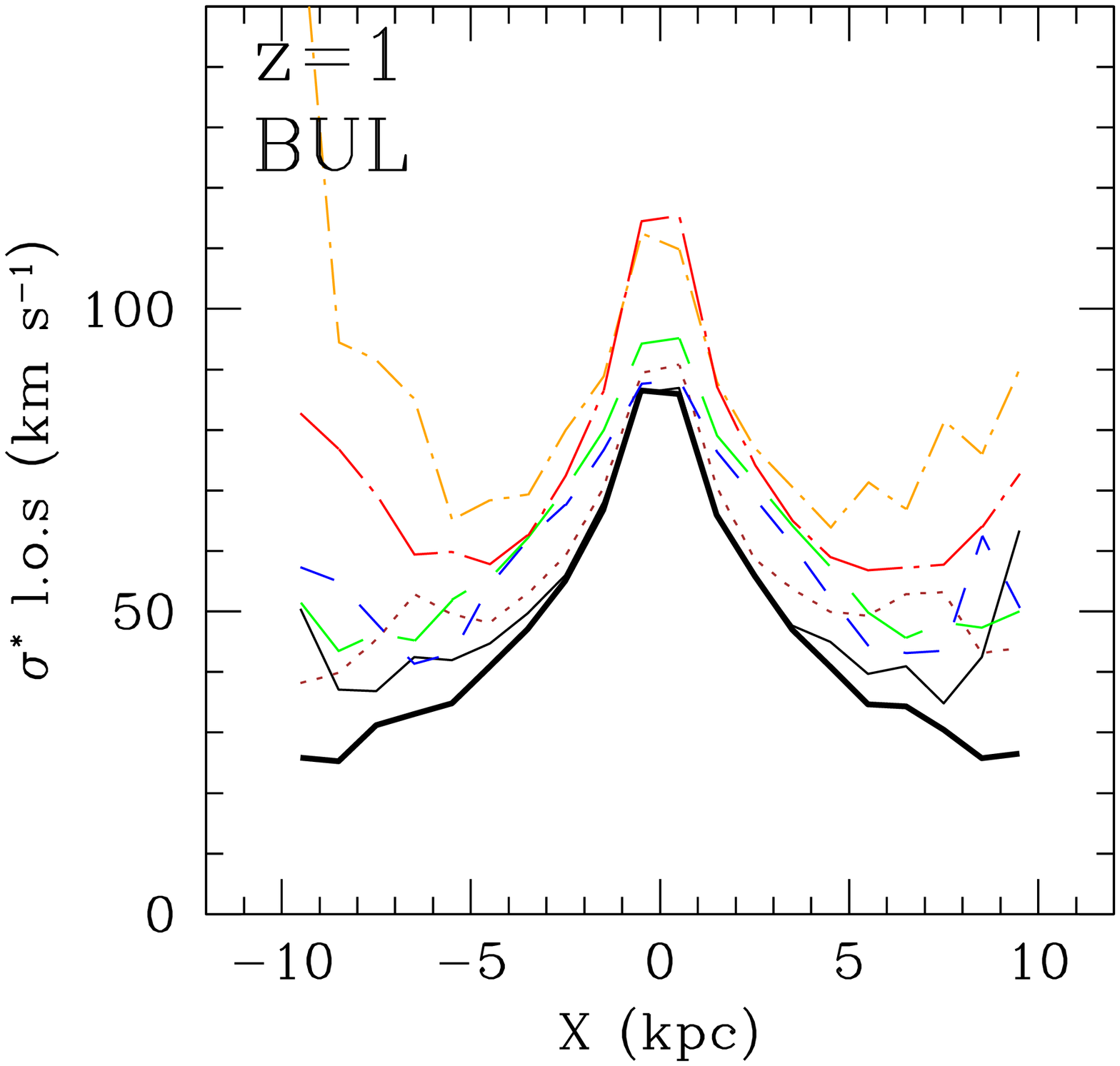}\hspace*{-1.36cm}
\includegraphics[width=50mm]{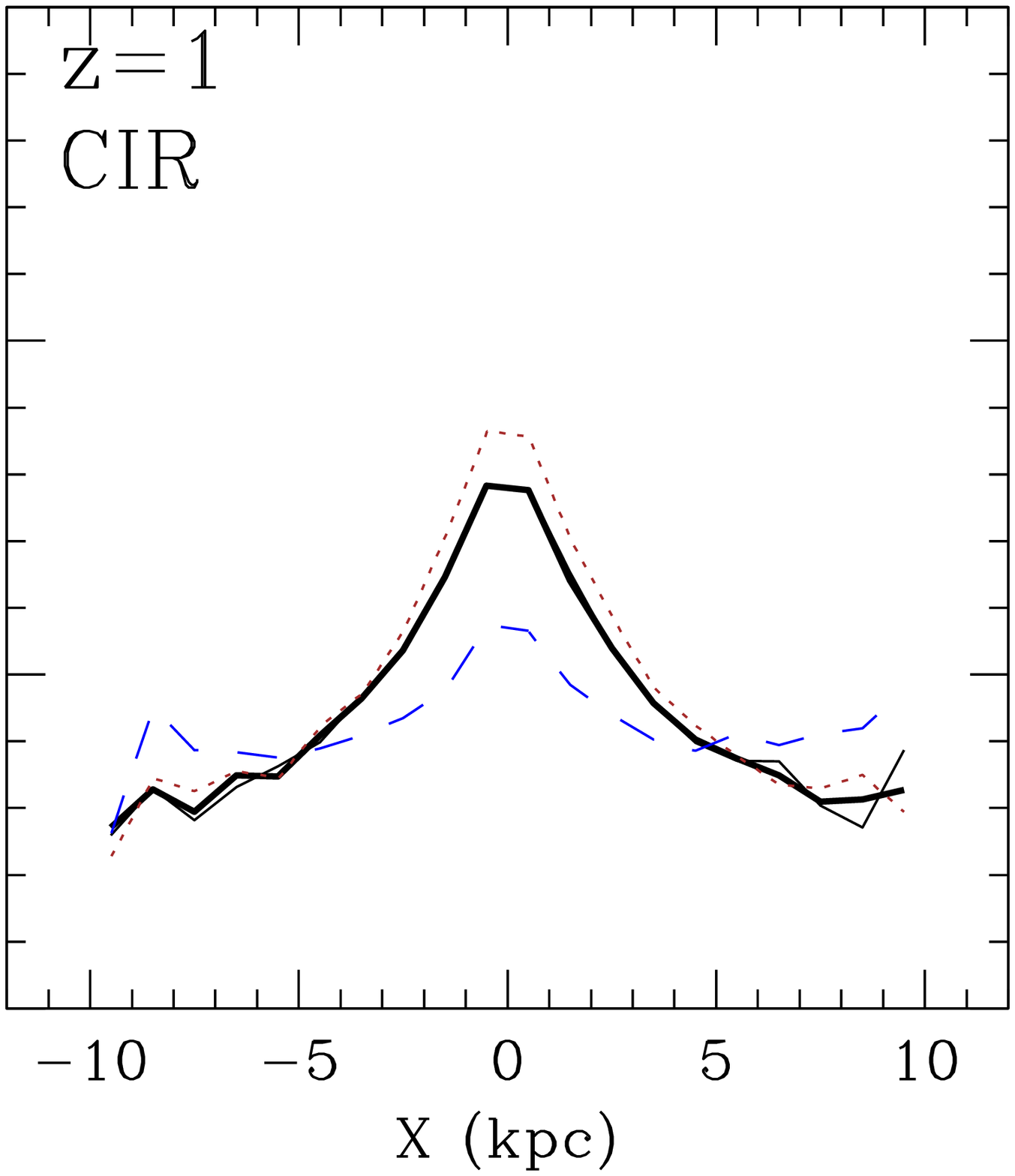}\hspace*{-1.36cm}
\includegraphics[width=50mm]{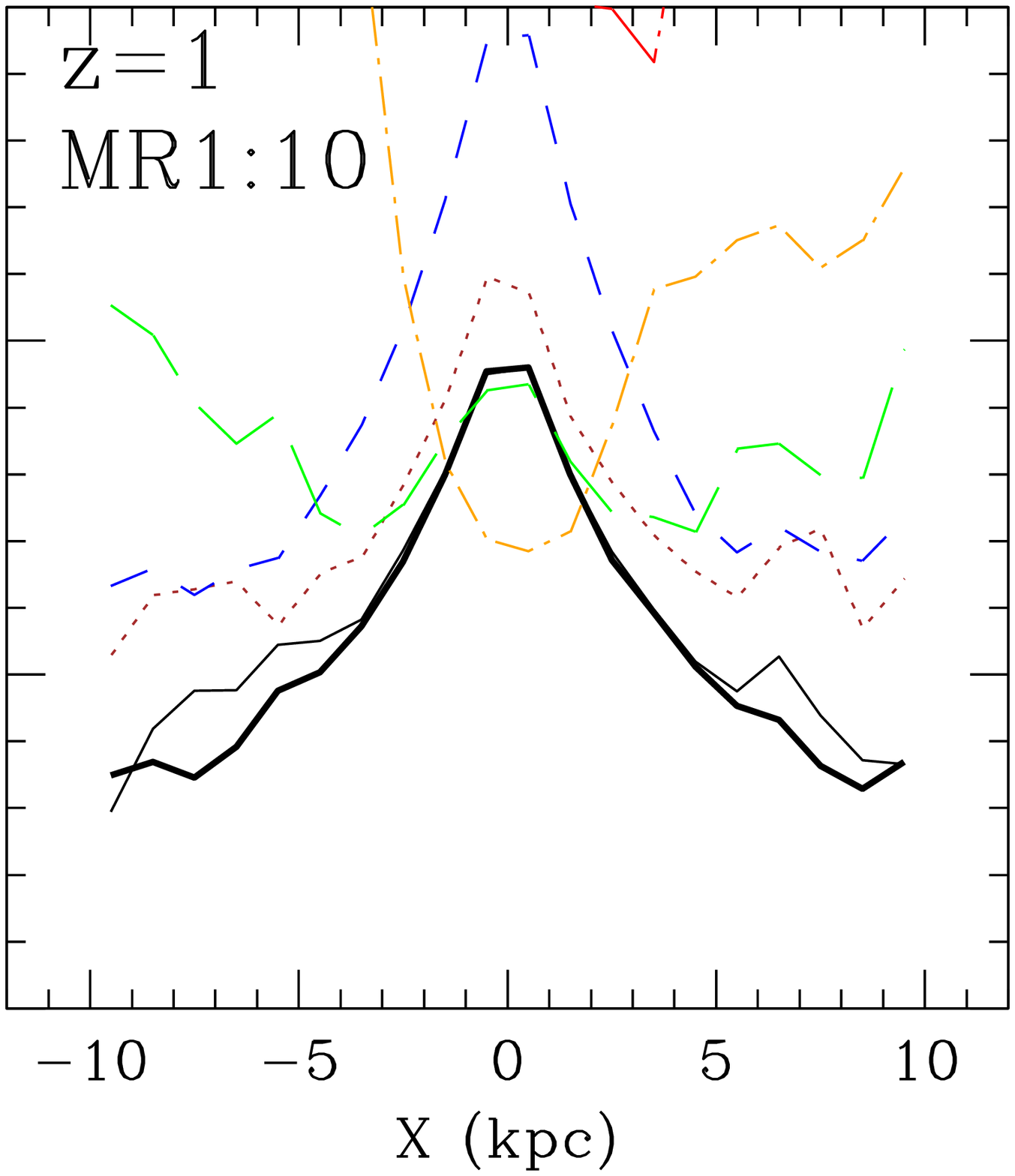}\hspace*{-1.36cm}
\includegraphics[width=50mm]{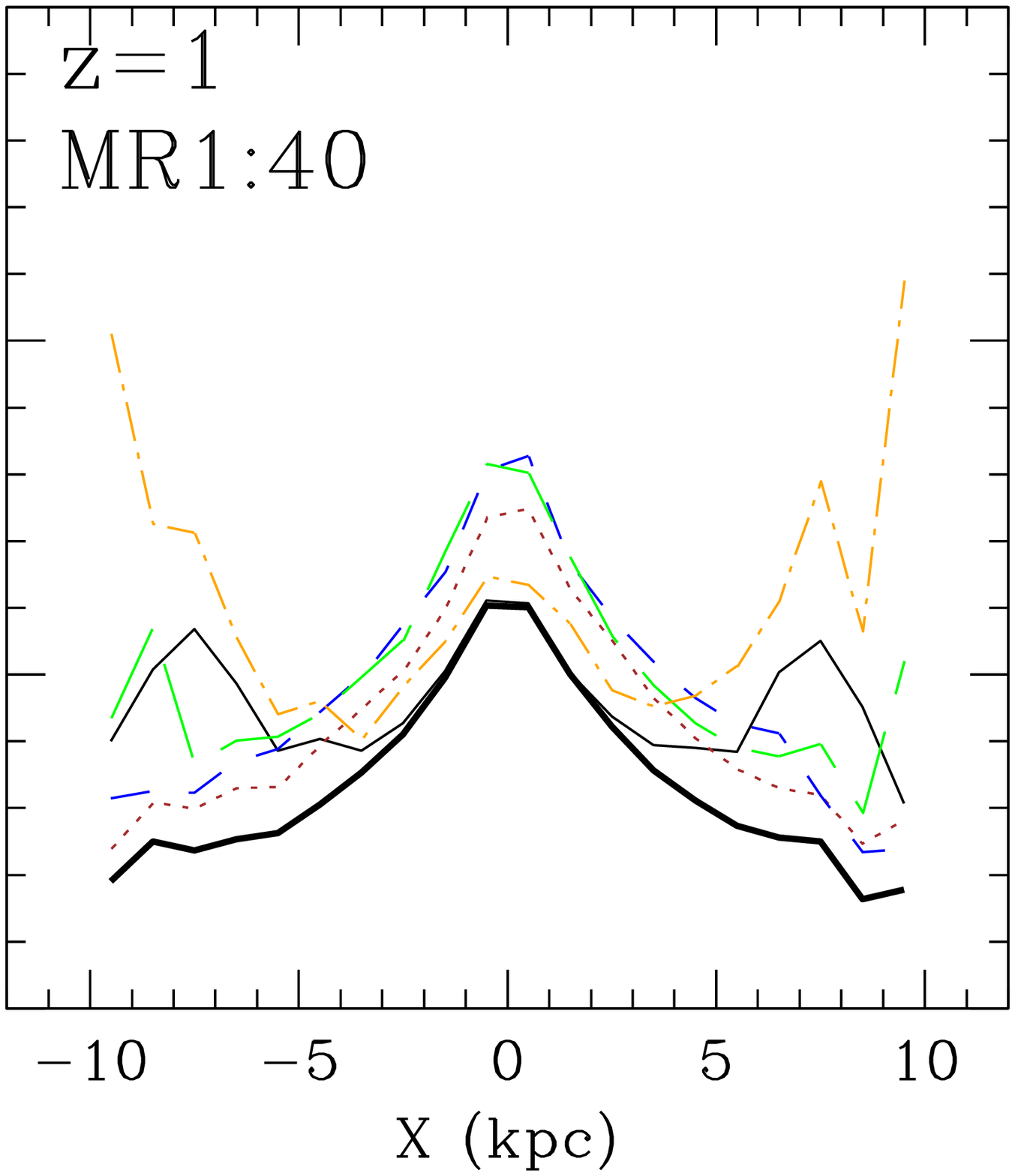}
\end{center}
\caption{Continuation of Figure~\ref{kinematics-evol-los}.
Evolution of the kinematics of disc galaxies in terms of their line-of-sight 
velocities and velocity dispersions, after each of their first six pericentric 
passages. All discs are placed edge-on, with their midplanes along the X-axis. 
The kinematics have been computed in 1~kpc-wide bins along the X-axis, considering 
\emph{all} disc stars (both bound and unbound) located within 3~kpc from the 
midplane, and have been corrected by the evolution of the respective disc in 
isolation. The line types and colour code are as in Figures~\ref{app-f4} 
and~\ref{app-f5}.}
\label{app-f6}
\end{figure*}

\bsp

\label{lastpage}

\end{document}